\documentclass[12pt, a4paper, twoside]{report}
\usepackage{amsmath}
\usepackage{amssymb}
\usepackage[dvips]{graphics}
\usepackage{epsfig}
\usepackage{calc}
\usepackage{amsfonts}
\usepackage{graphicx}
\usepackage[ansinew]{inputenc}

\def\opone{\leavevmode\hbox{\small1\kern-3.8pt\normalsize1}}

\def\be{\begin{equation}}
\def\ee{\end{equation}}
\def\ba{\begin{eqnarray}}
\def\ea{\end{eqnarray}}

\def\w{\wedge}

\def\k{\kappa}

\def\a{\alpha}

\def\g{\gamma}
\def\G{\Gamma}

\def\e{\epsilon}

\def\m{\mu}
\def\n{\nu}

\def\l{\lambda}

\def\S{\Sigma}

\def\Ct{{\widetilde C}}

\def\St{{\widetilde S}}

\def\IR{\relax{\rm I\kern-.18em R}}

\def\inv{^{\raise.15ex\hbox{${\scriptscriptstyle -}$}\kern-.05em 1}}

\def\be{\begin{equation}}
\def\ee{\end{equation}}
\def\ba{\begin{eqnarray}}
\def\ea{\end{eqnarray}}

\def\tr{\,{\rm tr}\,}
\def\Tr{\,{\rm Tr}\,}

\def\a{\alpha}

\def\g{\gamma}
\def\G{\Gamma}

\def\e{\epsilon}

\def\m{\mu}
\def\n{\nu}

\def\l{\lambda}

\def\k{\kappa}

\def\o{\omega}

\def\ks{{k \kern-.5em /}}
\def\es{{\e \kern-.4em /}}
\def\ds{{\partial \kern-.5em /}}
\def\Ds{{D \kern-.7em /}}

\def\inv{^{\raise.15ex\hbox{${\scriptscriptstyle -}$}\kern-.05em 1}}

\begin{document}
\setcounter{footnote}{0}
\setcounter{page}{0} 
\pagenumbering{roman}
\pagestyle{headings}

\begin{titlepage}
\begin{center}
{\bf \Large M-THEORY COMPACTIFICATIONS, \\
\vspace{0.5cm}
$G_2$-MANIFOLDS AND ANOMALIES}\\

\vspace{1cm} {
DIPLOMARBEIT\\
VON\\
STEFFEN METZGER}\footnote{steffen.metzger@physik.uni-muenchen.de}\\
\vspace{2cm}
\begin{figure}[ht]
\centering
\includegraphics[width=0.4\textwidth]{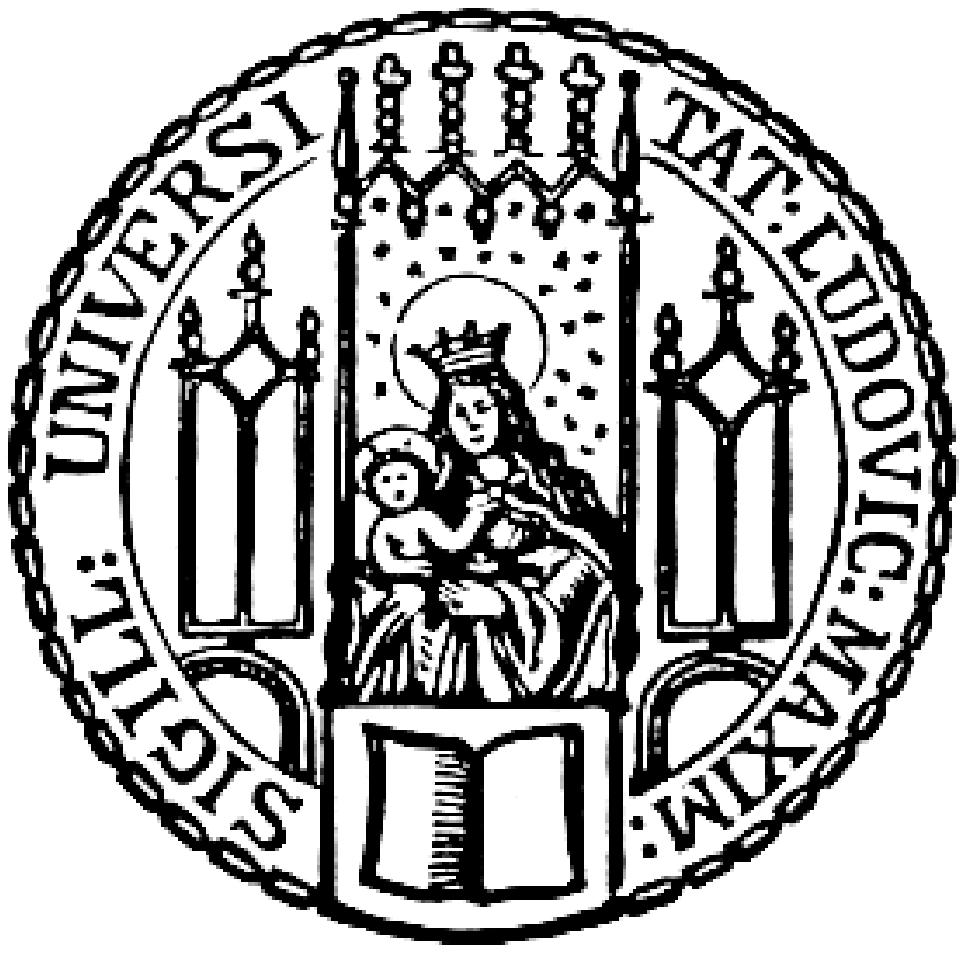}\\
\end{figure}

\vspace{1.5cm} {

\bigskip
LABORATOIRE DE PHYSIQUE TH\'{E}ORIQUE\\
PROF. DR. A. BILAL\\
ÉCOLE NORMALE SUPÉRIEURE\\
PARIS\\

\bigskip
und

\bigskip
SEKTION PHYSIK\\
LEHRSTUHL F\"UR MATHEMATISCHE PHYSIK\\
PROF. DR. J. WESS\\
LUDWIG-MAXIMILIANS-UNIVERSIT\"AT M\"UNCHEN}
\end{center}
\end{titlepage}

\cleardoublepage

\vspace*{2cm}
\begin{center}
{\bf Abstract}
\end{center}
This diploma thesis has three major objectives. Firstly, we give
an elementary introduction to M-theory compactifications, which
are obtained from an analysis of its low-energy effective theory,
eleven-dimensional supergravity. In particular, we show how the
requirement of $\mathcal{N}=1$ supersymmetry in four dimensions
leads to compactifications on $G_2$-manifolds. We also examine the
Freund-Rubin solution as well as the M2- and M5-brane. Secondly,
we review the construction of realistic theories in four
dimensions from compactifications on $G_2$-manifolds. It turns out
that this can only be achieved if the manifolds are allowed to
carry singularities of various kinds. Thirdly, we are interested
in the concept of anomalies in the framework of M-theory. We
present some basic material on anomalies and examine three cases
where anomalies play a prominent role in M-theory. We review
M-theory on $\mathbb{R}^{10}\times S^1/\mathbb{Z}_2$ where
anomalies are a major ingredient leading to the duality between
M-theory and the $E_8\times E_8$ heterotic string. A detailed
calculation of the tangent and normal bundle anomaly in the case
of the M5-brane is also included. It is known that in this case
the normal bundle anomaly can only be cancelled if the topological
term of eleven-dimensional supergravity is modified in a suitable
way. Finally, we present a new mechanism to cancel anomalies which
are present if M-theory is compactified on $G_2$-manifolds
carrying singularities of codimension seven. In order to establish
local anomaly cancellation we once again have to modify the
topological term of eleven-dimensional supergravity as well as the
Green-Schwarz term.

\cleardoublepage

\vspace*{5cm} {\hspace{7.5cm} F\"ur meine Eltern} \cleardoublepage

\tableofcontents

\cleardoublepage

\pagenumbering{arabic}

\chapter{Introduction}

The fundamental degrees of freedom of M-theory are still unknown.
Nevertheless, over the last few years indication has been found
that M-theory is a consistent quantum theory that contains all
known string theories as a certain limit of its parameter space.
Like string theory, M-theory comprises both general relativity and
quantum field theory and therefore might well be a major step
towards a unified theory of
all the forces in nature. \\
As is well known, M-theory needs to be formulated in eleven
dimensions. So in order to make this model realistic we have to
ask whether there is a vacuum of the theory that contains only
four macroscopic dimensions, with the other seven dimensions
compact
and small and hence invisible.\\
Experimental data - for instance the huge difference in energy
between the electroweak scale and the Planck scale, also known as
the hierarchy problem - tell us that our world can most probably
be described by a quantum theory with $\mathcal{N}=1$
supersymmetry \cite{WZ74a}, \cite{WZ74b}, \cite{WB92}. At some
other scale - that might even be reached by present day's
accelerators - this supersymmetry has to be broken again, of
course, as we do not observe any superpartners of the
known particles.\\
So what we want to study are M-theory vacua with $\mathcal{N}=1$
supersymmetry and four macroscopic space-time dimensions. There
are two known ways to obtain such theories from M-theory. The
first possibility is to compactify M-theory on a space
$\mathbb{R}^{3,1}\times X$, where $X$ is a manifold with boundary
$\partial X$, and $\partial X$ is a Calabi-Yau manifold
\cite{HW95}, \cite{HW96}. The basic setup of this approach will
be described in chapter 4. The second possibility is to take X to
be a manifold with holonomy group $G_2$. This approach will be
analyzed in detail in chapters 5 to 8.\\
The low-energy effective action of M-theory is eleven-dimensional
supergravity \cite{Wi95}. So if $X$ is large compared to the
Planck scale, and smooth, supergravity will give a good
approximation for low energies. Thus, in order to study M-theory
on a given space $\mathbb{R}^{3,1}\times X$, it will be sufficient
for most purposes to consider the compactification of its
low-energy limit. These sorts of compactification have been
studied for quite some time \cite{DNP86}. They are well
understood, as eleven-dimensional
supergravity is well-defined at least on the classical level.\\
Yet, once we have obtained a four-dimensional theory, we certainly
want to go further. We want to reproduce the field content of the
standard model in its very specific form. In particular, there
should be non-Abelian gauge groups and charged chiral fermions. It
was shown in \cite{Wi83}, however, that this field content cannot
be achieved by compactifying on smooth manifolds. Indeed, the
compactification on a smooth $G_2$-manifold, which will be
performed explicitly in chapter 5, gives only Abelian gauge groups
and neutral chiral multiplets. Nevertheless, there is a
possibility to derive interesting theories from compactifications.
This is achieved by using an idea that has been popular in string
theory over the last years. Usually physicists only work with
smooth manifolds, but it turns out that this approach is too
restrictive and that it is useful to admit spaces carrying
singularities. When these singularities are present, new effects
occur, and we will show that it is possible to obtain both charged
chiral fermions and non-Abelian gauge groups from
compactifications on singular spaces. As will be explained in
chapters 7 and 8, conical singularities in the compact
seven-manifold $X$ lead to chiral fermions, whereas $ADE$
singularities, singularities of codimension four, in $X$ yield
non-Abelian gauge groups. It is clear that the concept of a
manifold is no longer valid for these singular spaces, which
complicates the mathematical description. Often one needs to leave
the familiar grounds of differential geometry and resort to the
methods of algebraic geometry so as to give a mathematically
precise analysis of these spaces.\\
It turns out that one of the most important tools which can be
used in M-theory calculations are anomalies. It is well-known that
anomalies of local gauge symmetries have to vanish in order to
render the theory unitary and hence well-defined. Thus, we get
rather strong conditions on the theories under consideration. This
is particularly useful because anomalies are an infrared effect,
implying that an anomaly of the low-energy effective theory
destroys the consistency of the full M-theory. That way we can
infer information about the full quantum theory.\\
The main objective of this work is to understand M-theory
compactifications on singular spaces carrying holonomy $G_2$. For
that purpose we provide some mathematical background material in
chapter 2, where the mathematics of $G_2$-manifolds is described
in
detail and examples of both compact and non-compact $G_2$-manifolds are given.\\
Chapter 3 lists some background material from physics, namely the
basics of eleven-dimensional supergravity, anomaly theory and
Kaluza-Klein theory. \\
In chapter 4 we perform our first M-theory calculation by
explaining the duality between M-theory and $E_8\times E_8$
heterotic string theory. This chapter emphasizes the importance of
anomalies in M-theory and explains some ideas and techniques which
are also useful in the different context of compactifications on
$G_2$-manifolds.\\
The general tack to find M-theory vacua is described in chapter 5,
where we give several solutions of the equations of motion of
eleven-dimensional supergravity. In particular, we show that the
direct product of Minkowski space and a $G_2$-manifold is a
possible vacuum with $\mathcal{N}=1$ supersymmetry. We also
describe the $M2$- and $M5$-brane solutions and comment on their
basic properties. Finally, we perform the explicit
compactification of eleven-dimensional supergravity on a smooth
compact manifold with holonomy $G_2$.\\
Chapter 6 gives the details of anomaly cancellation in M-theory in
the presence of $M$-branes. Again the techniques of this chapter
are important for understanding the case of $G_2$-manifolds.\\
In chapter 7 we review how realistic theories can arise from
compactifications on singular spaces. In order to do so, we
explain the duality of M-theory on $K3$ and the heterotic string
on the torus $T^3$. We show that at certain points in moduli space
the $K3$ surface develops $ADE$ singularities which lead to
enhanced gauge symmetries. These singularities can be embedded
into a $G_2$-manifold leading to non-Abelian gauge groups in four
dimensions. Chiral fermions arise from
compactifications on spaces containing conical singularities.\\
To confirm these results, we perform an anomaly analysis of
M-theory on these singular spaces in chapter 8. We find that the
theory is anomaly-free if chiral fermions and non-Abelian gauge
groups are present. Details of the methods and ideas used in this
chapter
can be found in \cite{BM03a} and \cite{BM03b}.\\
We have added a number of appendices, mainly to fix our notation.
After a short presentation of the general notation we give some
details on Clifford algebras and spinors in appendix B. Appendix C
provides the basic formulae for general gauge theories and in
appendix D we derive some relations in the vielbein formalism of
gravity. Some basic results on index theorems are given in
appendix E, while the $ADE$ subgroups of $SU(2)$ are listed in
appendix F.

\chapter{Preliminary Mathematics}

In this chapter we collect the results from mathematics that are
necessary to understand the physical picture of M-theory
compactified on $G_2$-manifolds. Our definitions and notations
follow closely those of \cite{Jo00}, which is the general
reference for this chapter.

\section{Some Facts from Algebraic Geometry}
In this section we present some basic definitions from algebraic
geometry. The aim is to understand the concept of the blow-up of a
singularity.\\
First of all, however, we want to give the definition of an
orbifold, which we will need many times.

\bigskip
{\bf Definition 2.1}\\
A real {\em orbifold} is a topological space which admits an open
covering $\{U_i\}$, such that each patch $U_i$ is isomorphic to
$\mathbb{R}^n/G_i$, where the $G_i$ are finite subgroups of
$GL(n,\mathbb{R})$.\\
A complex orbifold is a topological space
with coordinate patches biholomorphic to $\mathbb{C}^n/G_i$ and
holomorphic transition functions. Here the $G_i$ are finite
subgroups of $GL(n,\mathbb{C})$.

\bigskip
{\bf Definition 2.2}\\
An {\em algebraic set} in $\mathbb{CP}^n$ is the set of common
zeros of a finite number of homogeneous polynomials
$h:\mathbb{CP}^n\rightarrow \mathbb{C}$. An algebraic set in
$\mathbb{CP}^n$ is said to be {\em irreducible} if it is not the
union of two algebraic sets in $\mathbb{CP}^n$. An irreducible
algebraic set is called a {\em projective algebraic variety}. An
{\em algebraic variety} or simply {\em variety} is an
open\footnote{See \cite{GH78} and \cite{Jo00} for a detailed
analysis of the topology.} subset of a projective algebraic
variety.

\bigskip
{\bf Definition 2.3}\\
Let $X$ be a variety in $\mathbb{CP}^n$ and let $x\in X$. $x$ is
called a {\em nonsingular} point if $X$ is a complex submanifold
of $\mathbb{CP}^n$ in a neighbourhood of $x$. If $x$ is not
nonsingular it is called {\em singular}. The variety $X$ is
called singular if it contains at least one singular point,
otherwise it is nonsingular.

\bigskip
Next we present the definition of a {\em blow-up}. This can be
given for points or subspaces in manifolds as well as in algebraic
varieties. The definitions will not be very precise but should
give an idea of the basic mechanism. Details can be found in
\cite{GH78}.

\bigskip
{\bf Definition 2.4}\\
We start with the blow-up of the origin in a disc $\Delta$ in
$\mathbb{C}^n$. Let $z:=(z_1,\ldots ,z_n)$ be Euclidean
coordinates in $\Delta$ and $l:=[l_1,\ldots ,l_n]$ homogeneous
coordinates in $\mathbb{CP}^{n-1}$. Define
$\widetilde{\Delta}\subset\Delta\times \mathbb{CP}^{n-1}$ as
\begin{equation}
\widetilde{\Delta}:=\{(z,l):z_il_j=z_jl_i,\ \forall i,j\},
\end{equation}
and a projection $\pi:\widetilde{\Delta}\rightarrow\Delta$ by
$\pi(z,l):=z$. For $z\neq0$ this is an isomorphism and
$\pi^{-1}(0)$ is the projective space of lines in $\Delta$. The
pair $(\widetilde{\Delta},\pi)$ is
called the {\em blow-up} of $\Delta$ at 0.\\
\\
Next we define blow-ups of points of manifolds. Let $M$ be a
complex $n$-manifold, $x$ a point in $M$ and let $(U,\phi)$ be a
chart on $M$ centered around $x$, s.t. $\phi:U\rightarrow \Delta$.
Define $\widetilde{\Delta}$ as before then we have another
projection $\pi':\widetilde{\Delta}\rightarrow U$ given by
$\pi':=\phi^{-1}\circ\pi$. With $E:=\pi^{'-1}(x)$ we see that
$\pi'$ is an isomorphism on $\widetilde{\Delta}\backslash E$. The
blow-up of $M$ at $x$ is defined as
\begin{equation}
\widetilde{M}:=M\backslash \{x\}\cup\widetilde{\Delta},
\end{equation}
obtained by replacing $U\subset M$ by $\widetilde{\Delta}$,
together with the natural projection $\pi:\widetilde{M}\rightarrow M$.\\
\\
Finally, we give the definition of the blow-up of a disc along a
coordinate plane. Let $\Delta\subset\mathbb{C}^n$ be defined as
before and take $V\subset\Delta$, s.t. $V:=\{(z_1,\ldots
,z_n):z_{k+1}=\ldots =z_n=0\}$. Furthermore we take
$[l_{k+1},\ldots ,l_n]$ to be homogeneous coordinates in
$\mathbb{CP}^{n-k-1}$. Define
$\widetilde{\Delta}\subset\Delta\times\mathbb{CP}^{n-k-1}$ by
\begin{equation}
\widetilde{\Delta}:=\{(z,l):z_il_j=z_jl_i,\ k+1\leq i,j\leq n\}.
\end{equation}
As before the projection
$\pi:\widetilde{\Delta}\rightarrow\Delta$ is an isomorphism away
from $V$, while $\pi^{-1}(x)\cong\mathbb{CP}^{n-k-1}$ for $x\in
V$. The pair $(\widetilde{\Delta},\pi)$ is called the blow-up of
$\Delta$ along $V$. This definition can be extended to the
blow-up of a manifold $M$ along a submanifold $N$ in $M$ \cite{GH78}.\\
\\
Furthermore, it is possible to define the blow-up of an algebraic
variety $X$ either at a point or along a subset $Y$ of $X$. The
result will again be an algebraic variety. We will not give any
details concerning the blow-up of varieties but the example of
the blow-up of $\mathbb{C}^2/\mathbb{Z}_2$ given below should
clarify the procedure. For singular varieties the blow-up is
particularly useful, as is shown by the following theorem.

\bigskip
{\bf Theorem 2.5}\\
Let $X$ be a singular variety. Then there exists a nonsingular
variety $\widetilde{X}$, which is the result of a finite sequence
of blow-ups of $X$.\\
\\
The blow-up is defined along a subvariety of $X$. If $Y$ is the
set of singularities of $X$ this subvariety is naturally taken to
be $Y$. The theorem states that either the blow-up of $X$ along
$Y$ is already nonsingular or we can continue this process until
we get a nonsingular variety. The physical picture is that the
singularities are cut out of the variety and a smooth manifold is
glued in instead.

\begin{figure}[h]
\centering
\includegraphics[width=0.6\textwidth]{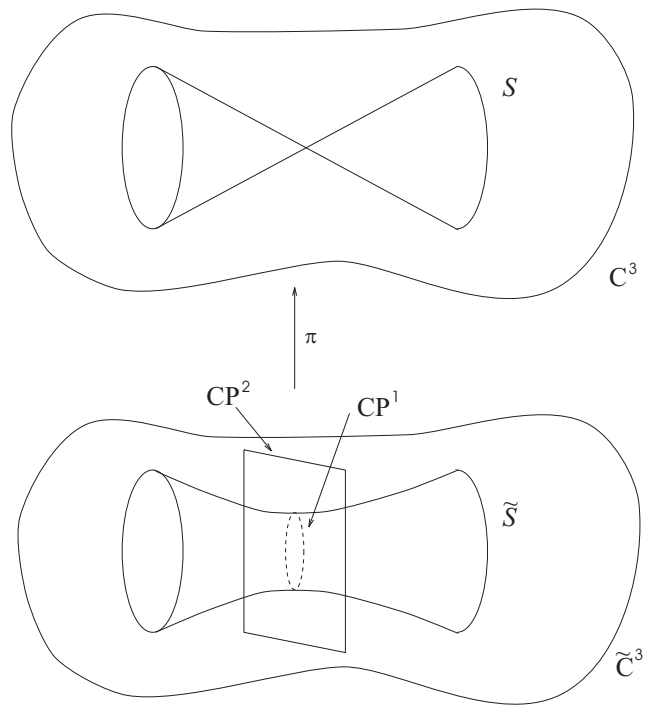}\\
\caption[]{The blow-up of $\mathbb{C}^2/\mathbb{Z}_2$.}
\label{blow-up}
\end{figure}

\bigskip
{\bf Example: The blow-up of $\mathbb{C}^2/\mathbb{Z}_2$ at 0}\\
The following illustrative example is taken from \cite{As96}.
Consider $\sigma:\mathbb{C}^2\rightarrow \mathbb{C}^2$ with
$\sigma(z^1,z^2):=(-z^1,-z^2)$. Then
$\mathbb{C}^2/\langle\sigma\rangle\cong\mathbb{C}^2/\mathbb{Z}_2$
is a complex orbifold and can be understood as a singular
algebraic variety. It can be embedded into $\mathbb{C}^3$ as
\begin{equation}
\mathbb{C}^2/\mathbb{Z}_2\cong S:=\{(z_0,z_1,z_2)\in \mathbb{C}^3:
z_0z_1-z_2^2=0\}.
\end{equation}
The hypersurface $S$ is singular at the origin of $\mathbb{C}^3$
and smooth otherwise. $S$ can be parameterized by $z_0=\zeta^2$,
$z_1=\eta^2$ and $z_2=\zeta\eta$. Then $(\zeta,\eta)$ and
$(-\zeta,-\eta)$ denote the same point, hence $S$ describes the
orbifold $\mathbb{C}^2/\mathbb{Z}_2$.\\
According to the prescription given above we consider the
following subspace of $\mathbb{C}^3\times\mathbb{CP}^2$,
\begin{equation}
\widetilde{\mathbb{C}^3}:=\{((z_0,z_1,z_2),[l_0,l_1,l_2])\in
\mathbb{C}^3\times\mathbb{CP}^2:z_il_j=z_jl_i,\ \forall i,j\}.
\end{equation}
Fixing a non-zero point in $\mathbb{C}^3$ gives a single point in
$\mathbb{CP}^2$. On the other hand for $(z_0,z_1,z_2)=(0,0,0)$ we
get an entire $\mathbb{CP}^2$. Defining
$\pi:\widetilde{\mathbb{C}^3}\rightarrow\mathbb{C}^3$ in the
natural way we found the blow-up $(\widetilde{\mathbb{C}^3},\pi)$
of $\mathbb{C}^3$ in 0. What we did so far is to excise the
origin in $\mathbb{C}^3$ and glue in a copy of $\mathbb{CP}^2$
instead.\\
Now let us consider what happens to our hypersurface $S$ as we
blow up the origin. It is natural to define
$\widetilde{S}:=\overline{\pi^{-1}(S\backslash 0)}$ as the
blow-up of $S$ at $0$. Consider following a path in S towards the
origin. In the blow-up, the point we land on in $\mathbb{CP}^2$
depends on the angle at which we approach the origin. The line
given by $(z_0t,z_1t,z_2t),\ t\in \mathbb{C},\ z_0z_1-z_2^2=0$
will land on the point $[l_0,l_1,l_2]$ in $\mathbb{CP}^2$ where
again $l_0l_1-l_2^2=0$. So we find that in $\widetilde{S}$ the
origin of $S$ is substituted by the points of $\mathbb{CP}^2$
subject to the condition $l_0l_1-l_2^2=0$. This space can be
shown to be a sphere $S^2\cong \mathbb{CP}^1$. Hence, we found
the smooth blow-up $\widetilde{S}$ of the singular space
$\mathbb{C}^2/\mathbb{Z}_2$ in which the origin is blown up to a
sphere. This process is shown in figure \ref{blow-up}.

\bigskip
As the example of the resolution of $\mathbb{C}^2/\mathbb{Z}_2$
is very important, we want to look at it from yet another
perspective. To do so we need to give the definition of a cone.

\bigskip
{\bf Definition 2.6}\\
Let $(M,g)$ be a Riemannian $n$-manifold. A point $p\in M$ is said
to be a {\em conical singularity} in $M$ if there is a
neighbourhood $U_p$ of $p$ such that on $U_p\backslash \{p\}$ the
metric takes the form
\begin{equation}
ds^2=dr^2+r^2d\Omega_N^2.\label{cone}
\end{equation}
Here $d\Omega_N^2$ is a metric on an $(n-1)$-dimensional manifold
$N$. If $g$ can be globally written in the form (\ref{cone}) $M$
is said to be a {\em cone} on $N$. Note that $\mathbb{R}^{n+1}$
can be regarded as a cone on $S^n$ with $d\Omega_N^2$ the standard
round metric on $S^n$. In that case, of course, we only have a
coordinate singularity at $r=0$. In all other cases we have a real
singularity in $M$.

\begin{figure}[ht]
\centering
\includegraphics[width=0.6\textwidth]{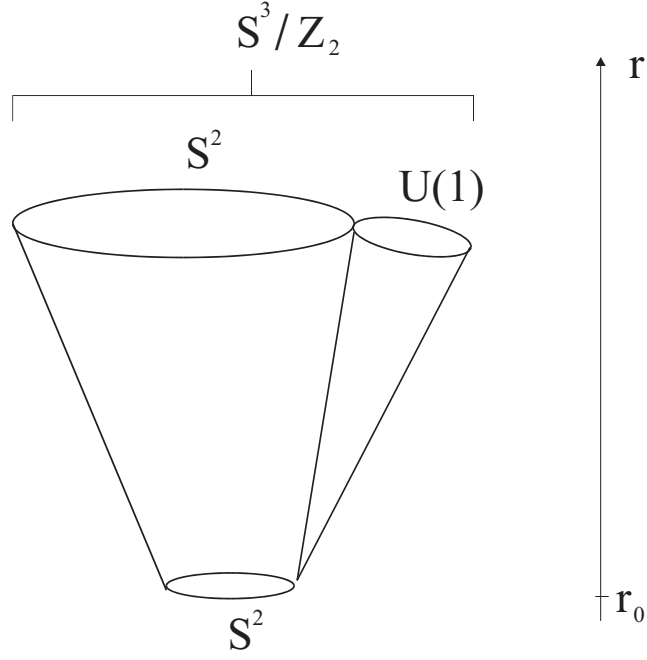}\\
\caption[]{Another picture of the resolved space $\widetilde{S}$.}
\label{asympcone}
\end{figure}

\bigskip
Certainly, $S\cong\mathbb{C}^2/\mathbb{Z}_2$ is a cone over
$S^3/\mathbb{Z}_2$, as $\mathbb{C}^2$ is a cone over $S^3$. With
the special $\mathbb{Z}_2$ defined above, $S^3/\mathbb{Z}_2$ is
actually smooth, as the $\mathbb{Z}_2$ does not have fixed points
in $S^3$. In fact, it identifies two antipodal points of the
sphere $S^3$, so we see that $S^3/\mathbb{Z}_2$ is actually
isomorphic to $SO(3)$. Let us introduce a coordinate $r$ given by
$r^2:=\sum_{i=0}^2|z_i|^2$. It is important to note that on the
blow-up $\widetilde{S}$ there is a smallest value of $r$, which is
bigger than zero. It is the radius of the sphere $S^2$ sitting in
the center of $\widetilde{S}$. We denote this radius $r_0$. Of
course, the resolution $\widetilde{S}$ no longer is a cone over
$S^3/\mathbb{Z}_2$, but for large $r$ $\widetilde{S}$ tends to
$S$, so $\widetilde{S}$ is {\em asymptotically conical}. Now
consider $SO(3)$ as a $U(1)$-bundle over $S^2$. Then the resolved
space can be viewed as shown in figure \ref{asympcone}.
We see that the $U(1)$ fibre collapses as $r$ goes to $r_0$ but we
are left with an uncollapsed $S^2$. In the singular space this
$S^2$ collapses as well.\\
We mentioned above that the physical picture of a blow-up is that
we first cut out the singularity and then glue in a smooth space
instead. In our example this smooth space is called {\em
Eguchi-Hanson space} $\mathbb{EH}$. Its metric is given by
\begin{equation}
ds^2_{\mathbb{EH}}={1\over{1-\left({r_0\over
r}\right)^4}}dr^2+r^2\left(\sigma_x^2+\sigma_y^2+\left(1-\left({r_0\over
r} \right)^4\right)\sigma_z^2\right),\label{EH}
\end{equation}
with $r_0\in \mathbb{R}^+$ a parameter and
\begin{eqnarray}
\sigma_x&=&\cos\psi\, d\theta+\sin\psi \sin\theta\, d\phi,\nonumber\\
\sigma_y&=&-\sin\psi\, d\theta+\cos\psi \sin\theta\, d\phi,\\
\sigma_z&=&d\psi+\cos\theta\, d\phi.\nonumber
\end{eqnarray}
These $\sigma_i$ are invariant under the left action of $SU(2)$
on $S^3\cong SU(2)$. Furthermore, it is easy to check that they
satisfy $d\sigma_i=-\epsilon_{ijk}\sigma_j\wedge\sigma_k$.\\
The range of coordinates in (\ref{EH}) is $r_0\leq r<\infty$,
$0\leq \psi<2\pi$, $0\leq\theta\leq\pi$ and $0\leq\phi<2\pi$.
Note that the metrics on $S^2$ and $S^3$ can be written as
\begin{eqnarray}
ds^2_{S^2}&=&\sigma_x^2+\sigma_y^2,\\
ds^2_{S^3}&=&\sigma_x^2+\sigma_y^2+\sigma_z^2.
\end{eqnarray}
The entire sphere $S^3$ is covered if $\theta$ and $\phi$ range as
above, but $0\leq\psi<4\pi$. Thus, we see that for large $r$ the
Eguchi-Hanson space is asymptotic to a cone on $S^3/\mathbb{Z}_2$,
as the $\mathbb{Z}_2$ acts on $S^3$ via $\psi\rightarrow
\psi+2\pi$. So topologically a surface of constant $r$ in
Eguchi-Hanson space is $S^3/\mathbb{Z}_2$ and it is possible to
excise the singularity of $\mathbb{C}^2/\mathbb{Z}_2$ and glue in
an Eguchi-Hanson space. Details on this procedure can be found in
\cite{Jo00}. Note also that for $r=r_0$ we find the $S^2$ sitting
on the tip of the Eguchi-Hanson space.\\
Finally we want to mention already at this point that
$\mathbb{EH}$ is Ricci flat and its holonomy group is $SU(2)$.

\section{The $K3$ Surface}
A complex manifold that turns up again and again in string and
M-theory is the $K3$ surface. In this section we give its
definition and basic properties as well as some examples.

\bigskip
{\bf Definition 2.7}\\
A vector bundle $E$ whose fibre $F$ is one-dimensional is called a
{\em line bundle}.\\
A holomorphic vector bundle $E$ on a complex manifold $M$ with
fibre $\mathbb{C}$ is called a {\em holomorphic line bundle}.\\
Let $M$ be a complex manifold of dimension $m$, then
$\Lambda^{m,0}M$ is a holomorphic line bundle, called the {\em
canonical bundle} $K_M$. It is the bundle of complex volume forms
on $M$.

\bigskip
{\bf Definition 2.8}\\
A {\em $K3$ surface} is a compact, complex surface\footnote{We did
not use the term ``manifold" as we want to allow for
singularities.} $(X,J)$ with
$h^{1,0}(X)=0$ and trivial canonical bundle.\\
\\
A simple example of a $K3$ surface is given by the so called {\em
Fermat quartic}
\begin{equation}
FQ:=\{ [z_0,\ldots
,z_3]\in\mathbb{CP}^3:z_0^4+z_1^4+z_2^4+z_3^4=0\}.
\end{equation}
The proof can be found in \cite{Jo00}.\\
\\
Another interesting example is the following. Let
$\Lambda\cong\mathbb{Z}^4$ be a lattice in $\mathbb{C}^2$. Then
$\mathbb{C}^2/\Lambda\cong T^4$. Define a map
$\sigma:T^4\rightarrow T^4$ by $\sigma(z^1,z^2):=(-z^1,-z^2)$.
Obviously this map has 16 fixed points. Thus
$T^4/\langle\sigma\rangle$ is a complex orbifold with 16
singularities which are locally isomorphic to
$\mathbb{C}^2/\mathbb{Z}_2$. Each of these singularities can be
resolved as we showed above. Define
$\widetilde{T^4/\langle\sigma\rangle}$ to be the space resulting
from blowing up all the singularities. Then
$\widetilde{T^4/\langle\sigma\rangle}$ is a $K3$ surface
\cite{Jo00}.

\bigskip
Next we list some important properties of $K3$ surfaces, which are
crucial to understand various duality conjectures.

\bigskip
{\bf Proposition 2.9}\\
Let $X$ be a $K3$ surface. Then its Betti numbers are $b^0=b^4=1$,
$b^1=b^3=0$ and $b^2=22$. Its Hodge numbers are
$h^{0,2}=h^{2,0}=1$ and $h^{1,1}=20$.\\
\\
From our resolution of a singularity using an Eguchi-Hanson space
the number of two-cycles of $K3$ can easily be determined. The
number of two-cycles of $T^4$ is 6 and these are not effected by
the $\mathbb{Z}_2$. $T^4/\langle\sigma\rangle$ has 16
singularities which are substituted by an Eguchi-Hanson space
with a sphere $S^2$ on its tip. This increases the number of
two-cycles by 16 and we get $b_2=b^2=22$. This result holds for
any $K3$ surface, as one can show that all $K3$ surfaces are
homeomorphic.

\bigskip
{\bf Proposition 2.10}\\
The moduli space $\mathcal{M}(K3)$ of Einstein metrics on a K3
surface $S$ is given by \cite{Wi95}, \cite{As96}
\begin{equation}
\mathcal{M}(K3)=\mathbb{R}^+\times
\left(SO(19,3;\mathbb{Z})\backslash
SO(19,3;\mathbb{R})/(SO(19)\times SO(3))\right).\label{ModspaceK3}
\end{equation}
Any Einstein metric on K3 is Ricci flat \cite{Hi74}.

\section{Holonomy Groups}
The topic of holonomy is a rich one. Holonomy groups can be
defined on vector bundles as well as principal bundles, they are
related to concepts as different as the topology and the curvature
of a manifold. Given the limitations of space, we only present the
definition for vector bundles. A case of particular interest is
the tangent bundle of a Riemannian manifold, equipped with the
Levi-Civita connection. In this special case we speak of
Riemannian holonomy groups.

\bigskip
{\bf Definition 2.11}\\
Let $M$ be a manifold, $E$ a vector bundle over $M$, and
$\nabla^E$ a connection on $E$. Fix a point $p\in M$. If $\gamma$
is a loop based at $p$ (i.e. $\gamma(0)=\gamma(1)=p$), then the
parallel transport map $P_{\gamma}:E_p\rightarrow E_p$ is an
invertible linear map, so that $P_{\gamma}$ lies in $GL(E_p)$, the
group of invertible linear transformations of $E_p$. Define the
holonomy group $Hol_p(\nabla^E)$ of $\nabla^E$ based at $p$ to be
\begin{equation*}
Hol_p(\nabla^E):=\lbrace P_{\gamma}|\  \gamma \ \mbox{is a loop
based at} \ p\rbrace \subset GL(E_p).
\end{equation*}

{\bf Proposition 2.12}\\
Let $M$ be a connected manifold. Then the holonomy group is
independent of $p \in M$ and we denote
$Hol(\nabla^E):=Hol_p(\nabla^E)$.

\bigskip
{\bf Definition 2.13}\\
Let $(M,g)$ be a Riemannian manifold with Levi-Civita connection
$\nabla$. Define the holonomy group $Hol(g)$ of $g$ to be
$Hol(\nabla)$. Then $Hol(g)$ is a subgroup of $O(n)$, called the
{\em Riemannian holonomy group}.

\bigskip
The following proposition - which can be easily understood
geometrically - clarifies the intimate relation between holonomy
and curvature.

\bigskip
{\bf Proposition 2.14}\\
Let $M$ be a manifold, $E$ a vector bundle
over $M$ and $\nabla^E$ a connection on $E$. If $\nabla^E$ is
flat, so that $R(\nabla^E)=0$ then
$Hol(\nabla^E)=\lbrace1\rbrace$.

\bigskip
Another proposition relating curvature and holonomy that will be
useful later on is the following.

\bigskip
{\bf Proposition 2.15}\\
Let $M$ be a manifold, $E$ a vector bundle over $M$, and
$\nabla^E$ a connection on $E$. Then for each $p\in M$ the
curvature $R(\nabla^E)_p$ of $\nabla^E$ at $p$ lies in
$\verb"hol"_p(\nabla^E)\otimes\Lambda^2T^*_pM$, where
$\verb"hol"_p(\nabla^E)$ is the Lie algebra of the holonomy group
$Hol_p(\nabla^E)$.\\
\\
This proposition is the mathematical formulation of formula
(\ref{curvgeneral}) given in the appendix. In particular, we see
that $R_{MNAB}\Gamma^{AB}$ is the generator of the holonomy group
acting on the spin bundle.

\bigskip
In order to understand the definition of a $G_2$-manifold, that
will be given below, we need to introduce the concept of
$G$-structures.

\bigskip
{\bf Definition 2.16}\\
Let $M$ be a manifold of dimension $n$, and $F$ the frame bundle
over $M$. Then $F$ is a principal bundle over $M$ with fibre
$GL(n,\mathbb{R})$. Let $G$ be a Lie subgroup of
$GL(n,\mathbb{R})$. Then a {\em G-structure} on $M$ is a
principal subbundle $P$ of $F$, with fibre $G$.

\bigskip
\begin{center}
{\bf The classification of Riemannian holonomy groups}
\end{center}

{\bf Definition 2.17}\\
Let $M$ be a manifold with $p\in M$ and let $G$ be a Lie group
acting on $M$ from the left. The {\em orbit} of $p$ under this
action is defined as
\begin{equation*}
Gp:=\lbrace gp:g\in G\rbrace.
\end{equation*}

{\bf Definition 2.18}\\
Let $M$ be a manifold and let $G$ be a Lie group acting on $M$
from the left. The action is called {\em transitive} if $\forall
p_1,
p_2 \in M \ \exists g\in G$, such that $gp_1=p_2$.\\

\bigskip
{\bf Definition 2.19}\\
Let $M$ be a manifold with $p\in M$ and let $G$ be a Lie group
acting on M from the left. The {\em isotropy group} ({\em little
group, stabilizer}) of $p$ is a subgroup of $G$ defined by
\begin{equation*}
H(p):=\lbrace g\in G:gp=p\rbrace.
\end{equation*}
Often the isotropy group is independent of $p$ and will be denoted
$H$ in that case.

\bigskip
{\bf Proposition 2.20}\\
Let $G$ be a Lie group and $H$ a Lie subgroup of $G$. The coset
space $G/H$ is a manifold called a {\em homogeneous space}.\\
Its dimension is given by ${\rm dim}\ (G/H)={\rm dim}\ G-{\rm
dim}\ H$.

\bigskip
Let $(M,g)$ be a symmetric Riemannian manifold, let $G$ be (a
subgroup of) the group of isometries acting transitively on $M$
from the left and let $H$ be its isotropy group. Then we have
$M\cong G/H$. Note that we sometimes have to take a subgroup of
the isometry group to establish the isomorphism. Precise
definitions of symmetric spaces and the required properties of $G$
can be found in \cite{Jo00}. In that case the holonomy group of
$M$ can be shown to be
\begin{equation}
Hol(g)=H.
\end{equation}
As an example let us consider the sphere $S^n$. Clearly its
isometry group is $SO(n+1)$, its isotropy group is $SO(n)$, we
have the isomorphism $S^n\cong SO(n+1)/SO(n)$ and the holonomy
group is $SO(n)$.\\
Given these properties we can construct spaces carrying particular
holonomy groups. In fact, a classification of the holonomy groups
of symmetric spaces was found by Cartan. However, the
classification of the holonomy groups of nonsymmetric spaces is
rather interesting. The question which subgroups of $O(n)$ can be
the holonomy group of a non-symmetric Riemannian $n$-manifold $M$
was answered by Berger \cite{Be55}, who proved the following
theorem.

\bigskip
{\bf Theorem 2.21 (Berger)}\\
Let $M$ be a simply connected manifold of dimension $n$, $g$ a
Riemannian metric on $M$ that is irreducible and
nonsymmetric\footnote{"Irreducible" basically means that we should
not allow for direct product spaces.}. Then exactly one of the
following cases holds\footnote{$Sp(m)Sp(1):=(Sp(m)\times
Sp(1))/\mathbb{Z}_2$}.
\begin{center}
\mbox{
\begin{tabular}{ll}
(i)  & $Hol(g)=SO(n)$\\
(ii) & $n=2m$ with $m\geq2$, and $Hol(g)=U(m)$ in $SO(2m)$,\\
(iii)& $n=2m$ with $m\geq2$, and $Hol(g)=SU(m)$ in $SO(2m)$,\\
(iv) & $n=4m$ with $m\geq2$, and $Hol(g)=Sp(m)$ in $SO(4m)$,\\
(v)  & $n=4m$ with $m\geq2$, and $Hol(g)=Sp(m)Sp(1)$ in $SO(4m)$,\\
(vi) & $n=7$ and $Hol(g)=G_2$ in $SO(7)$,\\
(vii)& $n=8$ and $Hol(g)=Spin(7)$ in $SO(8)$.\\
\end{tabular}}
\end{center}
$K3$ is an example for spaces with holonomy\footnote{See
\cite{Jo00} for a precise version of this statement.} $SU(2)$.
Because the groups $G_2$ and $Spin(7)$ are the exceptional cases
in this classification, they are referred to as {\em exceptional
holonomy groups}. The first examples of complete metrics with
holonomy $G_2$ and $Spin(7)$ on non-compact manifolds were given
by Bryant and Salomon in \cite{BS89} and by Gibbons, Page and Pope
in \cite{GPP90}. Compact manifolds with holonomy $G_2$ and
$Spin(7)$ were constructed by Joyce \cite{Jo96a}, \cite{Jo96b},
\cite{Jo96c}, \cite{Jo00}, however, no explicit metric is known.

\section{Spinors and Holonomy Groups}

The objective of compactifying M-theory is to obtain a
four-dimensional gauge theory with $\mathcal{N}=1$ supersymmetry.
This is equivalent to the existence of a covariantly constant
spinor on the compact manifold, as will be shown below. The
theorem in this subsection points out that in order to satisfy
this requirement the compact manifold must have holonomy group
$G_2$. This is the reason why physicists have been studying
$G_2$-manifolds over the
last years.\\
The concept of a spinor is defined in appendix B, the spin
connection $\nabla^S$ is introduced in appendix D. Contrary to the
discussion in the appendix we now take the metric on the base
manifold $M$ to have signature $+,\ldots ,+$, as we want to
compactify on Euclidean manifolds later on. $S$ is the spin bundle
over $M$ and $C^{\infty} (S)$ denotes the set of sections of $S$.

\bigskip
{\bf Definition 2.22}\\
Take a spinor $\eta \in C^{\infty} (S)$. $\eta$
is called a {\em parallel spinor} or {\em (covariantly) constant
spinor}, if $\nabla^S\eta=0$.

\bigskip
In chapter 5 we will relate the number of unbroken supersymmetry
after compactification to the number of covariantly constant
spinors of the compact space. Hence, the following theorem is of
fundamental importance.

\bigskip
{\bf Theorem 2.23}\\
Let $M$ be an orientable, connected, simply connected spin
$n$-manifold for $n\geq3$, and $g$ an irreducible Riemannian
metric on $M$. Define $N$ to be the dimension of the space of
parallel spinors on $M$. If $n$ is even, define $N_{\pm}$ to be
the dimension of the space of parallel spinors in
$C^{\infty}(S_{\pm})$, so that
$N=N_++N_-$.\\
Suppose $N\geq 1$. Then, after making an appropriate choice of
orientation for $M$, exactly one of the following holds:
\begin{center}
\mbox{
\begin{tabular}{ll}
(i)  & $n=4m$ for $m\geq1$ and $Hol(g)=SU(2m)$, with $N_+=2$ and $N_-=0$,\\
(ii) & $n=4m$ for $m\geq2$ and $Hol(g)=Sp(m)$, with $N_+=m+1$ and $N_-=0$,\\
(iii)& $n=4m+2$ for $m\geq1$ and $Hol(g)=SU(2m+1)$, with $N_+=1$ and $N_-=1$,\\
(iv) & $n=7$ and $Hol(g)=G_2$, with $N=1$,\\
(v)  & $n=8$ and $Hol(g)=Spin(7)$ with $N_+=1$ and $N_-=0$.\\
\end{tabular}}
\end{center}
With the opposite orientation, the values of $N_{\pm}$ are
exchanged.

\bigskip
In particular, we note that the requirement of a one-dimensional
space of covariantly constant spinors on a seven-manifold leads to
a manifold with $Hol(g)=G_2$.

\section{The Group $G_2$ and the Concept of $G_2$-Manifolds}
In this section we give the definition of $G_2$-manifolds and
collect a number of their properties. We define some useful
concepts related to them and study three explicit examples.

\bigskip
{\bf Definition 2.24}\\
Let $(x_1,\ldots ,x_7)$ be coordinates on $\mathbb{R}^7$. Write
$d{\bf x}_{ij\ldots l}$ for the exterior form $dx_i\wedge
dx_j\wedge\ldots \wedge dx_l$ on $\mathbb{R}^7$. Define a
three-form $\varphi_0$ on $\mathbb{R}^7$ by
\begin{equation}
\varphi_0:=d\textbf{x}_{123}+d\textbf{x}_{145}+d\textbf{x}_{167}+
d\textbf{x}_{246}-d\textbf{x}_{257}-d\textbf{x}_{347}-d\textbf{x}_{356}.\label{G2form}
\end{equation}
The subgroup of $GL(7,\mathbb{R})$ preserving $\varphi_0$ is the
exceptional Lie group $G_2$. It is compact, connected, simply
connected, semisimple and 14-dimensional, and it also fixes the
four-form
\begin{equation}
*\varphi_0=d{\bf x}_{4567}+d{\bf x}_{2367}+d{\bf x}_{2345}+d{\bf
x}_{1357}-d{\bf x}_{1346}-d{\bf x}_{1256}-d{\bf x}_{1247},
\end{equation}
the Euclidean metric $g_0=dx_1^2+\ldots dx_7^2$, and the
orientation on $\mathbb{R}^7$.

\bigskip
{\bf Definition 2.25}\\
A {\em $G_2$-structure} on a seven-manifold $M$ is a principal
subbundle of the frame bundle of $M$ with structure group $G_2$.
Each $G_2$-structure gives rise to a 3-form $\varphi$ and a metric
$g$ on $M$, such that every tangent space of $M$ admits an
isomorphism with $\mathbb{R}^7$ identifying $\varphi$ and $g$ with
$\varphi_0$ and $g_0$, respectively. We will refer to
$(\varphi,g)$ as
a $G_2$-structure.\\
Let $\nabla$ be the Levi-Civita connection, then $\nabla\varphi$
is called the {\em torsion} of $(\varphi,g)$. If $\nabla\varphi=0$
then $(\varphi,g)$ is
called {\em torsion free}.\\
A {\em $G_2$-manifold} is defined as the triple $(M,\varphi,g)$,
where $M$ is a seven-manifold, and $(\varphi,g)$ a torsion-free
$G_2$-structure on $M$.

\bigskip
{\bf Proposition 2.26}\\
Let $M$ be a seven-manifold and $(\varphi,g)$ a $G_2$-structure on
$M$.
Then the following are equivalent:\\
\\
\mbox{
\begin{tabular}{ll}
(i)  & $(\varphi,g)$ is torsion-free,\\
(ii) & $\nabla\varphi=0$ on $M$, where $\nabla$ is the Levi-Civita connection of $g$,\\
(iii)& $d\varphi=d*\varphi=0$ on $M$,\\
(iv) & $Hol(g)\subseteq G_2$, and $\varphi$ is the induced three-form.\\
\end{tabular}}\\
\\
Note that the holonomy group of a $G_2$-manifold is not
necessarily $G_2$.

\bigskip
\begin{center}
{\bf Properties of $G_2$-manifolds}
\end{center}

{\bf Proposition 2.27}\\
Let $(M,g)$ be a Riemannian manifold with $Hol(g)=G_2$. Then $M$
is a spin manifold and the space of parallel spinors has dimension
one, as stated above.

\bigskip
{\bf Proposition 2.28}\\
Let $(M,g)$ be a Riemannian seven-manifold. If $Hol(g)\subseteq
G_2$,
then $g$ is Ricci-flat.\\
\\
To prove this we note that we have at least one covariantly
constant spinor $\nabla^S \eta=0$. But then we know that
$0=[\nabla^S_m,\nabla^S_n]\eta={1\over4}R_{mnpq}\Gamma^{pq}\eta$
where we used (\ref{curv}). Now multiply by $\Gamma^n$ to get
\begin{eqnarray}
R_{mnpq}\Gamma^n\Gamma^{pq}\eta&=&0,\nonumber\\
\Rightarrow\ \ R_{m(npq)}\Gamma^n\Gamma^{pq}\eta&=&0,\ \
\mbox{as}\ \
R_{m[npq]}=0\nonumber\\
\Rightarrow\ \ \ \ \ \ \ \ \ \mathcal{R}_{mn}\Gamma^n\eta&=&0,\nonumber\\
\Rightarrow\ \ \ \ \ \ \ \ \ \ \ \ \ \
\mathcal{R}_{mn}&=&0,\nonumber
\end{eqnarray}
where we used
$\Gamma^n\Gamma^{pq}=\Gamma^{npq}-\Gamma^p\delta^{nq}+\Gamma^q\delta^{pn}$.

\bigskip
{\bf Proposition 2.29}\\
Let $(M,g)$ be a compact Riemannian manifold with $Hol(g)=G_2$,
then $H^1(M,\mathbb{R})=\lbrace 0\rbrace$, so that $b^1(M)=0$.

\bigskip
This proposition together with the connectedness of $(M,g)$ and
the Poincar\'{e} duality enables us to write down the Betti
numbers of a compact manifold with holonomy $G_2$. They are
$b^0=b^7=1$, $b^1=b^6=0$ and $b^2=b^5$ and $b^3=b^4$ arbitrary.

\bigskip
{\bf Proposition 2.30}\\
Let $(M,\varphi,g)$ be a compact $G_2$-manifold. Then $Hol(g)=G_2$
if and only if $\pi_1(M)$ is finite.

\bigskip
{\bf Proposition 2.31}\\
Let $M$ be a compact $G_2$-manifold with $Hol(g)=G_2$. Then the
isometry group of $M$ is trivial.

\bigskip
This will be proved in section 5.3.

\bigskip
{\bf Proposition 2.32}\\
Let $(M,\varphi, g)$ be a $G_2$-manifold. Then $\Lambda^kT^*M$
splits orthogonally into components. In particular,
\begin{equation}
\Lambda^3T^*M=\Lambda^3_1\oplus\Lambda^3_7\oplus\Lambda^3_{27},
\label{splitting}
\end{equation}
where $\Lambda^k_l$ corresponds to an irreducible representation
of $G_2$ of dimension $l$. This splitting is the only one we need.
The other splittings and more details can be found in \cite{Jo00}.

\bigskip
{\bf Definition 2.33}\\
Let $M$ be an oriented seven-manifold. For each $p\in M$, define
$\mathcal{P}^3_pM$ to be the subset of three-forms $\varphi\in
\Lambda^3T^*M$ for which there exists an oriented isomorphism
between $T_pM$ an $\mathbb{R}^7$ identifying $\varphi$ and
$\varphi_0$ of (\ref{G2form}). Define
\begin{equation}
\mathcal{X}:=\lbrace\varphi\in
C^{\infty}(\mathcal{P}^3M):d\varphi=d*\varphi=0\rbrace.
\end{equation}
Let $\mathcal{D}$ be the group of all diffeomorphisms $\Psi$ of
$M$ isotopic to the identity. Then $\mathcal{D}$ acts naturally on
$C^{\infty}(\mathcal{P}^3M)$ and $\mathcal{X}$ by
$\varphi\stackrel{\Psi}{\rightarrow}\Psi_*(\varphi)$. Define the
{\em moduli space} of torsion free $G_2$-structures on M to be
$\mathcal{M}:=\mathcal{X}/\mathcal{D}$.

\bigskip
{\bf Proposition 2.34}\\
Let $M$ be a compact seven-manifold, and
$\mathcal{M}=\mathcal{X}/\mathcal{D}$ the moduli space of
$G_2$-structures on $M$. Then $\mathcal{M}$ is a smooth manifold
of dimension $b^3(M)$.

\vspace{1.5cm}
\begin{center}
{\bf Calibrated geometry and $G_2$-manifolds}
\end{center}

{\bf Definition 2.35}\\
Let $(M,g)$ be a Riemannian manifold. An {\em oriented tangent
$k$-plane} $V$ on $M$ is a vector subspace $V$ of some tangent
space $T_pM$ to $M$ with dim$V=k$, equipped with an orientation.
If $V$ is an oriented tangent $k$-plane
 on $M$ then $g|_V$ is a Euclidean metric on $V$, so combining
 $g|_V$ with the orientation on $V$ gives a natural volume form
 ${\rm vol}_V$ on $V$, which is a $k$-form on $V$. \\
Now let $\varphi$ be a closed $k$-form on $M$. We say that
$\varphi$ is a {\em calibration} on $M$ if for every oriented
$k$-plane $V$ on $M$ we have $\varphi|_V\leq {\rm vol}_V$. Here
$\varphi|_V=\alpha\cdot {\rm vol}_V$ for some $\alpha \in
\mathbb{R}$ and $\varphi|_V\leq {\rm vol}_V$ if $\alpha \leq 1$.
Let $N$ be an oriented submanifold of $M$ with dimension $k$. Then
each tangent space $T_pN$ for $p\in N$ is an oriented tangent
$k$-plane. We say that $N$ is a {\em calibrated submanifold}
\footnote{Physicists usually call these manifolds {\em
supersymmetric cycles}.} if $\varphi|_{T_pN}={\rm vol}_{T_pN}$ for
all $p\in N$.

\bigskip
{\bf Proposition 2.36}\\
Let $(M,g)$ be a Riemannian manifold, $\varphi$ a calibration on
$M$, and $N$ a compact calibrated submanifold in $M$. Then $N$ is
volume minimizing in its homology class.

\bigskip
To prove this we denote $k:=dim(N)$, and let $[N]\in
H_k(M,\mathbb{R})$ and $[\varphi]\in H^k(M,\mathbb{R})$ be the
homology and cohomology classes of $N$ and $\varphi$. Then
\begin{equation*}
[\varphi]\cdot [N]=\int_{x\in N} \varphi|_{T_xN}=\int_{x\in N}
{\rm vol}_{T_xN}={\rm vol}(N),
\end{equation*}
since $\varphi_{T_xN}={\rm vol}_{T_xN}$. If $N'$ is any other
compact $k$-submanifold of $M$ with $[N']=[N]$ in
$H_k(M,\mathbb{R})$, then
\begin{equation*}
[\varphi]\cdot [N]=[\varphi]\cdot [N']=\int_{x\in
N'}\varphi|_{T_xN'}\leq\int_{x\in N'}{\rm vol}_{T_xN'}={\rm
vol}(N'),
\end{equation*}
since $\varphi|_{T_xN'}\leq {\rm vol}_{T_xN'}$ because $\varphi$
is a calibration. Thus, ${\rm vol}(N')\leq {\rm vol}(N)$.

\bigskip
A Riemannian manifold $(M,g)$ with holonomy $G_2$ defines a
three-form $\varphi$ and a four-form $\star\varphi$, as given
above. These are both calibrations. Calibrated submanifolds with
respect to $\varphi$ are called {\em associative three-folds},
calibrated submanifolds with respect to $\star\varphi$ are called
{\em
co-associative four-folds}.\\
The general relation between calibrations and holonomy groups can
be found in \cite{Jo00}.

\bigskip
\begin{center}
{\bf Weak $G_2$}
\end{center}

There is yet another notion which is important for M-theory
compactifications, namely that of weak $G_2$-holonomy, or a nearly
parallel $G_2$-structure \cite{FKMS97}, \cite{Gr71}. These are
$G_2$-structures $(\varphi,g)$ with a Killing spinor, rather than
a constant spinor. That is we have
$\nabla^S\eta=i{\lambda\over2}\gamma\eta$. On such manifolds the
three-form satisfies $d\varphi=4\lambda *\varphi$ and
$d*\varphi=0$, for some $\lambda \in \mathbb{R}$ and they are
automatically Einstein with non-negative scalar curvature
$6\lambda^2$. In the case of $\lambda=0$ we get a $G_2$-manifold.
Explicit metrics of compact weak $G_2$-manifolds with conical
singularities were first constructed in \cite{BM03a}.

\newpage
\begin{center}
{\bf Examples of $G_2$ manifolds}
\end{center}
Compact $G_2$-manifolds are complicated objects and have been
constructed only recently. We present the simplest construction
of such a manifold following the general construction mechanism
presented by Joyce \cite{Jo00}. The basic tack is a follows. We
start with a torus $T^7$, equipped with a flat $G_2$-structure
$(\varphi_0,g_0)$ and a finite group $\Gamma$ of automorphisms of
$T^7$ preserving $(\varphi_0,g_0)$. Then $T^7/\Gamma$ is an
orbifold with a flat $G_2$-structure. This orbifold can be
resolved and deformed to a smooth compact manifold of holonomy
$G_2$. The resulting manifold depends on the choice of the group
$\Gamma$ and the way of resolving the orbifold. Various different
$G_2$-manifolds can be constructed in that way.\\
After having analyzed the simplest example we will move on to the
more complicated manifolds which will play an important role in
M-theory compactifications.

\bigskip
{\bf Example 1}\\
For our first example let $(x_1,x_2,x_3,x_4,x_5,x_6,x_7)$ be
coordinates on $\mathbb{R}^7$, and let $(\varphi_0,g_0)$ be the
flat $G_2$-structure on $\mathbb{R}^7$, i.e.
\begin{eqnarray}
\varphi_0&:=&d\textbf{x}_{123}+d\textbf{x}_{145}+d\textbf{x}_{167}+
d\textbf{x}_{246}-d\textbf{x}_{257}-d\textbf{x}_{347}-d\textbf{x}_{356},\nonumber\\
g_0&=&dx_1^2+\ldots +dx_7^2.\nonumber
\end{eqnarray}
Let $\mathbb{Z}^7$ act on $\mathbb{R}^7$ by translation in the
obvious way, and let $T^7:=\left(\mathbb{R}/\mathbb{Z}\right)^7$.
Then $(\varphi_0,g_0)$ can be pulled back on $T^7$. We take
$([x_1],[x_2],[x_3],[x_4],[x_5],[x_6],[x_7])$ to be coordinates on
$T^7$, where $[x_i]:=x_i+\mathbb{Z}$.\\
To proceed we define the maps $\alpha,\beta,\gamma:T^7\rightarrow
T^7$ by
\begin{eqnarray}
\alpha([x_1],[x_2],[x_3],[x_4],[x_5],[x_6],[x_7])&=&([x_1],[x_2],[x_3],[-x_4],[-x_5],[-x_6],[-x_7]),\nonumber\\
\beta([x_1],[x_2],[x_3],[x_4],[x_5],[x_6],[x_7])&=&([x_1],[-x_2],[-x_3],[x_4],[x_5],[{1\over 2}-x_6],[-x_7]),\nonumber\\
\gamma([x_1],[x_2],[x_3],[x_4],[x_5],[x_6],[x_7])&=&([-x_1],[x_2],[-x_3],[x_4],[{1\over2}-x_5],[x_6],[{1\over2}-x_7])\nonumber.
\end{eqnarray}
Obviously these maps preserve $\varphi_0$ and $g_0$, therefore,
$\Gamma:=\langle\alpha,\beta,\gamma\rangle$ is a group of
isometries of $T^7$ preserving the flat $G_2$-structure
$(\varphi_0,g_0)$. Furthermore they certainly are involutions,
$\alpha^2=\beta^2=\gamma^2=1$, and $\alpha, \beta , \gamma$
commute, e.g.
\begin{eqnarray}
(\alpha\circ\beta&-&\beta\circ \alpha)
([x_1],[x_2],[x_3],[x_4],[x_5],[x_6],[x_7])\nonumber\\
&=&\alpha([x_1],[-x_2],[-x_3],[x_4],[x_5],[{1\over2}-x_6],[-x_7])\nonumber\\
&&-\beta([x_1],[x_2],[x_3],[-x_4],[-x_5],[-x_6],[-x_7])\nonumber\\
&=&([x_1],[-x_2],[-x_3],[-x_4],[-x_5],[x_6-{1\over2}],[x_7])\nonumber\\
&&-([x_1],[-x_2],[-x_3],[-x_4],[-x_5],[{1\over2}+x_6],[x_7])\nonumber\\
&=&([0],[0],[0],[0],[0],[-1],[0])=0,\nonumber
\end{eqnarray}
and similarly for all other commutators. Collecting all these
results we find the isomorphism $\Gamma\cong\mathbb{Z}_2^3$.\\
Next we want to analyze the set $T^7/\Gamma$, in particular we are
interested in its singularities. To do so we list the elements of
$\Gamma$ explicitly,
\begin{eqnarray}
\Gamma&=&\lbrace1,\alpha,\beta,\gamma,\alpha\beta,\alpha\gamma,\beta\gamma,\alpha\beta\gamma\rbrace,\nonumber\\
\alpha\beta&=&([x_1],[-x_2],[-x_3],[-x_4],[-x_5],[x_6-{1\over2}],[x_7]), \nonumber\\
\alpha\gamma&=& ([-x_1],[x_2],[-x_3],[-x_4],[x_5-{1\over2}],[-x_6],[x_7-{1\over2}]),\nonumber\\
\beta\gamma&=& ([-x_1],[-x_2],[x_3],[x_4],[{1\over2}-x_5],[{1\over2}-x_6],[x_7-{1\over2}]),\nonumber\\
\alpha\beta\gamma&=&([-x_1],[-x_2],[x_3],[-x_4],[x_5-{1\over2}],[x_6-{1\over2}],[{1\over2}-x_7]).\nonumber
\end{eqnarray}
In particular, we see that $\alpha\beta$, $\alpha\gamma$,
$\beta\gamma$ and $\alpha\beta\gamma$ have no fixed points, as
e.g. $[x_6]\stackrel{\alpha\beta}{\rightarrow}[x_6-{1\over2}]$,
$[x_5]\stackrel{\alpha\gamma}{\rightarrow}[x_5-{1\over2}]$,
$[x_7]\stackrel{\beta\gamma}{\rightarrow}[x_7-{1\over2}]$ and
$[x_5]\stackrel{\alpha\beta\gamma}{\rightarrow}[x_5-{1\over2}]$.
Next we analyze the fixed point sets of $\alpha,\beta,\gamma$.
Clearly for $x_4,\ x_5,\ x_6,\
x_7\in\lbrace\mathbb{Z},\mathbb{Z}+{1\over2}\rbrace$
$([x_1],[x_2],[x_3],[x_4],[x_5],[x_6],[x_7])$ are fixed points of
$\alpha$ for any value of $x_1,\ x_2,\ x_3$. So the fixed point
set of
$\alpha$ is the union of 16 disjoint copies of $T^3$.\\
Similarly $\beta$ and $\gamma$ have have fixed point sets of 16
tori $T^3$. Let us now study the action of $\beta$ on the fixed
point set of $\alpha$. Clearly
$[x_4]\stackrel{\beta}{\rightarrow}[x_4]$,
$[x_5]\stackrel{\beta}{\rightarrow}[x_5]$,
$[x_6]\stackrel{\beta}{\rightarrow}[x_6+{1\over2}]$ and
$[x_7]\stackrel{\beta}{\rightarrow}[x_7]$ for $x_4,\ x_5,\ x_6,\
x_7\in\lbrace\mathbb{Z},\mathbb{Z}+{1\over2}\rbrace$. Similarly
for $\gamma$, $[x_4]\stackrel{\gamma}{\rightarrow}[x_4]$,
$[x_5]\stackrel{\gamma}{\rightarrow}[x_5+{1\over2}]$,
$[x_6]\stackrel{\gamma}{\rightarrow}[x_6]$ and
$[x_7]\stackrel{\gamma}{\rightarrow}[x_7+{1\over2}]$ for $x_4,\
x_5,\ x_6,\ x_7\in\lbrace\mathbb{Z},\mathbb{Z}+{1\over2}\rbrace$
and $\beta\gamma$,
$[x_4]\stackrel{\beta\gamma}{\rightarrow}[x_4]$,
$[x_5]\stackrel{\beta\gamma}{\rightarrow}[x_5+{1\over2}]$,
$[x_6]\stackrel{\beta\gamma}{\rightarrow}[x_6+{1\over2}]$ and
$[x_7]\stackrel{\beta\gamma}{\rightarrow}[x_7+{1\over2}]$ for
$x_4,\ x_5,\ x_6,\
x_7\in\lbrace\mathbb{Z},\mathbb{Z}+{1\over2}\rbrace$. This
implies that the group $\langle\beta,\gamma\rangle$ acts
freely\footnote{The action of a group is said to be free if the
only element that has a fixed point is the unit element.} on the
fixed point set of $\alpha$. Following the same line of arguments
we can show that $\langle\alpha, \gamma\rangle$ acts freely on the
fixed point set of $\beta$ and $\langle\alpha, \beta\rangle$ acts
freely on the fixed point set of $\gamma$. Now we are ready to
formulate the following lemma.

\bigskip
{\bf Lemma 2.37}\\
The singular set $S$ of $T^7/\Gamma$ is a disjoint union of 12
copies of $T^3$, and the singularity at each $T^3$ is locally
modelled by $T^3\times \mathbb{C}^2/\mathbb{Z}_2$.

\bigskip
Let $S'$ be the set of fixed points of some non-identity element
in $\Gamma$. We know that $S'$ is the union of three sets of 16
tori $T^3$. To show that $S'$ in fact is the disjoint union of 48
tori suppose that two of the sets intersect, say those of $\alpha$
and $\beta$. Then the intersection is fixed by $\alpha$ and
$\beta$. But this would imply that $\alpha\beta$ has fixed
points, which is wrong.\\
Now $S=S'/\Gamma$ and $\langle\beta,\gamma\rangle$ acts freely on
the 16 tori $T^3$ fixed by $\alpha$. Therefore these contribute 4
$T^3$ to $S$. The same is true for $\beta$ and $\gamma$, so we
deduce that $S$ consists of 12 disjoint tori $T^3$.\\
It can be shown \cite{Jo00} that $T^7/\Gamma$ is simply connected
and has Betti numbers $b^0=b^7=1$, $b^1=b^2=b^5=b^6=0$ and
$b^3=b^4=7$. As was shown by Joyce, $T^7/\Gamma$ has a resolution
$\widetilde{T^7/\Gamma}$ carrying torsion free $G_2$-structures.
The basic idea is that near a singular $T^3$ the orbifold
$T^7/\Gamma$ looks like $T^3\times \mathbb{C}^2/\mathbb{Z}_2$. For
each singularity a copy of $T^3\times \mathbb{C}^2/\mathbb{Z}_2$
is excised and substituted by a space $T^3\times \mathbb{EH}$,
where $\mathbb{EH}$ is an Eguchi-Hanson space.\footnote{In general
this procedure is not unique. There might exist many different
resolutions for one orbifold. For details see \cite{Jo00}.} We saw
above that $\mathbb{EH}$ is diffeomorphic to the blow-up at 0 of
$\mathbb{C}^2/\mathbb{Z}_2$. It has Betti numbers $b^0=b^2=1$ and
$b^1=b^3=b^4=0$. One can show that
$\pi(\widetilde{T^7/\Gamma})\cong\pi(T^7/\Gamma)$, so we conclude
that $\widetilde{T^7/\Gamma}$ is simply connected. From
proposition 2.30 we deduce that the manifold
$\widetilde{T^7/\Gamma}$ has holonomy $G_2$. Finally, let us look
at the Betti numbers. We have
\begin{equation}
b^k(\widetilde{T^7/\Gamma})=b^k(T^7/\Gamma)-12
b^k(T^3\times\mathbb{C}^2/\mathbb{Z}_2)+12 b^k(T^3\times
\mathbb{EH}),
\end{equation}
which is quite intuitive keeping in mind the excision procedure.
With $b^2(T^3\times \mathbb{EH})=b^3(T^3\times \mathbb{EH})=4$,
$b^2(T^3\times \mathbb{C}^2/\mathbb{Z}_2)=3$ and $b^3(T^3\times
\mathbb{C}^2/\mathbb{Z}_2)=1$ we get
\begin{eqnarray}
b^2(\widetilde{T^7/\Gamma})&=&12,\\
b^3(\widetilde{T^7/\Gamma})&=&43.
\end{eqnarray}
So we constructed a compact manifold $\widetilde{T^7/\Gamma}$
with holonomy group
$G_2$ and Betti numbers $b^2=12$ and $b^3=43$.\\
This manifold was used in a duality conjecture in string theory
in \cite{Ac95}.

\bigskip
{\bf Example 2}\\
The second example we want to consider has $T^7$
and $(\varphi_0, g_0)$ as before but this time we define
\begin{eqnarray}
\alpha([x_1],[x_2],[x_3],[x_4],[x_5],[x_6],[x_7])&=&([x_1],[x_2],[x_3],[-x_4],[-x_5],[-x_6],[-x_7]),\nonumber\\
\beta([x_1],[x_2],[x_3],[x_4],[x_5],[x_6],[x_7])&=&([x_1],[-x_2],[-x_3],[x_4],[x_5],[-x_6],[-x_7]),\nonumber\\
\gamma([x_1],[x_2],[x_3],[x_4],[x_5],[x_6],[x_7])&=&([-x_1],[x_2],[{1\over2}-x_3],[x_4],[-x_5],[x_6],[-x_7])\nonumber.
\end{eqnarray}
As before one can show that
$\Gamma:=\langle\alpha,\beta,\gamma\rangle$ is isomorphic to
$\mathbb{Z}_2^3$ and that it preserves $(\varphi_0,g_0)$. An
explicit calculation similar to the one given in our first
example yields the following results \cite{Jo00}.\\
The elements $\beta\gamma$ and $\alpha\beta\gamma$ of $\Gamma$
have no fixed points on $T^7$. The fixed points of $\alpha$,
$\beta$, $\alpha\beta$, $\gamma$ and $\alpha\gamma$ in $T^7$ are
each 16 copies of $T^3$. Moreover $\langle\beta,\gamma\rangle$
acts trivially on the set of 16 $T^3$ fixed by $\alpha$, and
$\alpha$ acts trivially on the sets of 16 $T^3$ fixed by $\beta$,
$\alpha\beta$, $\gamma$ and $\alpha\gamma$. The fixed points of
$\alpha$, $\beta$ and $\alpha\beta$ intersect in 64 $S^1$ in
$T^7$, the fixed point set of $\langle\alpha,\beta\rangle$.
Similarly, the fixed points of $\alpha$, $\gamma$ and
$\alpha\gamma$ intersect in 64 $S^1$ in $T^7$, the fixed point set
of
$\langle\alpha,\gamma\rangle$.\\
The fixed point set $S$ of $T^7/\Gamma$ is the union of\\
\mbox{
\begin{tabular}{ll}
(i)  & 16 $T^3/\mathbb{Z}^2_2$ from the $\alpha$ fixed points,\\
(ii) & 8 $T^3/\mathbb{Z}^2_2$ from the $\beta$ fixed points,\\
(iii)& 8 $T^3/\mathbb{Z}^2_2$ from the $\alpha\beta$ fixed points,\\
(iv) & 8 $T^3/\mathbb{Z}^2_2$ from the $\gamma$ fixed points,\\
(v)  & 8 $T^3/\mathbb{Z}^2_2$ from the $\alpha\gamma$ fixed points.\\
\end{tabular}}\\
This union is not disjoint. Instead, the sets (i), (ii) and (iii)
intersect in 32 $S^1$ in $T^7/\Gamma$ from the fixed points of
$\langle\alpha,\beta\rangle$ and the sets (i), (iv) and (v)
intersect in 32 $S^1$ in $T^7/\Gamma$ from the fixed points of
$\langle\alpha,\gamma\rangle$.\\
The fixed points of $\langle\alpha,\beta\rangle$ are
$\{([x_1],\ldots ,[x_7]): x_2,\ldots
,x_7\in\{\mathbb{Z},\mathbb{Z}+{1\over2}\}\}$. These are 64 copies
of $S^1$ in $T^7$. As $\gamma$ acts freely upon them, their image
in $T^7/\Gamma$ is 32 $S^1$. We shall describe the singularities
of $T^7/\Gamma$ close to one of these $S^1$, say that with
$[x_2]=\ldots =[x_7]=\mathbb{Z}$. Near this $S^1$ identify $T^7$
with $S^1\times \mathbb{C}^3$ by equating $([x_1],[y_2],\ldots
,[y_7])$ with $([x_1],y_2+iy_3,\ldots ,y_6+iy_7)$, where $y_i$ is
small. Then the action of $\alpha$ and $\beta$ on $T^7$ can be
understood as an action on $S^1\times\mathbb{C}^3$
\begin{eqnarray}
\alpha([x_1],(z_1,z_2,z_3))&=&([x_1],(z_1,-z_2,-z_3)),\\
\beta([x_1],(z_1,z_2,z_3))&=&([x_1],(-z_1,z_2,-z_3)).
\end{eqnarray}
Thus $T^7/\Gamma$ is locally isomorphic to $S^1\times
\mathbb{C}^3/\mathbb{Z}_2^2$.\\
This kind of singularity can be resolved in various ways. One
possibility is to resolve it in two steps. In the first stage we
resolve $T^7/\langle\alpha\rangle=T^3\times T^4/\mathbb{Z}_2$. The
resolution of $T^4/\mathbb{Z}_2$ is the K3 surface, as we
discussed already below definition 2.8, so we get $T^3\times K3$.
This can be done in a way such that the action of
$\langle\beta,\gamma\rangle$ lifts to $T^3\times K3$, thus
$T^3\times K3/\langle\beta,\gamma\rangle$ is an orbifold. Then, in
a second step one can resolve the orbifold $T^3\times
K3/\langle\beta,\gamma\rangle$ to get a smooth $G_2$-manifold. To
do so one has to analyze the singularities of $T^3\times
K3/\langle\beta,\gamma\rangle$. It turns out that near the
singular points the orbifold is modelled by\footnote{See appendix
F for the definition of $\Gamma_{A_1}$.} $\mathbb{R}^3\times
\mathbb{R}^4/\mathbb{Z}_2=\mathbb{R}^3\times
\mathbb{C}^2/\mathbb{Z}_2=\mathbb{R}^3\times
\mathbb{C}^2/\Gamma_{A_1}$. The details of the resolution can be
found in \cite{Jo00} and we will not need them in what follows.
However, it is important to keep in mind that there exists a
singular limit of a smooth compact $G_2$-manifold which looks
locally like $\mathbb{R}^3\times \mathbb{C}^2/\Gamma_{A_1}$.

\bigskip
{\bf Example 3: A non-compact $G_2$-manifold}\\
Non-compact $G_2$-manifolds were constructed in \cite{BS89} and
\cite{GPP90}. Their structure is less difficult than the one of
compact manifolds. In particular, the metric can be written down
explicitly.

\begin{figure}[h]
\centering
\includegraphics[width=0.6\textwidth]{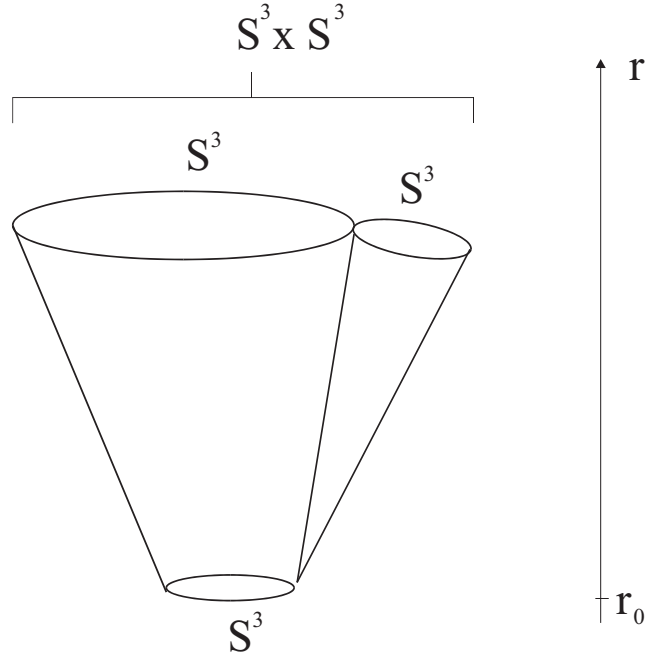}\\
\caption[]{A non-compact $G_2$-manifold.} \label{S3S3}
\end{figure}

\bigskip
One example is a space that is topologically $\mathbb{R}^4\times
S^3\cong\mathbb{C}^2\times S^3$ and which carries the metric
\begin{equation}
ds_7^2={1\over 1-\left({r_0\over
r}\right)^3}dr^2+{1\over9}r^2\left(1-\left({r_0\over
r}\right)^3\right)(\nu_1^2+\nu_2^2+\nu_3^2)+{r^2\over12}(\sigma_1^2+\sigma_2^2+\sigma_3^2).\label{noncompG2}
\end{equation}
Here $\nu_i:=\Sigma_i-{1\over2}\sigma_i$ and $\sigma_i$ and
$\Sigma_i$ are left invariant one-forms on two different $S^3$'s.
For large $r$ the space tends to a cone on $S^3\times S^3$, so it
is asymptotically conical. However, it is not a cone on $S^3\times
S^3$ as for $r=r_0$ we get one uncollapsed $S^3$, thus there is no
singularity. The space is depicted in figure \ref{S3S3}, its
structure is similar to the one of Eguchi-Hanson space. We see
that globally it is a fibre bundle over $S^3$ with fibres
$\mathbb{C}^2$. Note that the metric does not involve the standard
metric on $S^3\times S^3$ but rather the homogeneous metric on
$SU(2)^3/SU(2)_{diag}$. In particular this allows us to define an
action of $SU(2)$ on our manifold which keeps fixed the base $S^3$
and acts on the fibre $\mathbb{C}^2$ in a natural way. These
properties will be important in chapter 7.\\
For a proof that the Levi-Civita connection of this metric really
has holonomy $G_2$ see \cite{GPP90}. A detailed description of
this space can be found in \cite{GPP90} and \cite{AtW01}.

\chapter{Preliminary Physics}

\section{Eleven-dimensional Supergravity}
It is current wisdom in string theory \cite{Wi95} that the low
energy limit of M-theory is eleven-dimensional supergravity
\cite{CJS78}. Therefore, M-theory results can be found using this
well understood supergravity theory. In this section we review the
basic field content, the Lagrangian and its symmetries as well as
the equations of motion. More details can be found in \cite{WB92},
\cite{JPD02}, \cite{Du99}, \cite{DNP86} and \cite{West98}.\\

\begin{center}
{\bf The action of eleven-dimensional supergravity}
\end{center}
The field content of eleven-dimensional supergravity is remarkably
simple. It consists of the metric $g_{MN}$, a Majorana
spin-${3\over 2}$ fermion $\psi_M$ and a three-form $C={1\over
{3!}}C_{MNP}dz^M\wedge dz^N\wedge dz^P$, where $z^M$ is a set of
coordinates on the space-time manifold $M_{11}$. These fields can
be combined to give the unique $\mathcal{N}=1$ supergravity theory
in eleven dimensions. The full action is\footnote{We define
$\bar\psi_M:=i\psi_M^{\dagger}\Gamma^0$, see appendix B.}

\bigskip
\fbox{\parbox{14cm}{
\begin{eqnarray}
S&=&{1\over {2\kappa_{11}^2}}\int
[\mathcal{R}*1-{1\over2}G\wedge*G-{1\over 6}
C\wedge G\wedge G ]\nonumber\\
&& +{1\over 2\kappa_{11}^2}\int d^{11}z
\sqrt{g}\bar{\psi}_M\Gamma^{MNP}\nabla^S_N\left({\omega+\hat{\omega}\over
2}\right)\psi_P\nonumber\\
&&-{1\over 2\kappa_{11}^2}{1\over 192}\int d^{11}z
\sqrt{g}\left(\bar{\psi}_M\Gamma^{MNPQRS}\psi_N+12\bar{\psi}^P\Gamma^{RS}\psi^Q\right)(G_{PQRS}+\hat{
G}_{PQRS})\nonumber.
\end{eqnarray}}\hfill\parbox{8mm}{\begin{eqnarray}\label{SUGRAaction}\end{eqnarray}}}

\bigskip
The general notation and conventions adopted are given in the
appendix. To explain the contents of the action, we start with the
commutator of the vielbeins, which defines the {\em anholonomy
coefficients} $\Omega_{AB}^{\ \ \ \ C}$
\begin{equation}
[e_A,e_B]:=[e^{\ \ M}_A\partial_M,e^{\ \
N}_B\partial_N]=\Omega_{AB}^{\ \ \ \ C}e_C.
\end{equation}
Relevant formulae for the spin connection are
\begin{eqnarray}
\omega_{MAB}(e)&=&{1\over2}(-\Omega_{MAB}+\Omega_{ABM}-\Omega_{BMA})\nonumber\\
\omega_{MAB}&=&\omega_{MAB}(e)+{1\over 8}
[-\bar{\psi}_P\Gamma_{MAB}^{\ \ \ \ \ \
PQ}\psi_Q+2(\bar{\psi}_M\Gamma_B\psi_A-\bar{\psi}_M\Gamma_A\psi_B+\bar{\psi}_B\Gamma_M\psi_A
)]\nonumber\\
\hat{\omega}_{MAB}&:=&\omega_{MAB}+{1\over
8}\bar{\psi}_P\Gamma_{MAB}^{\ \ \ \ \ \ PQ}\psi_Q.
\end{eqnarray}
$\hat{\omega}$ is the supercovariant connection, whose variation
does not involve derivatives of the infinitesimal Grassmann
parameter. $\psi_M$ is a Majorana vector-spinor. The Lorentz
covariant derivative reads
\begin{equation}
\nabla^S_M(\omega)\psi_N=\partial_M\psi_N+{1\over
4}\omega_{MAB}\Gamma^{AB}\psi_N.
\end{equation}
For further convenience we define
\begin{equation}
\widetilde{\nabla}^S_M(\omega)\psi_N:=\nabla^S_M(\omega)\psi_N-{1\over288}(\Gamma_M^{\
\  PQRS}-8\delta_M^P\Gamma^{QRS})\hat{G}_{PQRS}\psi_N\nonumber.
\end{equation}
\begin{equation}
G:=dC \ \ \mbox{i.e.}\ \ G_{MNPQ}=4\partial_{[M}C_{NPQ]}.
\end{equation}
The supercovariantization $\hat{G}_{MNPQ}$ is defined as a term
without derivatives of the infinitesimal parameter in its
supersymmetry variation,
\begin{equation}
\hat{G}_{MNPQ}:=G_{MNPQ}+3\bar{\psi}_{[M}\Gamma_{NP}\psi_{Q]}.
\end{equation}

\bigskip
\begin{center}
{\bf Symmetries of eleven-dimensional supergravity}
\end{center}
The action and equations of motion are invariant under the
following symmetries.\\

\bigskip
a, {\em $d=11$ general covariance} (parameter $\xi^M$)
\begin{eqnarray}
\delta e_{\ \ M}^A&=&e_{\ \ N}^A\partial_M\xi^N+\xi^N\partial_N e_{\ \ M}^A\nonumber\\
\delta C_{MNP}&=&3C_{Q[MN}\partial_{P]}\xi^Q+\xi^Q\partial_Q
C_{MNP}\nonumber\\
\delta\psi_M&=&\psi_N\partial_M\xi^N+\xi^N\partial_N\psi_M
\end{eqnarray}

\bigskip
b, {\em Local $SO(1,10)$ Lorentz transformations} (parameter
$\alpha_{AB}=-\alpha_{BA}$)
\begin{eqnarray}
\delta e_{\ \ M}^A&=&-\alpha_{\ \ B}^Ae_{\ \ M}^B\nonumber\\
\delta C_{MNP}&=&0\nonumber\\
\delta\psi_M&=&-{1\over 4}\alpha_{AB}\Gamma^{AB}\psi_M
\end{eqnarray}

\bigskip
c, {\em $\mathcal{N}=1$ Supersymmetry} (parameter $\eta$,
anti-commuting)
\begin{eqnarray}
\delta e_{\ \ M}^A&=&-{1\over2}\bar{\eta}\Gamma^A\psi_M\nonumber\\
\delta C_{MNP}&=&-{3\over2}\bar{\eta}\Gamma_{[MN}\psi_{P]}\nonumber\\
\delta\psi_M&=&\widetilde{\nabla}^S_M(\hat{\omega})\eta
\end{eqnarray}

\bigskip
d, {\em Abelian gauge transformations} (parameter $\Lambda={1\over
2}\Lambda_{MN}dz^M\wedge dz^n$)
\begin{eqnarray}
\delta e_{\ \ M}^A&=&0\nonumber\\
\delta C_{MNP}&=&3\partial_{[M}\Lambda_{NP]}\Leftrightarrow\delta
C=d\Lambda\nonumber\\
\delta\psi_M&=&0
\end{eqnarray}

\bigskip
e, Odd number of space or time reflections together with
\begin{equation}
C_{MNP}\rightarrow -C_{MNP}
\end{equation}

\bigskip
\begin{center}
{\bf Field equations of eleven-dimensional supergravity}
\end{center}
We will only need solutions of the equations of motion with the
property that $\psi_M\equiv0$. Hence we can set $\psi_M$ to zero
before varying the equations of motion. This leads to an enormous
simplification of the calculations. The equations of motion with
vanishing fermion field read\footnote{We use the definition
$\nabla_MG^{MPQR}:={1\over\sqrt{g}}\partial_M(\sqrt{g}G^{MPQR})$.}
\begin{eqnarray}
\mathcal{R}_{MN}(\omega)-{1\over2}g_{MN}\mathcal{R}(\omega)={1\over
12}\left(G_{MPQR}G_N^{\ \ PQR}-{1\over8}g_{MN}G_{PQRS}G^{PQRS}\right)\label{beom1}\\
\nabla_MG^{MPQR}+{1\over
2\cdot4!\cdot4!}\epsilon^{PQRSTUVWXYZ}G_{STUV}G_{WXYZ}=0\label{beom2}.
\end{eqnarray}The last equation can be rewritten more conveniently
in terms of differential forms
\begin{equation}
d*G+{1\over2}G\wedge G=0 \label{beomforms}.
\end{equation}
In addition to those field equations we also know that G is
closed, as it is exact,
\begin{equation}
dG=0\label{Gclosed}.
\end{equation}
We note in passing that these equations enable us to define two
conserved charges
\begin{equation}
Q_e:=\int_{\partial M_8}\left (*G+{1\over2}C\wedge G\right),
\end{equation}
\begin{equation}
Q_m:=\int_{\partial M_5}G,
\end{equation}
where the integrations are over the boundary at infinity of a
space-like subspace of eight and five dimensions. Note that these
subspaces do not fill out the ten-dimensional space, so the
situation is different from Maxwell's theory in four dimensions.

\section{Anomalies}
In this section we will give the main results on anomalies that
will be needed at various places in our analysis of M-theory. We
will focus on general aspects of anomalies without reference to
their explicit calculation from perturbative quantum field theory.
General references for this section are \cite{AG85}, \cite{AGG84},
\cite{AGW84}, \cite{Wb96} and \cite{We86}. Details on the concept of
anomaly inflow and anomalies in M-theory can be found in \cite{BM03c}.\\
\\
In order to construct a quantum field theory one usually starts
from a classical theory which is quantized by following one of
several possible quantization schemes. Therefore, a detailed
analysis of the classical theory is a crucial prerequisite for
understanding the dynamics of the quantum theory. In particular,
the symmetries and the related conservation laws should be
mirrored on the quantum level. However, it turns out that this is
not always true. If the classical theory possesses a symmetry
that cannot be maintained on the quantum level we speak of an
{\em anomaly}.\\
To see explicitly how an anomaly can arise it is useful to look
at the individual steps involved in the process of quantization.
As is well known, many quantum field theories lead to divergences
if naive calculations are performed. To get rid of these the
theory has to be carefully regularized. If this regularized
theory still has the same symmetries as the classical theory no
anomalies can occur. This changes, however, if some symmetries
cannot be maintained by any regularization scheme. Then we can no
longer expect that the corresponding conservation laws hold on
the quantum level after the regulator is removed. An explicit
check has to be made, using the methods of perturbation theory.

\bigskip
Before discussing anomalies in detail we want to point out the
connection between symmetries and conserved currents. A theory
containing a massless gauge field $A$ is only consistent if the
action is invariant under the infinitesimal local gauge
transformation
\begin{equation}
A'(x)=A(x)+D\epsilon(x).\label{localgaugetrafo}
\end{equation}
The invariance of the action can be written as
\begin{equation}
D_M(x){{\delta S[A]}\over{\delta A_{aM}(x)}}=0,\label{gaugeinv}
\end{equation}
where $A=A_aT_a=A_{aM}T_adx^M$. Then we can define a current
corresponding to this symmetry,
\begin{equation}
J_a^M(x):={\delta S[A]\over\delta A_{aM}(x)},\label{localcurrent}
\end{equation}
and gauge invariance (\ref{gaugeinv}) of the action tells us that
this current is conserved,
\begin{equation}
D_MJ^M_a(x)=0.\label{classcurr}
\end{equation}
If on the other hand a symmetry is violated on the quantum level
we can no longer expect that the corresponding current is
conserved. Suppose we consider a theory containing
massless\footnote{Massive fermions cannot contribute to any
anomaly.} fermions $\psi$ in the presence of an
external gauge field $A$.\\
In such a case the expectation value of an operator is defined
as\footnote{We work in Euclidean space after having performed a
Wick rotation. Our conventions in the Euclidean are as follows:
$S_M=iS_E$, $ix^0_M=x^1_E$, $x^1_M=x^2_E,\ldots x^{d-1}_M=x^d_E$;
$i\Gamma^0_M=\Gamma^1_E,\ \Gamma^1_M=\Gamma^2_E,\ldots
\Gamma^{d-1}_M=\Gamma^d_E$;
$\Gamma_E:=i^{d\over2}\Gamma^1_E,\ldots ,\Gamma^d_E$. For details
on conventions in Euclidean space see \cite{BM03c}.}
\begin{equation}
\langle\mathcal{O}\rangle={\int D\psi D\bar\psi \ \mathcal{O}
\exp(-S[\psi,A])\over\int D\psi D\bar\psi \ \exp(-S[\psi,A])}.
\end{equation}
We define the quantity
\begin{equation}
\exp(-X[A]):=\int D\psi D\bar\psi \
\exp(-S[\psi,A]),\label{generating functional}
\end{equation}
where $S[\psi,A]$ is the fermion action
\begin{equation}
S[\psi,A]=\int d^dx\ \bar\psi i\Gamma^M(\partial_M+A_M)\psi.
\end{equation}
In particular we have the free fermion action
\begin{equation}
S[\psi]=\int d^dx\ \bar\psi i\Gamma^M\partial_M\psi,
\end{equation}
and a term proportional to $A$. But according to
(\ref{localcurrent}) this can be rewritten as
\begin{equation}
S[\psi,A]=S[\psi]+\int d^dx\ J^M_a(x)A_{aM}(x).
\end{equation}
Now it is easy to see that
\begin{equation}
\langle J^M_a(x)\rangle={\delta X[A]\over\delta A_{aM}(x)},
\end{equation}
as
\begin{eqnarray}
{\delta X[A]\over\delta A_{aM}(x)}&=&{\delta\over\delta
A_{aM}(x)}(-\ln(\int D\psi D\bar\psi \ \exp(-S[\psi,A])))\nonumber\\
&=&-{\int D\psi D\bar\psi \ {\delta\over\delta A_{aM}(x)}
\exp(-S[\psi,A])\over\int D\psi D\bar\psi \
\exp(-S[\psi,A])}\nonumber\\
&=&-{\int D\psi D\bar\psi \ \exp(-S[\psi,A])(-J^M_a(x))\over\int
D\psi D\bar\psi \
\exp(-S[\psi,A])}\nonumber\\
&=&\langle J^M_a(x)\rangle.
\end{eqnarray}
An anomaly occurs if a symmetry is broken on the quantum level.
This means that its corresponding quantum current will no longer
be conserved. In such a case we get a generalized version of
(\ref{classcurr}),
\begin{equation}
D_M\langle J^M_a(x)\rangle=iG_a[x;A].\label{localcurrentquantum}
\end{equation}
$G_a[x;A]$ is called the {\em anomaly}.

\bigskip
Not every symmetry of an action has to be a local gauge symmetry.
Sometimes there are global symmetries of the fields
\begin{equation}
\Phi'=\Phi+i\epsilon\Delta\Phi.\label{global symmetry}
\end{equation}
These symmetries lead to a conserved current as follows. As the
action is invariant under (\ref{global symmetry}), for
\begin{equation}
\Phi'=\Phi+i\epsilon(x)\Delta\Phi
\end{equation}
we get a transformation of the form
\begin{equation}
\delta S[\Phi]=-\int d^dx\ J^M(x)\partial_M\epsilon(x).
\label{dSglobal}
\end{equation}
If the fields $\Phi$ now are taken to satisfy the field equations
then (\ref{dSglobal}) has to vanish. Integrating by parts we find
\begin{equation}
\partial_M J^M(x)=0,
\end{equation}
the current is conserved on shell\footnote{This can be
generalized to theories in curved space-time, where we get
$\nabla_MJ^M(x)=0$, with the Levi-Civita connection $\nabla$.}.
Again this might no longer be true on the quantum level as we
will see in detail in the next section.

\bigskip
An anomaly of a global symmetry is not very problematic. It
simply states that the quantum theory is less symmetric than its
classical origin. If on the other hand a local gauge symmetry is
lost on the quantum level the theory is inconsistent. This comes
about as the gauge symmetry of a theory containing massless spin-1
fields is necessary to cancel unphysical states. In the presence
of an anomaly the quantum theory will no longer be unitary and
hence useless. This gives a strong constraint for valid quantum
theories as one has to make sure that all the local anomalies
vanish.

\subsection{The Chiral Anomaly}
In this section we calculate the Abelian anomaly in four flat
dimensions with Euclidean signature using Fujikawa's method
\cite{Fu79}. We will consider the specific example of non-chiral
fermions $\psi$ which are coupled to external gauge fields
$A=A_aT_a=A_{a \mu}T_adx^{\mu}$. The Lagrangian of the system is
given by
\begin{equation}
\mathcal{L}=\bar\psi i\gamma^{\mu}D_{\mu}\psi=\bar\psi
i\gamma^{\mu}(\partial_{\mu}+A_{\mu})\psi.\label{Lagr}
\end{equation}
This Lagrangian is invariant under the usual local gauge
transformation\\
\parbox{14cm}{
\begin{eqnarray}
\psi'(x)&=&g(x)^{-1}\psi(x),\nonumber\\
A'(x)&=&g(x)^{-1}(A(x)+d)g(x),\nonumber
\end{eqnarray}}\hfill\parbox{8mm}{\begin{eqnarray}\label{gaugetrafo1}\end{eqnarray}}
where
\begin{equation}
g(x)=\exp(\Lambda_a(x)T_a),\label{defg}
\end{equation}
and the $T_a$ are anti-Hermitian generators of the gauge group.
The corresponding classical current is given by
\begin{equation}
J^{\mu}_a(x)=i\bar{\psi}(x)T_a\gamma^{\mu}\psi(x),
\end{equation}
which is conserved, $D_{\mu}J^{\mu}_a=0$. The transformation
(\ref{gaugetrafo1}) will not lead to any anomalies on the quantum
level. To see this we consider the functional (\ref{generating
functional})
\begin{equation}
\exp(-X[A])=\int D\psi D\bar\psi \ \exp \left(-\int d^4x\ \bar\psi
i\gamma^{\mu}D_{\mu}\psi\right). \label{effact}
\end{equation}
The action is invariant under (\ref{gaugetrafo1}) but we still
need to check whether this is also true for the measure. In order
to do so we need to give a precise definition of the measure. As
we work in Euclidean space the Dirac operator
$i\gamma^{\mu}D_{\mu}$ is Hermitian, so we can find a basis of
orthonormal eigenfunctions with real eigenvalues,
\begin{equation}
i\gamma^{\mu}D_{\mu}\psi_i=\lambda_i\psi_i.
\end{equation}
The orthonormality conditions reads
\begin{equation}
\langle\psi_i|\psi_j\rangle=\int d^4x\
\psi_i^{\dagger}(x)\psi_j(x)=\delta_{ij}.
\end{equation}
Then we can expand $\psi$ and $\bar\psi$
\begin{eqnarray}
\psi&=&\sum_i a_i\psi_i,\\
\bar\psi&=& \sum_i \bar{b}_i \psi_i^{\dagger},
\end{eqnarray}
with $a_i$ and $\bar{b}_i$ Grassmann variables. The measure is
defined as
\begin{equation}
D\psi D\bar\psi:=\prod_ida_id\bar{b}_i.
\end{equation}
The infinitesimal version of (\ref{gaugetrafo1}) for $\psi$ is
given by
\begin{eqnarray}
\psi'(x)&=&(1-\Lambda(x))\psi(x)=\sum_ia_i'\psi_i,\\
{\bar\psi}'(x)&=&\bar\psi(1+\Lambda(x))=\sum_i\bar{b}_i'\psi^{\dagger}_i.
\end{eqnarray}
From orthonormality we obtain
\begin{eqnarray}
a_i'&=&\langle\psi_i|\psi'\rangle=\langle\psi_i|(1-\Lambda)|\psi\rangle=\sum_j\langle\psi_i|(1-\Lambda)|\psi_j\rangle a_j,\\
\bar{b}_i'&=&\langle\psi'|\psi_i\rangle=\langle\psi|(1+\Lambda)|\psi_i\rangle=\sum_j\bar{b}_j\langle\psi_j|(1+\Lambda)|\psi_i\rangle.
\end{eqnarray}
Now consider the transformation of the product
\begin{eqnarray}
\prod
da_i'&=&[\det(\langle\psi_k|(1-\Lambda)|\psi_l\rangle)]^{-1}\prod
da_i\nonumber\\
&=&\exp\left(-{\rm
tr}(\ln(\opone-\langle\psi_k|\Lambda|\psi_l\rangle))\right)\prod
da_i\nonumber\\
&\approx&\exp\left({\rm
tr}\langle\psi_k|\Lambda|\psi_l\rangle\right)\prod
da_i\nonumber\\
&=&\exp\left(\sum_k\langle\psi_k|\Lambda|\psi_k\rangle\right)\prod
da_i.
\end{eqnarray}
Similarly,
\begin{eqnarray}
\prod d\bar{b}_i'&=&\prod
d\bar{b}_i [\det(\langle\psi_k|(1+\Lambda)|\psi_l\rangle)]^{-1}\nonumber\\
&\approx&\prod d\bar{b}_i
\exp\left(-\sum_k\langle\psi_k|\Lambda|\psi_k\rangle\right),
\end{eqnarray}
and therefore the measure is invariant
\begin{equation}
\prod_ida_i'd\bar{b}_i'=\prod_ida_id\bar{b}_i.
\end{equation}

\bigskip
Thus, we showed that the right-hand side of (\ref{effact}) is
invariant under the transformation (\ref{gaugetrafo1}). But the
left-hand side gives
\begin{eqnarray}
\delta \exp(-X[A])&=&-\exp(-X[A])\int d^4x\left({\delta
X[A]\over\delta
A_{a\mu}(x)}(D_{\mu}\epsilon(x))_a\right)\nonumber\\
&=&\exp(-X[A])\int d^4x\left(D_{\mu}{\delta X[A]\over\delta
A_{a\mu}(x)}\epsilon_a(x)\right)\nonumber\\
&=&\exp(-X[A])\int d^4x\ D_{\mu}\langle
J^{\mu}_a(x)\rangle\epsilon_a(x).\label{deltaeX}
\end{eqnarray}
We conclude
\begin{equation}
D_{\mu}\langle J^{\mu}_a(x)\rangle=0,
\end{equation}
the symmetry is conserved at the quantum level.

\bigskip
However, (\ref{Lagr}) is also invariant under the global
transformation
\begin{equation}
\psi':=\exp(i\epsilon\gamma_5)\psi,\label{chiralsym}
\end{equation}
with $\epsilon$ an arbitrary real parameter. This symmetry is
called the {\em chiral symmetry}. The corresponding (classical)
current is
\begin{equation}
J_{5}^{\mu }(x)=\bar{\psi}(x)\gamma^{\mu}\gamma_5\psi(x)\nonumber,
\end{equation}
and it is conserved $\partial_{\mu}J^{\mu}_5=0$, by means of the
equations of motion. To proceed we analyze how (\ref{effact})
transforms under infinitesimal {\em local} chiral
transformations\\
\parbox{14cm}{
\begin{eqnarray}
\psi'(x)&=&(1+i\epsilon(x)\gamma_5) \psi(x)=\sum_ia_i'\psi_i,\nonumber\\
{\bar\psi}'(x)&=&\bar\psi(x)(1+i\epsilon(x)\gamma_5)=\sum_i\bar{b}_i'\psi^{\dagger}_i,\nonumber
\end{eqnarray}}\hfill\parbox{8mm}{\begin{eqnarray}\label{infchiral}\end{eqnarray}}
where we take $\epsilon(x)$ to be a smooth function of $x$. Note
that the properties of $\gamma_5$ lead to the same factor
$1+i\epsilon(x)\gamma_5$ for $\psi$ and $\bar\psi$. The current is
defined in equation (\ref{dSglobal}), so we know that under
(\ref{infchiral}) the action transforms as
\begin{equation}
\delta S=-\int d^4x\ J^{\mu}_5(x)\partial_{\mu}\epsilon(x).
\end{equation}
Once again we need to analyze the transformation of the measure,
\begin{eqnarray}
\prod
da_i'&=&\exp\left(-i\sum_k\langle\psi_k|\epsilon\gamma_5|\psi_k\rangle\right)\prod
da_i,\\
\prod d\bar{b}_i'&=&\prod d\bar{b}_i\
\exp\left(-i\sum_k\langle\psi_k|\epsilon\gamma_5|\psi_k\rangle\right),
\end{eqnarray}
hence, in that case we find a transformation
\begin{equation}
\prod_ida_i'd\bar{b}_i'=\prod_ida_id\bar{b}_i\
\exp\left(-2i\sum_k\langle\psi_k|\epsilon\gamma_5|\psi_k\rangle\right).
\end{equation}
This can be rewritten in the form
\begin{equation}
D\psi' D\bar{\psi}'=\exp\left(i\int d^4x\
\epsilon(x)G[x;A]\right)D\psi D\bar{\psi}\label{measuretrafo}
\end{equation}
with
\begin{equation}
G[x;A]=-2\sum_k \psi^{\dagger}_k(x)\gamma_5 \psi_k(x).
\end{equation}
Now let us consider the variation of the functional
(\ref{generating functional}) under (\ref{infchiral}). Clearly
the variation of the left-hand side vanishes\footnote{This is
actually one version of the Ward-Takahashi identity.} and we get
\begin{eqnarray}
0&=&\delta \exp(-X[A])\nonumber\\
&=&\delta\int D\psi D\bar\psi \exp(-S)\nonumber\\
&=&\int d^4x \int D\psi
D\bar\psi[iG[x;A]\epsilon(x)+J^{\mu}_5(x)\partial_{\mu}\epsilon(x)]\exp(-S)\label{traf}.
\end{eqnarray}
Integration by parts leads to
\begin{equation}
\partial_{\mu}\langle J^{\mu}_5(x)\rangle=i
G[x;A].
\end{equation}
So we see already at this point that the theory is anomalous if
$G[x;A]$ does not vanish. Let us work out its explicit structure.
To do so we have to introduce a regulator, as the integral in
(\ref{measuretrafo}) is ill defined. We write
\begin{eqnarray}
\int d^4x\  G[x;A]&=&-2\int
d^4x\sum_k\psi_k^{\dagger}(x)\gamma_5\psi_k(x)\exp[-(\lambda_k/M)^2]\Big\vert_{M\rightarrow\infty}\nonumber\\
&=&-2\sum_k\langle\psi_k|\gamma_5\exp\left(-(i\gamma^{\mu}D_{\mu}/M)^2\right)|\psi_k\rangle\Big\vert_{M\rightarrow\infty}\label{regG}.
\end{eqnarray}
Using $[D_{\mu},D_{\nu}]=F_{\mu\nu}$ (see appendix C) we find
that
\begin{eqnarray}
(i\gamma^{\mu}D_{\mu})^2&=&-\gamma^{\mu}\gamma^{\nu}D_{\mu}D_{\nu}\nonumber\\
&=&-(\eta^{\mu\nu}+{1\over2}[\gamma^{\mu},\gamma^{\nu}]){1\over2}(\{D_{\mu},D_{\nu}\}+F_{\mu\nu})\nonumber\\
&=&-D_{\mu}D^{\mu}-{1\over4}[\gamma^{\mu},\gamma^{\nu}]F_{\mu\nu}.
\end{eqnarray}
Then
\begin{equation}
G[x;A]=-2\sum_k\langle\psi_k|x\rangle\langle
x|\gamma_5\exp\left((D^2+{1\over4}[\gamma^{\mu},\gamma^{\nu}]F_{\mu\nu})/M^2\right)|\psi_k\rangle\Big\vert_{M\rightarrow\infty},
\end{equation}
and after introducing a plane wave basis
\begin{eqnarray}
G[x;A]&=&-2\int {d^4p\over(2\pi)^4}\int
{d^4p'\over(2\pi)^4}\sum_k\langle\psi_k|p'\rangle\langle p'|
x\rangle
\gamma_5\cdot\nonumber\\
&&\cdot\
\exp\left((D^2+{1\over4}[\gamma^{\mu},\gamma^{\nu}]F_{\mu\nu})/M^2\right)\langle
x|p\rangle\langle p
|\psi_k\rangle\Big\vert_{M\rightarrow\infty}\nonumber\\
&=&-2\int {d^4p\over(2\pi)^4} {\rm
tr}\left\{\gamma_5\exp\left((-p^2+{1\over4}[\gamma^{\mu},\gamma^{\nu}]F_{\mu\nu})/M^2\right)\right\}\Big\vert_{M\rightarrow\infty}.
\end{eqnarray}
Introducing $\widetilde{p}:=p/M$ this becomes
\begin{equation}
G[x;A]=-2{\rm
tr}\left\{\gamma_5\exp\left({1\over4}[\gamma^{\mu},\gamma^{\nu}]F_{\mu\nu}/M^2\right)\right\}M^4\int
{d^4\widetilde{p}\over
(2\pi)^4}exp(-\widetilde{p}^2)\Big\vert_{M\rightarrow\infty}.
\end{equation}
Now we expand, take the limit and use\footnote{Note that we are
working in Euclidean space, where $\gamma_5$ is defined as
$\gamma_5=-\gamma^1\gamma^2\gamma^3\gamma^4$.}
\begin{eqnarray}
{\rm tr}\ \gamma_5={\rm tr}\ \gamma_5\gamma^{\mu}\gamma^{\nu}&=&0,\\
{\rm tr}\ \gamma_5
\gamma^{\mu}\gamma^{\nu}\gamma^{\rho}\gamma^{\sigma}&=&-4\epsilon^{\mu\nu\rho\sigma},\\
\int d^4\widetilde{p}\ \exp(-\widetilde{p}^2)&=&\pi^2,
\end{eqnarray}
to get the final result
\begin{equation}
G[x;A]={1\over16\pi^2}{\rm
tr}[\epsilon^{\mu\nu\rho\sigma}F_{\mu\nu}(x)F_{\rho\sigma}(x)]\label{G[x;A]}.
\end{equation}
We conclude that the chiral symmetry is broken on the quantum
level and we are left with the anomaly
\begin{eqnarray}
\partial_{\mu}\langle J_5^{\mu}(x)\rangle&=&{i\over16\pi^2}\epsilon^{\mu\nu\rho\sigma}{\rm tr}F_{\mu\nu}(x)F_{\rho\sigma}(x)\label{abeliananomaly}\\
&=&{i\over4\pi^2}\epsilon^{\mu\nu\rho\sigma}{\rm
tr}[\partial_{\mu}(A_{\nu}\partial_\rho
A_{\sigma}+{2\over3}A_{\nu}A_{\rho}A_{\sigma})].
\end{eqnarray}
This was first calculated by Adler \cite{Ad69} and Bell and Jackiw
\cite{BJ69} using perturbative quantum field theory. The right
hand side of (\ref{abeliananomaly}) is called the {\em chiral
anomaly}.

\subsection{The non-Abelian Anomaly}
Next we study a theory containing a Weyl spinor $\chi$ coupled to
an external gauge field $A=A_aT_a$. Again we take the base
manifold to be  flat and four-dimensional. The Lagrangian of this
theory is
\begin{equation}
\mathcal{L}=\bar\chi i\gamma^{\mu}D_{\mu}P_+\chi=\bar\chi
i\gamma^{\mu}(\partial_{\mu}+A_{\mu})P_+\chi.
\end{equation}
It is invariant under the transformations\\
\parbox{14cm}{
\begin{eqnarray}
\chi'&=&g^{-1}\chi,\nonumber\\
A'&=&g^{-1}(A+d)g,\nonumber
\end{eqnarray}}\hfill\parbox{8mm}{\begin{eqnarray}\label{gaugetrafo2}\end{eqnarray}}
with the corresponding current
\begin{equation}
J_a^{\mu}(x):=i\bar{\chi}(x)T_a\gamma^{\mu}P_+\chi(x).
\end{equation}
Again the current is conserved on the classical level, i.e. we
have
\begin{equation}
D_{\mu}J_a^{\mu}(x)=0.
\end{equation}
We now want to check whether this is true on the quantum level as
well. This can be done in various ways. First of all one might
check the conservation of the current at the one-loop level using
perturbation theory. The explicit calculation can be found in
\cite{Wb96}. Another approach is to proceed as we did to
calculate the Abelian anomaly and check the invariance of the
measure. The details of this calculation are given in
\cite{Na90}. As the calculations are rather involved and we do
not need them later on we only present the results.\\
For $g=e^{\epsilon}$ and $\epsilon(x)=\epsilon_a(x)T_a$
infinitesimal the gauge transformation (\ref{gaugetrafo2}) reads
$A'(x)=A(x)+D\epsilon(x)$. The action is invariant under this
transformation but as we saw above this is not necessarily true
for the measure. Suppose it transforms again as
\begin{equation}\label{trans}
D\psi D\bar\psi\rightarrow \exp\left(i\int d^4x\
\epsilon_a(x)G_a[x;A] \right)D\psi D\bar\psi,
\end{equation}
with some anomaly function $G_a[x;A]$. Then the variation of the
functional (\ref{generating functional}) gives
\begin{equation}
\exp(-X[A])\int d^4x\ D_{\mu}\langle
J^{\mu}_a(x)\rangle\epsilon_a(x)=\int d^4x \int D\psi
D\bar\psi[iG_a[x;A]\epsilon_a(x)]\exp(-S),
\end{equation}
where the variation of the left-hand side is calculated as in
(\ref{deltaeX}) and the result for the right-hand side is similar
to (\ref{traf}). But this gives once again
\begin{equation}
D_{\mu} \langle J^{\mu}_a(x)\rangle=iG_a[x;A].
\end{equation}
In principle $G_a[x;A]$ can be calculated using similar methods as
in the case of the Abelian anomaly. The result of this calculation
is\footnote{Note that this anomaly is actually purely imaginary as
it should be in Euclidean space, since it contains three factors
of $T_a=-it_a$.}
\begin{equation}
D_{\mu} \langle
J^{\mu}_a(x)\rangle={1\over24\pi^2}\epsilon^{\mu\nu\rho\sigma}{\rm
tr}[T_a\partial_{\mu}(A_{\nu}\partial_\rho
A_{\sigma}+{1\over2}A_{\nu}A_{\rho}A_{\sigma})].\label{nonabelian
anomaly}
\end{equation}
Later on we will need the result for chiral fermions coupled to
Abelian gauge fields. In that case the anomaly simplifies to
\begin{eqnarray}
D_{\mu} \langle
J^{\mu}_a(x)\rangle&=&-{i\over24\pi^2}\epsilon^{\mu\nu\rho\sigma}\partial_{\mu}A_{\nu}^b\partial_\rho
A_{\sigma}^c\cdot (q_aq_bq_c)\nonumber\\
&=&-{i\over96\pi^2}\epsilon^{\mu\nu\rho\sigma}F^b_{\mu\nu}F^c_{\rho\sigma}\cdot
(q_aq_bq_c).\label{nonabelianabelian}
\end{eqnarray}
Here we used $T_a=iq_a$ which leads to $D=d+iq_aA_a$, the correct
covariant derivative for Abelian gauge fields. The index $a$ now
runs from one to the number of Abelian gauge fields present in the
theory.

\subsection{Consistency Conditions and Descent Equations}
In this section we study anomalies related to local gauge
symmetries from a more abstract point of view.\\
As we saw above a theory containing massless spin-1 particles has
to be invariant under local gauge transformations to be a
consistent quantum theory. These transformations read in their
infinitesimal form $A_{\mu}(y)\rightarrow
A_{\mu}(y)+D_{\mu}\epsilon(y)$. This can be rewritten as $A_{\mu
b }(y)\rightarrow A_{\mu b }(y)-i\int
d^4x\epsilon_a(x)\mathcal{T}_a(x)A_{\mu b}(y)$, with
\begin{equation}
-i\mathcal{T}_a(x):=-{\partial\over\partial
x^{\mu}}{\delta\over\delta A_{\mu a}(x)}-C_{abc}A_{\mu
b}(x){\delta\over\delta A_{\mu c}(x)}.
\end{equation}
Using this operator we can rewrite the divergence of the quantum
current (\ref{localcurrentquantum})
\begin{equation}
D_M\langle J^M_a(x)\rangle=D_M{\delta X[A]\over\delta
A_{aM}(x)}=iG_a[x;A]
\end{equation}
as
\begin{equation}
\mathcal{T}_a(x)X[A]=G_a[x;A].\label{anomaly}
\end{equation}
It is easy to show that the generators $\mathcal{T}_a(x)$ satisfy
the commutation relations
\begin{equation}
[\mathcal{T}_a(x),\mathcal{T}_b(y)]=iC_{abc}\mathcal{T}_c(x)\delta(x-y).\label{comm}
\end{equation}
From (\ref{anomaly}) and (\ref{comm}) we derive the {\em
Wess-Zumino consistency condition} \cite{WZ71}
\begin{equation}
\mathcal{T}_a(x)G_b[y;A]-\mathcal{T}_b(y)G_a[x;A]=iC_{abc}\delta(x-y)G_c[x;A].\label{WZconsistency}
\end{equation}
This condition can be conveniently reformulated using the BRST
formalism. We introduce a ghost field $c(x):=c_a(x) T_a$ and
define the BRST operator by
\begin{eqnarray}
sA&:=&-Dc,\label{BRST}\\
sc&:=&-{1\over2}[c,c].
\end{eqnarray}
s is nilpotent, $s^2=0$, and satisfies the Leibnitz rule
$s(AB)=s(A)B\pm As(B)$, where the minus sign occurs if $A$ is a
fermionic quantity. Furthermore it anticommutes with the exterior
derivative, $sd+ds=0$.\\
Next we define the anomaly functional
\begin{eqnarray}
G[c;A]:=\int d^4x\ c_a(x)G_a[x;A].
\end{eqnarray}
For our example (\ref{nonabelian anomaly}) we get
\begin{eqnarray}
G[c;A] &=&-{i\over24\pi^2}\int {\rm tr}\left\lbrace c\
d\left[AdA+{1\over2}A^3\right]\right\rbrace\nonumber\\
&=&-{i\over24\pi^2}\int {\rm tr}\left\lbrace c\
d\left[AF-{1\over2}A^3\right]\right\rbrace.\label{G example}
\end{eqnarray}
Using the consistency condition (\ref{WZconsistency}) it is easy
to show that
\begin{equation}
sG[c;A]=0.\label{WZconditionBRST}
\end{equation}
Suppose $G[c;A]=sF[A]$ for some local functional $F[A]$. This
certainly satisfies (\ref{WZconditionBRST}) as $s$ is nilpotent.
However, it is possible to show that all these terms can be
cancelled by adding a local functional to the action. This
implies that anomalies of quantum field theories are
characterized by the cohomology groups of the BRST operator. They
are the local functionals $G[c;A]$ of ghost number one satisfying
the Wess-Zumino consistency condition (\ref{WZconditionBRST}),
which cannot be expressed as the BRST
operator acting on some local functional of ghost number zero.\\
\\
Solutions to the consistency condition can be constructed using
the {Stora-Zumino descent equations}. To explain this formalism
we take the dimension of space-time to be $2n$. Consider the
$(2n+2)$-form
\begin{equation}
{\rm ch}_{n+1}(A):={1\over(n+1)!}{\rm
tr}\left({iF\over2\pi}\right)^{n+1},
\end{equation}
which is called the (n+1)-th {\em Chern character}\footnote{A more
precise definition of the Chern character is the following. Let
Let $E$ be a complex vector bundle over $M$ with gauge group $G$,
gauge potential $A$ and curvature $F$. Then ${\rm ch}(A):={\rm
tr}\ \exp\left({iF\over2\pi}\right)$ is called the total Chern
character. The jth Chern character is ${\rm ch}_j(A):={1\over
j!}{\rm tr}\left({iF\over2\pi}\right)^j$.}. As $F$ satisfies the
Bianchi identity we have
\begin{equation}
dF=[A,F],
\end{equation}
and therefore, ${\rm tr}F^{n+1}$ is closed,
\begin{equation}
d\ {\rm tr}F^{n+1}=(n+1){\rm tr}\lbrace[A,F]F^n\rbrace={\rm
tr}\lbrace AF^{n+1}-FAF^n\rbrace=0.
\end{equation}
We now want to show that on any coordinate patch the Chern
character can be written as
\begin{equation}
{\rm ch}_{n+1}(A)=d\Omega_{2n+1},
\end{equation}
for some $\Omega_{2n+1}$. To proof this we need to note that the
Chern character does depend on the connection only up to a total
derivative\footnote{In the mathematical literature this statement
is known as the Chern-Weil theorem for invariant polynomials.}.
Let $A$ and $B$ be two connections defined on a given patch of our
base manifold and define the interpolating connection
\begin{equation}
A_t:=B+t(A-B)
\end{equation}
for $0\leq t\leq 1$. The respective curvature is calculated to be
\begin{eqnarray}
F_t&:=&dA_t+{1\over2}[A_t, A_t]\nonumber\\
&=&F_B+t(F_A-F_B)+{1\over2}(t^2-t)[(A-B),(A-B)].
\end{eqnarray}
The difference of the two Chern characters is
\begin{eqnarray}
{\rm ch}_{n+1}(A)-{\rm ch}_{n+1}(B)&=&\int_0^1dt\ {d\over dt}{\rm ch}_{n+1}(A_t)\nonumber\\
&=&{1\over(n+1)!}\left({i\over2\pi}\right)^{n+1}\int_0^1dt{d\over
dt}{\rm tr}
F^{n+1}_t\nonumber\\
&=&{1\over n!}\left({i\over2\pi}\right)^{n+1}\int_0^1dt \ {\rm
tr}(
D_t(A-B)F_t^n)\nonumber\\
&=&{1\over n !}\left({i\over2\pi}\right)^{n+1}d\int_0^1dt\ {\rm
tr}((A-B)F_t^n).
\end{eqnarray}
Here we used that ${d\over dt}F_t=D_t(A-B)$, the Bianchi identity
and the fact that for tensors the exterior derivative of the
invariant trace and the covariant derivative coincide. The term
${1\over n !}\left({i\over2\pi}\right)^{n+1}\int_0^1dt\ {\rm
tr}(A-B)F_t^{n-1}$ is known as the {\em transgression} of ${\rm
ch}_{n+1}(A)$. But now we can take a frame in which $B\equiv0$ on
the chosen patch and we get
\begin{equation}
{\rm ch}_{n+1}(A)=d\left\{{1\over n
!}\left({i\over2\pi}\right)^{n+1}\int_0^1dt\ {\rm tr}(
AF^n_t)\right\}=:d\Omega_{2n+1}(A).
\end{equation}
The term $\Omega_{2n+1}(A)$ is known as the {\em Chern-Simons
form} of ${\rm ch}_{n+1}(A)$. From the definition of the BRST
operator and the gauge invariance of ${\rm tr}F^{n+1}$ we find
that $s({\rm tr}F^{n+1})=0$. Hence
$d(s\Omega_{2n+1}(A))=-sd\Omega_{2n+1}(A)=-s({\rm
ch}_{n+1}(A))=0$, and from Poincar\'{e}'s lemma\footnote{The
descent equations can be derived more rigorously without making
use of Poincaré's lemma, see e.g. \cite{We86}.},
\begin{equation}
s\Omega_{2n+1}(A)=d\Omega_{2n}^1(c,A).\label{descent1}
\end{equation}
Similarly, $d(s\Omega_{2n}^1(c,A))=-s^2\Omega_{2n+1}(A)=0$, and
therefore
\begin{equation}
s\Omega_{2n}^1(c,A)=d\Omega_{2n-1}^2(c,A).\label{descent2}
\end{equation}
(\ref{descent1}) and (\ref{descent2}) are known as the {\em
descent equations}. They imply that the integral of
$\Omega_{2n}^1(c,A)$ over $2n$-dimensional space-time is BRST
invariant,
\begin{equation}
s\int_{M_{2n}}\Omega_{2n}^1(c,A)=0.
\end{equation}
But this is a local functional of ghost number one, so it is
identified (up to possible factors) with the anomaly $G[c;A]$.
Thus, we found a solution of the Wess-Zumino consistency condition
by integrating the two equations $d\Omega_{2n}^1(A)={\rm
ch}_{n+1}(A)$ and $d\Omega_{2n}^1(c,A)=s\Omega_{2n+1}(A)$. As an
example let us consider the case of four dimensions. We get
\begin{eqnarray}
\Omega_5(A)&=&{1\over2}\left(i\over2\pi\right)^3\int_0^1dt\  {\rm tr}(AF_t^2),\\
\Omega_4^1(c,A)&=&{i\over48\pi^3}{\rm tr}\left\lbrace c\
d\left[AF-{1\over2}A^3\right]\right\rbrace.
\end{eqnarray}
Comparison with our example of the non-Abelian anomaly (\ref{G
example}) shows that indeed
\begin{equation}
G[c;A]=-2\pi \int\Omega_4^1(c,A).\label{anomaly omega}
\end{equation}
Having established the relation between certain polynomials and
solutions to the Wess-Zumino consistency condition using the BRST
operators it is actually convenient to rewrite the descent
equations in terms of gauge transformations. Define
\begin{equation}
G[\epsilon;A]:=\int d^4x\ \epsilon_a(x)G_a[x;A].
\end{equation}
From (\ref{BRST}) it is easy to see that we can construct an
anomaly from our polynomial by making use of the descent
\begin{equation}\label{gaugedescent}
{\rm ch}_{n+1}(A)=d\Omega_{2n+1}(A)\ \ ,\ \
\delta_{\epsilon}\Omega_{2n+1}(A)=d\Omega^1_{2n}(\epsilon,A),
\end{equation}
where $\delta_{\epsilon}A=D\epsilon$. Clearly we find for our
example
\begin{equation}
\Omega_4^1(\epsilon,A)=-{i\over48\pi^3}{\rm tr}\left\lbrace
\epsilon\  d\left[AF-{1\over2}A^3\right]\right\rbrace.
\end{equation}
and we have
\begin{equation}\label{GOmega}
G[\epsilon,A]=2\pi\int\Omega_4^1(\epsilon,A).
\end{equation}
We close this section with two comments.\\
\\
$\bullet$ The Chern character vanishes in odd dimension and thus
we cannot get an anomaly in these cases.\\
\\
$\bullet$ The curvature and connections which have been used were
completely arbitrary. In particular all the results hold for the
curvature two-form $R$. Anomalies related to a breakdown of local
Lorentz invariance or general covariance are called
{gravitational anomalies}. Gravitational anomalies are only
present in $4m+2$ dimensions.

\subsection{Anomalies and Index Theory}
Calculating an anomaly from perturbation theory is rather
cumbersome. However, it turns out that the anomaly $G[x;A]$ is
related to the index of an operator. The index in turn can be
calculated from topological invariants of a given quantum field
theory using powerful mathematical theorems, the Atiyah-Singer
index theorem and the Atiyah-Patodi-Singer index
theorem\footnote{The latter holds for manifolds with boundaries
and we will not consider it here.}. This allows us to calculate
the anomaly from the topological data of a quantum field theory,
without making use of explicit perturbation theory calculations.
We conclude, that an anomaly depends only on the field under
consideration and the dimension and topology of space, which is a
highly non-trivial result.

\bigskip
Let us start by determining the relationship between the chiral
anomaly and the index theorem.\\
The eigenvalues of the Dirac operator $i\gamma^{\mu}D_{\mu}$
always come in pairs, since for $\psi_{i}$ s.t.
$i\gamma^{\mu}D_{\mu}\psi_{i}=\lambda_i\psi_{i}$ we also have
$\gamma_5\psi_{i}$ with
$i\gamma^{\mu}D_{\mu}\gamma_5\psi_{i}=-\lambda_i\gamma_5\psi_{i}$.
Hence, the sum in (\ref{regG}) only receives contributions from
the zero mode sector, i.e. from eigenfunctions $\psi_i$ with
eigenvalue $\lambda_i=0$. These are not generally paired. As
$\gamma_5$ anticommutes with the Dirac operator we can choose
these functions to be not only eigenfunctions of the Dirac
operator but also of $\gamma_5$ with eigenvalues $\pm1$. Then
(\ref{regG}) becomes
\begin{eqnarray}
\int d^4x\
G[x;A]&=&-2\sum_k\langle\psi_k|\gamma_5\exp[-(i\gamma^{\mu}D_{\mu}/M)^2]|\psi_k\rangle\Big\vert_{M\rightarrow\infty}\nonumber\\
&=&-2\left[\sum_{i_+}\langle\psi_{i_+}|\psi_{i_+}\rangle-\sum_{i_-}\langle\psi_{i_-}|\psi_{i_-}\rangle\right]\nonumber\\
&=&-2(n_+-n_-)\nonumber\\
&=&-2\ {\rm ind}\left(i\gamma^{\mu}D_{\mu}\right),
\end{eqnarray}
where we chose $i_+$ $(i_-)$ to label the eigenstates of the Dirac
operator with positive  (negative) eigenvalue of $\gamma_5$. The
Atiyah-Singer index theorem (\ref{ASI}) gives the index of the
Dirac operator
\begin{equation}
{\rm ind}(i\gamma^{\mu}D_{\mu})=\int_M[{\rm ch}(F)\hat A(M)]_{{\rm
vol}}.
\end{equation}
For the trivial background geometry of section 3.2.1 we get $\hat
A(M)=\opone$. Using (\ref{ch(F)}) we find
\begin{equation}
{\rm ind}\left(i\gamma^{\mu}D_{\mu}\right)=-{1\over8\pi^2}\int
{\rm tr}F^2.
\end{equation}
and
\begin{equation}
G[x;A]={1\over16\pi^2}{\rm
tr}[\epsilon^{\mu\nu\rho\sigma}F_{\mu\nu}(x)F_{\rho\sigma}(x)],
\end{equation}
which is the same result as (\ref{G[x;A]}). So it was possible to
determine the structure of $G[x;A]$ using the index theorem.

\bigskip
Unfortunately, in the case of the non-Abelian or gravitational
anomaly the calculation is not that easy. The anomaly can be
calculated from the index of an operator in these cases as well.
However, the operator no longer acts on a $2n$-dimensional space
but on a space with $2n+2$ dimensions, where $2n$ is the dimension
of space-time of the quantum field theory. Hence, non-Abelian and
gravitational anomalies in $2n$ dimensions can be calculated from
index theorems in $2n+2$ dimensions. As we will not need the
elaborate calculations we only present the results. They were
derived in \cite{AGW84} and \cite{AGG84} and they are reviewed in
\cite{AG85}.\\
In section 3.2.3 we saw that it is possible to construct solutions
of the Wess-Zumino condition, i.e. to find the structure of the
anomaly of a quantum field theory, using the descent formalism.
Via descent equations the anomaly $G[c;A]$ in dimension $2n$ is
related to a unique $2n+2$-form, known as the {\em anomaly
polynomial}. It is this $2n+2$-form which contains all the
important information of the anomaly and which can be calculated
from index theory. Furthermore, the $2n+2$-form is unique, but the
anomaly itself is not. This can be seen from the fact that if the
anomaly $G[c;A]$ is related to a $2n+2$-form $I$, then
$G[c,A]+sF[A]$, with a $2n$-form $F[A]$ of ghost number zero, is
related to the same anomaly polynomial $I$. Thus, it is very
convenient, to work with anomaly polynomials instead of
anomalies.\\
The only fields which can lead to anomalies are spin-${1\over2}$
fermions, spin-${3\over2}$ fermions and also forms with
(anti-)self-dual field strength. Their anomalies were first
calculated in \cite{AGW84} and were related to index theorems in
\cite{AGG84}. The result is expressed most easily in terms of the
non-invariance of the Euclidean quantum effective action $X$. The
master formula for all these anomalies reads
\begin{equation}
\delta X=-2\pi i \int \hat I_{2n}^1
\end{equation}
where
\begin{equation}\label{gendescent}
d\hat I_{2n}^1=\delta\hat I_{2n+1}\ \ ,\ \ d\hat I_{2n+1}=\hat
I_{2n+2}
\end{equation}
The $2n+2$-forms for the three possible anomalies
are
\begin{eqnarray}
\hat I_{2n+2}^{(1/2)}&=&\left[\hat A(M_{2n})\ {\rm
ch}(F)\right]_{2n+2}\label{Ihat1/2}\\
\hat I_{2n+2}^{(3/2)}&=&\left[\hat A(M_{2n})\ \left(\tr
\exp\left({i\over2\pi}R\right)-1\right)\ {\rm
ch}(F)\right]_{2n+2}\label{Ihat3/2}\\
\hat I_{2n+2}^{A}&=&\left[\left(-{1\over2}\right){1\over4}\
L(M_{2n})\right]_{2n+2}.\label{IhatA}
\end{eqnarray}
To be precise these are the anomalies of spin-${1\over2}$ and
spin-${3\over 2}$ particles of positive chirality and a self-dual
form in Euclidean space under the gauge transformation $\delta
A=D\epsilon$ and the local Lorentz transformations
$\delta\omega=D\epsilon$. All the objects which appear in these
formulae are explained in appendix E.\\
\\
Let us see whether these general formula really give the correct
result for the non-Abelian anomaly. From (\ref{trans}) we have
$\delta X=-i\int \e(x)G[x;A]=-iG[\e;A]$. Next we can use
(\ref{GOmega}) to find $\delta X=-2\pi i\Omega_4^1(\epsilon,A)$.
But $\Omega_4^1(\epsilon,A)$ is related to ${\rm
ch}_{n+1}(A)=[{\rm ch}(F)]_{2n+2}$ via the descent
(\ref{gaugedescent}) which is the same as (\ref{gendescent}).
Finally $[{\rm ch}(F)]_{2n+2}$ is exactly (\ref{Ihat1/2}) as we
are working in flat space where $\hat A(M)$=1.\\
\\
The spin-${1\over2}$ anomaly\footnote{From now on the term
``anomaly" will denote both $G[x;A]$ and the corresponding
polynomial $\hat I$.} is often written as a sum
\begin{equation}
\hat I^{(1/2)}=\hat I^{(1/2)}_{gauge}+\hat I^{(1/2)}_{mixed}+n\hat
I^{(1/2)}_{grav},
\end{equation}
with the pure gauge anomaly
\begin{equation}
\hat I^{(1/2)}_{gauge}:=[{\rm ch}(A)]_{2n+2}={\rm
ch}_{n+1}(A),\label{Igauge}
\end{equation}
a gravitational anomaly
\begin{equation}
\hat I^{(1/2)}_{grav}=[\hat A(M)]_{2n+2},\label{I1/2grav}
\end{equation}
and finally all the mixed terms
\begin{equation}
\hat I^{(1/2)}_{mixed}:=\hat I^{(1/2)}-\hat
I^{(1/2)}_{gauge}-n\hat I^{(1/2)}_{grav}.\label{Imixed}
\end{equation}
$n$ is the dimension of the representation of the gauge group
under which $F$ transforms.\\
\\
We do not want to write the factor $-2\pi$ all the time. For any
polynomial $\hat I$ we define
\begin{equation}\label{IIhat}
I:=-2\pi \hat I.
\end{equation}
Next we want to present the explicit form of the polynomials $I$
in various dimensions.

\bigskip
\begin{center}
{\bf Anomalies in four dimensions}
\end{center}
There are no purely gravitational anomalies in four dimensions.
The only particles which might lead to an anomaly are chiral
spin-1/2 fermions. The anomaly polynomials are six-forms and they
read for a positive chirality spinor in Euclidean
space\footnote{Note that the polynomials are real, as we have, as
usual $A=A_aT_a$ and $T_a$ is anti-Hermitian.}
\begin{eqnarray}
I_{gauge}^{({1/2})}(F)&=&-2\pi \ {\rm ch}_3(A)\nonumber\\
&=&{i\over{{(2\pi)}^23!}}{\rm tr}F^3.\label{I4gauge}
\end{eqnarray}
The mixed anomaly polynomial of such a spinor is only present for
Abelian gauge fields as ${\rm tr}(T_a)F_a$ vanishes for all simple
Lie algebras. It reads
\begin{equation}
I_{mixed}^{({1/2})}(R,F)=-{i\over{{(2\pi)}^23!}}{1\over 8}{\rm
tr}R^2{\rm tr}F={1\over{{(2\pi)}^23!}}{1\over 8}{\rm tr}R^2
F^aq_a\label{I4mixed}.
\end{equation}

\bigskip
\begin{center}
{\bf Anomalies in ten dimensions}
\end{center}
In ten dimensions there are three kinds of fields which might lead
to an anomaly. These are chiral spin-3/2 fermions, chiral spin-1/2
fermions and self-dual or anti-self-dual five-forms. The twelve
forms for gauge and gravitational anomalies are calculated using
the general formulae (\ref{Ihat1/2}) - (\ref{IhatA}) and
(\ref{IIhat}), together with the explicit expressions for
$\hat{A}(M)$ and $L(M)$ given in appendix E. One obtains the
result
\begin{eqnarray}
I_{gauge}^{({1/2})}(F)&=&{1\over{{(2\pi)}^56!}}{\rm Tr}F^6\nonumber\\
I_{mixed}^{({1/2})}(R,F)&=&{1\over{{(2\pi)}^56!}}\left({1\over
16}{\rm tr}R^4 {\rm Tr}F^2+{5\over 64}({\rm tr}R^2)^2 {\rm
Tr}F^2-{5\over8}{\rm tr}R^2
{\rm Tr}F^4\right)\nonumber\\
I_{grav}^{(1/2)}(R)&=&{1\over{{(2\pi)}^56!}}\left(-{1\over504}{\rm
tr}R^6-{1\over
384}{\rm tr}R^4 {\rm tr}R^2-{5\over 4608}({\rm tr}R^2)^3\right)\nonumber\\
I_{grav}^{({3/2})}(R)&=&{1\over{{(2\pi)}^56!}}\left({55\over56}{\rm
tr}R^6-{75\over
128}{\rm tr}R^4  {\rm tr}R^2+{35\over 512}({\rm tr}R^2)^3\right)\nonumber\\
I_{grav}^{(5-form)}(R)&=&{1\over{{(2\pi)}^56!}}\left(-{496\over504}{\rm
tr}R^6+{7\over 12}{\rm tr}R^4 {\rm tr}R^2-{5\over 72}({\rm
tr}R^2)^3\right)\label{anomalypolynomials10}.
\end{eqnarray}
 The
Riemann tensor $R$ is regarded as an $SO(9,1)$ valued two-form,
the trace ${\rm tr}$ is over $SO(1,9)$ indices.\\
It is important that these formulae are additive for each
particular particle type. For Majorana-Weyl spinors an extra
factor of $1\over2$ must be included, negative chirality spinors
(in the Euclidean) carry an extra minus sign.

\begin{center}
{\bf Anomalies in six dimensions}
\end{center}
Six-dimensional field theories also involve three types of fields
which contribute to anomalies. These are chiral spin 3/2 fermions,
chiral spin 1/2 fermions and self-dual or anti-self-dual
three-forms. The anomaly polynomials are eight-forms, which have
been calculated to be
\begin{eqnarray}
I_{gauge}^{({1/2})}(F)&=&{1\over{{(2\pi)}^34!}}\left(-{\rm Tr}F^4\right)\nonumber\\
I_{mixed}^{({1/2})}(R,F)&=&{1\over{{(2\pi)}^34!}}\left({1\over
4}{\rm tr}R^2  {\rm Tr}F^2\right)\nonumber\\
I_{grav}^{(1/2)}(R)&=&{1\over{{(2\pi)}^34!}}\left(-{1\over240}{\rm
tr}R^4-{1\over
192}({\rm tr}R^2)^2\right)\nonumber\\
I_{grav}^{({3/2})}(R)&=&{1\over{{(2\pi)}^34!}}\left(-{49\over48}{\rm tr}R^4+{43\over 192}({\rm tr}R^2)^2\right)\nonumber\\
I_{grav}^{(3-form)}(R)&=&{1\over{{(2\pi)}^34!}}\left(-{7\over60}{\rm
tr}R^4+{1\over 24}({\rm
tr}R^2)^2\right)\label{anomalypolynomials6}.
\end{eqnarray}
Conventions are as above except that $R$ now is a $SO(5,1)$
valued two-form.

\subsection{Anomalies in Effective Supergravity Theories}
It is interesting that the effective supergravity theories of the
five known string theories are free of anomalies. We comment on
them one by one.\\
\\
$\bullet$ {\bf IIA Supergravity}\\
This theory is parity conserving and therefore free of (local)
anomalies.\\
\\
$\bullet$ {\bf IIB Supergravity}\\
IIB supergravity in ten dimensions contains a self-dual five-form
field strength, a pair of chiral spin-3/2 Majorana-Weyl gravitinos
and a pair of antichiral Majorana-Weyl spin-1/2 fermions. Thus the
total anomaly is given by
\begin{equation}
I_{12}=I_{grav}^{(5-form)}(R)+2{1\over2}I_{grav}^{(3/2)}(R)-2{1\over2}I_{grav}^{(1/2)}(R).
\end{equation}
The two factors of 1/2 come from the fact that all the spinors are
Majorana-Weyl. Adding up the terms we find $I_{12}=0$. IIB
supergravity
is anomaly free.\\
$\bullet$ {\bf Type I Supergravity coupled to d=10 super-Yang-Mills}\\
Type I supergravity is parity violating and in general gives rise
to anomalies. However, as was shown in a seminal paper by Green
and Schwarz \cite{GS84} the anomalies vanish provided Type I
supergravity is coupled to super-Yang-Mills theory with gauge
group $E_8\times E_8$ or $SO(32)$. The basic ideas are as follows.
The field content of $\mathcal{N}=1$ supergravity in ten
dimensions consists of a chiral Majorana-Weyl spin-3/2 gravitino
and an antichiral Majorana-Weyl spin 1/2 dilatino. This theory is
coupled to super-Yang-Mills which contains chiral Majorana-Weyl
spin-1/2 gauginos living in the adjoint representation of the
relevant gauge group $G$. The total anomaly of this theory is
\begin{equation}
I_{12}={1\over2}\left(I_{grav}^{(3/2)}(R)-I_{grav}^{(1/2)}(R)\right)+{1\over2}\left(nI_{grav}^{(1/2)}(R)+I_{mixed}^{(1/2)}(R,F)+I_{gauge}^{(1/2)}(F)\right),
\end{equation}
where\footnote{To be more precise $n$ is the dimension of the
representation of $G$, but as $F$ transforms in the adjoint
representation these two numbers coincide.} $n:={\rm dim}\ G$. If
we make use of the explicit formulas given in
(\ref{anomalypolynomials10}), we get
\begin{eqnarray}
I_{12}&=&{1\over2(2\pi)^56!}\left({{496-n}\over504}{\rm tr}R^6-{{224+n}\over384}{\rm tr}R^4{\rm tr}R^2+{5\over4608}(64-n)({\rm tr}R^2)^3\right.\nonumber\\
&&+\left.{1\over16}{\rm tr}R^4 {\rm Tr}F^2+{5\over64}({\rm
tr}R^2)^2 {\rm Tr}F^2-{5\over8}{\rm tr}R^2{\rm Tr}F^4+{\rm
Tr}F^6)\right).
\end{eqnarray}
To cancel this anomaly via a {\em Green-Schwarz mechanism}, i.e.
by adding a local counter term to the action, the anomaly
polynomial has to factorize into a four-form and an eight-form
\cite{GS84}. But the ${\rm tr}R^6$ term does not allow such a
factorization and therefore it has to vanish. This gives a first
condition on the structure of the gauge group, namely
\begin{equation}
n=496.
\end{equation}
Then we are left with
\begin{eqnarray}
I_{12}&=&{1\over2(2\pi)^56!}\left(-{15\over8}{\rm tr}R^4 {\rm
tr}R^2-{15\over32}({\rm tr}R^2)^3+{1\over16}{\rm tr}R^4
{\rm Tr}F^2\right.\nonumber\\
&&+\left. {5\over64}({\rm tr}R^2)^2 {\rm Tr}F^2-{5\over8}{\rm
tr}R^2 {\rm Tr}F^4+{\rm Tr}F^6\right).
\end{eqnarray}
In order for this to factorize we need
\begin{equation}
{\rm Tr}F^6={1\over48}{\rm Tr}F^4{\rm Tr}F^2-{1\over14400}({\rm
Tr}F^2)^3.
\end{equation}
There are only two 496-dimensional groups with this property,
$SO(32)$ and $E_8\times E_8$. For these groups the anomaly
polynomial reads
\begin{equation}
I_{12}={1\over16\pi^2}\left({1\over30}{\rm Tr}F^2-{\rm
tr}R^2\right) \widehat{X}_8,\label{ISUGRAYManomaly}
\end{equation}
with
\begin{equation}
\widehat{X}_8={1\over(2\pi)^34!}\left({1\over8}{\rm
tr}R^4+{1\over32}({\rm tr}R^2)^2-{1\over240}{\rm tr}R^2{\rm
Tr}F^2+{1\over24}{\rm Tr}F^4-{1\over7200}({\rm
Tr}F^2)^2\right).\label{hatX}
\end{equation}
For $SO(32)$ we have ${\rm Tr}=30{\rm tr}$ and for $E_8\times E_8$
we define ${\rm tr}:=1/30{\rm Tr}$, and thus
\begin{equation}
\widehat{X}_8={1\over(2\pi)^34!}\left({1\over8}{\rm
tr}R^4+{1\over32}({\rm tr}R^2)^2-{1\over8}{\rm tr}R^2{\rm
tr}F^2+{5\over4}{\rm tr}F^4-{1\over8}({\rm tr}F^2)^2\right).
\end{equation}
It is remarkable that the anomaly of the coupled
supergravity-super-Yang-Mills system cancels for the gauge groups
which play such an important role in string theory. In particular
we showed that all the low energy effective actions of the five
known string theories are anomaly free. As anomalies are an
infrared effect this is sufficient to tell us that string theory
is a consistent quantum theory.

\subsection{Anomaly Inflow}
The concept of anomaly inflow in effective theories was pioneered
in \cite{CH84} and further studied in \cite{Nak87}. Here we study
the extension of these ideas in the context of M-theory.\\
Consider once again the derivation of the non-Abelian anomaly as
it was given in section 3.2.2. The variation of (\ref{generating
functional}) gave us
\begin{equation*}
\exp(-X[A])\int (d^dx)_ED_{M}\langle
J^{M}_a(x)\rangle\epsilon_a(x)=\int (d^dx)_E \int D\psi
D\bar\psi[iG_a[x;A]\epsilon_a(x)]\exp(-S^E),
\end{equation*}
where $(d^dx)_E$ denotes the Euclidean measure and we used the
invariance of the Euclidean action $S^E$ under local gauge
transformations. It turns out that this formalism has to be
generalized as we often encounter problems in M-theory in which
the classical action is not fully gauge invariant. One might
argue that in this case the term ``anomaly" loses its meaning,
but this is in fact not true. The reason is that in many cases we
study theories on manifolds with boundary which are gauge
invariant in the bulk, but the non-vanishing boundary contributes
to the variation of the action. So in a sense, the variation does
not vanish because of global geometrical properties of a given
theory. If we studied the same Lagrangian density on a more
trivial base manifold the action would be perfectly gauge
invariant. This is why it still makes sense to speak of an
anomaly. Of course, if we vary the functional (\ref{generating
functional}) in theories which are not gauge invariant we obtain
an additional contribution on the right-hand side. This
contribution is called an {\em anomaly inflow term} for reasons
which will become clear presently.

\bigskip
Consider for example a theory which contains the topological term
of eleven-dimensional supergravity. In fact, all the examples we
are going to study involve either this term or terms which can be
treated similarly. Clearly $\delta\int_{M_{11}}C\wedge dC\wedge
dC$ is invariant as long as $M_{11}$ has no boundary. In the
presence of a boundary we get the non-vanishing result
$\int_{\partial{M_{11}}}\Lambda\wedge dC\wedge dC$.  Let us study
what happens in such a case to the variation of our functional. To
do so we first need to find out how our action can be translated
to Euclidean space. The rules are as follows (see also
\cite{BM03c})
\begin{eqnarray}
x^1_E&:=&ix^0_M,\ \ x^2_E:=x^1_M,\ldots\nonumber\\
(d^{11}x)_E&:=&id^{11}x,\nonumber\\
C_{1MN}^E&:=&-iC_{0MN},\nonumber\\
\epsilon^E_{123\ldots 11}&=&+1.
\end{eqnarray}
We know that $S^M=iS^E$, where $S^M$ is the Minkowski action, but
explicitly we have\footnote{Recall the definition of the
$\e$-tensor given in appendix A.}
\begin{eqnarray}
S_{kin}^M&\propto&-{1\over2}\int d^{11}x\sqrt{g}\  {1\over4!}G_{MNPQ}G^{MNPQ}\nonumber\\
&=&\ {i\over2}\int
(d^{11}x)_E\sqrt{g}\ {1\over4!}G^E_{MNPQ}(G^E)^{MNPQ},\nonumber\\
S_{top}^M&\propto&{1\over6}\int d^{11}x\sqrt{g}\
{1\over3!4!4!}\epsilon^{M_1\ldots M_{11}}C_{M_1M_2M_3}\partial_{M_4}C_{M_5M_6M_7}\partial_{M_8}C_{M_9M_{10}M_{11}}\nonumber\\
&=&-{1\over6}\int (d^{11}x)_E\sqrt{g}\
{1\over3!4!4!}(\epsilon^E)^{M_1\ldots
M_{11}}C^E_{M_1M_2M_3}\partial_{M_4}C^E_{M_5M_6M_7}\partial_{M_8}C^E_{M_9M_10M_{11}}.\nonumber
\end{eqnarray}
But then we can read off
\begin{eqnarray}
S^E&\propto&{1\over2}\int(d^{11}x)_E\sqrt{g}\ {1\over4!}G^E_{MNPQ}(G^E)^{MNPQ}\nonumber\\
&&-i\left(-{1\over6}\right)\int (d^{11}x)_E\sqrt{g}\
{1\over3!4!4!}(\epsilon^E)^{M_0\ldots
M_{10}}C^E_{M_0M_1M_2}\partial_{M_3}C^E_{M_4M_5M_6}\partial_{M_7}C^E_{M_8M_9M_{10}}\nonumber,
\end{eqnarray}
where a crucial factor of $i$ turns up. We write
$S^E=S^E_{kin}+S^E_{top}=S^E_{kin}-i\widetilde{S}^E_{top}$,
because $S^E_{top}$ is imaginary, so $\widetilde{S}^E_{top}$ is
real.

\bigskip
After having seen how the supergravity action translates into
Euclidean space let us calculate the variation of (\ref{generating
functional}) for a slightly more general case. Suppose we have a
theory $S^E_N[A]$ on a $d$-submanifold $N$ of a $D$-manifold $M$
which is invariant under Abelian gauge transformations,
$A'=A+d\epsilon$. Furthermore, let
$S^E_M[A]:=S^E_{M,kin}-i\widetilde{S}^E_{M,top}$ be a theory on
the manifold $M$. The total action is given by
$S^E[A]=S^E_N[A]+S^E_M[A]$. We use that $S_N^E$ and $S^E_{M,kin}$
are gauge invariant and find
\begin{eqnarray}
\int_N (d^dx)_E\ D_{P}\langle J^{P}(x)\rangle\epsilon(x)&=&\int_N
(d^dx)_E\ iG[x;A] \epsilon(x)+i\int_M(d^Dx)_E\  {\delta
\widetilde{S}^E_{M,top}\over\delta
A_P}\partial_P\epsilon(x),\nonumber\\
\end{eqnarray}
or after integration by parts
\begin{eqnarray}
\int_N (d^dx)_E\ D_{P}\langle J^{P}(x)\rangle\epsilon(x)&=&\int_N
(d^dx)_E\ iG[x;A]
\epsilon(x)\nonumber\\
&&+i\int_{\partial M}(d^{D-1}x)_E\ {\delta
\widetilde{S}^E_{M,top}\over\delta
A_P}\epsilon(x)\hat{n}_P\nonumber\\
&&-i\int_M(d^Dx)_E\ \epsilon(x)
\partial_P\left({\delta
\widetilde{S}^E_{M,top}\over\delta A_P}\right).\nonumber\\
\end{eqnarray}
Clearly, we get possible contributions to the anomaly from the new
terms of the right-hand side. These terms come from a theory which
lives on the manifold $M$ and they ``flow into" the manifold $N$
which justifies their name. This picture is particularly nice in
the case in which $N=\partial M$ and $\partial_P\left({\delta
\widetilde{S}^E_{M,top}\over\delta A_P}\right)=0$. Then we are
left with
\begin{equation}
D_{P}\langle J^{P}(x)\rangle=i\left(G[x;A]+ {\delta
\widetilde{S}^E_{M,top}\over\delta A_P}\hat n_P\right).
\end{equation}
Very often the geometrical anomaly inflow term can be used to
cancel anomalies $G[x;A]$ present in the theory on $N$. Sometimes
a similar mechanism works in a case in which we do not have a
boundary in our space but in which $\partial_P\left({\delta
\widetilde{S}^E_{M,top}\over\delta A_P}\right)$ does not vanish on
the lower dimensional manifold $N$. This happens for example in
the case of the M5-brane or in the special setup of M-theory on
singular $G_2$-manifolds considered in chapter 8.\\
After these general considerations we want to explain how anomaly
cancellation from inflow works in in practice. Suppose one has a
theory with $\delta S^M\neq0$. Then our master formula for the
anomaly is generalised to
\begin{equation}\label{master}
\delta X=\delta S^E-2\pi i\int \hat I_{2n}^1=-i\delta S^M+i\int
I_{2n}^1.
\end{equation}
The theory is anomaly free if and only if the right-hand side
vanishes. The following recipe is quite convenient to calculate
the anomaly of a theory in $4k+2$ dimensions. One first calculates
$\delta S^M=\int I_{2n}^{1,inflow}$ and the corresponding
$I_{2n+2}^{inflow}$. Then we {\it add} to this polynomial the
$2n+2$-forms $I$ (read off from (\ref{anomalypolynomials10}) and
(\ref{anomalypolynomials6})) that correspond to the fields which
are present in the {\it Minkowskian} theory (e.g. if a
ten-dimensional theory contains a spin-${3\over2}$ field of
positive chirality we add the fourth line of
(\ref{anomalypolynomials10}) to $I_{2n+2}^{inflow}$). The sum has
to vanish in an
anomaly free theory.\\
A detailed derivation of this recipe is given in \cite{BM03c}. The
main idea is that with our conventions in $d=4k+2$ we have
$\Gamma_E=-\Gamma_M$. This gives an additional sign if we continue
from Minkowskian to Euclidean space. In (\ref{master}) it seems as
if we had to subtract the inflow from the polynomial, but taking
into account this additional sign we have to add the two. If the
reader is not satisfies with this shortcut he can, of course,
always continue everything to Euclidean space and see whether
$\delta X$ vanishes.

\section{Kaluza-Klein Compactification}
The main idea of Kaluza and Klein was that a complicated quantum
field theory in a given dimension might be explained by a
dimensional reduction of a simple theory living in a
higher-dimensional space. As we want to compactify M-theory to
four dimensions it is worth studying how Kaluza-Klein reduction
can be done in general. We will perform an explicit
compactification of M-theory on a compact and smooth
seven-manifold in section 5.3. The general mechanism of
Kaluza-Klein compactification can be described as follows
\cite{DNP86}.

\bigskip
$\bullet$ We start from a theory in dimension $d=4+k$ on a
Riemannian manifold $(M_{4+k},g)$ with signature ($-,+,\ldots ,+$)
and coordinates $z^M$, containing gravity $g_{MN}$ and matter
fields $\Phi$, where $M,N,\ldots \in\lbrace0,1,\ldots
,4+k-1\rbrace$. The theory is described by the $d$-dimensional
Einstein-Hilbert action
\begin{equation}
S=\int_{M_{4+k}} \mathcal{R} *1+ \ldots \,.
\end{equation}

$\bullet$ Next one looks for stable {\em ground state}
solutions of the field equations, $\langle g_{MN}\rangle$ and
$\langle \Phi \rangle$, such that $(M_{4+k},\langle g\rangle)$ is
a Riemannian product $(M_{4+k},\langle g\rangle)=(M_4\times M_k,
g_1\times g_2)$,\footnote{At this point it seems as if we put in
by hand the condition of a macroscopic space with 1+3 dimensions.
However, it turns out that the 4+7 split of eleven-dimensional
supergravity is an output of the theory. For more details see
\cite{DNP86} and the discussion of the Freund-Rubin solution given
below.}
\begin{equation}
\langle g_{MN}(x,y)\rangle=\left(
     \begin{array}{cc}
        {g_1}_{\mu\nu}(x) & 0 \\
        0 & {g_2}_{mn}(y)
\end{array}
     \right).
\end{equation}
$(M_4,g_1)$ is supposed to be four-dimensional space-time with
signature ($-,+,+,+$), coordinates $x^{\mu}$ and
$\mu,\nu,\ldots\in\lbrace 0,1,2,3,\rbrace$. $(M_k,g_2)$ is a
$k$-dimensional space with Euclidean signature, coordinates $y^m$
and
$m,n,\ldots \in\lbrace 1,2,\ldots ,k\rbrace$.\\
In addition we impose the condition of {\em maximal
symmetry}\footnote{A $d$ dimensional manifold is maximally
symmetric if it admits ${1\over2}{d(d+1)}$ Killing vectors.} for
the $d=4$ space-time $(M_4,g_1)$. This requirement restricts the
curvature of the vacuum to be of the form
\begin{equation}
R_{\mu\nu\rho\sigma}(g_1)={\Lambda\over
3}({g_1}_{\mu\rho}{g_1}_{\nu\sigma}-{g_1}_{\mu\sigma}{g_1}_{\nu\rho}).
\end{equation}
This is an Einstein space with $\mathcal{R}_{\mu\nu}(g_1)=\Lambda
{g_1}_{\mu\nu}$. Maximally symmetric spaces are either {\em de
Sitter space} $dS$, Minkowski space or {\em anti-de Sitter space}
$AdS$. However, of those three possibilities only Minkowski and
$AdS$ ground states admit supersymmetry and a positive energy
theorem \cite{DNP86}. Therefore, we restrict to cases in which
$\Lambda\leq0$.\footnote{Current experimental data seem to
indicate, however, that the cosmological constant is small but
non-zero and positive, which would lead to de
Sitter space.}\\

\bigskip
We note at this point that the condition of a Riemannian product
space may be relaxed. A metric that is compatible with the
condition of maximal symmetry can be written as
\begin{equation}
\langle g_{MN}(x,y)\rangle=\left(
     \begin{array}{cc}
        \Delta(y){g_1}_{\mu\nu}(x) & 0 \\
        0 & {g_2}_{mn}(y)
\end{array}
     \right).
\end{equation}
The function $\Delta(y)$ is called the {\em warp-factor}. For the
time being we will restrict ourselves to spaces with warp-factor
one.

\bigskip
There are various restrictions that are imposed on $(M_k,g_2)$.
First of all it certainly must satisfy the field equations,
secondly it should lead to interesting non-Abelian gauge groups
and finally it should be compact in order to guarantee a discrete
mass spectrum in $d=4$. Typically, this is achieved by taking
$(M_k,g_2)$ to be Einstein $\mathcal{R}_{mn}(g_2)=c{g_2}_{mn}$, as
in that case we can refer to two important propositions.

\bigskip
{\bf Proposition 3.1}\\
Complete Einstein spaces with $c>0$ are always compact
\cite{My41}.

\bigskip
{\bf Proposition 3.2}\\
Compact Einstein spaces with $c<0$ have no continuous symmetries
\cite{Ya70}.

\bigskip
So by choosing a metric with $c>0$ we get what is called {\em
spontaneous compactification}. This means we found a solution of
the field equations which is a Riemannian product $(M_4\times
M_k, g_1\times g_2)$ with $M_k$ a compact manifold. Of course not
all higher-dimensional spaces admit such a spontaneous
compactification, but - interestingly enough - eleven-dimensional
supergravity does.

\bigskip
$\bullet$ To determine the spectrum of the
four-dimensional theory, we consider small fluctuations of the
$d$-dimensional fields about their ground their ground state
values
\begin{eqnarray}
g_{MN}(x,y)&=&\langle g_{MN}(x,y)\rangle+h_{MN}(x,y)\\
\Phi(x,y)&=&\langle\Phi(x,y)\rangle + \phi(x,y)
\end{eqnarray}
These equations are substituted into the equations of motion and
the terms linear in $h$ and $\phi$ are kept. The fluctuations are
decomposed as a sum, e.g.
\begin{equation}
\phi(x,y)=\sum_i\widetilde{\phi}^{i}(x)
\omega^{i}(y),\label{expansion}
\end{equation}
where $\omega^{i}(y)$ are eigenfunctions of the mass operator
\begin{equation}
M^2 \omega^{i}(y)=m_i^2 \omega^{i}(y). \label{mass eq}
\end{equation}
In this way we obtain an effective $d=4$ theory with an infinite
tower of massive states with masses $m_i$ together with a finite
number of massless states, coming from the zero-eigenvalue
modes.\\
Having found a spontaneous compactification one must check
whether the vacuum is stable, i.e. whether all states have
positive energy. In Minkowski space we need $m_i^2\geq0$, in $AdS$
the problem is more complicated.\footnote{Often the analysis is
facilitated by the fact that supersymmetric vacua are
automatically stable.}

\bigskip
$\bullet$ If $(M_k,g_2)$ has a symmetry group $G$, i.e. if it
admits Killing vectors $K_m^{(i)}$ $(i=1,\ldots ,\ {\rm dim}\ G)$,
\begin{equation}
\nabla_{m}K^{(i)}_{n}+\nabla_{n}K^{(i)}_{m}=0,\label{Killingcond}
\end{equation}
then the massless states will include Yang-Mills gauge fields
with gauge group $G$. To see this we look at $g_{\mu m}=h_{\mu m}$
and expand
\begin{equation}
g_{\mu m}(x,y)=\sum_i A^{(i)}_{\mu}(x)K^{(i)}_m(y)+\ldots\ \
\mbox{(massive modes).}
\end{equation}
This is exactly expansion (\ref{expansion}), as the Killing
vectors are zero-eigenvalue eigenfunctions of the Laplacian on
$M_k$\footnote{A proof of this statement can be found above
equation (\ref{LapV0}).},
\begin{equation}
\Delta_7 K^{(i)}_m=0.
\end{equation}
We know that the Lie bracket of Killing vectors gives another
Killing vector,
\begin{equation}
[ K^{(i)}, K^{(j)}]=\sum_kf^{ijk}K^{(k)}.
\end{equation}
Obviously this is the Lie algebra of of the symmetry group $G$ with structure constants $f^{ijk}$.\\
Now consider a general (passive) coordinate transformation
\begin{equation}
{z^M}'=z^M+\xi^M(z)
\end{equation}
which implies the transformation of the metric
\begin{equation}
\delta
g_{MN}(z)=\mathcal{L}_{\xi}g_{MN}=g_{NP}(z)\nabla_M\xi^P(z)+g_{MP}(z)\nabla_N\xi^P(z)+\xi^P(z)\nabla_P
g_{MN}(z),
\end{equation}
where $\mathcal{L}_{\xi}g$ is the Lie derivative of $g$ with
respect to $\xi$. If we focus on the special transformation
\begin{equation}
\xi^M(x,y):=\left(0,\sum_i\epsilon^{(i)}(x)K^{(i)m}(y)\right)\label{Spectrafo}
\end{equation}
the transformation law for $g_{\mu m}$ reads
\begin{equation}
\delta g_{\mu m}(x,y)=g_{mp}(x,y)\nabla_{\mu}\xi^p(x,y)+g_{\mu p
}(x,y)\nabla_m\xi^p(x,y)+\xi^p(x,y)\nabla_p g_{\mu m }(x,y).
\end{equation}
Substituting (\ref{Spectrafo}) and $g_{\mu m}(x,y)=\sum_i
A^{(i)}_{\mu}(x)K^{(i)}_m(y)$ gives
\begin{eqnarray}
\delta g_{\mu m}(x,y)&=&\sum_i \delta
A^{(i)}_{\mu}(x)K^{(i)}_m(y)+\sum_i A^{(i)}_{\mu}(x)\delta
K^{(i)}_m(y)\nonumber\\
&=&\sum_i\partial_{\mu}\epsilon^{(i)}(x)K^{(i)}_m(y)\nonumber\\
&&+\sum_{i,j}A_{\mu}^{(j)}(x)K^{(j)}_p(y)\nabla_m\left(\epsilon^{(i)}(x)K^{(i)p}(y)\right)\nonumber\\
&&+\sum_{i,j}\epsilon^{(i)}(x)K^{(i)p}(y)\nabla_p
\left(A_{\mu}^{(j)}(x)K^{(j)}_m(y)\right)\nonumber\\
&=&\sum_i\partial_{\mu}\epsilon^{(i)}(x)K^{(i)}_m(y)\nonumber\\
&&+\sum_{i,j}\epsilon^{(i)}(x)A_{\mu}^{(j)}(x)\left(K^{(i)p}(y)\nabla_pK_m^{(j)}(y)+K^{(j)}_p(y)\nabla_mK^{(i)p}(y)\right)\nonumber\\
&=&\sum_i\partial_{\mu}\epsilon^{(i)}(x)K^{(i)}_m(y)\nonumber\\
&&+\sum_{i,j}\epsilon^{(i)}(x)A_{\mu}^{(j)}(x)\left(K^{(i)p}(y)\nabla_pK_m^{(j)}(y)-K^{(j)p}(y)\nabla_pK^{(i)}_m(y)\right)\nonumber\\
&=&\sum_i\partial_{\mu}\epsilon^{(i)}(x)K^{(i)}_m(y)+\sum_{i,j}\epsilon^{(i)}(x)A_{\mu}^{(j)}(x)[K^{(i)}(y),K^{(j)}(y)]_m\nonumber\\
&=&\sum_i\partial_{\mu}\epsilon^{(i)}(x)K^{(i)}_m(y)+\sum_{i,j,k}\epsilon^{(i)}(x)A_{\mu}^{(j)}(x)f^{ijk}K^{(k)}_m(y)\nonumber\\
&=&\sum_{i}\left
(\partial_{\mu}\epsilon^{(i)}(x)+\sum_{j,k}\epsilon^{(j)}(x)A^{(k)}_{\mu}(x)f^{jki}
\right)K^{(i)}_m(y),
\end{eqnarray}
where we used (\ref{Killingcond}) and the fact that
$[X,Y]_m=X^p\nabla_pY_m-Y^p\nabla_pX_m$. So we have
\begin{eqnarray}
\delta K^{(i)}_m(y)&=&0,\\
\delta
A^{(i)}_{\mu}&=&\partial_{\mu}\epsilon^{(i)}(x)+\sum_{j,k}f^{ijk}
\epsilon^{(j)}(x)A^{(k)}_{\mu}(x).
\end{eqnarray}
This is precisely the transformation law for a Yang-Mills field
with gauge group $G$.\\
We note at this point that the gauge group is determined not by
the topology of $M_k$ but by the metric $g_2$.\\
\\
Substitution of the ansatz $g_{\mu m}(x,y)=\sum_i
A^{(i)}_{\mu}(x)K^{(i)}_m(y)$ into the $d$-dimensional equations
of motion yields the Yang-Mills equations of $A^{(i)}_{\mu}(x)$.

\bigskip
$\bullet$ In most realistic Kaluza-Klein theories the massive
states are very heavy and cannot be observed. Therefore, only the
massless sector of the reduction is relevant. The hope is that
the states which are massless at tree-level acquire mass via
quantum effects.\\
Unfortunately neglecting the massive sector sometimes causes
problems. For example what appears to be massive in a false vacuum
might be massless in a true vacuum and vice-versa. So we really
need to make sure that $\langle g_{MN}\rangle$ and
$\langle\Phi\rangle$ describe the true ground state.

\begin{center}
{\bf The standard model from Kaluza-Klein theory}
\end{center}
We noted already that the gauge group which arises in four
dimensions after Kaluza-Klein reduction is determined by the
isometry group of the compact space $(M_k,g_2)$. If $M_k$ is a
homogeneous space $G/H$ this group can be found very easily to be
$G$. In fact, the maximum possible symmetry groups for manifolds
of a given
dimension always arise for homogeneous spaces.\\
It is interesting to determine the dimension of a compactification
space $M_k$ that gives the gauge group $Q:=SU(3)\times SU(2)\times
U(1)$ of the standard model. Suppose $Q$ acts transitively on
$M_k$ and let the little group of a point be a subgroup $Q_0$.
$Q_0$ should not contain all of one of the factors $SU(3)$,
$SU(2)$, $U(1)$ as if so this factor would leave $M_k$ invariant
and would no longer be a symmetry group. The maximal subgroup that
does not contain any of these factors is $SU(2)\times U(1)\times
U(1)=:Q_0$. Thus the smallest dimension of a space $M_k$ with
gauge group $Q$ is ${\rm dim}Q- {\rm dim}Q_0=12-5=7$. So this
simple consideration led us to Kaluza-Klein theory in 4+7
dimensions and therefore to eleven-dimensional supergravity.

\bigskip
After having collected a lot of material from both mathematics
and physics we are ready to tackle the problem we set out to
solve. We want to construct realistic theories from M-theory
compactifications. The obvious thing to do is to solve the
equations of motion of eleven-dimensional supergravity and to
look for realistic vacua. Yet, before we do so in chapter 5, we
want to present another formalism which is useful for various
reasons. We will consider the relation between M-theory and the
$E_8\times E_8$ heterotic string, which makes clear that M-theory
is not an independent theory but it is strongly related to string
theory. This duality enables us to reproduce many string theory
results from M-theory. In particular we know that the
compactification of string theories on a Calabi-Yau manifold
leads to a four-dimensional theory with $\mathcal{N}=1$
supersymmetry. We will see how we can reproduce these theories
coming from eleven dimensions. Many of the concepts that will
turn up in the calculations of the next chapter will be very
useful later on. Let us then turn to the conjectured duality
between M-theory and the heterotic string.

\chapter{M-theory on $\mathbb{R}^{10}\times S^1/\mathbb{Z}_2$}
After having presented some prerequisites from mathematics and
from physics we now turn to a first M-theory calculation. As we
mentioned already in the introduction the fundamental degrees of
freedom of M-theory are still unknown. What we do know, however,
is that the low energy effective field theory of M-theory is
eleven-dimensional supergravity. Furthermore, over the last years
various duality relations between different string theories and
supergravity in eleven dimensions have been established,
confirming the evidence for a single underlying theory. One of the
conjectured duality relations discovered by Ho\v{r}ava and Witten
relates M-theory on the orbifold $\mathbb{R}^{10}\times
S^1/\mathbb{Z}_2$ to $E_8\times E_8$ heterotic string theory. Here
we will study the basic features of this duality concentrating on
the occurring anomalies and their cancellation. The explicit
calculations involve eleven-dimensional supergravity on
$\mathbb{R}^{10}\times S^1/\mathbb{Z}_2$. We will analyze the
anomalies of this specific theory and show that they cancel if we
add a pair of $E_8$ vector multiplets in a certain way. This
procedure is viable as anomalies are an infrared effect. That is,
it is sufficient to show that the low energy effective theory is
free of anomalies to
make sure that the full theory is consistent.\\
The main references for this chapter are \cite{HW95}, \cite{HW96},
\cite{BDSa99} and \cite{BM03c}. Contrary to our usual convention
for indices which is given in the appendix, we take $M,N,\ldots
\in \lbrace0,1,\ldots ,10\rbrace$ and $\mu,\nu,\ldots
\in\lbrace0,1,\ldots 9\rbrace$ in this chapter.

\begin{center}
{\bf The orbifold $\mathbb{R}^{10}\times S^1/\mathbb{Z}_2$}
\end{center}
Let the eleven-dimensional manifold $M_{11}$ be the Riemannian
product of ten-dimensional Minkowski space
$(\mathbb{R}^{10},\eta)$ and a circle $S^1$ with its standard
metric. The coordinates on the circle are taken to be
$\phi\in[-\pi,\pi]$ with the two endpoints identified\footnote{Of
course to define proper coordinates on $S^1$ we need two
charts.}. In particular the radius of the circle will be taken to
be one. The equivalence classes in $S^1/\mathbb{Z}_2$ are the
pairs of points with coordinate $\phi$ and $-\phi$, i.e.
$\mathbb{Z}_2$ acts as $\phi\rightarrow-\phi$. This map has the
fixed points 0 and $\pi$, thus the space $\mathbb{R}^{10}\times
S^1/\mathbb{Z}_2$ contains two
singular ten-dimensional planes.\\
In section 3.1 we noticed that $\phi\rightarrow-\phi$ is a
symmetry of the action provided $C\rightarrow -C$. This implies\\
\parbox{14cm}{
\begin{eqnarray}
C_{\mu\nu\rho}&\rightarrow& -C_{\mu\nu\rho},\nonumber\\
C_{\mu\nu 10}&\rightarrow& C_{\mu\nu 10}.\nonumber
\end{eqnarray}}\hfill\parbox{8mm}{\begin{eqnarray}\label{Ccond}\end{eqnarray}}\\
As $\phi$ and $-\phi$ are identified under $\mathbb{Z}_2$ we find
that $C_{\mu\nu\rho}$ is projected out and $C$ can be written as $C=\tilde B\w d\phi$.\\
Following \cite{BDSa99} we define for further convenience\\
\parbox{14cm}{
\begin{eqnarray}
\delta_1&:=&\delta(\phi)d\phi,\nonumber\\
\delta_2&:=&\delta(\phi-\pi)d\phi,\nonumber\\
\epsilon_1(\phi)&:=&{\rm sig}(\phi)-{\phi\over\pi},\nonumber\\
\epsilon_2(\phi)&:=&\epsilon_1(\phi-\pi),\nonumber
\end{eqnarray}}\hfill\parbox{8mm}{\begin{eqnarray}\end{eqnarray}}
which satisfy
\begin{equation}
d\epsilon_i=2\delta_i-{d\phi\over\pi}.\label{depsilon}
\end{equation}
After regularization we get \cite{BDSa99}
\footnote{Of course $\delta_{ij}$ is the usual Kronecker symbol, not to be confused with $\delta_i$.}\\
\begin{equation}\label{reg}
\delta_i\epsilon_j\epsilon_k\rightarrow{1\over3}(\delta_{ji}\delta_{ki})\delta_i.
\end{equation}

\begin{center}
{\bf Anomalies of M-theory on $\mathbb{R}^{10}\times
S^1/\mathbb{Z}_2$}
\end{center}
Anomalies are a low energy effect and should be computable from
the low energy limit of M-theory. Hence, we must determine the
field content of eleven-dimensional supergravity on the given
orbifold. As long as we are on the bulk we get the usual fields
presented in section 3.1, subject to condition (\ref{Ccond}). On
the fixed planes, however, new effects occur and we need to
analyze how the supergravity fields behave on these planes. If we
compactify on a circle $S^1$ the eleven-dimensional
Rarita-Schwinger field reduces to a sum of infinitely many massive
modes, which are anomaly free and the massless chiral
ten-dimensional gravitino. This massless mode leads to a
gravitational anomaly in ten dimensions. However, in the present
case we do not have a circle but an orbifold with two
ten-dimensional fixed planes and it is not clear a priori how the
fields can be reduced in that case. What we do know, however, is
that the two planes fixed under $\mathbb{Z}_2$ are completely
symmetric, thus it is natural to assume that both should carry the
same anomaly. Furthermore, in the case in which the radius of
$S^1$ reduces to zero we get the usual ten-dimensional anomaly of
a massless gravitino field. We conclude that in the case of finite
radius the anomaly which is situated on the ten-dimensional planes
is given by exactly one half of the usual gravitational anomaly in
ten dimensions. Then, the anomaly on a single fixed plane is given
by the polynomial of these so-called {\em untwisted fields}
\begin{equation}
I_{12(i)}^{(untwisted)}={1\over2}\left\{{1\over2}\left(I_{grav}^{(3/2)}(R)-I_{grav}^{(1/2)}(R)\right)\right\}.\label{I12untwisted}
\end{equation}
The second factor of ${1\over2}$ arises because the fermions are
Majorana-Weyl. The $(-1)$ of the second term comes from the fact
that the spinor is anti-chiral and $i=1,2$ denotes the two planes.\\
So M-theory on $\mathbb{R}^{10}\times S^1/\mathbb{Z}_2$ is
anomalous and has to be modified in order to be a consistent
theory. An idea that has been very fruitful in string theory over
the last years is to introduce new fields which live on the
singularities of the space under consideration. Following this
general tack we introduce massless modes living only on the fixed
planes of our orbifold. These so-called {\em twisted fields} have
to be ten-dimensional vector multiplets because the vector
multiplet is the only ten-dimensional supermultiplet with all
spins $\leq 1$. In particular there will be gaugino fields
contributing to the pure and mixed gauge anomalies, as well as to
the gravitational ones. This gives an additional term in our
anomaly polynomial,
\begin{equation}
I_{12(i)}^{(twisted)}={1\over2}\left(n_iI_{grav}^{(1/2)}(R)+I_{mixed}^{(1/2)}(R,F_i)+I_{gauge}^{(1/2)}(F_i)\right).\label{I12twisted}
\end{equation}
Here $n_i$ is the dimension of the adjoint representation of the
gauge group $G_i$. Adding all the pieces gives
\begin{eqnarray}
I_{12(i)}^{(fields)}&=&I_{12(i)}^{(untwisted)}+I_{12(i)}^{(twisted)}\nonumber\\
&=&{1\over{2(2\pi)^5 6!}}\left[{{496-2n_i}\over1008}{\rm
tr}R^6+{{-224-2n_i}\over768}{\rm tr}R^4
{\rm tr}R^2 +{{320-10n_i}\over{9216}}({\rm tr}R^2)^3\right. \nonumber\\
&&+\left. {1\over{16}} {\rm tr}R^4 {\rm Tr}F_i^2 +{5\over64}({\rm
tr}R^2)^2 {\rm Tr}F_i^2-{5\over8}{\rm tr}R^2{\rm Tr}F_i^4+{\rm
Tr}F_i^6\right],
\end{eqnarray}
where now $\Tr$ denotes the adjoint trace. To derive this formula
we made use of the general form of the anomaly polynomial that was
given in chapter 3. The anomaly cancels only if several conditions
are met. First of all it is not possible to cancel the ${\rm
tr}R^6$ term by a Green-Schwarz type mechanism. Therefore, we get
a restriction on the gauge group $G_i$, namely
\begin{equation}
n_i=248.
\end{equation}
Then we are left with
\begin{eqnarray}
I_{12(i)}^{(fields)} &=&{1\over{2(2\pi)^5
6!}}\left[-{15\over16}{\rm tr}R^4 {\rm tr}R^2 -{15\over{64}}({\rm
tr}R^2)^3+{1\over{16}} {\rm tr}R^4
{\rm Tr}F_i^2\right. \nonumber\\
&&+\left. {5\over64}({\rm tr}R^2)^2 {\rm Tr}F_i^2-{5\over8}{\rm
tr}R^2{\rm Tr}F_i^4+{\rm Tr}F_i^6\right].
\end{eqnarray}
In order to cancel this remaining part of the anomaly we will
apply a sort of Green-Schwarz mechanism. This is possible if and
only if the anomaly polynomial factorizes into the product of a
four-form and an eight-form. For this factorization to occur we
need
\begin{equation}
{\rm Tr}F_i^6={1\over24}{\rm Tr}F_i^4{\rm
Tr}F_i^2-{1\over3600}({\rm Tr}F^2_i)^3.
\end{equation}
There is exactly one non-Abelian Lie group with this property,
which is the exceptional group $E_8$. Defining ${\rm
tr}:={1\over30}{\rm Tr}$ for $E_8$ and making use of the
identities
\begin{eqnarray}
{\rm Tr}F^2&=:&30\ {\rm tr}F^2,\\
{\rm Tr}F^4&=&{1\over100}({\rm Tr}F^2)^2,\\
{\rm Tr}F^6&=&{1\over7200}({\rm Tr}F^2)^3,
\end{eqnarray}
which can be shown to hold for $E_8$, we can see that the anomaly
factorizes,
\begin{equation}
I_{12(i)}^{(fields)}={\pi\over3}(I_{4(i)})^3+I_{4(i)}\wedge
X_8\label{I12i},
\end{equation}
with
\begin{eqnarray}
I_{4(i)}:={1\over
16\pi^2}\left({\rm tr}F_i^2-{1\over2}{\rm tr}R^2\right),\\
X_8:={1\over(2\pi)^34!}\left({1\over8}{\rm tr}R^4-{1\over32}({\rm
tr}R^2)^2\right)\label{X8}.
\end{eqnarray}
It is important to note at this point that $X_8$ differs from
$\widehat{X}_8$ given in (\ref{hatX}) even if the gauge fields of
the super-Yang-Mills theory of section 3.2.5 are set to zero.
$X_8$ is related to the forms $X_7$ and $X_6$ via the usual
descent mechanism
\begin{eqnarray}
X_8&=&dX_7,\\
\delta X_7&=&dX_6.
\end{eqnarray}

\begin{center}
{\bf The modified Bianchi identity}
\end{center}
So far we saw that M-theory on $S^1/\mathbb{Z}_2$ is anomalous
and we added new fields onto the fixed planes to cancel part of
that anomaly. But now the theory has changed. It no longer is
pure eleven-dimensional supergravity on a manifold with boundary
but we have to couple this theory to ten-dimensional
super-Yang-Mills theory, with the action
\begin{equation}
S_{SYM}=-{1\over4\lambda^2}\int d^{10}x \sqrt{g_{10}}\
F_{\mu\nu}F^{\mu\nu},
\end{equation}
with an unknown coupling constant $\l$. The explicit coupling of
these two theories was determined in \cite{HW96}. The crucial
result of this calculation is that the Bianchi identity needs to
be modified. It reads\footnote{This differs by a factor 2 from
\cite{HW96} which comes from the fact that our $\k_{11}$ is the
``downstairs" $\k$. \cite{HW96} use its ``upstairs" version and
the relation between the two is
$2\k_{downstairs}\equiv2\k_{11}=\k_{upstairs}$.}
\begin{eqnarray}
dG&=&-{2\kappa_{11}^2\over\lambda^2}\sum_i\delta_i\wedge\left({\rm tr}F^2_i-{1\over2}{\rm tr}R^2\right)\nonumber\\
&=&-(4\pi)^2{2\kappa_{11}^2\over\lambda^2}\sum_i\delta_i\wedge
I_{4(i)}.
\end{eqnarray}
Since $\delta_i$ has support only on the fixed planes and is a
one-form $\sim d\phi$, only the values of the smooth four-form
${I_4}_{(i)}$ on this fixed plane are relevant and only the
components not including $d\phi$ do not vanish. The gauge part
${\rm tr}F_i^2$ always satisfies these conditions but for the
${\rm tr}R^2$ term this is non-trivial. In the following a bar on
a form will indicate that all components containing $d\phi$ are
dropped and the argument is set to $\phi=\phi_i$. Then the
modified Bianchi identity reads
\begin{equation}
dG=\gamma\sum_i\delta_i\wedge \bar{I}_{4(i)},\label{BI}
\end{equation}
where we introduced
\begin{equation}
\gamma:=-(4\pi)^2{2\kappa_{11}^2\over\lambda^2}.\label{gamma}
\end{equation}
Define the Chern-Simons form
\begin{equation}
\bar{\omega}_i:= {1\over(4\pi)^2}\left({\rm tr}(A_i
dA_i+{2\over3}A_i^3)-{1\over2}{\rm tr}(\bar{\Omega}_i
d\bar{\Omega}_i +{2\over3}\bar{\Omega}_i^3 )\right),
\end{equation}
so that
\begin{equation}
d\bar{\omega}_i=\bar{I}_{4(i)}.
\end{equation}
Under a gauge and local Lorentz transformation with parameters
$\Lambda^g$ and $\Lambda^L$ independent of $\phi$ one has
\begin{equation}
\delta\bar{\omega}_i=d\bar{\omega}_i^1,
\end{equation}
where
\begin{equation}
\bar{\omega}_i^1:={1\over(4\pi)^2}\left({\rm tr}\Lambda^g
dA_i-{1\over2}{\rm tr}\Lambda^Ld\bar{\Omega_i}\right).
\end{equation}
Making use of (\ref{depsilon}) we find that the Bianchi identity
(\ref{BI}) is solved by
\begin{equation}
G=dC-(1-b)\gamma\sum_i\delta_i\wedge\bar{\omega_i}
+b\gamma\sum_i{\epsilon_i\over2}\bar{I}_{4(i)}
-b\gamma\sum_i{d\phi\over2\pi}\wedge\bar{\omega_i}.\label{G}
\end{equation}
As $G$ is a physical field it is taken to be gauge invariant,
$\delta G=0$. Hence we get the transformation law of the
$C$-field,
\begin{equation}
\delta
C=dB_2^1-\gamma\sum_i\delta_i\wedge\bar{\omega}_i^1-b\gamma\sum_i
{\epsilon_i\over2}d\bar{\omega}_i^1,
\end{equation}
with some two-form $B_2^1$. Recalling that $C_{\mu\nu\rho}$ is
projected out, this equation can be solved, as $C_{\mu\nu\rho}=0$
is only reasonable if we also have $\delta C_{\mu\nu\rho}=0$. This
gives
\begin{equation}
(dB_2^1)_{\mu\nu\rho}={b\gamma\over2}\sum_i(\epsilon_i
d\bar{\omega}_i^1)_{\mu\nu\rho},
\end{equation}
which is solved by
\begin{equation}
(B_2^1)_{\mu\nu}=\gamma{b\over2}\sum_i\epsilon_i(\bar{\omega}_i^1)_{\mu\nu}.
\end{equation}
So we choose
\begin{equation}
B_2^1=\gamma{b\over2}\sum_i\epsilon_i\bar{\omega}_i^1,
\end{equation}
and get
\begin{equation}
\delta
C=\gamma\sum_i\left[(b-1)\delta_i\wedge\bar{\omega}_i^1-{b\over2\pi}d\phi\wedge\bar{\omega}_i^1\right].\label{deltaC}
\end{equation}

\begin{center}
{\bf Inflow terms and anomaly cancellation}
\end{center}
In the last sections we saw that introducing a vector
supermultiplet cancels part of the gravitational anomaly that is
present on the ten-dimensional fixed planes. Furthermore the
modified Bianchi identity led to a very special transformation law
for the $C$-field. In this section we show that this modified
transformation law leads to an anomaly free theory.\\
Before we proceed let us introduce some nomenclature. So far we
worked on the space $\mathbb{R}^{10}\times S^1$ with an additional
$\mathbb{Z}_2$-projection imposed. This is what is called the
``upstairs" formalism. In particular when we wrote down the
modified Bianchi identity we assumed implicitly that we work in
this space that does not have boundaries. Equivalently, one might
work on the manifold $\mathbb{R}^{10}\times I$ with $I=[0,\pi]$.
This is referred to as the ``downstairs" approach. It is quite
intuitive to work downstairs on the interval but for calculational
purposes it is more convenient to work on manifolds without
boundary. Otherwise one would have to impose boundary conditions
for $G$ instead of our modified Bianchi identity. Starting from
supergravity on $\mathbb{R}^{10}\times I$ it is easy to obtain the
action in the upstairs formalism. Because of
$\int_I\ldots={1\over2}\int_{S^1}\ldots$ we have
\begin{equation}
S_{top}=-{1\over12\k^2_{11}}\int_{\mathbb{R}^{10}\times I} C\w
dC\w dC=-{1\over24\k^2_{11}}\int_{\mathbb{R}^{10}\times S^1} C\w
dC\w dC.
\end{equation}
However, we no longer have $G=dC$ and thus it is no longer clear
whether the correct Chern-Simons term is $CdCdC$ or rather $CGG$.
It turns out that the correct term is the one which maintains the
structure $\Ct d\Ct d\Ct$ everywhere except on the fixed planes.
However, it is not $C$ but $\Ct$, a modified version of $C$ which
is relevant.\footnote{This modification of the topological term
was motivated from considerations in \cite{FHMM98} which will be
reviewed in chapter 6. A similar modification was used in
\cite{BM03b} to cancel anomalies on singular $G_2$-manifolds, as
will be explained in chapter 8.} To be concrete let us study the
structure of $G$ in more detail. It is given by
\begin{equation}
G=d\left(C+{b\over2}\g\sum_i\e_i\bar\o_i\right)-\g\sum_i\delta_i\w\bar\o_i=:d\Ct-\g\sum_i\delta_i\w\bar\o_i.
\end{equation}
That is we have $G=d\Ct$ except on the fixed planes where we get
an additional contribution. Thus, in order to maintain the
structure of the Chern-Simons term almost everywhere we postulate
it to read
\begin{eqnarray}
\St_{top}&=&-{1\over24\k^2_{11}}\int_{\mathbb{R}^{10}\times S^1}
\Ct\w G\w
G\nonumber\\
&=&-{1\over24\k^2_{11}}\int_{\mathbb{R}^{10}\times S^1} \Ct\w
\left(d\Ct-\g\sum_i\delta_i\w\bar\o_i\right)\w
\left(d\Ct-\g\sum_i\delta_i\w\bar\o_i\right)\nonumber\\
&=&-{1\over24\k^2_{11}}\int_{\mathbb{R}^{10}\times S^1}\left(
\Ct\w d\Ct\w d\Ct-2\Ct\w d\Ct\w\g\sum_i\delta_i\w\bar\o_i\right).
\end{eqnarray}
To see that this is reasonable let us calculate its variation
under gauge transformations. From (\ref{deltaC}) we find that
\begin{equation}
\delta \Ct=d\left({\g
b\over2}\sum_i\e_i\bar\o_i^1\right)-\gamma\sum_i\delta_i\w\bar\omega_i^1,
\end{equation}
and therefore
\begin{eqnarray}
\delta
\St_{top}&=&-{1\over24\kappa_{11}^2}\int_{\mathbb{R}^{10}\times
S^1}\delta \Ct \wedge G
\wedge G\nonumber\\
&=&-{1\over24\kappa_{11}^2}\int_{\mathbb{R}^{10}\times
S^1}\left[d\left({\g
b\over2}\sum_i\e_i\bar\o_i^1\right)-\gamma\sum_i\delta_i\w\bar\omega_i^1\right]
\wedge G
\wedge G\nonumber\\
&=&-{1\over24\kappa_{11}^2}\int_{\mathbb{R}^{10}\times
S^1}\left[-\gamma\sum_i\delta_i\w\bar\omega_i^1\w d\Ct\w d\Ct\right.\nonumber\\
&&\left.+d\left({\g
b\over2}\sum_i\e_i\bar\o_i^1\right)\w2d\Ct\w\left(-\g\sum_k\delta_k\w\bar\o_k\right)\right]\nonumber\\
&=&-{1\over24\kappa_{11}^2}\int_{\mathbb{R}^{10}\times
S^1}\left[-\gamma\sum_i\delta_i\w\bar\omega_i^1\w \left({\gamma b\over2}\sum_j\e_j\bar I_{4(j)}\right)\w \left({\gamma b\over2}\sum_k\e_k\bar I_{4(k)}\right)\right.\nonumber\\
&&\left.+d\left({\g
b\over2}\sum_i\e_i\bar\o_i^1\right)\w2\left({\gamma
b\over2}\sum_j\e_j\bar I_{4(j)}\right)\w\left(-\g\sum_k\delta_k\w\bar\o_k\right)\right]=\nonumber\\
&=&{\g^3b^2\over96\k_{11}^2}\int_{\mathbb{R}^{10}\times
S^1}\sum_{i,j,k}(\delta_i\e_j\e_k+2\e_i\e_j\delta_k)\bar\o_i^1\w\bar
I_{4(j)}\w\bar I_{4(k)}.
\end{eqnarray}
Using (\ref{reg}) we find
\begin{equation}
\delta\St={\g^3b^2\over96\k^2_{11}}\sum_i\int_{}\bar\o_i^1\w \bar
I_{4(i)}\w \bar I_{4(i)}
\end{equation}
which corresponds to the anomaly polynomial
\begin{equation}
I_{12}^{(top)}=\sum_i{\g^3b^2\over96\k^2_{11}}(\bar
I_{4(i)})^3=:\sum_iI_{12(i)}^{(top)}.\label{I12top}
\end{equation}
If we choose $\g$ to be
\begin{equation}\label{g1}
\g=-\left(32\pi\k^2_{11}\over b^2\right)^{1/3}
\end{equation}
this cancels the first part of the anomaly (\ref{I12i}) through
inflow, as described in chapter 3. Note that this amounts to
specifying a certain choice for the coupling constant $\l$.\\

As for the second part of the anomaly we need to introduce yet
another term which is a higher order correction to the
supergravity action. This so called {\em Green-Schwarz term} reads
\cite{VW95, DLM95, BM03c}
\begin{equation}
S_{GS}:=-{1\over(4\pi\k^2_{11})^{1/3}}\int_{\mathbb{R}^{10}\times
I} G\wedge
X_7=-{1\over2(4\pi\k^2_{11})^{1/3}}\int_{\mathbb{R}^{10}\times
S^1} G\wedge X_7,\label{Green-Schwarz term}
\end{equation}
where $dX_7=X_8$ and $X_8$ is given in (\ref{X8}). This term is
not present in classical supergravity and can be regarded as a
first quantum correction. It is necessary to establish anomaly
cancellation not only in Ho\v{r}ava-Witten theory but also in
M-theory in the presence of an M5-brane, which will be analysed in
detail in chapter 6. The variation of the Green-Schwarz term gives
the final contribution to our anomaly,
\begin{eqnarray}
\delta
S_{GS}&=&-{1\over2(4\pi\k^2_{11})^{1/3}}\int_{\mathbb{R}^{10}\times
S^1} G\wedge \delta
X_7=-{1\over2(4\pi\k^2_{11})^{1/3}}\int_{\mathbb{R}^{10}\times
S^1} G\wedge
dX_6^1\nonumber\\
&=&{1\over2(4\pi\k^2_{11})^{1/3}}\int_{\mathbb{R}^{10}\times S^1}
dG\wedge X_6^1
={\g\over2(4\pi\k^2_{11})^{1/3}}\int_{\mathbb{R}^{10}\times S^1}
\sum_i\delta_i\wedge \bar I_{4(i)}\wedge
X_6\nonumber\\
&=&\sum_i{\g\over2(4\pi\k^2_{11})^{1/3}}\int_{\mathbb{R}^{10}}
\bar I_{4(i)}\wedge \bar X_6.
\end{eqnarray}
Here we used the descent equations $X_8=dX_7$ and $\delta
X_7=dX_6^1$. The corresponding anomaly polynomial is
\begin{equation}
I_{12}^{(GS)}=\sum_i {\g\over2(4\pi\k^2_{11})^{1/3}}\ \bar
I_{4(i)}\wedge \bar X_8=:\sum_iI_{12(i)}^{(GS)}.\label{I12GS}
\end{equation}
which cancels the second part of our anomaly if
\begin{equation}\label{g2}
\g=-(32\pi\k^2_{11})^{1/3}.
\end{equation}
Happily, this is consistent with our first condition for anomaly
cancellation and selects
\begin{equation}
b=1.
\end{equation}
This value for $b$ was suggested in \cite{BDSa99} from general
considerations unrelated to anomaly cancellation.\\
Choosing $\g$ (and thus the corresponding value for $\l$) as in
(\ref{g2}) leads to a local cancellation of the anomalies. Indeed
let us collect all the contributions to the anomaly of a single
fixed plane, namely (\ref{I12i}), (\ref{I12top}) and (\ref{I12GS})
\begin{equation}
I_{12(i)}=I_{12(i)}^{(untwisted)}+I_{12(i)}^{(twisted)}+I_{12(i)}^{(top)}+I_{12(i)}^{(GS)}=0.
\end{equation}

\bigskip
We succeeded in constructing an anomaly free theory on the
orbifold $\mathbb{R}^{10}\times S^1/\mathbb{Z}_2$. Note in
particular, that the anomalies cancelled on each of the two
ten-dimensional fixed planes separately. We are thus led to the
conjecture that anomalies always have to cancel {\em locally}.
The concepts of local and
global anomaly cancellation will be very important later on.\\
Much more could be said about Ho\v{r}ava-Witten theory. Its most
interesting feature is the fact that it builds a link between
M-theory and the $E_8\times E_8$ heterotic string theory. This
can be found in the limit when the radius of $S^1$ goes to zero.
It is conjectured that M-theory on $\mathbb{R}^{10}\times
S^1/\mathbb{Z}_2$ is the strong coupling limit of this heterotic
string theory.\\
Furthermore, we want to take up a comment made in the
introduction. Having obtained the heterotic string invites us to
go even further, namely to compactify this theory on a Calabi-Yau
threefold. Thus we would be led to a four-dimensional theory with
$\mathcal{N}=1$ supersymmetry.  This sort of compactification of
ten-dimensional string theory has been studied extensively
\cite{CHSW85} before the advent of M-theory. In what follows we
adopt a different method to obtain four-dimensional supersymmetric
theories by compactifying directly on seven-dimensional spaces.\\
We will not pursue these ideas any further but it will be useful
to keep in mind the basic techniques and features of
Ho\v{r}ava-Witten theory which will be needed in the following.

\chapter{M-Theory on $G_2$-Manifolds}
After having given a first M-theory calculation in chapter 4 we
now turn to our main interest, the compactification of M-theory on
spaces which lead to four-dimensional supersymmetric field
theories. In section 5.1 we study a number of M-theory vacua by
solving the equations of motion of eleven-dimensional
supergravity. Section 5.2 describes how supersymmetry comes into
the game and how the condition of supersymmetry of the vacuum
leads to additional constraints. We show that the direct product
of Minkowski space and a compact manifold $X$ preserves
$\mathcal{N}=1$ supersymmetry provided $X$ is a $G_2$-manifold.
In section 5.3 we determine the field content by performing an
explicit Kaluza-Klein reduction of supergravity on a smooth
$G_2$-manifold.

\section{Vacua of Eleven-dimensional Supergravity}
In this section we describe various solutions of the equations of
motion of eleven-dimensional supergravity. Each solution is a
possible vacuum and might therefore describe our world.
Supergravity is defined on a manifold\footnote{Sometimes $M_{11}$
is a manifold with boundary, see e.g. \cite{HW95}, \cite{HW96}.}
$M_{11}$ which we will often take to be a product manifold
$M_{11}:=M_4\times M_7$. The motivation is, of course, that we
want to interpret $M_4$ as the four-dimensional space we perceive.
$M_{11}$ is supposed to carry a pseudo-Riemannian metric $g$ of
signature ($-+\ldots +$). We only consider the Levi-Civita
connection on the tangent bundle $TM_{11}$ and induced connections
on various other bundles. Note, however, that the fact that
$M_{11}$ is a product manifold does not imply that is is a {\em
Riemannian product}, i.e. that $g$ is a product metric
$g=g_1\times g_2$ (although this will actually be the case for
most of the solutions that we will study). Even if $M_{11}$ is a
Riemannian product it is not clear that the metric $g_1$ on $M_4$
will have signature ($-+++$) and $g_2$ on $M_7$ signature (+\ldots
+). In fact there is the solution $AdS_7\times S^4$ with
non-definite signature in the seven-dimensional space.

\bigskip
{\bf Definition 5.1}\\
A {\em vacuum}\footnote{Sometimes a vacuum is referred to as a
{\em background}. If $\langle\psi_M\rangle=0$ we speak of a
bosonic background.} is a tuple $(M,\langle g\rangle, \langle
C\rangle, \langle\psi\rangle)$, such that $\langle g\rangle,
\langle C\rangle$ and $\langle\psi\rangle$ satisfy the equations
of motion of eleven-dimensional supergravity\footnote{Of course
this is a classical concept. It is possible to avoid classical
statements, see \cite{CR84}.}.

\subsection{Ricci-flat Solutions}
The simplest solutions of the equations of motion can be found by
imposing the following conditions.

\bigskip
$\bullet$ $(M_{11},g)$ is a Riemannian product
$(M_{11},g):=(M_4\times
M_7,g_1\times g_2)$. This implies\\
\parbox{14cm}{
\begin{eqnarray}
\langle g_{\mu m}(x,y)\rangle&=&0,\nonumber\\
\langle g_{\mu\nu}(x,y)\rangle&=&\langle g_{\mu\nu}(x)\rangle, \ \mbox{and}\nonumber\\
\langle g_{mn}(x,y)\rangle&=&\langle\nonumber
g_{mn}(y)\rangle\label{Riemannian product}.
\end{eqnarray}}\hfill\parbox{8mm}{\begin{eqnarray}\end{eqnarray}}\\
In particular we do not have a warp factor.\\
It is useful to note at this point that because of
(\ref{Riemannian product}) the mixed components of the Christoffel
symbols, as $\Gamma^{\mu}_{m\nu}$, vanish. This implies that
$\mathcal{R}_{\mu\nu}:=R^M_{\mu M\nu}=R^{\rho}_{\mu\rho\nu}$ is
the Ricci tensor of $(M_4,g_1)$ and does not depend on $g_2$.
Similarly $\mathcal{R}_{mn}$ is the Ricci tensor of $(M_2,g_2)$.

\bigskip
$\bullet$ $(M_4,g_1)$ is taken to be maximally symmetric.

\bigskip
$\bullet$ As we want to include spinors into our theory
$(M_{11},g)$ needs to be a spin manifold. Because of the maximal
symmetry, $(M_4,g_1)$ is spin and this implies that $(M_7,g_2)$
needs to be spin as well.

\bigskip
$\bullet$ The vacuum is supposed to be invariant under $SO(3,1)$
in $M_4$, yielding\\
\parbox{14cm}{
\begin{eqnarray}
\langle\psi_M\rangle&=&0,\nonumber\\
\langle G_{\mu mnp}\rangle&=&\langle G_{\mu\nu mn} \rangle=\langle
G_{\mu\nu\rho m}\rangle=0,\nonumber\\
\langle
G_{\mu\nu\rho\sigma}\rangle&=&f\sqrt{g_1}\ \widetilde{\epsilon}_{\mu\nu\rho\sigma},\nonumber\\
\langle G_{mnpq}\rangle &\ &\mbox{arbitrary}.\nonumber
\end{eqnarray}}\hfill\parbox{8mm}{\begin{eqnarray}\end{eqnarray}}
Together with the fact that $G$ is closed, $dG=0$, i.e.
$\partial_{[M}G_{NPQR]}=0$ we get the stronger conditions
$\partial_{m}G_{\mu\nu\rho\sigma}(x,y)=0$ and
$\partial_{\mu}G_{mnpq}(x,y)=0$. Thus we get\footnote{Note that
this does not only hold for the ground state.}
\parbox{14cm}{
\begin{eqnarray}
\langle
G_{\mu\nu\rho\sigma}(x,y)\rangle&=&\langle G_{\mu\nu\rho\sigma}(x)\rangle,\nonumber\\
\langle G_{mnpq}(x,y)\rangle&=&\langle
G_{mnpq}(y)\rangle\nonumber.
\end{eqnarray}}\hfill\parbox{8mm}{\begin{eqnarray}\end{eqnarray}}

\bigskip
$\bullet$ Furthermore, in the case of Ricci-flat solutions we
want to constrain the $C$-field even further by imposing $\langle
C\rangle=0$\footnote{This is only possible on manifolds with
$\lambda(M_{11}):={{p_1(M_{11})}\over2}$, s.t.
${{\lambda}\over2}\in \mathbb{Z}$ \cite{Wi96}. This is always the
case for manifolds $(\mathbb{R}^4\times M_7,\eta\times g_2)$ where
$(M_7,g_2)$ is a manifold of holonomy $G_2$ \cite{HM99}.}. After
what we said before it is clear that this final condition amounts
to saying that $\langle G_{mnpg}\rangle$ and $f$ have to vanish,
as well. We will relax the latter condition for non-Ricci-flat
solutions later on.

Having imposed all these conditions the equation of motion (\ref{beom2}) is satisfied trivially and (\ref{beom1}) reduces to\\
\parbox{14cm}{
\begin{eqnarray}
\mathcal{R}_{\mu\nu}&=&0,\nonumber\\
\mathcal{R}_{mn}&=&0\nonumber\label{Ricci-flat}.
\end{eqnarray}}\hfill\parbox{8mm}{\begin{eqnarray}\end{eqnarray}}\\
So both $(M_4,g_1)$ and $(M_7,g_2)$ have to be Ricci-flat.\\
From the requirement of maximal symmetry we know that $(M_4,g_1)$
is either $dS$, $AdS$ or Minkowski space. As only Minkowski space
is Ricci-flat we conclude $(M_4,g_1)=(\mathbb{R}^4,\eta)$. The
most trivial solution of the equations of motion is
eleven-dimensional Minkowski space\footnote{The first line specifies the topology.},\\
\parbox{14cm}{
\begin{eqnarray}
M_{11}&=&\mathbb{R}^{11}\nonumber\\
\langle\psi\rangle&=&0,\nonumber\\
\langle C\rangle&=&0,\nonumber\\
\langle g_{MN}\rangle&=&\eta_{MN}\nonumber.
\end{eqnarray}}\hfill\parbox{8mm}{\begin{eqnarray}\end{eqnarray}}
This is not very interesting, of course, as we certainly do not
observe additional macroscopic dimensions. However, it is easy to
find other solutions satisfying (\ref{Ricci-flat}). In particular
any $(M_4\times M_7,\eta^{(4)}\times g_2)$ with $M_7$ compact and
$(M_7,g_2)$ Ricci-flat is a possible vacuum. Seven-dimensional
compact Ricci-flat metrics exist and as we are free to take the
volume of $M_7$ to be small we found a vacuum that might be
relevant for our real world. Supersymmetry considerations will
give us a much deeper insight into the structure of these
solutions in the next section. Known examples of this type are
the solutions $M_7=T^7$ \cite{CJ79} and $M_7=K3\times T^3$
\cite{DNP83} as well as compact manifolds with $G_2$-holonomy,
which will be studied in detail below.\\
At this point we also need to come back to Ho\v{r}ava-Witten
theory and check whether the vacuum that was chosen in that case
is a solution of the equations of motion. But this can be seen
easily as we have neither $G$-flux nor a fermion background and
$\mathbb{R}^6\times S^1/\mathbb{Z}_2$ clearly is Ricci-flat.
Another interesting solution is $\mathbb{R}^{10}\times S^1$, which
gives the relation between M-theory and type IIA string
theory\footnote{The compactification of eleven-dimensional
supergravity on $S^1$ leads to type IIA supergravity in ten
dimensions. It is conjectured that M-theory on $S^1$ is the strong
coupling limit of IIA string theory.}. Finally $CY_3\times S^1$
and $CY_3\times S^1/\mathbb{Z}_2$, with $CY_3$ a Calabi-Yau
three-fold, are Ricci-flat. M-theory on these spaces is
conjectured to be the strong coupling limit of type IIA or
$E_8\times E_8$ heterotic string theory on $CY_3$.

\subsection{The Freund-Rubin Solutions}
In order to find the Ricci-flat solutions we had to impose rather
severe constraints on the vacuum. In this section we want to find
more general solutions \cite{FR80}. The conditions we impose are
the following.

\bigskip
$\bullet$ As above we want $(M_{11},g)$ to be a Riemannian
product.

\bigskip
$\bullet$ The four-dimensional theory is taken to be maximally
symmetric.

\bigskip
$\bullet$ $(M_{11},g)$ is spin.

\bigskip
$\bullet$ The vacuum is taken to be invariant under $SO(3,1)$.

\bigskip
That is, we impose the same constraints as in the Ricci flat case
except that we no longer demand vanishing $G$-flux.\\

\bigskip
The Freund-Rubin ansatz is\\
\parbox{14cm}{
\begin{eqnarray}
\langle G_{\mu\nu\rho\sigma}(x)\rangle&:=&f\sqrt{\langle g_1\rangle}\ \widetilde{\epsilon}_{\mu\nu\rho\sigma},\nonumber\\
\langle G_{mnpq}(y)\rangle&:=&0,\nonumber
\end{eqnarray}}\hfill\parbox{8mm}{\begin{eqnarray}\label{FR}\end{eqnarray}}
where f is a real constant. Then the field equation for $C$ reads
\begin{equation}
\partial_{\mu}\left(\sqrt{\langle g\rangle}\ f{1\over{\sqrt{\langle g_1\rangle}}}\widetilde{\epsilon}^{\ \mu\nu\rho\sigma}\right)=f\widetilde{\epsilon}^{\ \mu\nu\rho\sigma}\partial_{\mu}\sqrt{\langle
g_2(y)\rangle}=0,
\end{equation}
and hence is trivially satisfied. To check Einstein's equations
it is convenient to rewrite them as\footnote{We omit the brackets
$\langle\ \rangle$ in equations (\ref{Einstein2}) - (\ref{FR7}). }
\begin{equation}
\mathcal{R}_{MN}=T_{MN}-{1\over 9}g_{MN}T^K_{\ \
K}=:\widetilde{T}_{MN}.\label{Einstein2}
\end{equation}
The ansatz (\ref{FR}) gives
\begin{eqnarray}
T_{\mu\nu}&=&-{1\over4}f^2g_{\mu\nu},\\
T_{mn}&=&{1\over 4}f^2 g_{mn},\\
T^K_{\ \ K}&=&{3\over4}f^2,
\end{eqnarray}
and we find
\begin{eqnarray}
\mathcal{R}_{\mu\nu}&=&-{1\over3}f^2g_{\mu\nu},\label{FR4}\\
\mathcal{R}_{mn}&=&{1\over6}f^2g_{mn}. \label{FR7}
\end{eqnarray}
So we find solutions with non-vanishing G-flux provided
$(M_4,g_1)$ is anti-de Sitter space and $(M_7,g_2)$ is a
seven-dimensional Einstein space. Note that (\ref{FR7}) implies
that $M_7$ is compact, so we get what is called {\em spontaneous
compactification}. In a sense this is very satisfying, as we do
not have to impose the condition that $M_7$ is compact, but it is
an output of the theory. However, the problem of the Freund-Rubin
solution lies somewhere else. If we consider (\ref{FR4}) and
(\ref{FR7}) we see that the curvature of the two Einstein spaces
is of the same magnitude. This means that if we want to have a
tiny compact space which is highly curved, the curvature of $AdS$
has to be large as well.\\
A particular solution is, of course, $AdS_4\times S^7$. To
summarize, the Freund-Rubin solution reads\footnote{Recall that the topology of $AdS_4$ is $S^1\times \mathbb{R}^3$.}\\
\parbox{14cm}{
\begin{eqnarray}
M_{11}&=&S^1\times\mathbb{R}^{3}\times M_7, \ \ M_7\ \mbox{compact}\nonumber\\
\langle\psi\rangle&=&0,\nonumber\\
\langle g_1\rangle&=& g(AdS_4)\nonumber\\
\langle g_2\rangle&\ \ & \mbox{Einstein, s.t. $\mathcal{R}_{mn}={1\over6}f^2\langle {g_2}_{mn}\rangle$}\nonumber\\
\langle G_{\mu\nu\rho\sigma}\rangle&=&f\sqrt{\langle g_1\rangle}\
\widetilde{\epsilon}_{\mu\nu\rho\sigma}\nonumber.
\end{eqnarray}}\hfill\parbox{8mm}{\begin{eqnarray}\end{eqnarray}}
Similarly one can show that there is yet another solution, where
$M_{11}$ has the structure $AdS_7\times S^4$ \cite{FR80}.

\subsection{M-branes}
In 1990 and 1991  two other solution of the equations of motion
were found, namely the {\em $M2$-brane} \cite{DS90} and the {\em
$M5$-brane} \cite{G91}. We will analyze the $M2$-brane in detail,
the M5-brane solution can be found analogously (see \cite{BM03c}
for some details).
For reviews see \cite{Du99}, \cite{St98}.\\
To motivate the $M2$-brane solution consider Maxwell's theory.
There the $A$ field couples to a one-dimensional world-line
$\gamma$ that is swept out by an electron via $\int_{\gamma} A$.
In eleven-dimensional supergravity the three-form $C$ can couple
to a (2+1)-dimensional object, the M2-brane, $\int_{M2}C$. So we
are looking for solutions of the full equations of motion with
symmetry $P(2,1)\times G$, where
$P(2,1)$ is the Poincaré group in (2+1) dimensions and $G$ is
the symmetry group of the space transverse to the brane.\\
A possible ansatz for such a solution is\footnote{For a more
general version of this ansatz see \cite{St98}.}
\begin{eqnarray}
\langle g\rangle&=&\left(
\begin{array}{cccc}
        -H(r)^{-{2\over3}} &             0          & 0          &  0 \\
        0                     & H(r)^{-{2\over3}}\opone_2  &  0 &0\\
        0                     &             0          &
        H(r)^{1\over3}&0\\
        0                    &              0          &0&r^2H(r)^{1\over3}g^{(7)}
\end{array}\right),\\
\langle
C_{012}\rangle&=&H(r)^{-1},\\
\langle \psi_M\rangle&=&0.
\end{eqnarray}
Here $\lbrace x^0,x^1,x^2\rbrace$ denote coordinates on the
$M2$-brane, $r$ measures essentially the distance to the brane and
$\lbrace y^1,\ldots y^7\rbrace$ are coordinates on a
seven-manifold $X$ with metric $g^{(7)}$. As always we want
$(M_{11},g)$ to be a spin manifold. The line element reads
\begin{equation}
ds^2=H(r)^{-{2\over3}}(-(dx^0)^2+(dx^1)^2+(dx^2)^2)+H(r)^{1\over3}(dr^2+r^2ds_7^2).
\end{equation}
This ansatz can be visualized as a 2+1 dimensional object embedded
in eleven-dimensional space-time. Now we check whether this
solution really satisfies (\ref{beom1}) and \ref{beom2}. In this
subsection we adopt the index conventions $M,N,\ldots
\in\lbrace0,\ldots 10\rbrace$, $\hat M,\hat N,\ldots \in\lbrace
0,\ldots 10\rbrace$ (tangent space), $\mu,\nu,\ldots \in\lbrace
0,\ldots 3\rbrace$, $\hat {\mu},\hat{\nu},\ldots \in\lbrace
0,\ldots 3\rbrace$ (tangent space), $a,b,\ldots \in\lbrace
0,\ldots 2\rbrace$, $\widehat{a},\widehat{b},\ldots \in\lbrace
0,\ldots 2\rbrace$ (tangent space), $m,n,\ldots \in\lbrace
4,\ldots 10\rbrace$, $\hat m,\hat n,\ldots \in\lbrace 4,\ldots
10\rbrace$ (tangent space), $dx^3:=dr$. That is indices without a
hat are indices in a coordinate basis, indices with hat are the
corresponding tangent space indices. We will also introduce a hat
on numbers in tangent space. This is necessary as we need to tell
apart $\gamma^0$ from $\gamma^{\hat 0}$, etc. These index
conventions are used in sections 5.1.3 and 5.2.3 only.
\begin{center}
{\bf Field equation for C}
\end{center}
To see that the field equation for the C field is satisfied we
note that $G_{0123}={H^{\prime}\over H^2}$, $G\wedge G=0$ and
$G^{0123}=-H^{-{1\over3}}H^{\prime}$. The field equation reads
\begin{equation}
\partial_M\left(\sqrt{g}G^{MNPQ}\right)=0,
\end{equation}
which gives
\begin{equation}
H^{\prime\prime}r^7+7H^{\prime}r^6=0.\label{cond0}
\end{equation}
This is solved by
\begin{equation}
H(r)=H_{\infty}+{k\over{r^6}}.
\end{equation}
Now we need to check whether this is consistent with Einstein's
equations.

\begin{center}
{\bf Einstein equation}
\end{center}
We make use of the form (\ref{Einstein2}) of Einstein's equations.
From (\ref{beom1}) we know that
\begin{equation}
T_{MN}={1\over 12}\left[G_{MNPQ}G_N^{\ \
PQR}-{1\over8}g_{MN}G_{PQRS}G^{PQRS}\right].
\end{equation}
With
\begin{eqnarray}
G_{PQRS}G^{PQRS}&=&-24\left({{H^{\prime}}\over
H}\right)^2H^{-{1\over3}},\\
G_{\mu PQR}G_{\nu}^{\ \ PQR}&=&-6\left({{H^{\prime}}\over H}
\right)^2H^{-{1\over3}}g_{\mu\nu},
\end{eqnarray}
we obtain\\
\parbox{14cm}{
\begin{eqnarray}
\widetilde{T}_{mn}&=&{1\over6}\left({{H^{\prime}}\over H
}\right)^2 H^{-{1\over3}}g_{mn},\nonumber\\
\widetilde{T}_{m\mu}&=&0,\nonumber\\
\widetilde{T}_{\mu\nu}&=&-{1\over3}\left({{H^{\prime}}\over H
}\right)^2 H^{-{1\over3}}g_{\mu\nu}.\nonumber
\end{eqnarray}}\hfill\parbox{8mm}{\begin{eqnarray}\label{M2Energy}\end{eqnarray}}\\
In order to evaluate the curvature we use the vielbein formalism
and write\footnote{A hat on numbers indicates that they are
indices in tangent space.}
\begin{eqnarray}
ds^2&=&\sum_{\hat{\mu}=\hat 0}^{\hat 3}e^{\hat \mu}\otimes
e^{\hat\mu}+\sum_{\hat m=\hat 4}^{\hat{10}}e^{\hat m}\otimes
e^{\hat m}\nonumber\\
&=&\sum_{\hat \mu=\hat 0}^{\hat 3}e^{\hat\mu}\otimes
e^{\hat\mu}+\sum_{\hat m=\hat 4}^{\hat{10}}(H^{1\over6}r
\widetilde{e}^{\ \hat m})\otimes (H^{1\over6}r\widetilde{e}^{\
\hat m}),
\end{eqnarray}
where $ds_7^2=\sum_{\hat m}\widetilde{e}^{\hat
m}\otimes\widetilde{e}^{\hat m}$ is the
metric on the seven-dimensional space. We read off\\
\parbox{14cm}{
\begin{eqnarray}
e^{\hat{a}}&=& H^{-{1\over3}}\delta^{\hat a}_{b}dx^{b},\nonumber\\
e^{\hat 3}&=& H^{{1\over6}}dr,\nonumber\\
e^{\hat m}&=&H^{1\over6}r\widetilde{e}^{\ \hat m}.\nonumber
\end{eqnarray}}\hfill\parbox{8mm}{\begin{eqnarray}\label{vielbeins}\end{eqnarray}}
Using the Maurer-Cartan structure equation (\ref{MC1}) for
vanishing torsion, we get the non-vanishing connection
coefficients\\
\parbox{14cm}{
\begin{eqnarray}
\omega^{\hat{a}}_{\ \
\hat3}&=&-{1\over3}H^{-{7\over6}}H^{\prime}e^{\hat{a}}=-{1\over3}H^{-{3\over2}}H^{\prime}\delta^{\hat a}_{b}dx^{b},\nonumber\\
\omega^{\hat3}_{\ \
\hat m}&=&-\left({H^{\prime} \over{6H}}+{1\over r}\right)H^{-{1\over6}}e_{\hat m}=-\left({{H^{\prime}r} \over{6H}}+1\right)\widetilde{e}_{\hat m},\nonumber\\
\omega^{\hat m}_{\ \ \hat n}&=&\widetilde{\omega}^{\hat m}_{\ \
\hat n}.\nonumber
\end{eqnarray}}\hfill\parbox{8mm}{\begin{eqnarray}\label{omegas}\end{eqnarray}}
Applying the second Maurer-Cartan equation (\ref{MC2}) gives the
curvature two-form
\parbox{14cm}{
\begin{eqnarray}
R^{\hat a}_{\
\hat b}&=&-{1\over 9}\left({{H^{\prime}}\over H}\right)^2H^{-{1\over3}}e^{\hat a}\wedge e_{\hat b},\nonumber\\
R^{\hat a}_{\
\hat3}&=&\left({{H''}\over{3H}}-{1\over2}\left({{H^{\prime}}\over
H}\right)^2\right)H^{-{1\over3}}e^{\hat a}\wedge e_{\hat
3},\nonumber\\
R^{\hat a}_{\ \hat m}&=&\left({1\over18}\left({{H^{\prime}}\over
H}\right)^2+{{H'}\over{3r
H}}\right)H^{-{1\over3}}e^{\hat a}\wedge e_{\hat m},\nonumber\\
R^{\hat 3}_{\ \hat3}&=&0,\nonumber\\
R^{\hat3}_{\ {\hat m}
}&=&-\left({{H''}\over{6H}}-{1\over6}\left({{H^{\prime}}\over
H}\right)^2+{{H^{\prime}}\over {6r H
}}\right)H^{-{1\over3}}e^{\hat3}\wedge e_{\hat m},\nonumber\\
R^{\hat m}_{\ \ \hat n}&=&\widetilde{R}^{\hat m}_{\ \ \hat n}
-\left({{H'}\over{6H}}+{1\over r }\right)^2H^{-{1\over3}} e^{\hat
m}\wedge e_{\hat n},\nonumber
\end{eqnarray}}\hfill\parbox{8mm}{\begin{eqnarray}\end{eqnarray}}
from which we finally derive the Ricci tensor,\\
\parbox{14cm}{
\begin{eqnarray}
\mathcal{R}_{\hat a}^{\ \hat b
}&=&\left({{H''}\over{3H}}+{{7H'}\over{3r H
}}-{1\over3}\left({{H'}\over H}\right)^2\right)H^{-{1\over3}}\delta_{\hat a}^{\hat b},\nonumber\\
\mathcal{R}_{\hat 3}^{\
\hat{3}}&=&\left(-{{H''}\over{6H}}-{{7H'}\over{6r H
}}-{1\over3}\left({{H'}\over H}\right)^2\right)H^{-{1\over3}},\nonumber\\
\mathcal{R}_{\hat m}^{\ \ \hat
n}&=&\left(-{{H''}\over{6H}}-{{7H'}\over{6r H
}}+{1\over6}\left({{H'}\over
H}\right)^2-{6\over{r^2}}+{3\over8}{{\lambda^2}\over{r^2}}\right)H^{-{1\over3}}\delta_{\hat
m}^{\hat n}.\nonumber
\end{eqnarray}}\hfill\parbox{8mm}{\begin{eqnarray}\label{M2Ricci}\end{eqnarray}}
Here we used the fact that the seven-dimensional space $X$ is an
Einstein space,
\begin{equation}
\widetilde{\mathcal{R}}_{\hat m}^{\ \ \hat
n}={3\over8}\lambda^2\delta_{\hat m}^{\hat n}.
\end{equation}
Comparing (\ref{M2Ricci}) to (\ref{M2Energy}) we get the
conditions
\begin{eqnarray}
H''+{{7H'}\over r}&=&0,\label{cond1a}\\
\lambda&=&4.\label{cond2}
\end{eqnarray}
(\ref{cond1a}) is in fact the same condition as (\ref{cond0}) and
we see that the field equation (\ref{cond2}) can be easily
satisfied by choosing the seven-dimensional cosmological constant
appropriately. Therefore, we proved that the M2-brane is a
possible vacuum of eleven-dimensional supergravity with the
following properties:
\begin{eqnarray}
\langle \psi_M\rangle&=&0\nonumber\\
\langle G_{0123}\rangle&=&{{H'}\over{H^2}}\nonumber\\
ds^2&=&\left(H_{\infty}+{k\over{r^6}}\right)^{-{2\over3}}(-(dx^0)^2+(dx^1)^2+(dx^2)^2)+\left(H_{\infty}+{k\over{r^6}}\right)^{1\over3}(dr^2+r^2ds_7^2)\nonumber\\
\widetilde{\mathcal{R}}_{\hat m}^{\ \ \hat n}&=&6\delta_{\hat
m}^{\hat n}\label{M2-brane}
\end{eqnarray}
Having found this vacuum some comments are in order. First of all
the metrics looks as if it had a singularity at $r=0$. However,
this is a mere coordinate singularity. In fact the metric does not
cover the entire space-time, but it can be extended \cite{St98}.
The extended metric contains a horizon together with a time like
singularity in its interior. This may be understood by adding a
$\delta$-function source term to the solution. Indeed, there
exists a membrane action \cite{BST87}
that can be coupled to $d=11$ supergravity.\\

\begin{figure}[ht]
\centering
\includegraphics[width=0.6\textwidth]{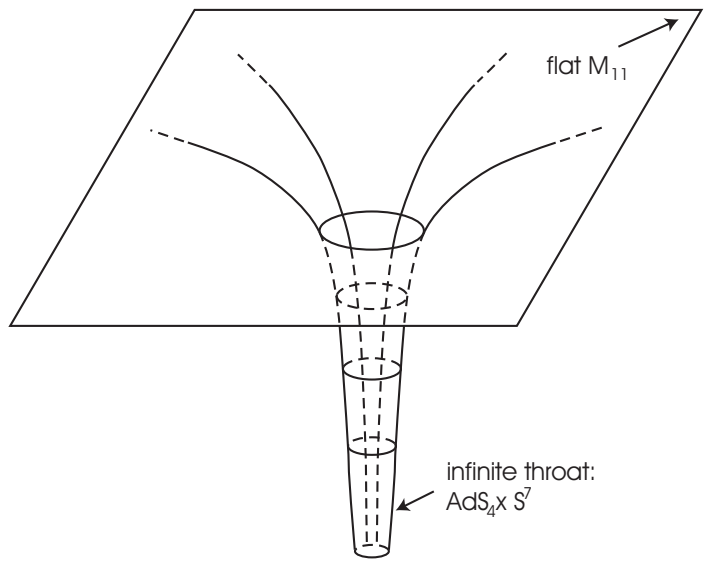}\\
\caption[]{The M2-brane interpolates between
$\mathcal{M}_{11}:=(\mathbb{R}^{11},\eta)$ and $AdS_4\times
S^7$.} \label{M2-interpol}
\end{figure}

Another interesting aspect of the $M2$-brane solution is its
interpolating character. It is obvious, that its limit for
$r\rightarrow\infty$ is eleven-dimensional Minkowski space,
provided $ds_7^2$ is the standard metric on the seven-sphere
$S^7$. In that case the metric tends to $AdS_4\times S^7$ for
$r\rightarrow 0$. So the membrane interpolates between two
different vacua as is shown in figure \ref{M2-interpol}. If $X$ is
not the seven-sphere we still have the limit $AdS_4\times X$ for
$r\rightarrow 0$, but for $r\rightarrow \infty$ we no longer
obtain Minkowski space. What we get is $\mathcal{M}_3\times
C(X)$, where $C(X)$ is the cone on $X$ and $\mathcal{M}_3$ is
three-dimensional Minkowski space.
These spaces will be analyzed in more detail below.\\
Recently a new type of solution was found \cite{CLP00} in which a
self-dual harmonic four-form was added to the $G$-flux of
(\ref{M2-brane}). These kinds of solutions no longer have
singularities. A variety of solutions of intersecting or wrapped
M-branes can be found in the literature.

\bigskip
Next to the solutions we studied so far there is also the so
called {\em $pp$-wave} solution \cite{H84} which has found
renewed interest recently.

\section{Supersymmetry in M-Theory - the Arisal of $G_2$}
After having discussed several solutions of the equations of
motion we now want to study whether these vacua preserve
supersymmetry.

\bigskip
{\bf Definition 5.2}\\
A vacuum $(M, \langle g\rangle, \langle C\rangle,
\langle\psi\rangle)$ is said to be {\em supersymmetric} if it
satisfies
\begin{eqnarray}
\delta e_{\ \ M}^A&:=&-{1\over2}\bar{\eta}\Gamma^A\psi_M\nonumber=0,\\
\delta C_{MNP}&:=&-{3\over2}\bar{\eta}\Gamma_{[MN}\psi_{P]}=0,\\
\delta\psi_M&:=&\widetilde{\nabla}^S_M(\hat{\omega})\eta=0\nonumber,
\end{eqnarray}
where the variations have to be calculated at the point $e^A_{\ \
M}=\langle e^A_{\ \ M}\rangle,\ C_{MNP}=\langle C_{MNP}\rangle,\
\psi_M=\langle\psi_M\rangle$. All the vacua we are going to study
have vanishing fermionic background, $\langle\psi_M\rangle=0$, so
the first two equations are trivially satisfied and the last one
reduces to
\begin{equation}
\widetilde{\nabla}^S_M(\omega)\eta=0,
\end{equation}
evaluated at $C_{MNP}=\langle C_{MNP}\rangle$, $e^A_{\ \
M}=\langle e^A_{\ \ M} \rangle$ and $\psi_M=0$. We see that
$e^A_{\ \ M}$ and $C_{MNP}$ are automatically invariant and we
find that the vacuum is supersymmetric if and only if there exists
a spinor $\eta$ s.t. $\forall M$
\begin{equation}
\nabla^S_M\eta-{1\over 288}\left(\Gamma_M^{\ \
PQRS}-8\delta_M^P\Gamma^{QRS}\right)G_{PQRS}\eta=0.\label{SUSYvac}
\end{equation}
We will now analyze the implications of this equation for the
vacua we studied above. See \cite{BDS01}, \cite{D02} and
\cite{DLPS95} for a discussion of some of the topics covered in
this section.

\subsection{Ricci-flat Solutions and $G_2$}
In this section we study the amount of supersymmetry that is
preserved by Ricci-flat solutions. As we saw above, these are
Riemannian products of Minkowski space and a seven-dimensional
space that is Ricci-flat. $G$-flux and the fermionic background
are taken to
vanish.\\
Given these conditions of the vacuum we get a further
simplification of (\ref{SUSYvac}), which now reads
\begin{equation}
\nabla^S\eta=0.\label{covcon}
\end{equation}
To be able to determine the number of supersymmetries in a given
vacuum we need to comment on supersymmetry in various dimensions.
A supersymmetric theory in $d$ dimensions is one in which the
Poincar\'{e} algebra is extended to a superalgebra by adding
spinorial generators $Q^I_{\alpha}$. $I$ is a label with
$I\in\{1,\ldots ,\mathcal{N}\}$, $\mathcal{N}\in \mathbb{N}$ and
$\mathcal{N}$ gives the number of supersymmetries of the theory.
For $\mathcal{N}=1$ we speak of unextended supersymmetry,
otherwise supersymmetry is said to be extended. The index $\alpha$
runs over the spin components of the spinor $Q^I$, but this means
that the range of $\alpha$ depends on the dimension $d$, because
the number of components of a spinor depends on $d$ (see appendix
B). At the moment we are particularly interested in the case
$d=11$ where a spinor has 32 components and the case $d=4$ with 4
component spinors. Suppose we start from a theory with
$\mathcal{N}=1$ in eleven dimensions and we compactify to four
dimensions. If the compactification is done in a way that all the
32 generators are generators of supersymmetry in the
four-dimensional theory we obtain $\mathcal{N}=8$. So we conclude
that starting from $\mathcal{N}=1$ in $d=11$ we can have
$0\leq\mathcal{N}\leq 8$ in $d=4$. We note in passing that there
are no consistent four-dimensional theories with $\mathcal{N}>8$
as this would lead to particles of spin bigger than two for which
no consistent field theories are known. Similarly, in eleven
dimensions
$\mathcal{N}=1$ is the only possibility.\\
Let us be more precise. A supersymmetric vacuum in eleven
dimensions is one that admits 32 linearly independent solutions of
equation (\ref{covcon}) corresponding to 32 generators of
supersymmetry. After compactification the original Poincar\'{e}
group $P(10,1)$ is broken to $P(3,1)\times P(7)$. The ${\bf 32}$
of $SO(10,1)$ decomposes as ${\bf 32}={\bf 4}\otimes {\bf 8}$,
thus, for a spinor in the compactified theory we have
\begin{equation}
\eta(x,y)=\epsilon(x)\otimes \theta(y),
\end{equation}
with $\epsilon$ a spinor in four and $\theta$ a spinor in seven
dimensions. The $\Gamma$-matrices can be rewritten as\\
\parbox{14cm}{
\begin{eqnarray}
\Gamma^{a}&=&\gamma^{a}\otimes \opone,\nonumber\\
\Gamma^m&=&\gamma_5\otimes \gamma^m,\nonumber
\end{eqnarray}}\hfill\parbox{8mm}{\begin{eqnarray}\label{gammasplit}\end{eqnarray}}
with $\{\gamma^m\}$ the generators of a Clifford algebra in seven
dimensions. This leads to a nice decomposition of
$\nabla^S=\nabla_M^Sdz^M$,
\begin{eqnarray}
\nabla^S&=&\nabla_{\mu}^Sdx^{\mu}+\nabla^S_{m}dy^m\nonumber\\
&=&(\partial_{\mu}+{1\over4}\omega_{\mu AB}\Gamma^{AB})dx^{\mu}+(\partial_m+{1\over4}\omega_{mAB}\Gamma^{AB})dy^m\nonumber\\
&=&(\partial_{\mu}+{1\over4}\omega_{\mu ab}\Gamma^{ab})dx^{\mu}+(\partial_m+{1\over4}\omega_{m\hat n\hat p}\Gamma^{\hat n\hat p})dy^m\nonumber\\
&=&(\partial_{\mu}+{1\over4}\omega_{\mu ab}\gamma^{ab}\otimes\opone)dx^{\mu}+(\partial_m+{1\over4}\omega_{m\hat n\hat p}\opone\otimes\gamma^{\hat n\hat p})dy^m\nonumber\\
&=&\nabla^S_4\otimes\opone+\opone\otimes\nabla_7^S.
\end{eqnarray}
Here we used the fact that our vacuum is a Riemannian product, and
thus the connection coefficients with mixed indices, as
$\omega_{m\hat na}$, vanish. But then (\ref{covcon}) reads
\begin{equation}
(\nabla^S_4\otimes\opone+\opone\otimes\nabla_7^S)\epsilon(x)\otimes
\theta(y)=\nabla^S_4\epsilon(x)\otimes\theta(y)+\epsilon(x)\otimes\nabla_7^S\theta(y)=0.
\end{equation}
We are looking for the number of linearly independent solutions of
this equations for a given vacuum. Any maximally symmetric space
admits the maximum number of covariantly constant spinors, in
particular, on Minkowski space we can find a basis of four
constant spinors $\epsilon^i$. The condition we are left with is
\begin{equation}
\nabla^S_7\theta(y)=0.\label{covcon7}
\end{equation}
Thus, the number of solutions of (\ref{covcon}) is four times the
number of covariantly constant spinors on the compact seven
manifold. This in turn implies that the number of supersymmetries
in four dimensions is given by the number of covariantly constant
spinors on the seven-manifold.

\bigskip
After these comments it is easy to determine the amount of
supersymmetry preserved by various vacua. The first vacuum we
found was eleven-dimensional Minkowski space. Obviously, in this
case we can find 32 covariantly constant spinors, we have a vacuum
with $\mathcal{N}=1$ supersymmetry. The spaces which are more
interesting are $\mathbb{R}^4\times T^7$ and $\mathbb{R}^4\times
T^3\times K3$. $T^7$ is flat, hence, we expect to find 8 solutions
of (\ref{covcon7}). This gives the maximal $\mathcal{N}=8$
supersymmetry in four dimensions. The situation gets more
interesting for $\mathbb{R}^4\times T^3\times K3$. The reason is
that the K3 surface has the reduced holonomy $SU(2)$, and from
theorem 2.23 we read off that we are left with only two
covariantly constant spinors. We have the decomposition ${\bf
32}={\bf 4}\otimes {\bf 2}\otimes {\bf 4}$ and as we get two
solutions for $T^3$ we are left with sixteen solutions altogether,
and thus $\mathcal{N}=4$ supersymmetry.

\bigskip
Our goal is to find a vacuum of M-theory which is a Riemannian
product of Minkowski space and a compact seven-manifold, s.t. the
effective field theory in four dimensions contains the field
content of the standard model with $\mathcal{N}=1$ supersymmetry.
After the discussions of this chapter it is easy to derive the
consequences of the latter condition. $\mathcal{N}=1$
supersymmetry means that the compact space has exactly one
covariantly constant spinor. Referring once again to theorem 2.23
we conclude that {\em the holonomy group of the compact
seven-dimensional space has to be $G_2$.} These spaces are
possible vacua because they are automatically Ricci-flat. So to
find realistic field theories we should concentrate on
compactifications on $G_2$-manifolds.\\
Before we proceed we want to understand the relation between a
covariantly constant spinor and $G_2$ holonomy from a more
physical point of view. It is clear that $\nabla^S_7\theta=0$
implies $R_{mn\hat p\hat q}\Gamma^{\hat p\hat q}\theta=0$. But
$R_{mn\hat p\hat q}\Gamma^{\hat p\hat q}$ are the generators of
the holonomy group on $M_7$. So we deduce $Hol(M_7)\theta=\theta$
for exactly one spinor $\theta$. Using a suitable basis this
spinor can be written as $\theta=(0,\ldots ,0,1)^{\tau}$ and
$Hol(M_7)$ is the subgroup of $SO(7)$ which fixes this spinor,
i.e. we must have ${\bf 8}\rightarrow{\bf 7}\oplus{\bf 1}$ under
$Hol(M_7)\subset SO(7)$. But this is true only for
$Hol(M_7)=G_2$.\\

Yet, we also need to mention at this point that compactification
on $G_2$-manifolds is not the only possible method to obtain
$\mathcal{N}=1$ supersymmetry from M-theory. The reason is that
the requirements of theorem 2.23 are rather strong. In particular,
we want the metric to be irreducible. This basically excludes the
Riemannian product of two spaces. If we allow for manifolds $M_7$
which are Riemannian products a number of other interesting
solutions with the right amount of supersymmetry exists. We
mentioned above that the compactification of M-theory on $S^1$ and
on $S^1/\mathbb{Z}_2$ gives the strong coupling limit of type IIA
and $E_8\times E_8$ heterotic string theory. But it is well known
\cite{CHSW85} that the compactification of the heterotic string on
Calabi-Yau three-folds, three-dimensional complex Ricci-flat
manifolds with holonomy $SU(3)$, leads to $\mathcal{N}=1$
supersymmetry, as well. In fact the analysis of this case is very
similar to what we did for M-theory in this chapter. Similarly,
compactifying the IIA theory on a Calabi-Yau three-fold leads to
$\mathcal{N}=2$ in four dimensions. We conclude that the desired
amount of supersymmetry can be obtained from both
compactifications of M-theory on $G_2$-manifolds as well as on
$CY_3\times S^1/\mathbb{Z}_2$.

\bigskip
So far we only considered Ricci-flat solutions, which have to
satisfy very restrictive conditions. However, it was shown  in
\cite{CR84} that any supersymmetric compactification of
eleven-dimensional supergravity to $(\mathbb{R}^4\times
M_7,\eta\times g)$, with $(\mathbb{R}^4,\eta)$ Minkowski space and
$(M_7,g)$ arbitrary but compact, leads to vanishing G-flux,
$\langle C\rangle=0$, and a Ricci-flat compact space
$\mathcal{R}_{mn}=0$.\\
If one relaxes the constraint of compactifications to Minkowski
space there are other supersymmetric vacua, as we will see in the
next section.

\subsection{The Freund-Rubin Solutions and Weak $G_2$}
Next we want to analyze the consequences of (\ref{SUSYvac}) for
the Freund-Rubin solutions. Remember that we have $\langle
G_{mnpq}\rangle=0$ and $\langle
G_{\mu\nu\rho\sigma}\rangle=f\epsilon_{\mu\nu\rho\sigma}$ in this
case. With\footnote{Remember
$\epsilon_{\mu\nu\rho\sigma}=\sqrt{g}\
\widetilde{\epsilon}_{\mu\nu\rho\sigma}$ and that hats denote
indices in the tangent space.}
\begin{equation}
\gamma_5:=-i\gamma^{\hat0}\gamma^{\hat1}\gamma^{\hat2}\gamma^{\hat3}=-{i\over
4!}\gamma^{\mu}\gamma^{\nu}\gamma^{\rho}\gamma^{\sigma}\epsilon_{\mu\nu\rho\sigma}
\end{equation}
and
\begin{equation}
\epsilon_{\mu\nu\rho\sigma}\gamma^{\nu}\gamma^{\rho}\gamma^{\sigma}=6i\gamma_{\mu}\gamma_5
\end{equation}
we get\\
\parbox{14cm}{
\begin{eqnarray}
\nabla^S_{\mu}\eta&=&-{if\over6}(\gamma_{\mu}\gamma_5\otimes\opone)\eta,\nonumber\\
\nabla^S_m\eta&=&{if\over12}(\opone\otimes\gamma_m)\eta,\nonumber
\end{eqnarray}}\hfill\parbox{8mm}{\begin{eqnarray}\label{Deta}\end{eqnarray}}
where we used the same decomposition of spinors and
$\Gamma$-matrices as in the Ricci-flat case. Again we have the
decomposition ${\bf 32}={\bf 4}\otimes {\bf 8}$ and hence
\begin{equation}
\eta(x,y)=\epsilon(x)\otimes \theta(y).
\end{equation}
Then (\ref{Deta}) reduce to
\begin{eqnarray}
\nabla^S_{\mu}\epsilon&=&-{if\over6}\gamma_{\mu}\gamma_5\epsilon,\label{D4e}\\
\nabla^S_m\theta&=&{if\over12}\gamma_m\theta\label{D7t}.
\end{eqnarray}
From standard supergravity theory in four dimensions \cite{JPD02}
one can show that on $AdS$ there are always four spinors
satisfying (\ref{D4e}). Therefore, the number of spinors $\eta$,
satisfying (\ref{Deta}) is four times the number of spinors
$\theta$ which are solutions of (\ref{D7t}). But this implies
that the number of supersymmetry being conserved is given by the
number of solutions of (\ref{D7t}).

\bigskip
Let us pause for a moment and check whether the condition of a
supersymmetric vacuum is consistent. To do so we calculate
\begin{eqnarray}
\ [\nabla^S_{\mu},\nabla^S_{\nu}]\epsilon&=&-{f^2\over36}(\gamma_{\mu}\gamma_{\nu}-\gamma_{\nu}\gamma_{\mu})\epsilon,\\
\
[\nabla^S_m,\nabla^S_n]\theta&=&{f^2\over144}(\gamma_m\gamma_n-\gamma_n\gamma_m)\theta,
\end{eqnarray}
and using (\ref{curv}) we obtain
\begin{eqnarray}
\mathcal{R}_{\mu\nu}&=&-{1\over3}f^2g_{\mu\nu},\\
\mathcal{R}_{mn}&=&{1\over6}f^2g_{mn}.
\end{eqnarray}
But this is exactly (\ref{FR4}) and (\ref{FR7}), thus it is
consistent to look for supersymmetric Freund-Rubin type solutions.

\bigskip
So to find Freund-Rubin solutions with $\mathcal{N}=k$
supersymmetry we need to find compact seven-dimensional Einstein
spaces with positive curvature and exactly $k$ Killing spinors.
One possible space is the seven-sphere which admits eight Killing
spinors, leading to maximal supersymmetry in four
dimensions.\footnote{One can show \cite{DNP86} that $S^7$ and
$T^7$ are in fact the only spaces
that lead to maximal supersymmetry.}\\
Another interesting example was found in \cite{ADP82}. There the
compact space has the topology of a sphere, but the metric has
been ``squashed" and it no longer is the standard metric. In fact,
it admits exactly one Killing spinor, leading to $\mathcal{N}=1$
supersymmetry. Here we can take up a comment from chapter 2, as
this sphere actually has {\em weak $G_2$ holonomy}. The reader is
referred to the literature for further details.

\bigskip
In fact, the Freund-Rubin ansatz is not the most general ansatz to
find solutions which are a Riemannian product of a maximally
symmetric space in four dimensions and a compact seven-manifold.
In \cite{E82} a solution with $\langle G_{mnpq}\rangle\neq0$ was
found. However, it was shown in \cite{DFN82} and \cite{ERS83} that
it breaks all supersymmetry. This is actually the case for all
solutions with $\langle G_{mnpq}\rangle\neq0$, therefore they are
not interesting for our purposes.

\subsection{M2-branes and Supersymmetry}
We now want to turn to the question of how much supersymmetry is
conserved in the case of the M2-brane solution. We saw already
that the amount of supersymmetry which is conserved by a given
vacuum depends on the metric on this vacuum. Therefore, it is
natural to expect that preserved supersymmetry depends on the
metric of the Ricci-flat compact seven-manifold $X$, which has
not been specified so far. This will indeed turn
out to be true.\\
As always the fermionic background vanishes and we need to
analyze once again the consequences of (\ref{SUSYvac}). Because
of the specific structure of the M2-brane with a
(2+1)-dimensional Minkowski space, a radial coordinate $x^3:=r$
and a compact seven-manifold $X$ we look at the decomposition
${\bf 32}={\bf 4}\otimes{\bf 8}={\bf 2}\otimes{\bf 2 }\otimes{\bf
8}$. We write our spinors as
\begin{equation}
\eta=\epsilon\otimes \theta,
\end{equation}
with a four component spinor $\epsilon$ on the
(1+2+1)-dimensional space and an eight component spinor $\theta$
on the seven-manifold $X$. However, we decompose the
$\Gamma$-matrices as
\begin{eqnarray}
\Gamma^{\hat a}&=&\gamma^{\hat a}\otimes
\sigma_3\otimes\opone,\nonumber\\
\Gamma^{\hat 3}&=&\opone\otimes
\sigma_1\otimes\opone,\\
\Gamma^{\hat m}&=&\opone\otimes \sigma_2\otimes\gamma^{\hat
m}.\nonumber
\end{eqnarray}
Before we proceed any further we collect some useful relations.
From (\ref{omegas}) we can read off that the only non-vanishing
components of the spin connection are
\begin{eqnarray}
\omega_{\hat a\ \hat 3}^{\ \hat b}&=&-{1\over3}H'H^{-{7 \over6}}\delta_{\hat a}^{\hat b},\nonumber\\
\omega_{\hat m\ \hat n}^{\ \hat 3}&=&-\left({H'\over6H}+{1\over r}\right)H^{-{1 \over6}}\delta_{\hat m\hat n},\\
\omega_{\hat n\ \hat q}^{\ \hat p}&=&\widetilde{\omega}_{\hat n\
\hat q}^{\ \hat p}.\nonumber
\end{eqnarray}
Furthermore, from (\ref{vielbeins}) we get
\begin{eqnarray}
e_{\hat a}^{b}&=&H^{1\over3}\delta^b_{\hat a},\nonumber\\
e_{\hat 3}^{\ 3}&=&H^{-{1\over6}},\\
e_{\hat n}^{\ m}&=&{1\over r}H^{-{1\over6}}\widetilde{e}_{\hat
n}^{\ m} .\nonumber
\end{eqnarray}
$\Gamma$-matrix conventions on the three-dimensional Minkowski
space are such that $\gamma^{\hat 0}\gamma^{\hat 1}\gamma^{\hat 2
}=-\opone$. Then, the substitution of the M2-brane solution into
(\ref{SUSYvac}) gives \cite{DLPS95}
\begin{eqnarray}
0&=& \delta_{\hat a}^{b}H^{{1\over3}}\partial_{b}\eta+{1\over6}H'H^{-{7\over6}}\gamma_{\hat a}\otimes\sigma_1(\sigma_3-\opone)\otimes\opone\eta\nonumber,\\
0&=& H^{-{1\over6}}\partial_r\eta+{1\over6}H'H^{-{7\over6}}\opone\otimes\sigma_3\otimes\opone\eta\label{M2susy},\\
0&=&{1\over r }H^{-{1\over6}}\left(\widetilde{e}_{\hat m}^{\ \
n}\opone\otimes\opone\otimes\nabla_n^S\eta-{i\over2}\opone\otimes\sigma_3\otimes\gamma_{\hat
m}\eta\right)-{i\over12}H'H^{-{7\over6}}\opone\otimes(\sigma_3-\opone)\otimes\gamma_{\hat
m}\eta.\nonumber
\end{eqnarray}
Again we are looking for the number of solutions $\eta$ of this
set of equations. The problem simplifies if we recall its
symmetries. We have a $P(2,1)$ symmetry on the three-dimensional
Minkowski space and a radial coordinate which can be viewed as
parameterizing seven-manifolds $X$. The most general spinor that
is invariant under the symmetries of this setup has the form
\begin{equation}
\eta=F(r)\epsilon_0\otimes\theta,
\end{equation}
with $\epsilon_0$ a constant spinor. If we plug this ansatz into
(\ref{M2susy}) we get conditions for our spinor,
\begin{eqnarray}
F'=-{H'\over6H}F,\\
\opone\otimes\sigma_3\epsilon_0=\epsilon_0,\\
\nabla^S_{\hat m}\theta-{i\over2}\gamma_{\hat m}\theta=0.
\end{eqnarray}
The first equation can easily be integrated and gives
$F=H^{-{1\over6}}$. The second equation is more interesting as it
is satisfied by only two of the four constant spinors on the
(1+2+1)-dimensional space. The last equation tells us that
$\theta$ has to be a Killing spinor on the seven-dimensional space
$X$. This is all very nice. We conclude that the number of
solutions of (\ref{SUSYvac}) is two times the number of Killing
spinors on $X$. For example in the case of the seven-sphere,
$X=S^7$, we have eight Killing spinors and hence sixteen
generators of supersymmetry. We see that in this case an M2-brane
breaks half of the supersymmetry. Certainly we can find a solution
which breaks supersymmetry completely. All one has to do is to
look for a compact Ricci-flat seven-manifold with
$\mathcal{R}_{mn}=6\delta_{mn}$ that does not admit Killing
spinors. But there are also interesting spaces in between these
two cases. In particular let us consider a space with weak $G_2$
holonomy. We know that these spaces admit exactly one Killing
spinor and that they are Einstein spaces with positive scalar
curvature. In that case only two supersymmetric generators remain
unbroken, we speak of a {\em weak $G_2$-brane}.

\section{Kaluza-Klein Compactification on Smooth $G_2$-Manifolds}
The analysis of the structure of $G_2$-manifolds showed that
M-theory on $(\mathcal{M}_4\times X, \eta\times g)$ with $(X,g)$ a
$G_2$-manifold leads to an effective $\mathcal{N}=1$ supergravity
theory in four dimensions. In this section we want to analyze the
field content that arises from Kaluza-Klein compactification on
smooth $G_2$-manifolds. This field content was determined in
\cite{PT95}, the compactification procedure is reviewed in \cite{Ac02}.\\
We start from  supergravity (\ref{SUGRAaction}) on the Riemannian
product $(\mathcal{M}_4\times X, \eta\times g)$. Together with
$\langle\psi_M\rangle=\langle C\rangle=0$ this is a possible
vacuum of eleven-dimensional supergravity. We assume that $X$ is
large compared to the Planck length in order for supergravity to
be valid and small compared to macroscopic scales. In particular,
it is supposed to be so small that massive Kaluza-Klein states
cannot be observed.

\bigskip
{\bf Proposition 5.3}\\
The low energy effective theory of M-theory on
$(\mathcal{M}_4\times X, \eta\times g)$ with $(X,g)$ a smooth
$G_2$-manifold is an $\mathcal{N}=1$ supergravity theory coupled
to $b^2(X)$ Abelian vector multiplets and $b^3(X)$ massless
neutral chiral multiplets.

\bigskip
In order to show this we proceed according to the recipe given in
section 3.3. Consider small fluctuations around the vacuum,
\begin{eqnarray}
g_{MN}(x,y)&=&\langle g_{MN}(x,y)\rangle+\delta g_{MN}(x,y)\label{fluctuationsg},\\
C(x,y)&=&\langle C(x,y)\rangle +\delta C(x,y)=\delta C(x,y)\label{fluctuationsC},\\
\psi_M(x,y)&=&\langle\psi_M(x,y)\rangle+\delta\psi_M(x,y)=\delta
\psi_M(x,y)\label{fluctuationspsi}.
\end{eqnarray}
Next we need to substitute these fluctuations into the field
equations, keeping only linear terms.

\bigskip
\begin{center}
{\bf Kaluza-Klein expansion of the $C$-field}
\end{center}
After substitution of (\ref{fluctuationsC}) the equation for the
C-field reads
\begin{equation}
d*G=0.
\end{equation}
If we impose the gauge condition $d*C=0$ we get
\begin{equation}
\Delta_{11} C=0.
\end{equation}
We split $\Delta_{11}=\Delta_4+\Delta_7$,
\begin{equation}
\Delta_4 C +\Delta_7 C=0,
\end{equation}
to see that possible mass terms for four dimensional fields might
arise from $\Delta_7 C$, which is identified with the mass
operator of equation (\ref{mass eq}).\\
\\
Let $\lbrace\Omega^{i}\rbrace\subset\Lambda^3T^*X$ be the set of
three-forms, s.t.
\begin{equation}
\Delta_7\Omega^{i}=\lambda^i_{(3)}\Omega^{i},
\end{equation}
$\lbrace\omega^{j}\rbrace\subset\Lambda^2T^*X$ be the set of
two-forms, s.t.
\begin{equation}
\Delta_7\omega^{j}=\lambda^j_{(2)}\omega^{j},
\end{equation}
$\lbrace a^{k}\rbrace\subset\Lambda^1T^*X$ be the set of
one-forms, s.t.
\begin{equation}
\Delta_7 a^{k}=\lambda^k_{(1)} a^{k},
\end{equation}
and let $\lbrace g^{l}\rbrace\subset\Lambda^0T^*X$ be the set of
eigenfunctions, s.t.
\begin{equation}
\Delta_7 g^{l}=\lambda^l_{(0)} g^{l}.
\end{equation}
Then we have the general expression
\begin{eqnarray}
C(x,y)&=&\sum_i p^{i}(x)\wedge \Omega^{i}(y)+\sum_j
A^{j}(x)\wedge\omega^{j}(y)\nonumber\\
&&+\sum_k B^{k}(x)\wedge a^{k}(y)+\sum_l H^{l}(x)\wedge g^{l}(y).
\end{eqnarray}
This can be rewritten as
\begin{eqnarray}
C(x,y)&=&\sum_{i=1}^{b^3(X)} p^{i}(x)\wedge
\Omega^{i}(y)+\sum_{j=1}^{b^2(X)}A^{j}(x)\wedge\omega^{j}(y)\nonumber\\
&&+\sum_{k^1}^{b^1(X)} B^{k}(x)\wedge a^{k}(y)+\sum_{l=1}^{b^0(X)}
H^{l}(x)\wedge g^{l}(y)+\ldots \mbox{(massive)},
\end{eqnarray}
where from now on $\Omega^{i},\ \omega^{j},\ a^{k}$ and $g^{l}$
are taken to be harmonic forms on $X$. Given the Betti numbers of
a $G_2$-manifold this simplifies to
\begin{equation}
C(x,y)=\sum_{i=1}^{b^3(X)} p^{i}(x)\wedge
\Omega^{i}(y)+\sum_{j=1}^{b^2(X)} A^{j}(x)\wedge\omega^{j}(y)+
H(x)+\ldots \ \mbox{(massive)}.\label{KKCfield}
\end{equation}
We see that we obtain $b^2(X)$ Abelian gauge fields and $b^3(X)$
pseudo-scalar\footnote{Recall the transformation of the $C$-field
under parity.} fields from compactifications of the
$C$-field\footnote{The four-dimensional three-form $H(x)$ is not
dynamical.}. Note that a necessary requirement was that $\langle
G\rangle=\langle\psi_M\rangle=0$, so this kind of expansion of
the $C$-field is only valid for vacua with vanishing $G$-flux.

\bigskip
\begin{center}
{\bf Kaluza-Klein expansion of the metric}
\end{center}
Compactifying the $C$-field gave us $b^3(X)$ pseudo-scalars in
four dimensions. In eleven dimensions the bosonic superpartner of
$C$ is $g$, so we should expect that superpartners of the
pseudo-scalars in four dimensions can be obtained from the metric,
which therefore should give another $b^3(X)$ scalars. To see this
explicitly we substitute the fluctuations into the Einstein
equation. As the energy momentum tensor is quadratic in the
fluctuations we are left with the condition of Ricci-flatness,
\begin{equation}
\mathcal{R}_{MN}=0,\label{RMN0}
\end{equation}
where $\mathcal{R}_{MN}$ is evaluated using the full metric of
(\ref{fluctuationsg}). Suppose furthermore that $\delta g_{\mu
m}(x,y)=0$, which (together with the Riemannian product structure
of our eleven-dimensional manifold) implies that Christoffel
symbols with mixed components vanish trivially. Then (\ref{RMN0})
reduces to $\mathcal{R}_{\mu\nu}=\mathcal{R}_{mn}=0$.
Substitution of $g_{mn}(x,y)=\langle g_{mn}(x,y)\rangle+\delta
g_{mn}(x,y)$ leads to the Lichnerowicz equation for $\delta g
_{mn}$,
\begin{equation}
\Delta_L\delta g_{mn}:=-\Delta_{11}\delta g_{mn}-2R_{mpnq}\delta
g^{pq}+2\mathcal{R}^{\ \ p}_{(m}\delta g_{n)p}=0,
\end{equation}
where now $R_{mpnq}$ and $\mathcal{R}_{pm}$ are calculated from
$\langle g_{mn}\rangle$. Again we write
$\Delta_{11}=\Delta_4+\Delta_7$ and obtain
\begin{equation}
\Delta_4 \delta g_{mn}+\Delta_7 \delta g _{mn}+2R_{mpnq} \delta
g^{pq}-2\mathcal{R}_{(m}^{\ \ p}\delta g_{n)p}=0.
\end{equation}
Next we make the Kaluza-Klein ansatz
\begin{equation}
\delta g_{mn}(x,y)=\sum_i s^{i}(x) h^{i}_{mn}(y),
\end{equation}
where $h^{i}_{mn}$ are eigenfunctions of the Lichnerowicz operator
\begin{equation}
\Delta_L h^{i}_{mn}(y)=\lambda^ih^{i}_{mn}(y).
\end{equation}
Then we are left with
\begin{eqnarray}
0&=&\left(\Delta_4 s^{i}(x)\right)h^{i}_{mn}(y)-s^{i}(x)\Delta_L
h^{i}_{mn}(y)\nonumber\\
&=&\left(\Delta_4 s^{i}(x)-\lambda^is^{i}(x)\right)h^{i}_{mn}(y).
\end{eqnarray}
We conclude that massless scalar fields in four dimensions are
given by the zero modes of the Lichnerowicz operator. It remains
to determine the number of zero modes. To do so we define the
three-form $\omega$ s.t.\footnote{The symmetric tensor $h_{mn}$
transforms as a ${\bf 27}$ on a seven-manifold, which is
irreducible under $G_2$. The {\bf 35} of a three-form decomposes
as {\bf 35}={\bf 1}+{\bf 7}+{\bf 27}, see proposition 2.32. As
$\varphi$ is in the trivial representation of $G_2$, $\omega$ and
$h$ are in the same representation. This is why there is an
isomorphism between $h_{mn}$ and three-forms $\omega_{mnp}$ on
$X$.}
\begin{equation}
\omega_{mnp}:=\varphi_{q[mn}h^{\ \ q}_{p]},
\end{equation}
where $\varphi$ is the $G_2$-invariant three-form on $X$. It can
be shown \cite{Ac02} that $h$ is a zero mode of the Lichnerowicz
operator if and only if $\omega$ is harmonic,
\begin{equation}
\Delta_L h=0\ \ \ \Leftrightarrow\ \ \ \Delta \omega=0.
\end{equation}
To prove this we need to note that
\begin{equation}
\varphi_{q[mn}R_{p]k\ l}^{\ \ \ q}h^{kl}=0,
\end{equation}
which gives
\begin{equation}
\Delta\omega_{mnp}=-\varphi_{q[mn}\Delta_Lh_{p]}^{\ \ q}
\end{equation}
and thus the desired result. We conclude that there are $b^3(X)$
scalars in the four dimensional theory
coming from the metric, as expected.\\
In fact, this could have been anticipated from the general rule
that if there is a $k$-dimensional family of $G_2$-holonomy
metrics on $X$, i.e. if the moduli space of these metrics has
dimension $k$, then there will be correspondingly $k$ massless
scalars in four dimensions. From proposition 2.34 we know that the
moduli space of $G_2$ metrics on $X$ is a manifold of
dimension $b^3(X)$.\\
Now, the scalars coming from the metric combine with those coming
from the $C$-field to give $b^3(X)$ complex massless scalars
$\Phi^{i}(x):=s^{i}+ip^{i}$ which are the lowest components of
massless chiral superfields in four dimensions. These scalar
fields can be rewritten in a very nice form by means of a basis
$\alpha_{i}$ for the third homology group of $X$,
$H_3(X,\mathbb{R})$. The basis is chosen s.t.
\begin{equation}
\int_{\alpha_{j}}\Omega^{i}=\delta^{ij}.
\end{equation}
Next we note that a fluctuation of the metric can be rewritten by
means of the three-form $\varphi$ characterizing a $G_2$-manifold,
\begin{equation}
\varphi'=\varphi+\delta\varphi=\varphi+\sum_i
s^{i}(x)\Omega^{i}(y).
\end{equation}
This gives
\begin{equation}
\Phi^{i}=\int_{\alpha_{i}}\left( \varphi'+iC\right).
\end{equation}

\bigskip
It remains to consider the fluctuations of the metric on
four-dimensional Minkowski space. As above the substitution of
$g_{\mu\nu}(x,y)=\langle g_{\mu\nu}(x,y)\rangle+\delta
g_{\mu\nu}(x,y)$ into $\mathcal{R}_{\mu\nu}(g_{\rho\sigma})$ leads
to the Lichnerowicz equation for $\delta g _{\mu\nu}$. But on
Minkowski space this equation simplifies to
\begin{equation}
\Delta_{11}\delta g_{\mu\nu}=0.
\end{equation}
The Kaluza-Klein ansatz
\begin{equation}
\delta g_{\mu\nu}(x,y)=\sum_i h^{i}_{\mu\nu}(x) t^{i}(y),
\end{equation}
with $t^{i}$ eigenfunctions of $\Delta_7$ gives
\begin{eqnarray}
0&=&\left(\Delta_4
h^{i}_{\mu\nu}(x)\right)t^{i}(y)+h^{i}_{\mu\nu}(x)\Delta_7
t^{i}(y)\nonumber\\
&=&\left(\Delta_4
h^{i}_{\mu\nu}(x)+\lambda^ih^{i}_{\mu\nu}(x)\right)t^{i}(y).
\end{eqnarray}
The massless particle corresponding to the zero modes of
$\Delta_7$ is the four-dimensional graviton.

\bigskip
The isometry group of a compact $G_2$-manifold is trivial, as we
will prove below. This implies that we do not get gauge fields
from the Kaluza-Klein expansion of the metric.

\bigskip
We could perform a Kaluza-Klein expansion of the gravitino as
well. However, this is not necessary, as we know from
considerations in the last section that the effective
four-dimensional theory has $\mathcal{N}=1$ supersymmetry. Hence,
there will be superpartners to all the bosonic particles we
considered so far. In particular, we will get the four-dimensional
gravitino and hence supergravity. But this supergravity theory is
coupled to a gauge theory with $b^2(X)$ Abelian vectors multiplets
and $b^3(X)$ neutral chiral multiplets. This means that the total
number of fermions is $b^2(X)+b^3(X)$ together with the gravitino.

\bigskip
Unfortunately the theory we obtained is not very interesting.
Neither did we find non-Abelian gauge groups which are a crucial
ingredient of the standard model, nor do we have charged chiral
fermions in our effective theory. Obviously the fermionic
superpartners of the Abelian gauge fields are neutral. The
effective action contains no term which couples the chiral
multiplets to the Abelian gauge fields, so we conclude that their
charges vanish as well. This means that we have to work harder to
get a realistic theory from compactifications on $G_2$-manifolds.
Note on the other hand that the theory is automatically free of
local gauge anomalies, as the fermions cannot couple to the gauge
fields.

\begin{center}
{\bf The isometry group of a manifold with $G_2$-holonomy}\\
\end{center}
In chapter 3 we discussed that non-Abelian gauge groups are
generated by the isometry group of the compact manifold. To prove
the statement that our Kaluza-Klein compactification does not lead
to non-Abelian gauge theories we must show that the isometry group
of a $G_2$-manifold is trivial.\\

\bigskip
{\bf Proposition 5.4}\\
A compact manifold $X$ with holonomy group $G_2$ does not have
continuous symmetries.

\bigskip
Continuous symmetries are generated by Killing vectors, so it is
sufficient to show, that there are no Killing vectors on $X$. A
Killing vector $V$ satisfies
\begin{equation}
\nabla_mV_n+\nabla_nV_m=0.
\end{equation}
Multiplying by $g^{mn}$ we get $\nabla^mV_m=0$. Furthermore
\begin{equation}
0=\nabla^m\nabla_mV_n+\nabla^m\nabla_nV_m.
\end{equation}
But using (\ref{curvgeneral})
\begin{equation}
\nabla^m\nabla_nV_m=\nabla_n\nabla^mV_m+[\nabla^m,\nabla_n]V_m=R^m_{\
nmp}V^p=\mathcal{R}_{np}V^p=0,
\end{equation}
as $X$ is Ricci flat. We are left with
\begin{equation}
\nabla^m\nabla_mV_n=0.\label{LapV0}
\end{equation}
This implies for compact manifolds $X$
\begin{equation}
0=\int_X V^n\nabla^m\nabla_mV_n=-\int_X
(\nabla^mV^n)(\nabla_mV_n),
\end{equation}
therefore, the Killing vector must be covariantly constant,
\begin{equation}
\nabla_mV_n=0.
\end{equation}
Thus we proved the following Lemma.

\bigskip
{\bf Lemma 5.5}\\
A Killing vector on a Ricci-flat compact manifold is covariantly
constant.

\bigskip
But for $G_2\subset SO(7)$ we have ${\bf 7}\rightarrow{\bf 7}$, a
vector always transforms in the fundamental representation of
$G_2$. In particular, there is no singlet of the holonomy group,
so covariantly constant vectors cannot exist on manifolds with
holonomy $G_2$. Thus, there are no Killing vectors and the
isometry group is trivial.\footnote{Note that this changes if the
holonomy group is a proper subgroup of $G_2$. For example in the
case of $SU(3)\subset G_2\subset SO(7)$ we have ${\bf 7}={\bf
7}={\bf 3}\oplus{\bf \bar 3}\oplus {\bf 1}$. There is a singlet
and thus a Killing vector. This occurs for example for $CY\times
S^1$ with $CY$ a Calabi-Yau manifold. Here the Killing vector
generates the isometry of $S^1$.}

\bigskip
\begin{center}
{\bf A final remark}
\end{center}
It is interesting to compare our result to the effective theory
obtained by compactifying the heterotic string on a Calabi-Yau
manifold. In this case the effective theory is again an
$\mathcal{N}=1$ supergravity theory in four dimensions. However,
the two compactifications cannot be equivalent, as the
compactification of the heterotic string leads to a chiral theory
with non-abelian gauge fields. Some attempt have been made
\cite{PT95} to establish the equivalence of the two
compactification schemes by compactifying the heterotic string on
a Calabi-Yau manifold with zero Euler number, as this leads to
non-chiral theories. If in addition the gauge group of the
heterotic string can be broken to $U(1)^{16}$ the effective theory
is $\mathcal{N}=1$ supergravity coupled to 16 $\mathcal{N}=1$
vector multiplets and $(h^{1,1}+h^{1,2}+1)$ $\mathcal{N}=1$
neutral chiral multiplets. The vanishing Euler number gives
$h^{1,1}=h^{1,2}$. So the two compactification mechanisms lead to
the same theory, provided $b^2(X)=16$ and $b^3(X)=2h^{1,1}+1$.
This can be established for example for a Calabi-Yau manifold with
$h^{1,1}=h^{1,2}=19$ and a $G_2$-manifold with $b^2(X)=16$ and
$b^3(X)=39$. Remarkably, these manifolds exist \cite{PT95},
\cite{Jo96b}, \cite{Jo00}.\\
The relation between the two compactification schemes will become
much clearer when we will consider $G_2$-manifolds carrying
singularities.\\

\chapter{M-theory Anomalies in the Presence of M-branes}
So far we have seen that M2-branes and M5-branes are solutions of
the equations of motions of eleven-dimensional supergravity. In
fact, the action of those objects is known \cite{APPS97},
\cite{BLNPST97}, \cite{BST87}, and they can be coupled to the
supergravity theory. This is similar to Maxwell's theory where
the action for electrons and magnetic monopoles can be coupled to
the free photon field. However, we need to check whether this new
theory is consistent, and in particular whether it is free of
anomalies.

\begin{center}
{\bf Five-branes in M-theory}
\end{center}
The five-brane has support on a 5+1-dimensional submanifold $W^6$
of eleven-dimensional space-time $M_{11}$. Both $W^6$ and $M_{11}$
are taken to be oriented and spin. For this particular setup the
notion of a normal bundle is useful.

\bigskip
{\bf Definition 6.1}\\
Let $M$ be an $m$-dimensional submanifold of $\mathbb{R}^{m+k}$
and let $p\in M$. Then $T_pM\subset
T_p\mathbb{R}^{m+k}\cong\mathbb{R}^{m+k}$. Define $N_pM\subset
T_p\mathbb{R}^{m+k}$ to be the vector space which is normal to
$T_pM$ in $T_p\mathbb{R}^{m+k}$, i.e. $u\cdot v=0$ $\forall u\in
N_pM,\ \forall v\in T_pM$, with respect to the metric on
$\mathbb{R}^{m+k}$. Then the {\em normal bundle} of $M$ is defined
as $NM:=\cup_{p\in M}N_pM$.

\bigskip
$TM_{11}$ and $TW^6$ denote the tangent bundle of $M_{11}$ and
$W^6$, respectively. If $TM_{11}|_{W^6}$ is the restriction of
$TM_{11}$ on $W^6$ it can be written as a Whitney sum
$TM_{11}|_{W^6}=TW^6\oplus NW^6$. Let $X^M(\xi^{\mu})$ be the
embedding functions of $W^6$ into $M_{11}$, where
$\mu,\n,\ldots\in\lbrace0,1,\ldots 5\rbrace$ (with corresponding
tangent space indices $a,b,\ldots$). This embedding breaks the
original $SO(10,1)$ Lorentz symmetry to $SO(5,1)\times SO(5)$ and
$GL(11,\mathbb{R})$ is broken to $GL(6,\mathbb{R})\times
GL(5,\mathbb{R})$. $M_{11}$ carries a metric $g_{MN}$ which can be
pulled back onto $W^6$. Then the components of the pullback
$h:=X^*g$ read
\begin{equation}
h_{\mu\nu}(\xi^{\rho})={\partial X^M(\xi^{\rho})\over\partial
\xi^{\mu}}{\partial X^N(\xi^{\rho})\over\partial
\xi^{\nu}}g_{MN}(X(\xi^{\rho})).
\end{equation}
From this metric we can construct the Levi-Civita connection and
the curvature tensor on $W^6$, which we will denote
$\widetilde{\Gamma}^{\rho}_{\mu\nu}$ and
$\widetilde{R}_{\mu\nu\rho\sigma}$. It is useful to introduce
coordinates on $M_{11}$, s.t. $W^6=\lbrace p\in
M_{11}|X^6=0,\ldots ,X^{10}=0 \rbrace$ and $g_{\mu
m}|_{_{W^6}}=\partial_Mg_{\mu m}|_{_{W^6}}=0$, where $m,n,\ldots$
label the coordinates orthogonal to the brane. These coordinates
reflect the broken symmetry of the system. Then,
$g_{\mu\nu}(X(\xi))$ equals $h_{\mu\nu}(\xi)$ up to coordinate
transformations on $W^6$. So in these coordinates we find
$\Gamma_{\mu\nu}^{\rho}|_{_{W^6}}=\widetilde{\Gamma}^{\rho}_{\mu\nu}$.
However, there is another interesting contribution of the
eleven-dimensional connection. Consider
\begin{equation}
\Gamma_{\mu m}^n|_{_{W^6}}={1\over2}g^{nM}(\partial_{\mu}
g_{mM}+\partial_m g_{M\mu}-\partial_M g_{\mu
m})|_{_{W^6}}={1\over2}g^{nk}\partial_{\mu} g_{mk}|_{_{W^6}}.
\end{equation}
We know, that $\omega^A_{MB}$ corresponding to $\Gamma_{MN}^K$
transforms as a connection under local Lorentz transformations,
$\omega'=g^{-1}(\omega+d)g$ with $g\in SO(5,1)\times SO(5)$.
Clearly, $\omega_{\mu \hat m}^{\hat n}$, with $\hat m,\hat n$
tangent space indices on $NW^6$, transforms trivially under the
$SO(5,1)$ part of this transformation. But for $g=(\opone, h)$,
with $h\in SO(5)$ we find
\begin{equation}
\omega_{\mu \hat m}^{'\hat n} =(h^{-1}(\omega_{\mu}+d)h)^{\hat n
}_{\hat m},
\end{equation}
thus, $\omega_{\mu \hat m}^{\hat n}$ transforms as an $SO(5)$
gauge field under local Lorentz transformations on the normal
bundle. This allows us to interpret $NW^6$ as a fibre bundle over
$W^6$ with structure group $SO(5)$ and a local connection one form
$A_{\mu}:=\omega_{\mu \hat m}^{\hat n}$. In this setup
$g_{mn}(X(\xi))$ can be interpreted as the fibre metric of the
normal bundle.\\
Clearly, the eleven-dimensional spin connection restricts to a
spin connection in six dimensions that reads
\begin{equation}
\nabla_\m^S=\partial_\m+{1\over4}\o_{\m
ab}\G^{ab}+{1\over4}\o_{\m\hat n\hat p}\G^{\hat n\hat p}.
\end{equation}
Using\\
\parbox{14cm}{
\begin{eqnarray}
\G^a&=&\g^a\otimes\opone\nonumber\\
\G^{\hat m}&=&\g\otimes \g^{\hat m},\nonumber
\end{eqnarray}}\hfill\parbox{8mm}{\begin{eqnarray}\end{eqnarray}}
where $\{\gamma^a\}$ are the $\g$-matrices in six dimensions with
$\g=-\g^{\hat 0}\g^{\hat 1}\ldots\g^{\hat 5}$ and $\{\g^{\hat
m}\}$ are the $\g$-matrices in five dimensions, we have
\begin{equation}\label{derW6}
\nabla_\m^S=\partial_\m+{1\over4}\o_{\m
ab}\g^{ab}\otimes\opone+{1\over4}\o_{\m\hat n\hat
p}\opone\otimes\g^{\hat n\hat p}.
\end{equation}
This operator acts on spinors which carry an additional
$SO(5)$-index in the $Spin(5)$ representation. The relevant index
for these fields is given in \ref{ASI} but one has to take care of
the fact that the gauge fields transforms in the spinor
representation (see below). Some more details can be found in
\cite{Wi96b} and \cite{BM03c}.

\bigskip
Next we want to analyze the fields which arise if M5-branes are
introduced. That is, we want to study the world-volume field
theory of these branes. To do so, we note that the M5-brane
solutions break half of the supersymmetry, leaving 16
supersymmetry generators\footnote{We saw this explicitly for the
M2-brane in chapter 5, the M5-brane can be analyzed similarly.
From now on we always use the brane solution that conserves the
maximal number of supersymmetries.}. This is equivalent to
$\mathcal{N}=2$ supersymmetry in six dimensions. The bosonic
degrees of freedom of the six-dimensional theory come about as
follows. The world-volume theory of the M5-brane contains five
scalar fields $X^m,\ m\in\lbrace 6,7,\ldots ,10\rbrace$ which
parameterize the position of the brane in eleven-dimensional
space. Furthermore, it was shown in \cite{KM95} and reviewed in
\cite{BM03c} that there is also a (Minkowskian) self-dual
three-form\footnote{This equation cannot hold globally, though, as
we have $dH=X^*G$, \cite{To95}, \cite{Wi95b}.} $H=dA_{(2)}$ on the
brane, which contains another three bosonic degrees of freedom.
The fermionic degrees of freedom are determined from supersymmetry
to be a pair of chiral spin-1/2 fermions.

\bigskip
After having discussed the geometry and the field content of the
five-brane setup we need to come back to the M-theory description.
If we allow for five-branes in M-theory four major changes occur
on the level of the low-energy effective action and the equations
of motion.

\bigskip
$\bullet$ Firstly, we need to introduce the explicit form of the
coupling of the M5-brane to the $G$-field. It is given by
\cite{To95}, \cite{Wi96b}
\begin{equation}
S_{coup}=\alpha_0 T_5\int_{W^6}|dA_{(2)}-C|^2+\alpha_1
T_5\int_{W^6}A_{(2)}\wedge dC,\label{Scoup}
\end{equation}
with some coefficients $\alpha_0$ and $\alpha_1$. $C$ is
understood to be the pullback of the $C$-field in eleven
dimensions, so if we want to be more precise we should write
$X^*C$ instead of $C$. The quantity $T_5$ is known as the {\em
five-brane tension} and it specifies the strength of the
coupling. It must be present for dimensional
reasons\footnote{Mass dimensions are as follows: $[C_{MNP}]=0$,
$[G_{MNPQ}]=1$, $[\kappa_{11}]=-9/2$, $[R_{MN}]=[F_{MN}]=2$.} and
its mass dimension is $[T_5]=6$. Similarly, for the M2-brane
there is a {\em membrane tension} $T_2$ with $[T_2]=3$. They are
related by \cite{Al96}
\begin{equation}
{T_5\over(T_2)^2}={1\over2\pi},\label{relTk1}
\end{equation}
and satisfy a quantization condition similar to the Dirac
quantization in the presence of magnetic monopoles,
\begin{equation}
2\kappa_{11}^2T_2T_5=2\pi n,\label{relTk2}
\end{equation}
for $n\in\mathbb{Z}$. This enables us to express them both in
terms of the fundamental constant $\kappa_{11}$.\\
The two-form $A_{(2)}$ transforms as $\delta A_{(2)}=\Lambda$, so
the first term in $S_{coup}$ is gauge invariant. However, we get a
variation from the second term
\begin{equation}
\delta S_{coup}=\alpha_1 T_5\int_{W^6}\Lambda\wedge dC.
\end{equation}
The corresponding anomaly polynomial is given by
\begin{equation}
I_8^{(coup)}=\alpha_1 T_5dC\wedge dC.
\end{equation}

\bigskip
$\bullet$ Secondly, the Bianchi identity has to be modified.
Five-branes are sources for the $G$-field just as magnetic
monopoles are sources for the field strength $F$ in Maxwell's
theory. That is, we have
\begin{equation}
dG={2\pi\over T_2}\delta_{W^6}^{(5)},\label{fivebranesource}
\end{equation}
where $\delta_{W^6}^{(5)}$ is a closed five-form with support
localized on the five-brane world-volume $W^6$, which integrates
to one in the directions normal to $W^6$. The factor containing
$T_2$ must be present for dimensional reasons. Such a five-brane
will contribute ${2\pi\over T_2}$ units of $G$-flux when
integrated over the five-dimensions transverse to the brane
bounded by a four-cycle $\omega_4$,\footnote{Some subtleties
related to this formula are discussed in \cite{Wi96} and
\cite{Wi96b}.}
\begin{equation}
\int_{\omega_4} G={2\pi\over T_2}.
\end{equation}
It turns out, that (\ref{fivebranesource}) has to be formulated
more carefully. Following \cite{FHMM98} we define a radial
distance $r$ from the brane and excise a neighbourhood
$D_{\epsilon}W^6$ of the brane of radius $\epsilon$. We are left
with a space $M_{11}\backslash D_{\epsilon}W^6$ with
boundary\footnote{Apparently, in \cite{FHMM98} the boundary of
$M_{11}\backslash D_{\epsilon}W^6$ was taken to be
$+S_{\epsilon}W^6$ but formula (\ref{eeep}) which is taken from
\cite{BC97} is only valid if the spheres carry their natural
orientation.} $-S_{\epsilon}W^6$, which is an $S^4$-bundle over
$W^6$ with its orientation reversed. Now instead of
(\ref{fivebranesource}) we write
\begin{equation}
dG={2\pi\over T_2}d\rho(r)\wedge e_4/2,\label{modfivebranesource}
\end{equation}
where $\rho$ is a smooth function which is $-1$ for small $r$ and
0 for large $r$, s.t. $\int_0^{\infty}dr (\partial_r\rho)=1$.
$e_4$ is the angular form, which is gauge invariant under $SO(5)$
gauge transformations and closed. Its detailed structure can be
found in \cite{FHMM98} and \cite{BB99}. One of its important
properties is
\begin{equation}
\int_{S_{\epsilon}} e_4=2,\label{inteis2}
\end{equation}
where $S_{\epsilon}$ is one of the fibres of $S_{\epsilon}W^6$.
In this setup integrals over $M_{11}$ are understood as
\begin{equation}
\int_{M_{11}} \mathcal{L}:=\lim_{\epsilon\to
0}\int_{M_{11}\backslash D_{\epsilon}W^6}\mathcal{L}
\end{equation}

\bigskip
$\bullet$ Thirdly, we need to add a higher derivative term to our
action (\ref{SUGRAaction}) in order to make the theory
consistent. This so-called {\em Green-Schwarz term} is given by
\begin{equation}
S_{GS}:=\alpha_2 T_2\int G\wedge X_7.
\end{equation}
$\alpha_2$ is a yet undetermined constant and $T_2$ is needed for
dimensional reasons. $X_7$ is defined as $X_8=dX_7$ and $X_8$ was
given in (\ref{X8}). It is very important to note that $X_8$
contains the curvature of $M_{11}$ rather than the one of $W^6$.
This new term in the low energy limit was discovered in
\cite{VW95} and it was used to cancel part of the anomaly in
\cite{DLM95}. We already used it to cancel the anomalies when we
discussed Ho\v{r}ava-Witten theory.\\
\\
The variation of the Green-Schwarz term gives\footnote{Note that
for $\epsilon\rightarrow 0$ the boundary contribution vanishes.}
\begin{equation}
\delta S_{GS}=\alpha_2 T_2\int G\wedge \delta X_7=\alpha_2 T_2\int
G\wedge dX_6^1=-\alpha_2 T_2\int dG\wedge X_6=-\alpha_2 (2\pi)
\int_{W^6}X_6^1.
\end{equation}
This leads to the anomaly polynomial
\begin{equation}
I_8^{(GS)}=-\alpha_2 (2\pi)X_8.
\end{equation}

\bigskip
$\bullet$ Finally, it turns out that the topological term in the
action has to be modified \cite{FHMM98} in order to cancel all the
anomalies. To do so we  solve the modified Bianchi identity
(\ref{modfivebranesource}) as we did in chapter 4. The general
solution is
\begin{equation}
G=dC-{2\pi\over T_2}dB\wedge d\rho+a{2\pi\over T_2} \rho
e_4/2+(a-1){2\pi\over T_2}d\rho\wedge e_3/2.
\end{equation}
The sign and coefficient of $B$ is chosen for later convenience.
$e_3$ is given by $e_4$ via descent equations, i.e. $e_4=de_3$ and
$\delta e_3=de_2^{(1)}$. However, $\rho e_4/2$ is singular at the
five-brane, whereas $d\rho e_3/2$ is regular because of the
definition of $\rho(r)$. As by imposing
(\ref{modfivebranesource}) we basically smoothed out the
five-brane we should expect regular behaviour at $r=0$. Therefore
we take $a=0$. If we require $G$ to be gauge invariant, $\delta
G=0$, we obtain the transformation behaviour of $B$ and $C$,
\begin{eqnarray}
\delta C&=&d\Lambda, \\
\delta B&=&e_2^{(1)}/2.
\end{eqnarray}
We see that the modified Bianchi identity leads to a non-trivial
relation between $C$ and $G$. In particular, it is no longer
clear how we should write the topological term of the
supergravity action. It turns out that all anomalies cancel if
this term is modified in a very special way. The modified
topological term in the presence of an M5-brane is postulated to
be \cite{FHMM98}
\begin{eqnarray}
\widetilde{S}_{top}&:=&-{1\over12\kappa_{11}^2}\lim_{\epsilon\to0}\int\left(C-{2\pi\over
T_2}B\wedge d\rho-{2\pi\rho e_3\over2T_2}\right)\wedge\nonumber\\
&&\wedge d\left(C-{2\pi\over T_2}B\wedge d\rho-{2\pi\rho
e_3\over2T_2}\right)\wedge
d\left(C-{2\pi\over T_2}B\wedge d\rho-{2\pi\rho e_3\over2T_2}\right).\nonumber\\
\end{eqnarray}
Let us comment a bit on the structure of this term. Firstly, we
note that it gives us the usual $CdCdC$-term together with a
variety of other contributions. It is interesting that the latter
are actually not present in the bulk but live ``close to the
brane'' as $d\rho$ and $\rho$ have support only there. One of the
basic properties of the topological term is that $\delta\int
CdCdC$ can at the most obtain boundary contributions, as $d^2C=0$.
This would no longer be the case if we had chosen
$\widetilde{S}_{top}\propto\int CGG$, because of the modified
Bianchi identity. In fact, we have $d(C-\ldots )=G+$corrections,
where the
corrections are chosen in such a way that $d(G+$corrections)=0.\\
To compare the coefficient to the one given in \cite{FHMM98} note
that using (\ref{relTk1}) and (\ref{relTk2}) with $n=1$ we find
\begin{equation}
-{1\over12\kappa_{11}^2}=-{2\pi\over6}\left({T_2\over2\pi}\right)^3.\label{relkT}
\end{equation}
Thus, the modified term can be rewritten as
\begin{eqnarray}
\widetilde{S}_{top}&:=&-{2\pi\over6}\lim_{\epsilon\to0}\int\left({T_2\over2\pi}C-B\wedge
d\rho-{\rho
e_3\over2}\right)\nonumber\\
&&\wedge d\left({T_2\over2\pi}C-B\wedge d\rho-{\rho
e_3\over2}\right)\wedge
d\left({T_2\over2\pi}C-B\wedge d\rho-{\rho e_3\over2}\right).\nonumber\\
\end{eqnarray}
Now let us calculate its variation,
\begin{eqnarray}
\delta\widetilde{S}_{top}&=&-{2\pi\over6}\lim_{\epsilon\to0}\int\left({T_2\over2\pi}d\Lambda-d\left({\rho
e_2^{(1)}\over2}\right)\right)\nonumber\\
&&\wedge d\left({T_2\over2\pi}C-B\wedge d\rho-{\rho
e_3\over2}\right)\wedge d\left({T_2\over2\pi}C-B\wedge d\rho-{\rho
e_3\over2}\right)\nonumber\\
&=&-{2\pi\over6}\lim_{\epsilon\to0}\int
{T_2\over2\pi}d\Lambda\wedge
d\left({T_2\over2\pi}C-B\wedge d\rho-{\rho e_3\over2}\right)\wedge d\left({T_2\over2\pi}C-B\wedge d\rho-{\rho e_3\over2}\right)\nonumber\\
&&+{2\pi\over6}\lim_{\epsilon\to0}\int d\left({\rho
e_2^{(1)}\over2}\right)\wedge
d\left({T_2\over2\pi}C-B\wedge d\rho-{\rho e_3\over2}\right)\wedge d\left({T_2\over2\pi}C-B\wedge d\rho-{\rho e_3\over2}\right)\nonumber\\
&=:&\delta\widetilde{S}_{top}^{(1)}+\delta\widetilde{S}_{top}^{(2)}.
\end{eqnarray}
We integrate by parts using the fact that the boundary of our
space is given by $-S_{\epsilon} W^6$ and note that $d\rho$ has
only components perpendicular to $S_{\epsilon}W^6$,
\begin{eqnarray}
\delta\widetilde{S}_{top}^{(1)}&=&{T_2\over6}\lim_{\epsilon\to0}\int_{S_{\epsilon}W^6}
\Lambda\wedge
d\left({T_2\over2\pi}C-B\wedge d\rho-{\rho e_3\over2}\right)\wedge d\left({T_2\over2\pi}C-B\wedge d\rho-{\rho e_3\over2}\right)\nonumber\\
&=&{T_2\over6}\lim_{\epsilon\to0}\int_{S_{\epsilon}W^6}
\Lambda\wedge
\left({T_2\over2\pi} dC-{\rho e_4\over2}\right)\wedge \left({T_2\over2\pi} dC-{\rho e_4\over2}\right)\nonumber\\
&=&{T_2\over6}\lim_{\epsilon\to0}\int_{S_{\epsilon}W^6}\Lambda\wedge
{T_2\over2\pi}dC\wedge
{T_2\over2\pi}dC\nonumber\\
&&+{T_2\over6}\lim_{\epsilon\to0}\int_{S_{\epsilon}W^6}\rho^2\Lambda\wedge{e_4\over2}
\wedge
{e_4\over2}\nonumber\\
&&-{T_2\over3}\lim_{\epsilon\to0}\int_{S_{\epsilon}W^6}\Lambda\wedge
{T_2\over2\pi}dC\wedge {\rho e_4\over2}\nonumber\\
&=&{(T_2)^2\over6\pi}\int_{W^6}\Lambda\wedge dC,
\end{eqnarray}
where we used (\ref{inteis2}), the fact that $\Lambda$ and $dC$
are regular on the brane and $\rho(0)=-1$ in the last line.
Similarly, we obtain
\begin{eqnarray}
\delta\widetilde{S}_{top}^{(2)}&=&-{2\pi\over6}\lim_{\epsilon\to0}\int_{S_{\epsilon}W^6}{\rho
e_2^{(1)}\over2}\wedge
d\left({T_2\over2\pi}C-B\wedge d\rho-{\rho e_3\over2}\right)\wedge d\left({T_2\over2\pi}C-B\wedge d\rho-{\rho e_3\over2}\right)\nonumber\\
&=&-{2\pi\over6}\lim_{\epsilon\to0}\int_{S_{\epsilon}W^6}{\rho
e_2^{(1)}\over2}\wedge
\left({T_2\over2\pi} dC-{\rho e_4\over2}\right)\wedge \left({T_2\over2\pi} dC-{\rho e_4\over2}\right)\nonumber\\
&=&-{2\pi\over6}\lim_{\epsilon\to0}\int_{S_{\epsilon}W^6}{
e_2^{(1)}\over2}\wedge {e_4\over2}\wedge
{e_4\over2}\rho^3\nonumber\\
&&-{2\pi\over6}\lim_{\epsilon\to0}\int_{S_{\epsilon}W^6}{\rho
e_2^{(1)}\over2}\wedge {T_2\over2\pi}dC\wedge
{T_2\over2\pi}dC\nonumber\\
&&+{2\pi\over3}\lim_{\epsilon\to0}\int_{S_{\epsilon}W^6}{
e_2^{(1)}\over2}\wedge {T_2\over2\pi}dC\wedge {e_4\over2}\rho^2\nonumber\\
&=&{2\pi\over48}\lim_{\epsilon\to0}\int_{S_\epsilon W^6}e_4\wedge
e_4\wedge e_2^{(1)}. \label{inflowtop}
\end{eqnarray}
Before we can write down the anomaly polynomials corresponding to
these variations we must present a formula which holds for our
very specific geometry. In \cite{BC97}, \cite{FHMM98} and
\cite{BB99}, it was shown that
\begin{equation}
{1\over2}\int_{S_{\epsilon}W^6} e_4\wedge e_4\wedge
e_2^{(1)}=\int_{W^6} (p_2(NW^6))^{(1)}.\label{eeep}
\end{equation}
Here $(p_2(NW^6))^{(1)}$ is obtained from $p_2(NW^6)$ via descent
equations and $p_2(NW^6)$ is the second Pontrjagin class given by
\begin{eqnarray}
p_2(NW^6)={1\over8}\left({1\over2\pi}\right)^4\left[({\rm
tr}R_{\bot}^2)^2-2\ {\rm tr}R_{\bot}^4\right],\label{P2}
\end{eqnarray}
where $R_{\bot}$ is the curvature of the $SO(5)$-gauge field
$A_{\mu}=\omega_{\mu \hat m}^{\hat n}$. Using this it is easy to
write down the anomaly polynomials
\begin{eqnarray}
I_8^{(top,1)}&=&{(T_2)^2\over6\pi}dC\wedge dC={T_5\over3}dC\wedge dC, \\
I_8^{(top,2)}&=&{2\pi\over24}p_2(NW^6).
\end{eqnarray}

\bigskip
\begin{center}
{\bf Anomaly cancellation for the M5-brane}
\end{center}
Having presented all the preliminaries we are now ready to tackle
the problem of anomaly cancellation. After the field content is
determined the anomaly can be written down immediately, making
use of the general formulae in (\ref{anomalypolynomials6}). The
gravitational anomaly reads
\begin{eqnarray}
I_8^{(tangent)}&=&I_{grav}^{(3-form)}(\widetilde{R})+2I_{grav}^{(1/2)}(\widetilde{R})\nonumber\\
&=&-{1\over(2\pi)^34!}\left({1\over8}{\rm
tr}\widetilde{R}^4-{1\over32}({\rm tr}\widetilde{R}^2)^2\right).
\end{eqnarray}
This anomaly is called the {\em tangent bundle} anomaly. But this
is not the only anomaly to occur. Diffeomorphisms of the normal
bundle are translated into $SO(5)$ gauge transformations of the
gauge field $A_{\mu}:=\omega_{\mu \hat m}^{\hat n}$. This leads to
an anomaly polynomial
\begin{equation}
I_8^{(normal)}={1\over(2\pi)^34!}\left({1\over8}{\rm
tr}R_{\bot}^4+{1\over16}{\rm tr}\widetilde{R}^2 {\rm
tr}R_\bot^2-{3\over32}({\rm tr}R_{\bot}^2)^2\right)
\end{equation}
which was calculated in \cite{Wi96b}. In fact this anomaly can be
understood as sort of a gauge and mixed anomaly in six dimensions
with the curvature given by $\widetilde{R}$ and the gauge field
given by $R_{\bot}$. The naive application of
(\ref{anomalypolynomials6}) gives us the structure of the first
two terms, although we do not get the coefficients right. The
third term seems more mysterious, as it is not contained in ${\rm
ch}(R_{\bot})$. The solution to this puzzle comes from the fact
that the ``gauge field" ${1\over4}\o_{\m\hat n\hat p}\g^{\hat
n\hat p}$ in (\ref{derW6}) transforms in the spinor
representation. Therefore ${\rm ch}(F)$ has to be replaced by $\tr
\exp\left({i\over2\pi}{1\over4}R_{\bot\hat m\hat n}\g^{\hat m\hat
n}\right)=:{\rm ch}(S(NW^6))$ in the relevant formulae for the
anomaly polynomials. That is the polynomial for the spinor fields
is $I_8^{(1/2)}=-2\pi\left[{1\over2} \hat A(W^6){\rm
ch}(S(NW^6)\right]$ where the factor of ${1\over2}$ comes from a
chirality projector \cite{AGG84, Wi96b, BM03c}. As usual the
polynomial for the self-dual three-form is given by
$I_8^{(3-form)}=-2\pi\left[-{1\over8}L(W^6)\right]$ and we find
indeed
\begin{equation}
-2\pi\left[{1\over2} \hat A(W^6){\rm
ch}(S(NW^6))-{1\over8}L(W^6)\right]_8=I_8^{(tangent)}+I_8^{(normal)}.
\end{equation}
More details can be found
in \cite{Wi96b} and \cite{BM03c}.\\

Now let us collect all the terms which might lead to an anomaly
and see whether anomaly cancellation can be established. We have
\begin{eqnarray}
I_8&=&I_8^{(coup)}+I_8^{(GS)}+I_8^{(top,1)}+I_8^{(top,2)}+I_8^{(tangent)}+I_8^{(normal)}\nonumber\\
&=&\alpha_1 T_5dC\wedge dC -\alpha_2 (2\pi)X_8+{T_5\over3}dC\wedge
dC+{2\pi\over24}p_2(NW^6)\nonumber\\
&&-{1\over(2\pi)^34!}\left({1\over8}{\rm tr}\widetilde{R}^4-{1\over32}({\rm tr}\widetilde{R}^2)^2\right)\nonumber\\
&&+{1\over(2\pi)^34!}\left({1\over8}{\rm
tr}R_{\bot}^4+{1\over16}{\rm tr}\widetilde{R}^2 {\rm tr}R_\bot^2
-{3\over32}({\rm tr}R_{\bot}^2)^2\right)\nonumber\\
&=&\left(\alpha_1 +{1\over3}\right)T_5dC\wedge dC -\alpha_2
(2\pi){1\over(2\pi)^34!}\left({1\over8}{\rm tr}R^4-{1\over32}({\rm tr}R^2)^2\right)\nonumber\\
&&+{2\pi\over24}{1\over8}\left({1\over2\pi}\right)^4\left[({\rm
tr}R_{\bot}^2)^2-2{\rm tr}R_{\bot}^4\right]
-{1\over(2\pi)^34!}\left({1\over8}{\rm tr}\widetilde{R}^4-{1\over32}({\rm tr}\widetilde{R}^2)^2\right)\nonumber\\
&&+{1\over(2\pi)^34!}\left({1\over8}{\rm
tr}R_{\bot}^4+{1\over16}{\rm tr}\widetilde{R}^2 {\rm
tr}R_\bot^2-{3\over32}({\rm tr}R_{\bot}^2)^2\right).
\end{eqnarray}
We see that a necessary requirement for anomaly cancellation is
\begin{equation}
\alpha_1=-{1\over3}.
\end{equation}
To proceed we need to rewrite $X_8$ in terms of Pontrjagin
classes. It reads\footnote{The first Pontrjagin class is given by
$p_1(TM_{11})=-{1\over2}\left(1\over2\pi\right)^2{\rm tr}R^2$ and
$p_2$ was given in (\ref{P2}).}
\begin{equation}
X_8={2\pi\over48}\left({p_1(TM_{11})^2\over4}-p_2(TM_{11})\right).
\end{equation}
Now we can use the properties of the Pontrjagin classes, namely
\begin{eqnarray}
p_1(TM_{11}|_{W^6})&=&p_1(TW^6\oplus NW^6)=p_1(TW^6)+p_1(NW^6),\\
p_2(TM_{11}|_{W^6})&=&p_2(TW^6\oplus
NW^6)=p_2(TW^6)+p_2(NW^6)+p_1(TW^6)p_1(NW^6),\nonumber\\
\end{eqnarray}
to rewrite $X_8$ as
\begin{equation}
X_8={2\pi\over48}\left({p_1(TW^6)^2\over4}-p_2(TW^6)+{p_1(NW^6)^2\over4}-p_2(NW^6)-{1\over2}p_1(TW^6)p_1(NW^6)\right),
\end{equation}
and hence
\begin{equation}
X_8={1\over(2\pi)^34!}\left\{\left({1\over8}{\rm
tr}\widetilde{R}^4-{1\over32}({\rm
tr}\widetilde{R}^2)^2\right)+\left({1\over8}{\rm
tr}R_{\bot}^4-{1\over32}({\rm
tr}R_{\bot}^2)^2\right)-{1\over16}{\rm tr}\widetilde{R}^2{\rm
tr}R_{\bot}^2\right\}.
\end{equation}
Using this we find indeed that the anomaly cancels,
\begin{equation}
I_8=0,
\end{equation}
provided we choose\footnote{With this choice of $\a_2$ the
Green-Schwarz term reads $S_{GS}=-{T_2\over2\pi}\int G\w X_7$.
Using (\ref{relTk1}) and (\ref{relTk2}) we get
$S_{GS}=-\left({1\over4\pi\k^2_{11}}\right)^{1/3}\int G\w X_7$.
This is exactly the Green-Schwarz term (\ref{Green-Schwarz term})
we used to cancel the anomalies in M-theory on $S^1/\mathbb{Z}_2$
in chapter 4.}
\begin{equation}
\alpha_2=-{1\over2\pi}.
\end{equation}
We conclude that M-theory in the presence of M5-branes is free of
local gauge and gravitational anomalies.\\

\begin{center}
{\bf Anomaly cancellation for the M2-brane}
\end{center}
There are no perturbative anomalies associated with membrane
zero-modes as the world-volume of an M2-brane is
three-dimensional and the world-volume theory is non-chiral. The
only possible anomalies related to membrane zero-modes are global
ones. Their cancellation was shown in \cite{Wi96}.

\chapter{Realistic Physics from Singularities}
Before taking the next step let us pause for a moment and
summarize our results. We have seen that the low energy effective
theory of M-theory, eleven-dimensional supergravity, can be
coupled both to M2- and M5-branes. We showed that M-theory
containing these objects is a consistent quantum theory in the
sense that it is anomaly free. Furthermore, we found a number of
vacua of M-theory, some of which preserve $\mathcal{N}=1$
supersymmetry. In particular the manifold $(\mathbb{R}^4\times
X,\eta \times g)$ with $(X,g)$ a $G_2$-manifold seemed to be a
promising candidate for a realistic four-dimensional theory.
However, the explicit Kaluza-Klein compactification of M-theory
on this manifold, which was performed in section 5.3, showed that
the resulting four-dimensional effective theory contains only
Abelian gauge groups and neutral chiral fermions. So if we want
to reproduce the basic features of the standard model we have to
work harder.\\
In this chapter we show that compactifications on singular spaces
lead to field theories with the desired properties. To be able to
do so we have to explain the mechanism of symmetry enhancement
and consider the notion of an $ADE$ singularity. Both of these
will be important to understand the duality between the heterotic
string on $T^3$ and supergravity on a $K3$ surface. This duality
in turn will be exploited to show that singular $G_2$-manifolds
lead to non-Abelian gauge groups. We conclude the chapter by
explaining how charged chiral fermions arise in such a model.

\section{Enhanced Gauge Symmetry}
Consider the compactification of a string theory on a space of
given topology. Usually this space can be equipped with a family
of metrics which are parameterized by points in the moduli space
of the theory. Suppose this theory possesses a $U(1)^k$ gauge
symmetry for generic points in moduli space. If for certain
special points in moduli space this symmetry group is enlarged to
a non-Abelian gauge group we speak of {\em enhanced gauge
symmetry}. This mechanism of symmetry enhancement occurs at
various places in string theory. In this section we will describe
the phenomenon by explaining the simplest example, namely symmetry
enhancement in the case of compactifying the closed bosonic
string on $\mathbb{R}^{25}\times S^1$.\\
General aspects of the bosonic string are discussed in
\cite{GSW87} and \cite{Pol}. The only thing that is important to
us is that after quantization a string is characterized by a set
of quantum numbers. These are its momentum and its excitation
levels. In the particular case of the closed string we have two
possible excitations, those of left-moving and of right-moving
modes. If the space in which the string moves is not simply
connected new quantum numbers occur. In particular, for the case
of $\mathbb{R}^{25}\times S^1$ we have one quantum number, which
indicates how often the string winds around the circle. It is
referred to as the winding number. So let $R$ be the radius of
$S^1$, $N$ and $\widetilde{N}$ the excitation levels for left- and
right-moving modes and $w$ the winding number of the string around
$S^1$. We also get a quantization of the momentum in the compact
direction. The corresponding quantum number is denoted $n$. In
this case the mass formula for the string is given by
\cite{GSW87}, \cite{Pol}
\begin{equation}
m^2={n^2\over
R^2}+{w^2R^2\over\alpha^{'2}}+{2\over\alpha'}(N+\widetilde{N}-2),
\end{equation}
subject to the constraint
\begin{equation}
0=nw+N-\widetilde{N}.\label{cons}
\end{equation}
This mass spectrum can be understood easily by looking at the
individual terms. The first term obviously gives the contribution
of the compact momentum, the second term gives the mass of the
winding modes and finally we also have the contribution of the
string oscillations together with a zero-point energy.\\
Now let us consider what kind of massless states we can possibly
get. For generic values of $R$ the only solutions are $n=w=0$ and
$N=\widetilde{N}=1$. If we denote the creation operators of left-
and right-moving modes by $\alpha_{-1}^M$ and
$\widetilde{\alpha}_{-1}^M$ with $M\in\{0,\ldots ,25\}$ we can
construct two massless vector fields,\footnote{In fact the vector
fields are the symmetric and anti-symmetric linear combination of
these states.}\\
\parbox{14cm}{
\begin{eqnarray}
&\alpha_{-1}^{\mu}\widetilde{\alpha}_{-1}^{25}|0;k\rangle,&\nonumber\\
&\alpha_{-1}^{25}\widetilde{\alpha}_{-1}^{\mu}|0;k\rangle,&\nonumber
\end{eqnarray}}\hfill\parbox{8mm}{\begin{eqnarray}\end{eqnarray}}
with $\mu\in\{0,1,\ldots ,24\}$. So for generic points in moduli
space, i.e. for generic radii of $S^1$, we expect a $U(1)\times
U(1)$ gauge symmetry. As always symmetry transformations are
generated by conserved charges. A careful analysis \cite{Pol}
shows that the charges of massive states under the $U(1)\times
U(1)$ symmetry are the compact
momentum $p_{25}$ and the winding number $w$.\\
The whole picture changes dramatically when we consider the
spectrum of massless fields at the specific radius
$R=\sqrt{\alpha'}$. In that case the condition for massless
fields reads
\begin{equation}
0=n^2+w^2+2N+2\widetilde{N}-4,
\end{equation}
where again we have to impose the constraint (\ref{cons}). Next to
the solutions presented above these conditions are satisfied by
the following combinations,\\
\parbox{14cm}{
\begin{eqnarray}
n=w=\pm1,\ N=0, \ \widetilde{N}=1,\nonumber\\
n=-w=\pm1,\ N=1,\ \widetilde{N}=0,\nonumber\\
n=\pm2,\ w=N=\widetilde{N}=0,\nonumber\\
w=\pm2,\ n=N=\widetilde{N}=0.\nonumber
\end{eqnarray}}\hfill\parbox{8mm}{\begin{eqnarray}\end{eqnarray}}
Consider the first two cases, from which we get four new gauge
bosons. The crucial point is that they also carry compact
momentum and winding number, so they are {\em charged gauge
bosons}. The only consistent theory of charged massless vectors
is non-Abelian gauge theory. Indeed the six sets of vector fields
combine to two sets of fields, each transforming under $SU(2)$.
The symmetry group gets enhanced from $U(1)\times U(1)$ to
$SU(2)\times SU(2)$ at this specific radius.\\
As we mentioned already this mechanism does not only occur for
the bosonic string. In particular we also get symmetry
enhancement in the case of a compactification of the heterotic
string on $\mathbb{R}^7\times T^3$. The details of this
enhancement are of no importance to us, however, the basic
mechanism is similar to the one just described.

\section{A Duality Conjecture}
It was understood a few years ago that the known string theories
are not independent but that they are related to each other and to
eleven-dimensional supergravity via duality relations. In one of
those relations non-Abelian gauge groups occurred in M-theory from
spaces carrying $ADE$ singularities. This is why we want to
consider the evidence that led to the following conjecture.

\bigskip
{\bf Conjecture}\\
M-theory on $\mathbb{R}^7\times K3$ is dual to the heterotic
string theory on $\mathbb{R}^7\times T^3$.\\
\\
The first piece of evidence confirming the conjecture concerns
the moduli spaces of the two theories. The moduli space of vacua
of the heterotic string on $\mathbb{R}^7\times T^3$ is
\begin{equation}
\mathcal{M}=\mathbb{R}^+\times\mathcal{M}_1
\end{equation}
with
\begin{equation}
\mathcal{M}_1=SO(19,3;\mathbb{Z})\backslash
SO(19,3;\mathbb{R})/(SO(19)\times SO(3)),
\end{equation}
the usual Narain moduli space \cite{Na85}. $\mathbb{R}^+$
parameterizes the possible values of the string coupling
constant.\\
Now let us consider the moduli space of vacua of
eleven-dimensional supergravity on $\mathbb{R}^7\times K3$. To be
more precise we consider the set of vacua of eleven-dimensional
supergravity with vanishing $G$-flux and $\langle\psi\rangle=0$.
The topology of the vacuum is taken to be a direct product of
$\mathbb{R}^7$ and the complex manifold $K3$. We specify the
metric on $\mathbb{R}^7$ to be flat, i.e. $(\mathbb{R}^7,\eta)$ is
seven-dimensional Minkowski space and we ask what metrics may be
allowed on $K3$. Given this setup the equations of motion reduce
to the condition of Ricci-flatness, exactly as in our
consideration of Ricci-flat solutions in section 5.1. Thus, we can
allow for any metric on $K3$ provided it is Ricci-flat. But we
know that in the case of $K3$ all Einstein metrics are Ricci-flat.
This amounts to saying that the moduli space of eleven-dimensional
supergravity on $\mathbb{R}^7\times K3$ is isomorphic to the
moduli space of Einstein metrics on $K3$. This space was given in
proposition 2.10 and it is in fact $\mathcal{M}$. In that case the
$\mathbb{R}^+$ parameterizes the volume of the $K3$ surface. So we
see that the moduli spaces of both theories are the same, even if
their origin is very different.

\bigskip
Next we want to show that both compactifications lead to the same
field content of the effective theory in seven dimensions. We
start by presenting the results for the heterotic string on $T^3$
without giving a detailed derivation which can be found in the
literature
\cite{GSW87}, \cite{Pol}, \cite{Ac02}.\\
To be able to determine the field content of the compactified
heterotic string we need to study its low energy effective
supergravity theory in ten dimensions. It is $\mathcal{N}=1$
supergravity coupled to ten-dimensional super-Yang-Mills theory
with gauge group $SO(32)$ or $E_8\times E_8$ depending on the
heterotic theory we started with. Its field content consists of a
metric $g$, a two-form $B$, a dilaton $\phi$, the non-Abelian
gauge fields and their fermionic
superpartners.\\
$\mathcal{N}=1$ supersymmetry in ten dimensions corresponds to
sixteen supersymmetry generators. To see how much supersymmetry is
conserved in the compactified theory we are looking for the number
of covariantly constant spinors on $\mathbb{R}^7\times T^3$,
similar to what we did in section 5.2. Compactifying on
$\mathbb{R}^7\times T^3$ implies the split ${\bf 16}={\bf
8}\otimes {\bf 2}$ and as $(\mathbb{R}^7,\eta)$ is flat we have
eight covariantly constant spinors on $\mathbb{R}^7.$ But $T^3$ is
flat as well, of course, so we get another two covariantly
constant spinors and hence we are left
with $\mathcal{N}=2$ supersymmetry in seven dimensions.\\
Now let us consider how the fields arise. We expand the two-form
\begin{equation}
B(x,y)=\sum_{i=1}^{b^2(T^3)} s^{(i)}(x)\wedge
\omega^{(i)}(y)+\sum_{i=1}^{b^1(T^3)} A^{(i)}(x)\wedge
\eta^{(i)}(y)+ H(x)+\ldots \ \mbox{(massive)}.\label{Bexpand}
\end{equation}
We have $b^2(T^3)=3$ and $b^1(T^3)=3$, so we get three Abelian
gauge fields and three scalars from the compactification of the
$B$-field. Another three gauge fields arise from the metric, as
the isometry group of $T^3$ is given by $U(1)^3$.\\
The family of flat metrics on $T^3$ is six-dimensional. The six
parameters are the radii of the three $S^1$ together with three
angles between them. The quantum fluctuations of these parameters
lead to another six scalar fields.\\
Furthermore, the fluctuations of the dilaton gives yet another
scalar. The origin of the remaining fields is a bit more technical
and we refer the reader to the literature \cite{GSW87},
\cite{Pol}, \cite{Ac02} for details. The result is that the
special structure of the heterotic string gives us another 16
Abelian gauge fields and 48
scalar fields.\\
The fermionic fields can be determined from supersymmetry. To
summarize, the effective field theory of the heterotic string on
$T^3$ is given by an $\mathcal{N}=2$ supergravity theory in seven
dimensions coupled to 58 scalar multiplets and 22 Abelian vector
multiplets.

\bigskip
Now let us see whether this field content can be reproduced from a
compactification of M-theory on $K3$. We start once again with the
amount of conserved supersymmetry. As in section 5.2 we are
looking for the number of covariantly constant spinors on
$\mathbb{R}^7\times K3$. As usual we split ${\bf 32}={\bf
8}\otimes{\bf 4}$. But we know that the holonomy group of $K3$ is
$SU(2)$, hence, according to theorem 2.23, it admits two
covariantly constant spinor. Thus, we are left with sixteen
supersymmetry generators or $\mathcal{N}=2$ in seven dimensions.
We see that the amount of
conserved supersymmetry is the same in both theories.\\
Next let us determine the field content of M-theory on $K3$. The
gauge fields come from an expansion of the $C$-field as in
(\ref{KKCfield}),
\begin{equation}
C(x,y)=\sum_{i=1}^{b^2(K3)} A^{(i)}(x)\wedge \omega^{(i)}(y)+
H(x)+\ldots \ \mbox{(massive)}.\label{Cexpans}
\end{equation}
The Betti number $b^2(K3)$ gives 22 Abelian gauge fields in seven
dimensions, so we get a gauge group $U(1)^{22}$.\\
The fluctuations of the metric are more complicated to derive.
They can be calculated explicitly with the methods of section 5.3,
using once again that for $K3$ surfaces an Einstein metric is
automatically Ricci flat. We will follow a different tack which is
somewhat more abstract but much quicker. In section 5.3 we noted
already that the result of $b^3(X)$ scalar fields might have been
obtained by studying the moduli space of the theory. In fact the
moduli space (\ref{ModspaceK3}) parameterizes all possible
Einstein metrics on $K3$. Writing $g=\langle g\rangle+h$, with $h$
the fluctuation of the metric, we want both $g$ and $\langle g
\rangle$ to be Einstein, so both are described by points in moduli
space. The fluctuations are infinitesimal so these points are
close to each other and can be connected by a continuous path in
moduli space. As the moduli space (\ref{ModspaceK3}) has dimension
58 it is parameterizes by 58 scalar fields $s$. These are the
scalar fields we observe in the seven-dimensional theory.\\
As always fermions are determined from supersymmetry.

\bigskip
Collecting all the listed fields we see that the field content of
both theories does indeed coincide. In fact, it was shown in
\cite{Wi95} that the low energy effective actions of M-theory on
$K3$ and the heterotic string on $T^3$ are also the same.

\bigskip
However, the situation turns out to be even richer. Above we only
listed massless fields for theories at generic points in the
moduli space. It is known that the moduli space of the heterotic
string on $T^3$ contains points at which new massless modes occur.
At these points we get an enhanced gauge symmetry through a
mechanism which is similar to the one we described in the last
section. If the two theories are really related to each other by a
duality transformation there must be points in the moduli space of
M-theory on $K3$ at which these enhanced gauge symmetries occur.
To see how this might be possible it is useful to recall the
construction of a $K3$ surface from blow-ups of singular spaces.
This construction was explained in detail in chapter 2, where we
excised the singularities and glued in an Eguchi-Hanson space
instead. It is exactly the limit in moduli space at which the
smooth $K3$ gets singular, i.e. when the two-sphere sitting on the
tip of the Eguchi-Hanson space collapses, where we have a chance
that new non-Abelian gauge groups occur. This will indeed be the
case a we will explain in the following section.

\section{A-D-E Singularities in $G_2$-Manifolds}
In one of the examples given in section 2.5 we saw that a smooth
compact $G_2$-manifold has a limit in which it looks locally like
$\mathbb{R}^3\times \mathbb{C}^2/\Gamma_{A_1}$. This was obtained
as a local model of the singularities in $T^3\times K3/\langle
\beta,\gamma\rangle$. It is natural to expect that depending on
the choice of $\langle\beta,\gamma\rangle$ we might arrive at
singularities that can be modelled locally as $\mathbb{R}^3\times
\mathbb{C}^2/A_n$. In fact, it can be shown \cite{Ac98}, that it
is possible to get singularities of the form $\mathbb{R}^3\times
\mathbb{C}^2/\Gamma_{ADE}$ with $\Gamma_{ADE}$ as given in
appendix F. We now want to analyze the physics of M-theory on
these singular spaces. All the effects of a singularity should be
local and independent of the structure of the manifold far away
from the singularity. Thus, it is sufficient to study M-theory on
the non-compact space $\mathbb{R}^4\times \mathbb{R}^3\times
\mathbb{C}^2/\Gamma_{ADE}$, where the first factor is ordinary
Minkowski space. The locus of the singularity in this space is
given by $\mathbb{R}^4\times \mathbb{R}^3\times \{0\}$. The basic
ideas of this section were developed in \cite{Ac98}.

\begin{center}
{\bf M-theory physics at the singularity}
\end{center}
Consider M-theory on the smooth non-compact space
$\mathbb{R}^4\times\mathbb{R}^3\times \mathbb{EH}$, where
$\mathbb{EH}$ is an Eguchi-Hanson space. This space is the blow-up
of $\mathbb{R}^4\times\mathbb{R}^3\times
\mathbb{C}^2/\Gamma_{A_1}$ and tends to it if the radius $r_0$ of
the two-sphere sitting on the tip of $\mathbb{EH}$ goes to zero.
If we interpret this sphere as a two-cycle in $\mathbb{EH}$ we can
construct a (compactly supported) dual harmonic two-form
$T_2\omega$. In fact, it is the only harmonic two-form in
$\mathbb{EH}$. Looking at the explicit structure of the metric
(\ref{EH}) we see that size and shape of the two-sphere sitting at
$r=r_0$ are parameterized by the three scalars $r_0$, $\theta$,
$\phi$. Of course in order for
$\mathbb{R}^4\times\mathbb{R}^3\times \mathbb{EH}$ to be a
consistent vacuum of eleven-dimensional supergravity we need to
have\footnote{Recall that the metric on $\mathbb{EH}$ is
Ricci-flat.} $\langle G\rangle=\langle\psi\rangle=0$, consistent
with the requirements
for compactifications on $G_2$-manifolds.\\
Next we want to perform a Kaluza-Klein analysis of
eleven-dimensional supergravity on $(\mathbb{R}^4\times
\mathbb{R}^3)\times\mathbb{EH}$. With the usual ansatz for the
$C$-field we find exactly one $U(1)$ gauge field in seven
dimensions. This gauge field combines with the fluctuations of the
three scalars parameterizing the sphere to the bosonic part of a
vector multiplet in seven dimensions. We conclude that the
spectrum of M-theory on $\mathbb{R}^4\times\mathbb{R}^3\times
\mathbb{EH}$
consists of an Abelian vector multiplet.\\
Now we turn to M-theory on the singular space
$\mathbb{R}^4\times\mathbb{R}^3\times \mathbb{C}^2/\Gamma_{A_1}$.
As we mentioned already we expect some sort of enhanced symmetry
to occur for these spaces from the conjectured duality to the
heterotic string. Let us explain the details of this mechanism. As
in our example given in section 7.1 we expect that the theory
contains some states which are massive as long as $r_0\neq 0$ and
the two-sphere is finite. For zero volume of the sphere the states
should become massless in order for enhanced symmetry to occur.
But there is a very natural way how this can happen in M-theory.
We have seen that among the fundamental constituents of M-theory
we have M2-branes, which are 2+1-dimensional objects. If these
wrap the two-sphere they appear as particles from the
seven-dimensional point of view. Furthermore, we know that the
M2-brane couples to the vector field $A$. The easiest way to see
this is to recall the standard coupling of the M2-brane to the
$C$-field, $\int_{M2}C$. With $C=A\omega$ and $M2=S^2\times\gamma$
for some path $\gamma$ in $\mathbb{R}^4\times\mathbb{R}^3$ we get
\begin{equation}
T_2\int_{M2}C=\int_{\gamma}A\int_{S^2}T_2\omega=\int_{\gamma}A.
\end{equation}
The M2-brane has a mass which is basically its tension times its
world-volume. Thus, it will have minimal energy if it wraps the
minimal two-sphere available in a given space.\footnote{In fact
the minimal two-sphere is a calibrated submanifold in the sense
of chapter 2. This has some interesting consequences, it ensures
for example that the mass value obtained from classical
calculations will not be corrected quantum mechanically. See
\cite{Ac02} for some details.}\\
The M2-brane wrapped around the cycle in the opposite orientation
has the opposing $U(1)$ charge to the previous one. So in the case
of a collapsing two-cycle we get two additional massless states of
opposite charge. Similar to the mechanism in 7.1 these combine
with the $U(1)$ gauge field to an enhanced symmetry group $SU(2)$.
In that case we expect that the effective theory in seven
dimensions is super-Yang-Mills theory with gauge group $SU(2)$.\\
More generally we expect the effective seven-dimensional theory of
M-theory on $\mathbb{R}^4\times\mathbb{R}^3\times
\mathbb{C}^2/\Gamma_{ADE}$ to be super-Yang-Mills theory with the
corresponding $ADE$ gauge group.

\begin{center}
{\bf $ADE$ singularities in $G_2$-manifolds and the M-theory
spectrum}
\end{center}
These considerations are very encouraging, as for the first time
we get non-Abelian gauge groups from M-theory calculations.
However, we started from compactifications on $G_2$-manifolds to
ensure that our theory is $\mathcal{N}=1$ supersymmetric. To see
that this really is still possible we have to study in more
detail, how the singularities are actually embedded into our
compact $G_2$-manifold. So far the part of the total space which
is not Minkowski was $\mathbb{R}^3\times
\mathbb{C}^2/\Gamma_{ADE}$. It is clear that this space will not
carry a $G_2$-holonomy metric. However, in order to be able to go
through the mechanism just described the singularities in $X$ need
to have codimension four. For $\mathbb{R}^3\times
\mathbb{C}^2/\Gamma_{ADE}$ the singular space is
$\mathbb{R}^3\times \{0\}$. So we should expect that if we want to
embed $ADE$ singularities into $G_2$-manifolds $X$ the space of
singularities $Q$ in $X$ has to be three-dimensional. Near this
singularity, i.e. near a point on $Q$ we have a local description
of the space as $\mathbb{R}^3\times \mathbb{C}^2/\Gamma_{ADE}$.
Globally, however, it must look differently to maintain the
$G_2$-structure. We are thus led to consider $X$ as a $K3$
fibration over $Q$ with a warped metric. It is not at all clear
that there exists a $G_2$-manifold $X$ with these properties but
it can be shown \cite{Ac98}, \cite{Ac02} that this is indeed the
case\footnote{For an analysis of the global structure of these
spaces see \cite{Liu98}.}. Explicit examples of compact
$G_2$-manifolds carrying $ADE$ singularities can be found in
\cite{Jo00} and \cite{Ac98}. They are the singular limits of
smooth $G_2$-manifolds which are constructed using Joyce's method.
The singular limit of our second example in chapter 2 is the
simplest of these cases. Non-compact manifolds with $ADE$
singularities are relatively easy to construct \cite{Ac02},
\cite{AtW01}. The idea is to take a smooth $G_2$-manifold which is
topologically $\mathbb{C}^2\times Q$. Suppose $\mathbb{C}^2\times
Q$ admits an action of $SU(2)$ which leaves $Q$ invariant and acts
on $\mathbb{C}^2$ in the natural way. Then the singular manifold
can be constructed as $\mathbb{C}^2/\Gamma_{ADE}\times Q$.
Interestingly, the manifold (\ref{noncompG2}) which we presented
in chapter 2 satisfies all these requirements with $Q=S^3$, so we
can construct the desired manifolds from it.\footnote{In the case
of $\mathbb{EH}$ we noted that the space on the tip of the cone is
a calibrated manifold. In fact this is true for the base $S^3$ of
the metric (\ref{noncompG2}) and more generally for any locus $Q$
of $ADE$ singularities in a $G_2$-manifold \cite{Ac98},
\cite{Ac02}. They are associative
three-folds as defined in chapter 2.}\\
Now it remains to ask for the structure of the effective theory in
four dimensional Minkowski space. But as we know that near the
singularity the space $X$ looks like $\mathbb{R}^3\times
\mathbb{C}^2/\Gamma_{ADE}$, we know that the effective theory in
seven dimensions is super-Yang-Mills with an $ADE$ gauge group on
$\mathbb{R}^4\times Q$. To obtain the four-dimensional theory we
just have to do another Kaluza-Klein compactification of this
seven-dimensional theory on $Q$. This has been done\footnote{In
seven dimensions we have three scalars and a gauge field together
with their superpartners. Compactification on $\mathbb{R}^4\times
Q$ leads to an $SO(3,1)\times SO(3)$ symmetry. The gauge fields
transform as $({\bf 1}\otimes {\bf 3})\oplus ({\bf 4\otimes {\bf
1}})$ and the scalars as $({\bf 1}\otimes {\bf 3})$. This gives a
four-dimensional gauge fields and two copies of the ${\bf 3}$ of
$SO(3)$. The latter can be interpreted as one-forms on $Q$. They
are massless if they are zero-modes of the Laplacian on $Q$, so
there are $b^1(Q)$ of them. Compactification of the fermion fields
gives all the superpartners. Thus we are left with
super-Yang-Mills in four dimensions, coupled to $b^1(Q)$ chiral
supermultiplets.} \cite{Ac98}, \cite{Ac02} and the resulting
effective theory in four dimensions is $\mathcal{N}=1$
super-Yang-Mills theory with $b^1(Q)$ massless adjoint chiral
supermultiplets. Thus, we reached our goal to construct
four-dimensional non-Abelian gauge theories by compactifying
M-theory on
$G_2$-manifolds which carry $ADE$ singularities. \\
This discussion has been rather abstract and we did not calculate
things explicitly. The aim was to show that effective non-Abelian
gauge theories occur if M-theory is compactified on singular
$G_2$-manifolds. In the next chapter we will confirm the results
of this section by an anomaly analysis of the theory which tells
us that it is only consistent with non-Abelian gauge fields on the
singularities.

\section{Chiral Fermions from $G_2$-Manifolds}
The presence of chiral fermions on singularities of
$G_2$-manifolds can be shown in three different ways. The first
possibility is to consider the duality with the heterotic string
\cite{AcW01}. As we have not given the necessary material to
understand this string theory approach, we will not present it
here. Secondly, one can examine M-theory dynamics on particular
spaces with conical singularities. The basic results were
obtained in \cite{AtW01} and we will provide one interesting
example in this section. Finally, one might check the results by
doing an anomaly analysis. We will leave this for the next
chapter.

\vspace{1cm}
\begin{center}
{\bf M-theory dynamics on asymptotically conical spaces}
\end{center}
Let $X$ be a smooth compact $G_2$-manifold with a metric such that
on some chart of $X$ the metric is asymptotically conical. This
means that we can deform the metric on this chart until a conical
singularity develops in $X$.\footnote{No explicit example of such
a space is known, however, it is probable that $G_2$-manifolds can
develop conical singularities. Examples of compact weak
$G_2$-manifolds with conical singularities were constructed in
\cite{BM03a}.} We expect all the effects of the singularity to be
local, so it is sufficient to consider only the small
neighbourhood covered by the chart. But this amounts to saying
that the part of the manifold which is not covered by the chart is
irrelevant for dynamics related to the singularity. Thus, it is
consistent to consider non-compact asymptotically conical spaces.
This is useful as metrics of $G_2$-holonomy on these spaces can be
written down explicitly. Examples of this kind were first
constructed in \cite{BS89} and \cite{GPP90} and one of them was
already presented in section 2.5.\\
Let us consider in more detail what it means to study M-theory on
these spaces. From now on $X$ denotes the asymptotically conical
non-compact manifold. According to the definitions of chapter 2 a
manifold is asymptotically conical if we have for large $r$
\begin{equation}
ds^2\approx dr^2+r^2d\Omega^2_Y.
\end{equation}
This means that for large values of $r$, far away from the tip of
the manifold, the metric looks like a cone on some six-manifold
$Y$.\\
We want to try to do some first steps towards the quantum
behaviour of our theory, so we must both specify a background and
quantum fluctuations. In our case the background is given by a
specific choice of the $G_2$-metric and the vanishing $C$-field,
$\langle C\rangle=0$, together with zero expectation value for the
gravitino. But as we want to study M-theory on a non-compact
manifold  we also need to specify boundary conditions at infinity.
This can be done by specifying a metric on $Y$, because the
metric is the only non-vanishing background field. Given both the
boundary conditions and the background the fields may fluctuate in
a way that all the fluctuations still satisfy the boundary
conditions. This means in particular, that we cannot change the
structure of $Y$, as it is kept fixed. We chose the vacuum to be
asymptotically conical, so the only parameter in the metric on
$X$ which is left to get quantum corrections is $r_0$, which
characterizes the difference of the metric on $X$ from a conical
metric. Of course, we also allow for fluctuations of $C$ and
$\psi_M$, provided they vanish on the boundary. To be able to
interpret the fluctuations as physical fields, the corresponding
kinetic energy terms have to be finite.

\newpage
\begin{center}
{\bf Symmetries of M-theory on asymptotically conical spaces}
\end{center}
It is very important to understand the symmetries of this
particular setup. The symmetries of the problem are given by the
symmetries of the fields at the boundary, that is on $Y$. As
always, the solutions of a theory may or may not have the same
symmetries. If a solution has less symmetry than the theory we
speak of {\em broken symmetry}. Applied to our case this means
that symmetries of fields on $Y$ do not necessarily extend to $X$.
An unbroken symmetry is a symmetry of both the fields on $Y$ and
on $X$, hence, it also leaves fixed the fields in the interior. In
general there are symmetries of the metric, which we call
geometric symmetries, and symmetries coming from the $C$-field.
Clearly we have $C'=C+d\Lambda$, but as $C$ vanishes at infinity
the symmetries are generated by two-forms $\Lambda$ with
$d\Lambda=0$ at infinity. This in turn tells us immediately that
the symmetries of the theory which come from the $C$-field are
given by $U(1)^{b^2(Y)}\cong H^2(Y;U(1))$. For an unbroken
symmetry we certainly need $0=\delta C=d\Lambda$ everywhere. Thus,
these are given by $U(1)^{b^2(X)}\cong H^2(X;U(1))$.\footnote{As
the discussion is quite abstract at this point the reader might
want to consult \cite{AtW01} for more details. Also, in
\cite{BM03a} the symmetries of weak $G_2$-manifolds are discussed
and might provide further intuition
for the case at hand.}\\

\begin{figure}[t]
\centering
\includegraphics[width=0.6\textwidth]{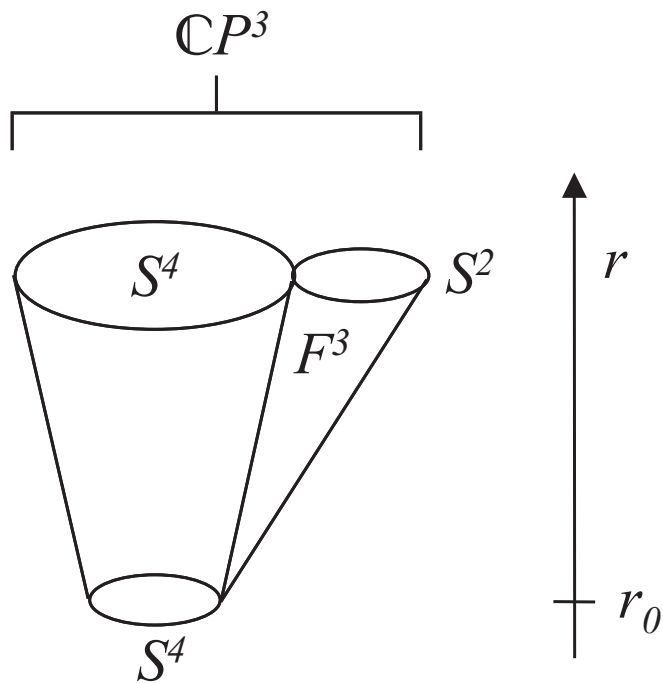}\\
\caption[]{Another non-compact $G_2$-manifold.} \label{S4S2}
\end{figure}

\begin{center}
{\bf Chiral fermions from singularities - an example}
\end{center}
Let us see how all this works in a specific example, which will
give us charged chiral fermions if M-theory is compactified on it.
The metric for this space is given by
\begin{equation}
ds^2_7={1\over1-\left({r_0\over
r}\right)^4}dr^2+{r^2\over4}(1-\left(r_0 \over
r\right)^4)(du_i+\epsilon_{ijk}A_ju_k)^2+{r^2\over
2}ds_4^2\label{CP3}.
\end{equation}
Here $ds_4^2$ is the standard metric on $S^4$, normalized in a way
that $\mathcal{R}_{rs}=3g_{rs}$, with $r,s\in\{1,2,3,4\}$. The
$u_i$ are any set of coordinates in $\mathbb{R}^3$ subject to the
additional condition
\begin{equation}
u_i^2=1.
\end{equation}
$r_0\in \mathbb{R}_0^+$ is a parameter. The $A_i$ are a set of
three local one-forms corresponding to the three almost complex
structure tensors which can be defined on $S^4$. See \cite{Jo00},
\cite{GPP90} and \cite{AtW01} for the details of this geometry.\\
We see that this metric is asymptotic to a cone on $Y$, with $Y$
an $S^2$-bundle over $S^4$. This bundle is isomorphic to
$\mathbb{CP}^3$, it is known as the {\em twistor space} on $S^4$.
The picture of these spaces is by now familiar, it is given in
figure \ref{S4S2}.
An important property of this metric is that if differs from a
conical metric by terms of order $(r_0/r)^4$, for $r\rightarrow
\infty$. The consequence of this fact is that fluctuations of
$r_0$ can be regarded as physical fields with finite kinetic
energy. This comes about as follows. Let $\langle g\rangle$ be the
metric as it is given above with fixed $\langle r_0\rangle$. Now
suppose $r_0$ fluctuates as $r_0=\langle r_0\rangle +\delta r_0$.
We write the corresponding metric as $g=\langle g\rangle+\delta
g$. If we define the norm of $\delta g$ as
\begin{equation}
|\delta g|^2:=\int_Xd^7x\sqrt{g}g^{ii'}g^{jj'}\delta g_{ij}\delta
g_{i'j'}
\end{equation}
we find that
\begin{equation}
|\delta g|^2<\infty,
\end{equation}
as $\delta g/g\sim r^{-4}$ and we integrate over a
seven-dimensional space. This means that in M-theory on
$\mathbb{R}^4\times X$ the kinetic energy associated with
fluctuations of $r_0$ is finite and they can be considered as a
physical field.\\
It is interesting to see, that the fields corresponding to a
fluctuation of $r_0$ really live only close to the tip of the
cone. Denote $g(\langle r_0\rangle):=g_{\langle
r_0\rangle}(x,y,r)$ the metric on the entire space
$\mathbb{R}^4\times X$. Then we get the variation of the metric
\begin{equation}
\delta g=g(\langle r_0\rangle+\delta r_0(x))-g(\langle
r_0\rangle)={\delta g\over \delta r_0}\Big\vert_{\langle
r_0\rangle}\delta r_0(x).
\end{equation}
But from the explicit form of the metric (\ref{CP3}) we see that
$\delta g$ gets small for large values of $r$ and is of order one
for $r\sim \langle r_0\rangle$. We conclude that in the case of a
singular space with $\langle r_0\rangle=0$ the fluctuations are
four-dimensional fields only present on the singularity. In
particular, if we start from a space with more than one
singularities there will be a four-dimensional field at each of
them. These fields do not interact, as they live in different
four-dimensional
spaces - their ``$r$-coordinate" is different.\\
As we ensured that our theory has $\mathcal{N}=1$ supersymmetry by
compactifying on a $G_2$-manifold, we expect that this fields
should have supersymmetric partners. Indeed there is a
contribution from the $C$-field which gives another scalar field
in four dimensions. The mechanism leading to this field is by now
familiar. On the space (\ref{CP3}) there is one $L^2$-harmonic
three-form $\Omega$ \cite{AtW01} and we get $C=\phi\Omega+\ldots $
. These two fields combine to a complex scalar as
\begin{equation}
z=V\exp\left(i{T_2}\int_{F^3}\widetilde{C}\right),
\end{equation}
where $V\propto r_0^4$ is the volume of the $S^4$ at the center
of our space and the integral is over one fibre of $X\rightarrow
S^4$. The field $\widetilde{C}$ will be described in detail in
the next chapter. It vanishes close to the tip of the cone where
the curvature is large and equals $C$ away from the tip of the
cone. Its most important property is its transformation under
gauge variation
$\delta\widetilde{C}=\Lambda\wedge\delta(r-R)dr+\ldots $ for some
positive parameter $R$.\\
The corresponding superpartner to this complex bosonic field is a
chiral fermion. Again it should be emphasized that if we start
with $\langle r_0\rangle=0$ we get a fermion living on the
singularity of our space. So we conclude that a
``compactification" of M-theory on (\ref{CP3}) leads to a chiral
multiplet in four dimensions, which carries the
additional label $\langle r_0\rangle$.\\
So far anything we did is familiar, as we should expect $b^3(X)$
neutral chiral multiplets in our manifold, according to the
considerations of section 5.3. However, something new will happen
in the case when we take $\langle r_0\rangle=0$, that is when we
really choose the vacuum on which we compactify to be singular. To
see this we need to study the symmetries of this model.\\
In principle we need to study both geometrical symmetries and
symmetries coming from the $C$-field. However, it turns out that
geometrical symmetries are not very important for this particular
model as all the symmetries of $Y$ extend over $X$ \cite{AtW01}.
The important symmetry in our case comes from the $C$-field. As we
have $b^2(Y)=1$ the symmetry of the theory is given by $U(1)\cong
H^2(Y;U(1))$. But $b^2(X)=0$ which tells us that this symmetry is
broken to nothing.\\
It is natural to expect that the Betti numbers of $Y$ and $X$ are
different as long as $X$ is smooth. This can be understood easily
by looking at a simple model. Consider the first homology group of
$S^1$ and the upper half of a sphere $S^2_+$ which is bounded by
$S^1$. Certainly $H_1(S^1;\mathbb{Z})=\mathbb{Z}$ but
$H_1(S^2_+;\mathbb{Z})=0$. This is clear as on $S^2_+$ any circle
is contractible. But now consider what happens if we deform
$S^2_+$ until it is a cone on $S^1$, $C(S^1)$. Then it is no
longer true that we can contract any circle to a point, as we have
a singularity at the tip of the cone. Hence, we can introduce a
winding number and $H_1(C(S^1);\mathbb{Z})=\mathbb{Z}$. But this
is exactly what happens in the case when we take $\langle
r_0\rangle$ to vanish. In this case the cohomology groups of $Y$
and $X$ are isomorphic and we conclude that the $U(1)$ symmetry is
restored. The crucial point is that the chiral multiplet is {\em
charged} under the corresponding gauge field \cite{AtW01}. To see
this we need to consider the Kaluza-Klein expansion of $C$. As
long as $X$ is non-singular we have no harmonic two-forms and thus
no massless gauge fields in four dimensions. The situation changes
if $X$ becomes singular. Then we have $H^2(X;U(1))\cong
H^2(Y;U(1))\cong U(1)$ and we can expand $C=A\wedge \omega+\ldots
$, so in the case of singular manifolds a new gauge field arises.
For the case of compact weak $G_2$-manifolds with conical
singularities the isomorphism $H^2(X;U(1))\cong H^2(Y;U(1))$ was
proven in \cite{BM03a}. Under a gauge transformation of this field
we have\footnote{The integral over $F^3$ of those terms of
$\delta\widetilde{C}$ which are not shown explicitly vanishes.
This can be seen easily from the detailed structure of $\delta
\widetilde{C}$ presented in the next chapter.}
\begin{equation}
z'=V\exp\left(i{T_2}\int_{F^3}(\widetilde{C}+\Lambda\wedge\delta(r-R)dr)\right)
=e^{i{\epsilon(x)}}V\exp\left(i{T_2}\int_{F^3}\widetilde{C}\right),
\end{equation}
where we used $\Lambda=\epsilon\omega+\ldots $ and
$\int_{S^2}T_2\omega=1$. But this is exactly the transformation
law of a charged chiral multiplet. Of course, in order to write
down a gauge invariant Lagrangian we need to introduce a covariant
derivative which couples the fermions to the new gauge field.
Hence, in the case of singular $G_2$-manifolds we really obtained
chiral multiplets which couple
to gauge fields.\\
Let us summarize what we have done so far. We saw that the
compactification of M-theory on a {\em smooth} $G_2$-manifold that
is asymptotic to a cone on $\mathbb{CP}^3$ gives us $b^2(X)=0$
Abelian vector multiplets and $b^3(X)=1$ {\em neutral} chiral
multiplets. This is exactly as expected from the analysis of
section 5.3. However, in the limit in which the manifolds becomes
singular a new four-dimensional Abelian gauge field arises and we
were able to show, that this gauge field couples to the chiral
multiplet present at the singularity. So we really obtained {\em
charged} chiral fermions from compactifications on {\em singular}
spaces. But of course these fields in general lead to anomalies.
So we have to study whether M-theory on singular $G_2$-manifolds
is anomaly free. This will be done in the next chapter.

\bigskip
Finally we want to mention that it is possible to get chiral
fermions from compactifications of M-theory on singular
$G_2$-manifolds which couple to non-Abelian gauge fields. We saw
already, that is necessary to have $ADE$ singularities localized
at $Q$ in $X$ to obtain non-Abelian gauge groups. In general $Q$
is a three-dimensional submanifold of $X$ but it might happen that
$Q$ itself develops a singularity. Close to these singularities
$X$ looks like a cone and we expect a chiral multiplet from its
fluctuations. In fact in this case the multiplet is charged under
the relevant $ADE$ gauge group. See \cite{AtW01}, \cite{Wi01} and
\cite{AcW01} for the details of this mechanism. In these
references an explicit example is given involving a cone on a
weighted projective space. We will comment on the anomaly
structure of these spaces in section 8.2.

\chapter{Anomaly Analysis of M-Theory on $G_2$-Manifolds}
In the last chapter we showed that it is possible to obtain
realistic theories from compactifications of M-theory on singular
$G_2$-manifolds. We obtained four-dimensional chiral fermions
which are charged under Abelian or non-Abelian gauge groups. So in
a sense we reached our goal of constructing realistic theories
from M-theory. However, once this is done we obviously need to
check whether this theory is really consistent. In general chiral
fermions give rise to anomalies and these have to cancel in a
well-defined theory. The idea of how this might occur was first
developed in \cite{Wi01}, however, we present the modified version
of \cite{BM03b}.

\section{Chiral Fermions from Anomaly Considerations}
Let $X$ be a compact $G_2$-manifold that is smooth except for
conical singularities. Up to now it is not clear whether compact
$G_2$-manifolds of that kind exist, however, examples of
non-compact spaces with conical singularities are known
\cite{GPP90}. The term manifold should be taken with a grain of
salt, as the space has a singularity and hence cannot be a
Riemannian manifold in the strict sense. However, it can be viewed
as a limiting point in the moduli space of $G_2$-metrics. For
example the non-compact $G_2$-metrics of (\ref{noncompG2}) and
(\ref{CP3}) develop a conical singularity in the limit in which
$r_0\rightarrow 0$.
\begin{figure}[h]
\centering
\includegraphics[width=0.6\textwidth]{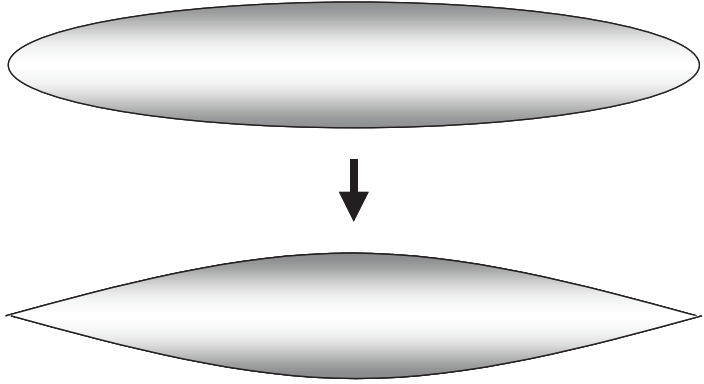}\\
\caption[]{A smooth $G_2$-manifold develops a conical
singularity.} \label{regsing}
\end{figure}
We saw in the last chapter that for certain geometries there are
four-dimensional charged chiral fermions sitting on the
singularities $P_{\alpha}$, with $\alpha$ a label running over the
number of singularities in $X$. From anomaly cancellation we will
learn that these fermions are of negative
chirality.\footnote{Recall that in our conventions we have
$\g^E_5=\g^M_5$ in four dimensions, in contrast to the conventions
chosen in \cite{BM03b, BM03c}. Note also, that because of this
relation our recipe to check for anomaly cancellation given in
chapter 3.2.6 no longer works. In this chapter we really have to
calculate the Euclidean inflow and anomaly and need to use our
master formula (\ref{master}).} These lead to a variation of the
Euclidean quantum effective action of the form
\begin{equation}
\delta X|_{anomaly}=-2\pi i \hat I_4^1\ \ \mbox{with} \ \ \hat
I_6=(-1)[\hat A(M_4){\rm ch}(F)]_6.
\end{equation}
The Euclidean gauge anomaly localized at $P_{\alpha}$ is (recall
that $I=-2\pi\hat I$ and $F=iF^iq^i$),
\begin{equation}\label{gaugeanomaly}
I_{\alpha}^{(gauge)}=-{1\over(2\pi)^23!}\sum_{\sigma\in
T_{\alpha}}\left(\sum_{i=1}^{b^2(X)}q_{\sigma}^iF^i\right)^3.
\end{equation}
$\sigma$ labels the four dimensional chiral multiplets
$\Phi_{\sigma}$ which are present at the singularity $P_\alpha$.
$T_{\alpha}$ is simply a set containing all these labels.
$q_{\sigma}^i$ is the charge of $\Phi_{\sigma}$ with respect to
the $i$-th gauge field $A^i$. As all the gauge fields come from a
Kaluza-Klein expansion of the $C$-field we have $b^2(X)$ of them.
Clearly we get a mixed anomaly as well,
\begin{equation}\label{mixedanomaly}
I_{\alpha}^{(mixed)}={1\over24}\sum_{\sigma\in
T_{\alpha}}\left(\sum_{i=1}^{b^2(X)}q_{\sigma}^iF^i\right)p_1'.
\end{equation}
Here $p_1'=-{1\over8\pi^2}\tr R\wedge R$ is the first Pontrjagin
class of four dimensional space-time $\mathbb{R}^4$.\\
As usual the effective four-dimensional theory is obtained by
integrating over the compact space $X$. But this means that the
effective theory has anomalies
\begin{eqnarray}
I^{(gauge)}&=&\sum_{\alpha}I_{\alpha}^{(gauge)},\label{globalanomaly1}\\
I^{(mixed)}&=&\sum_{\alpha}I_{\alpha}^{(mixed)}.\label{globalanomaly2}
\end{eqnarray}
Certainly, these anomalies have to cancel in a consistent theory.
However, we saw already at various places that it is not
sufficient to have global\footnote{To avoid confusion let us be
more precise about what we mean by a "global" anomaly. Each chiral
field which leads to an anomaly is localized on a four-dimensional
subspace of the eleven-dimensional manifold. If we speak of local
anomaly cancellation we want to cancel the anomalies on each of
these subspaces separately. This is different from the
cancellation of the global anomaly which is defined to be the sum
of all the localized anomalies as given in (\ref{globalanomaly1})
and (\ref{globalanomaly2}). Of course, the effective theory which
is obtained from integrating over the compact space only sees the
global anomaly. However, consistency of the eleven-dimensional
theory requires a local cancellation of all anomalies. An example
of this kind of anomaly cancellation was already given in chapter
4, where anomalies cancelled on each of the two ten-dimensional
planes separately. Note that this concept is completely unrelated
to local gauge versus global anomalies which were discussed in
chapter 3.} anomaly cancellation in M-theory. Instead anomalies
have to be cancelled locally. This means that we need to find a
mechanism, that cancels $I_{\alpha}^{(gauge)}$ and
$I_{\alpha}^{(mixed)}$ at each singularity $P_{\alpha}$
separately.

\bigskip
All the ``M-theory" calculations we performed so far actually used
the action of eleven-dimensional supergravity together with some
correction terms. However, in the neighbourhood of a conical
singularity the curvature of $X$ blows up. Close to the
singularity $X$ is a cone on some manifold $Y$. But as $X$ is
Ricci-flat $Y$ has to be Einstein with
$\mathcal{R}^Y_{mn}=5\delta_{mn}$. The Riemann tensor on $X$ and
$Y$ are related by $R^{Xmn}_{\ \ \ \ \ pq}={1\over r^2}
(R^{Ymn}_{\ \ \ \ \ pq}-
\delta^m_p\delta^n_q+\delta^m_q\delta^n_p)$, for $m\in\{1,2,\ldots
,6\}$. Thus, the supergravity description is no longer valid close
to a singularity and one has to resort to a full M-theory
calculation, a task that is currently not feasible. To avoid this
difficulty we could, following \cite{Wi01}, excise the
singularities of $X$, i.e. we could cut off the tips of all the
cones in our manifold. Then we were left with a manifold with
boundary $X'$ such that
\begin{equation}
\partial X'=-\cup_{\alpha}Y_{\alpha}.
\end{equation}
$Y_{\alpha}$ is the component of the boundary of $X'$ which comes
from cutting out $P_{\alpha}$ and we chose its orientation to be
opposite to the one induced by the boundary operator. In that way
we get rid of those points with arbitrary large curvature and the
supergravity approximation can be used for compactifications on
$X'$. So, the basic idea in \cite{Wi01} is to model the effect of
a singularity by a  boundary which arises from cutting off this
singularity.

\begin{figure}[h]
\centering
\includegraphics[width=0.6\textwidth]{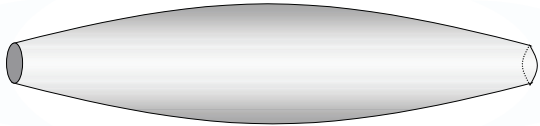}\\
\caption[]{Excising the singularities gives a manifold with
boundary.} \label{mfdwithboundary1}
\end{figure}

\begin{figure}[h]
\centering
\includegraphics[width=0.6\textwidth]{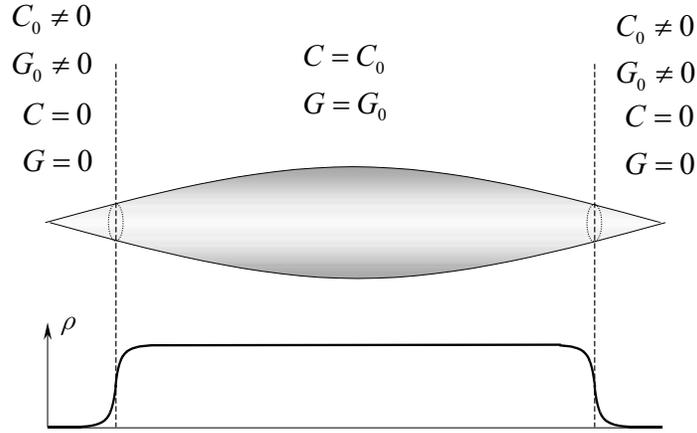}\\
\caption[]{Embedding a manifold with boundary into a manifold.}
\label{embedding}
\end{figure}

We want to try to follow a similar but different route to study
singular spaces. Suppose we start from M-theory on
$\mathbb{R}^4\times X$ with $X$ carrying conical singularities.
The expectation values of the $C$-field and the gravitino vanish
for compactifications on $G_2$-manifolds, but we allow for
fluctuations. We are only interested in fluctuations of the
three-form, which we will denote $C_0$. The corresponding field
strength is given by $G_0=dC_0$. As we discussed above we
cannot do supergravity calculations in this setup.\\
But now suppose that the fields can be shielded by some mechanism
close to the singularities. That is we introduce some arbitrary
parameter $R$ and a new set of fields $\{C, G\}$ on $X$ with two
important properties. Firstly we want them to vanish at points
which are closer to a singularity than $R$. Secondly, they are
given by the original fields $\{C_0, G_0\}$ for distances bigger
than $R$. Note that we do not introduce a new metric but we want
to keep fixed
the geometry of our setup. The aim is to understand the dynamics
of the new set of fields $\{C,G\}$ on $X$, which can be done
using supergravity.\\
The mathematical formulation of this is as follows. Choose a set
of charts on $X$ such that the neighbourhood of each singularity
is covered by a single chart. On this chart the metric is conical
and there is a natural coordinate $r_{\alpha}$ for each
$P_{\alpha}$. Next we introduce a set of smooth functions
$\rho_{\alpha}$ on the chart around $P_{\alpha}$ with the
properties
\begin{equation}
\rho_{\alpha}(r_{\alpha})= \left\{\begin{matrix}
                                                0&\mbox{for}&\ \ 0\leq r_{\alpha}\leq R-\epsilon\\
                                                1&\mbox{for}&r_{\alpha}\geq R+\epsilon
                                  \end{matrix}\right.
\end{equation}
where $\epsilon/R$ is small.\\
Using a partition of unity we can construct a smooth function
$\rho$ on $X$ from these $\rho_{\alpha}$ in such a way that $\rho$
vanishes for points with a distance to a singularity which is less
than $R-\epsilon$ and is one for distances larger than
$R+\epsilon$. We denote the points of radial coordinate $R$ in the
chart around $P_{\alpha}$ by $Y_{\alpha}$ and the subspace of $X$
which is bounded by these $Y_{\alpha}$ is denoted $X'$. All these
conventions are chosen in such a way that $\int_X(\ldots )\wedge
d\rho=\sum_{\alpha}\int_{Y_{\alpha}}(\ldots )$. This clarifies the
relation of the current approach to the one used in \cite{Wi01}.
Our model can be understood as the
embedding of a manifold with boundary into a manifold.\\
Now we want to impose the following boundary conditions on the
fields of our new theory\footnote{We choose the parameter $R$ in
such a way that it is not possible to be closer than $R$ to two
singularities.}
\begin{eqnarray}
C&=&\left\{\begin{matrix} 0 &\mbox{for one}& r_{\alpha}<R-\epsilon\\
                         C_0&\mbox{for all}&
                         r_{\alpha}>R+\epsilon
           \end{matrix}\right.\\
G&=&\left\{\begin{matrix} 0&\mbox{for one}&r_{\alpha}<R-\epsilon\\
                         G_0&\mbox{for all}&
                         r_{\alpha}>R+\epsilon\label{bcG}
           \end{matrix}\right.
\end{eqnarray}
So what did we gain from introducing this new theory on
$\mathbb{R}^4\times X$ with a new set of fields? The important
property of this theory is, that it reproduces the results of the
theory with the field content $\{C_0, G_0\}$ for distances from
the singularities larger than $R$.\footnote{Our basic idea is
familiar from standard electrodynamics. Suppose we are interested
in studying electrodynamics in a black hole geometry. Certainly,
close to the singularity the description is problematic, but if we
are only interested in what happens far away from the singularity
we might as well introduce a Faraday cage containing the black
hole. In this situation we have pure gravity in the interior of
the cage. Outside the cage, however, the theory does not change,
as the cage gives exactly the right boundary conditions. So what
we do in our model is to introduce a Faraday cage for M-theory.}
The basic setup is shown in figure (\ref{embedding}).

Let us analyze the consequences of this situation. Most
importantly, it is clear that we have to introduce a source for
the $G$-field at $R$. We need to have
\begin{equation}
dG=G_0\wedge d\rho.\label{Gsource}
\end{equation}
Note that the strength of the source is given by the value of
$G_0$ on $Y_{\alpha}$. If we worked with a manifold with boundary
we certainly would need to impose boundary conditions on
$Y_{\alpha}$. In our model, there is no boundary, but
(\ref{Gsource}) is necessary for the $G$-field to satisfy our
requirements. So our model leads to a modified Bianchi identity,
very much in the spirit of the chapters 4 and 6. The general
solution of this equation is given by
\begin{equation}
G=aG_0\rho+(1-a)C_0\wedge d\rho+bH_4+cdC_3+edB\wedge d\rho.
\end{equation}
Here $\{a,b,c,e\}$ are real coefficients, $H_4$ is a harmonic
four-form on $X$, $C_3$ is a three-form on $\mathbb{R}^4\times X$
and $B$ is an arbitrary two-form on $\mathbb{R}^4\times X$. Note,
however, that $d\rho$ has support only close to the $Y_{\alpha}$,
so one might have introduced a set of independent two-forms
$B_{\alpha}$ on $\mathbb{R}^4\times Y_{\alpha}$. In that case the
last term would read $\sum_{\alpha}dB_{\alpha}\wedge d\rho$.
Imposing the boundary conditions (\ref{bcG}) gives the simple
solution
\begin{equation}
G=G_0\rho.
\end{equation}
The gauge invariant kinetic term of our theory is constructed from
this field
\begin{equation}
S_{kin}=-{1\over4\kappa_{11}^2}\int G\wedge \star
G=-{1\over4\kappa_{11}^2}\int_{r\geq R} dC_0\wedge \star dC_0.
\end{equation}
To proceed we need to recall what we did in the presence of a
modified Bianchi identity in chapter 6. There the $G$-field which
occurred in the topological term in the supergravity action had to
be corrected in a way that $d(G+$corrections)=0. Applying this to
our model we define yet another field
\begin{eqnarray}
\widetilde{G}&:=&G-(C_0-dB)\wedge d\rho\nonumber\\
&=&G_0\rho-(C_0-dB)\wedge d\rho\nonumber\\
&=&d\left( C_0\rho + B\wedge d\rho\right)\nonumber\\
&=:&d\widetilde{C}
\end{eqnarray}
and postulate a modified topological term of the form
\begin{eqnarray}
\widetilde{S}_{top}&:=&-{1\over12\kappa_{11}^2}\int_{\mathbb{R}^4\times
X}\widetilde{C}\wedge\widetilde{G}\wedge\widetilde{G}\\
&=&-{2\pi\over6}\left(T_2\over2\pi\right)^3\int_{\mathbb{R}^4\times
X}\widetilde{C}\wedge\widetilde{G}\wedge\widetilde{G}.
\end{eqnarray}
Here we used the identity (\ref{relkT}) which is always valid in
M-theory. It is interesting to analyze the structure of this term,
\begin{eqnarray}
\widetilde{S}_{top}&=&-{2\pi\over6}\left(T_2\over2\pi\right)^3\int_{\mathbb{R}^4\times
X}\widetilde{C}\wedge\widetilde{G}\wedge\widetilde{G}\nonumber\\
&=&-{2\pi\over6}\left(T_2\over2\pi\right)^3\int_{\mathbb{R}^4\times
X}(C_0\rho+B\wedge d\rho)\wedge\nonumber\\
&&\wedge(G_0\rho-(C_0-dB)\wedge d\rho)\wedge(G_0\rho-(C_0-dB)\wedge d\rho)\nonumber\\
&=&-{2\pi\over6}\left(T_2\over2\pi\right)^3\int_{\mathbb{R}^4\times
X}\left(C_0\wedge G_0\wedge G_0\rho^3+\right.\nonumber\\
&&\left.+2C_0\wedge dB\wedge G_0
\wedge\rho^2d\rho+B\wedge G_0\wedge G_0\wedge\rho^2d\rho\right)\nonumber\\
&=&-{2\pi\over6}\left(T_2\over2\pi\right)^3\int_{\mathbb{R}^4\times
X}C_0\wedge G_0\wedge G_0\rho^3\nonumber\\
&&-{1\over3}\sum_{\alpha}{2\pi\over6}\left(T_2\over2\pi\right)^3\int_{\mathbb{R}^4\times
Y_{\alpha}}\left( 2C_0\wedge dB\wedge G_0+B\wedge G_0\wedge
G_0\right)\nonumber\\
&=&-{2\pi\over6}\left(T_2\over2\pi\right)^3\int_{\mathbb{R}^4\times
X}C_0\wedge G_0\wedge G_0\rho^3\nonumber\\
&&-\sum_{\alpha}{2\pi\over6}
\left(T_2\over2\pi\right)^3\int_{\mathbb{R}^4\times
Y_{\alpha}}B\wedge G_0\wedge G_0.
\end{eqnarray}
Here we used that $\rho^2d\rho\rightarrow {1\over3}d\rho$. So we
have the usual topological term on $X'$, nothing close to the
singularities and a contribution $BG_0G_0$ on each $Y_{\alpha}$.
This additional term is rather interesting and we will give some
speculative comments related to it in the next chapter. Of course
the whole construction is similar in spirit to what we did in
chapter 6 and clearly the structure of the modified topological
term is similar in both cases.\\
The field $\widetilde{G}$ is closed by construction but, because
of its position in the topological term, we also want it to be
gauge invariant\footnote{This is why $B$ has to be present in the
definition of $\widetilde{G}$.}. This gives us the transformation
law of $B$,
\begin{eqnarray}
0&=&\delta\widetilde{G}=d\left( d\Lambda \rho+\delta
B\wedge d\rho\right)\nonumber\\
&=&(d\delta B-d\Lambda)\wedge d\rho,
\end{eqnarray}
from which we read off
\begin{equation}
\delta B=\Lambda.
\end{equation}
Then it is easy to write down the transformation law of
$\widetilde{C}$,
\begin{eqnarray}
\delta\widetilde{C}&=&\delta C_0\rho+\delta
B\wedge d\rho\nonumber\\
&=&d\Lambda\rho+\Lambda\wedge d\rho\nonumber\\
&=&d(\Lambda\rho).
\end{eqnarray}
Now we are ready to calculate the gauge transformation of the
modified topological term. It reads
\begin{eqnarray}
\delta
\widetilde{S}_{top}&=&-{2\pi\over6}\left(T_2\over2\pi\right)^3\int_{\mathbb{R}^4\times
X}\delta\widetilde{C}\wedge\widetilde{G}\wedge\widetilde{G}\nonumber\\
&=&-{2\pi\over6}\left(T_2\over2\pi\right)^3\int_{\mathbb{R}^4\times
X}d(\Lambda \rho)\wedge(G_0\rho-(C_0-dB)\wedge d\rho)
\wedge(G_0\rho-(C_0-dB)\wedge d\rho)\nonumber\\
&=&-{2\pi\over6}\left(T_2\over2\pi\right)^3\int_{\mathbb{R}^4\times
X}d\Lambda \wedge G_0\wedge G_0
\rho^3\nonumber\\
&&+2{2\pi\over6}\left(T_2\over2\pi\right)^3\int_{\mathbb{R}^4\times
X}d\Lambda\wedge
G_0\wedge (C_0-dB)\wedge \rho^2d\rho-\nonumber\\
&&-{2\pi\over6}\left(T_2\over2\pi\right)^3\int_{\mathbb{R}^4\times
X}\Lambda\wedge G_0\wedge G_0\wedge \rho^2d\rho.
\end{eqnarray}
We substitute the Kaluza-Klein expansions
\begin{eqnarray}
C_0(x,y)&=&\sum_iA^i(x)\wedge\omega^i(y)+\ldots ,\\
\Lambda(x,y)&=&\sum_i\epsilon^i(x)\omega^i(y)+\ldots , \label{KKLambda}\\
B(x,y)&=&\sum_i\phi^i(x)\omega^i(y)+\ldots ,
\end{eqnarray}
and get
\begin{eqnarray}
\delta
\widetilde{S}_{top}&=&{2\over3}{2\pi\over6}\left(T_2\over2\pi\right)^3\int_{\mathbb{R}^4\times
X}d\epsilon^i F^j
(A^k-d\phi^k) \omega^i\wedge\omega^j\wedge\omega^k\wedge d\rho\nonumber\\
&&-{1\over3}{2\pi\over6}\left(T_2\over2\pi\right)^3\int_{\mathbb{R}^4\times X}\epsilon^i F^jF^k\omega^i\wedge\omega^j\wedge\omega^k\wedge d\rho+\ldots \nonumber\\
&=&{2\over3}\sum_{\alpha}{2\pi\over6}\left(T_2\over2\pi\right)^3\int_{\mathbb{R}^4\times
Y_{\alpha}}d\epsilon^iF^j
(A^k-d\phi^k) \omega^i\wedge\omega^j\wedge\omega^k\nonumber\\
&&-{1\over3}\sum_{\alpha}{2\pi\over6}\left(T_2\over2\pi\right)^3\int_{\mathbb{R}^4\times Y_{\alpha}}\epsilon^i F^jF^k\omega^i\wedge\omega^j\wedge\omega^k+\ldots \nonumber\\
&=&{2\over3}\sum_{\alpha}{2\pi\over6}\left(T_2\over2\pi\right)^3\int_{\mathbb{R}^4}d\epsilon^i
F^j
(A^k-d\phi^k) \int_{Y_{\alpha}}\omega^i\wedge\omega^j\wedge\omega^k\nonumber\\
&&-{1\over3}\sum_{\alpha}{2\pi\over6}\left(T_2\over2\pi\right)^3\int_{\mathbb{R}^4}\epsilon^i F^jF^k\int_{Y_{\alpha}}\omega^i\wedge\omega^j\wedge\omega^k+\ldots \nonumber\\
&=&-\sum_{\alpha}{1\over(2\pi)^23!}\int_{\mathbb{R}^4}\epsilon^i
F^jF^k\int_{Y_{\alpha}}(T_2)^3\omega^i\wedge\omega^j\wedge\omega^k+\ldots
\end{eqnarray}
Note that substituting the Kaluza-Klein expansions into
$d\Lambda\wedge G_0\wedge G_0$ gives zero. Furthermore, we
integrated by parts on $\mathbb{R}^4$ to obtain the last equation.
The result is a sum of terms which are localized at $Y_{\alpha}$.
The corresponding Euclidean anomaly polynomial is given by
\begin{equation}
I^{(top)}_E=\sum_\a I^{(top)}_{E,\a}=-i
\sum_{\alpha}I^{(top)}_{M,\alpha}=\sum_{\alpha}{i\over(2\pi)^23!}
F^iF^jF^k\int_{Y_{\alpha}}(T_2)^3\omega^i\wedge\omega^j\wedge\omega^k+\ldots
\end{equation}
This is very similar to the gauge anomaly $I_{\alpha}^{(gauge)}$
and we do indeed get a local cancellation of the anomaly provided
we have
\begin{equation}
\int_{Y_{\alpha}}(T_2)^3\omega^i\wedge\omega^j\wedge\omega^k=\sum_{\sigma\in
T_{\alpha}}q_{\sigma}^iq_{\sigma}^jq_{\sigma}^k.\label{chargecondition1}
\end{equation}
(Note that the condition of local anomaly cancellation is
$iI_{\alpha}^{(gauge)}+I^{(top)}_{E,\alpha}=0$, from
(\ref{master}).) In \cite{Wi01} it was shown that this equation
holds for all known examples of conical singularities. Note that
the dimensions of (\ref{chargecondition1}) are correct. We have
$[(C_0)_{\mu mn}]=0$ but $[A_{\mu}]=1$, therefore
$[\omega_{mn}]=-1$, hence we need a factor of dimension
nine on the left-hand side to give us a dimensionless quantity.\\
Turning things around, had we not known that there are charged
(anti-)chiral fermions living on the singularity we would be
forced to introduce them at this point in order to cancel the
anomalies of our theory. Their charges would be constraint to
satisfy the condition (\ref{chargecondition1}), thus we conclude
that it depends on the geometry of the manifold $X$ and in
particular on the properties of
the harmonic two-forms on $X$ whether chiral fermions are present or not.\\
It is particularly important that our modified topological term
gave us a sum of terms localized at $Y_{\alpha}$ without any
integration by parts on $X$. This is crucial, because local
quantities are no longer well-defined after an integration by
parts\footnote{Consider for example $\int_a^b
df=f(b)-f(a)=(f(b)+c)-(f(a)+c)$. It is impossible to infer the
value of $f$ at the boundaries $a$ and $b$.}. In fact let us
consider once again the variation of the topological term and let
us this time integrate by parts.
\begin{equation}
\delta
\widetilde{S}_{top}=-{2\pi\over6}\left(T_2\over2\pi\right)^3\int
d(\Lambda\rho)\wedge\widetilde{G}\wedge\widetilde{G}=2{2\pi\over6}\left(T_2\over2\pi\right)^3\int\Lambda\rho\wedge
d\widetilde{G}\wedge\widetilde{G}=0,\label{delS0}
\end{equation}
as $d\widetilde{G}=0$. Hence,
$I^{(top)}_E=\sum_{\alpha}I^{(top)}_{E,\alpha}=0$, the gauge
variation of the topological term vanishes. But this is a global
statement, as we saw explicitly that $\widetilde{S}_{top}$ is a
sum of terms located at $Y_{\alpha}$. This tells us that the
situation is rather subtle. We have the anomalies
$I_{\alpha}^{(gauge)}$, $I_{E,\alpha}^{(top)}$, $I^{(gauge)}$ and
$I^{(top)}_E$ with
$I^{(top)}_E=\sum_{\alpha}I_{E,\alpha}^{(top)}$,
$I^{(gauge)}=\sum_{\alpha}I_{\alpha}^{(gauge)}$ and
$I_{E,\alpha}^{(top)}=-iI_{\alpha}^{(gauge)}$. But we just showed
that $I^{(top)}_E=0$, so
$I^{(gauge)}=\sum_{\alpha}I_{\alpha}^{(gauge)}=i\sum_{\alpha}I_{E,\alpha}^{(top)}=iI^{(top)}_E=0$,
the global gauge anomaly vanishes as well.\\
Clearly in this situation the concept of local and global
anomalies is very important. We just saw that the sum of all the
anomalies in a theory with an unmodified topological term
vanishes, i.e. the anomalies vanish after integration over $X$.
Although this is what we will do in the end to obtain a
four-dimensional theory it is not sufficient. In fact the
anomalies are localized at the singularities as the chiral
fermions live only there. These localized anomalies have to be
cancelled {\em before} we integrate over the compact space. The
modification of the topological term is done in such a way that
the anomalies cancel independently at each singularity.

\bigskip
After having seen how anomaly cancellation works in the case of
gauge anomalies we turn to the mixed anomaly. We start by
considering the Green-Schwarz term in a theory on a smooth
manifold $\mathbb{R}^4\times X$. It can be rewritten as
\begin{equation}
S_{GS}=-{T_2\over2\pi}\int_{\mathbb{R}^4\times X}G_0\wedge
X_7=-{T_2\over2\pi}\int_{\mathbb{R}^4\times X}C_0\wedge X_8,
\end{equation}
with $X_8$ as in (\ref{X8}) and $X_8=dX_7$. The coefficient was
determined from the anomaly cancellation condition in chapter 6
and is consistent with the results of chapter 4.\\
Then, there is a natural modification of the Green-Schwarz term on
our singular manifold,
\begin{equation}
\widetilde{S}_{GS}:=-{T_2\over2\pi}\int_{\mathbb{R}^4\times
X}\widetilde{C}\wedge X_8.
\end{equation}
As before it is interesting to study the detailed structure of
this term,
\begin{eqnarray}
\widetilde{S}_{GS}&=&-{T_2\over2\pi}\int_{\mathbb{R}^4\times
X}(C_0\rho+B\wedge d\rho)\wedge
X_8\nonumber\\
&&=-{T_2\over2\pi}\int_{\mathbb{R}^4\times X}C_0\rho\wedge
X_8-{T_2\over2\pi}\sum_{\alpha}\int_{\mathbb{R}^4\times
Y_{\alpha}}B\wedge X_8.
\end{eqnarray}
Once again we get the usual term in the bulk, nothing close to the
singularities and a very special contribution from the
$Y_{\alpha}$.\\
The variation of the Green-Schwarz term is easily calculated,
\begin{eqnarray}
\delta\widetilde{S}_{GS}&=&-{T_2\over2\pi}\int_{\mathbb{R}^4\times
X}\delta\widetilde{C}\wedge X_8\nonumber\\
&=&-{T_2\over2\pi}\int_{\mathbb{R}^4\times X}d(\Lambda\rho)\wedge X_8\nonumber\\
&=&-{T_2\over2\pi}\int_{\mathbb{R}^4\times X}d\Lambda\rho\wedge
X_8-{T_2\over2\pi}\int_{\mathbb{R}^4\times X}\Lambda\wedge d\rho\wedge X_8\nonumber\\
&=&-{T_2\over2\pi}\int_{\mathbb{R}^4\times X}d\Lambda\rho\wedge
X_8-{T_2\over2\pi}\sum_{\alpha}\int_{\mathbb{R}^4\times
Y_{\alpha}}\Lambda\wedge
X_8\nonumber\\
&=&-{T_2\over2\pi}\sum_{\alpha}\int_{\mathbb{R}^4\times
Y_{\alpha}}\epsilon^i\omega^i\wedge X_8+\ldots
.\label{mixedinflow}
\end{eqnarray}
To obtain the last line we used the Kaluza-Klein expansion
(\ref{KKLambda}) and the structure of $X_8$. As we saw in chapter
6, $X_8$ can be rewritten in terms of the first and second
Pontrjagin class
\begin{eqnarray}
p_1&=&-{1\over2}\left({1\over2\pi}\right)^2{\rm tr}R^2,\\
p_2&=&{1\over8}\left(1\over2\pi\right)^4[({\rm tr}R^2)^2-2{\rm
tr}R^4],
\end{eqnarray}
as
\begin{equation}
X_8={\pi\over4!}\left[{p_1^2\over4}-p_2\right].
\end{equation}
The background we are working in is four-dimensional Minkowski
space times a $G_2$-manifold. In this special setup the Pontrjagin
classes can easily be expressed in terms of the Pontrjagin
classes $p_i'$ on $(\mathbb{R}^4,\eta)$ and those on $(X,g)$,
which we will write as $p_i''$. We get
\begin{eqnarray}
p_1&=&p_1'+p_1''\\
p_2&=&p_1'\wedge p_1''.
\end{eqnarray}
Using these relations we obtain a convenient expression for the
inflow (\ref{mixedinflow}),
\begin{eqnarray}
\delta\widetilde{S}_{GS}&=&-{T_2\over2\pi}\sum_{\alpha}\int_{\mathbb{R}^4\times
Y_{\alpha}}\epsilon^i\omega^i\wedge X_8\nonumber\\
&=&-{T_2\over2\pi}\sum_{\alpha}\int_{\mathbb{R}^4\times
Y_{\alpha}}\epsilon^i\omega^i\wedge{\pi\over
4!}\left[{p_1^2\over4}-p_2\right]\nonumber\\
&=&-{T_2\over2\pi}{\pi\over
4!}\sum_{\alpha}\int_{\mathbb{R}^4\times
Y_{\alpha}}\epsilon^i\omega^i\wedge\left[{1\over2}p_1'\wedge p_1''-p_1'\wedge p_1''\right]\nonumber\\
&=&{T_2\over2\pi}{\pi\over
48}\sum_{\alpha}\int_{\mathbb{R}^4}\epsilon^i
p_1'\int_{Y_{\alpha}}\omega^i\wedge p_1''.
\end{eqnarray}
The corresponding anomaly polynomial is given by
\begin{equation}
I^{(GS)}_E=\sum_{\alpha}I^{(GS)}_{E,\alpha}=-i\sum_{\alpha}{1\over24}F^ip_1'\int_{Y_{\alpha}}{T_2\over4}\omega^i\wedge
p_1'',
\end{equation}
and we see that the mixed anomaly cancels locally provided we have
\begin{equation}
\int_{Y_{\alpha}}{T_2\over4}\omega^i\wedge p_1''=\sum_{\sigma\in
T_{\alpha}}q_{\sigma}^i.\label{chargecondition2}
\end{equation}
All known examples satisfy this requirement. Once again turning
things around (\ref{chargecondition2}) might be seen as a second
condition which has to be satisfied by the charges of the chiral
fermions which have to be present on the singularities to have a
well-defined theory.\\
Note that from integration by parts,
\begin{equation}
\delta \widetilde{S}_{GS}=-{T_2\over2\pi}\int_{\mathbb{R}^4\times
X}d(\Lambda\rho)\wedge X_8={T_2\over2\pi}\int_{\mathbb{R}^4\times
X}\Lambda\rho\wedge dX_8=0.
\end{equation}
This tells us that
$0=I^{(GS)}_E=\sum_{\alpha}I^{(GS)}_{E,\alpha}=\sum_{\alpha}-iI^{(mixed)}_{\alpha}=-iI^{(mixed)}$.
The sum of all mixed anomalies vanishes.\\

\begin{center}
{\bf The example of $\mathbb{CP}^3$}
\end{center}
We saw that the anomalies of chiral fermions sitting on the
singularities of $G_2$-manifolds vanish provided their charges
satisfy certain geometrical conditions, namely
(\ref{chargecondition1}) and (\ref{chargecondition2}). In the
last chapter we studied the example of a manifold which is
asymptotic to a cone on $\mathbb{CP}^3$. For this manifold we know
both the geometry and the field content, thus we can check
whether our conditions hold in this case. The fields were
determined to consist of a single chiral multiplet with a charge
that can be normalized to one. So the right-hand side of our
conditions equals one. Let us check whether this is true for the
left-hand side as well. We need to calculate
\begin{equation*}
\int_{\mathbb{CP}^3}(T_2)^3\omega^i\wedge\omega^j\wedge\omega^k
\end{equation*}
and
\begin{equation}
{T_2\over4}\int_{\mathbb{CP}^3}\omega^i\wedge p_1''.
\end{equation}
We know that $b_2(\mathbb{CP}^3)=1$, which is obvious from the
fact that $\mathbb{CP}^3$ can be considered as an $S^2$-bundle
over $S^4$. From de Rham's theorem we derive that
$b^2(\mathbb{CP}^3)=1$ and we only have a single harmonic two-form
$\omega$. Poincar\'{e}'s duality tells us that there is a single
harmonic four-form $\Omega$ which can be normalized such that
\begin{equation}
T_2\int_{\mathbb{CP}^3}\omega\wedge\Omega=1.
\end{equation}
But of course $\omega\wedge \omega$ is both closed and co-closed
and thus $\omega\wedge\omega\propto\Omega$. This implies that
choosing the normalization of $\omega$ appropriately we get
\begin{equation}
\int_{\mathbb{CP}^3}(T_2)^3\omega\wedge\omega\wedge\omega=1.
\end{equation}
Hence, (\ref{chargecondition1}) is satisfied in this example. The
second condition can be verified easily as well, if one considers
the details of the geometry of $\mathbb{CP}^3$. In this space the
first two Chern classes are given by $c_1=4T_2\omega$,
$c_2=6(T_2)^2\omega\wedge\omega$ and the first Pontrjagin class is
$p_1''=c_1^2-2c_2$ \cite{EGH80}, \cite{Wi01}. Given this
geometrical data it is easy to do the integral, as we have
$p_1''=4(T_2)^2\omega\wedge\omega$ yielding
\begin{equation}
{T_2\over4}\int_{\mathbb{CP}^3}\omega\wedge
p_1''=\int_{\mathbb{CP}^3}(T_2)^3\omega\wedge\omega\wedge\omega=1,
\end{equation}
which is the required result. In \cite{Wi01} it was shown that the
equations (\ref{chargecondition1}) and (\ref{chargecondition2})
are in fact true for all known examples of singular
$G_2$-manifolds.

\section{Non-Abelian Gauge Groups and Anomalies}
Finally we also want to comment on anomaly cancellation in the
case of non-Abelian gauge groups. The calculations are relatively
involved in this case and we refer the reader to \cite{Wi01} and
\cite{BM03b} for the details. We only present the basic mechanism.
We saw in chapter 7 that chiral fermions which are charged under
non-Abelian gauge groups occur if the locus $Q$ of an $ADE$
singularity in a $G_2$-manifold $X$ develops a singularity. Close
to such a singularity $P_{\alpha}$ of $Q$ the space $X$ looks like
a cone on some $Y_\alpha$. If $U_{\alpha}$ denotes the
intersection of $Q$ with $Y_{\alpha}$ then close to $P_{\alpha}$
$Q$ is a cone on $U_{\alpha}$. We saw that in this case there are
$ADE$ gauge fields on $\mathbb{R}^4\times Q$ which reduce to
non-Abelian gauge fields on $\mathbb{R}^4$ if we perform a
Kaluza-Klein expansion on $Q$. On the $P_{\alpha}$ we have a
number of chiral multiplets $\Phi_{\sigma}$ which couple to both
the non-Abelian gauge fields and the Abelian ones, coming from the
Kaluza-Klein expansion of the $C$-field. Thus, we expect to get a
$U(1)^3$, $U(1)H^2$ and $H^3$ anomaly\footnote{The $U(1)^2G$
anomaly is not present as ${\rm tr}\  T_a$ vanishes for all
generators of $ADE$ gauge groups, and the $H^3$-anomaly is only
present for $H=SU(n)$.}, where $H$ is the relevant $ADE$ gauge
group. The relevant anomaly polynomial for this case is
\begin{equation}
I_6=-2\pi [(-1)\hat A(M){\rm ch}(F^{(Ab)}){\rm ch}(F)]_6
\end{equation}
where $F^{(Ab)}:=iq^iF^i$ denotes the Abelian and $F$ the
non-Abelian gauge field. Expansion of this formula gives four
terms namely (\ref{gaugeanomaly}), (\ref{mixedanomaly}) and
\begin{eqnarray}
I^{(H^3)}&=&-{i\over(2\pi)^23!}{\rm tr}F^3,\\
I^{(U(1)_iH^2)}&=&{1\over(2\pi)^22}q_iF_i{\rm tr}F^2.
\end{eqnarray}
The anomalies which are related to these polynomials via the
descent formalism are
\begin{eqnarray}
G^{H^3}[\e;A]&=&-{i\over(2\pi)^23!}\int_{\mathbb{R}^4}{\rm tr}\left\{\e\ d\left(AdA+{1\over2}A^3\right)\right\},\label{H^3}\\
G^{U(1)_iH^2}[\e,\e_i;A,A_i]&=&{q_i\over(2\pi)^22}\int_{\mathbb{R}^4}\left(\beta
\e_i{\rm tr}F^2+(1-\beta)F_i{\rm tr}(\e dA)\right).\label{U1H^2}
\end{eqnarray}
The anomalies (\ref{H^3}) and (\ref{U1H^2}) are cancelled from the
gauge variations of two terms which are present in our special
setup \cite{Wi01}, \cite{BM03b}. They read
\begin{eqnarray}
\widetilde{S}_1=-{i\over6(2\pi)^2}\int_{\mathbb{R}^4\times Q}K\wedge {\rm tr}(\widetilde{A}\widetilde{F}^2),\\
\widetilde{S}_2={T_2\over2(2\pi)^2}\int_{\mathbb{R}^4\times
Q}\widetilde{C}\wedge {\rm tr}\widetilde{F}^2.
\end{eqnarray}
Here $K$ is the curvature of a certain line bundle in our setup
described in \cite{Wi01} and $\widetilde{A}$ and $\widetilde{F}$
are modified versions of $A$ and $F$, the gauge potential and
field strength of the non-Abelian $ADE$ gauge field living on
$\mathbb{R}^4\times Q$. The gauge variation of these fields is
calculated in \cite{BM03b} to be exactly the one which is
necessary to achieve local cancellation of all anomalies. The main
steps are similar to what we did in the last chapter. The only
difficulty comes from the non-Abelian nature of the fields which
complicates the calculation.

\chapter{Summary and Conclusions}
We have come a far way by now. Starting from M-theory, which is
still far from being well-understood, we showed that there are
vacua of the theory carrying all the characteristic features of
the standard model. As it also contains gravity in a natural way,
M-theory may well be a unifying theory of all the known forces in
nature. The methods we used were sometimes rather complicated. The
main reason is that we had to use relatively unfamiliar
mathematics leaving the domain of smooth manifolds and
differential geometry, which is familiar to physicists. A crucial
ingredient of the model outlined are spaces carrying various
singularities. Depending on the singularity involved, one or
another feature of the standard model emerges. It is interesting
to see that charged chiral fermions can only arise in the very
special case of compactifications on spaces with conical
singularities. In a sense, only if the singularity has codimension
seven the $G_2$-structure can maintain its full power and force
supersymmetry to be $\mathcal{N}=1$. For singularities with a
smaller codimension more supersymmetry is conserved. But then we
get $CPT$ self-conjugate and hence non-chiral spectra. The
singularity required for non-Abelian gauge theories has a quite
different nature. Here we need it to have codimension four in the
$G_2$-manifold. Interestingly enough, these singularities can be
characterized using the classification scheme of simply laced
semi-simple Lie algebras.\\
At this point it should also be mentioned that almost all our
calculations are done using the low-energy effective theory of
M-theory, eleven-dimensional supergravity. It is amazing that we
can get this far, knowing so little about the fundamental
formulation of M-theory. One of the reasons why this is the case
is that practically everything we did was classical. There are
some first steps towards the quantum dynamics of M-theory on
$G_2$-manifolds, see for example \cite{AtW01} and \cite{Ac02}.
Given the limitations of space, time and the author's competence
we could not comment on these interesting developments. However,
it is probably fair to say that there is still much to be done
before a rigorous and hopefully simple theory of the
quantum dynamics on $G_2$-manifolds can be formulated.\\
The only real quantum phenomenon we encountered were anomalies.
These can be handled for two important reasons. First and most
importantly, it is not necessary to know the full quantum
behaviour of M-theory for being able to study its anomalies. The
reason is, of course, that anomalies are an infrared effect, which
can be read off from the low-energy effective action. Secondly,
the calculation of anomalies is facilitated by their relation to
topological quantities of the base manifolds of the theory.
Therefore, they can be evaluated from index theorems. As a
consequence, an anomaly only depends on the dimension of the base
space and the kind of field under consideration. Hence, anomalies
can be calculated for each field and dimension once and for all
and the results can be used without having to perform further
calculations. We have repeatedly seen that anomalies allow to
probe the quantum regime of a given theory. In chapter 4 they were
the basic reason why the duality between M-theory on
$\mathbb{R}^{10}\times S^1/\mathbb{Z}_2$ and the $E_8\times E_8$
heterotic string could be established. There we obtained the
Green-Schwarz term as a first quantum correction to the
supergravity action. This term again played a crucial role in the
case of anomaly cancellation of M-theory containing M5-branes. In
that case another new ingredient had to be introduced. The
topological term had to be modified to ensure the cancellation of
the normal bundle anomaly. Another interesting feature of anomaly
theory is the concept of local anomaly cancellation. M-theory is
formulated in eleven dimensions, where there are no local gauge
anomalies. The concept of local anomaly cancellation states that
the theory needs to be anomaly free on any even-dimensional
subspace of the eleven-dimensional base manifold which contains
four-dimensional Minkowski space. It is not sufficient to have an
anomaly-free theory after integration over the compact space. This
important concept makes it possible to show that four-dimensional
chiral fermions have to be present at conical singularities of a
$G_2$-manifold, as long as the geometry of the cone satisfies
certain requirements. This case is particularly subtle as,
contrary to Ho\v{r}ava-Witten theory, the sums of all gauge and
mixed anomalies vanish. Nevertheless, there is a local anomaly at
each singularity which can be cancelled only
if additional chiral fermions are included.\\
In various M-theory calculations another inconvenience was
encountered, namely that the space under consideration was not a
manifold, but a manifold with boundary. These are much less
comfortable as, for example, the Atiyah-Singer index theory is no
longer true, but we have to use the Atiyah-Patodi-Singer index
theorem. Hence, the anomaly calculations which are usually done
for manifolds without boundary need to be revised. In order to
avoid this difficulty, we adopted an approach to study singular
spaces which makes it not necessary to excise the singularities.
In a sense we embedded a manifold with boundary into a manifold.
All the fields were multiplied by a smooth function $\rho$, which
is supposed to model the boundary. In that way we were led to a
nice and natural mechanism of anomaly cancellation for singular
$G_2$-manifolds. This idea was is presented in detail in \cite{BM03b}.\\
At this point we cannot resist speculating on how this formalism
might be modified. In chapter 8 we saw that there is a source for
the $G$-field in the regime were the $\rho$-function changes
rapidly. One interpretation of this might be that we have
M5-branes in our theory, as they are the only sources for a
$G$-field that are known in M-theory. We might think of them as
being smeared over the ten-dimensional space-time to give a
``charge density", just as electrons can be distributed on a
conducting surface. Further evidence for this speculation comes
from the structure of the modified terms. We basically added two
interaction terms to our theory which read
$\int_{\mathbb{R}^4\times Y_{\alpha}}BG_0G_0$ and
$\int_{\mathbb{R}^4\times Y_{\alpha}}BX_8$, with a field $B$ that
transforms as $\delta B=\Lambda$. Comparing this to (\ref{Scoup}),
one might be tempted to identify the field $B$ with the two-form
$A_{(2)}$ living on the M5-brane. Taking up the idea of the
Faraday cage, our model may indeed be such a cage for M-theory.
The interpretation would be that M5-branes are distributed on the
surface of this cage, which is given by the union of the
$Y_{\alpha}$, in such a way that $G$ is zero close to the
singularities. However, then we would need to consider the tangent
and normal bundle anomaly of those M5-branes, a complicated task,
which has not been performed so far.\\
Finally, we should like to comment on possible criticism of the
approach we have taken. There are two main points which might cast
doubt on the model outlined being a promising candidate for a
realistic theory. One is that the space we constructed is one of
many possibilities, and one might ask why nature should have
chosen exactly this particular one. Secondly, one might argue
that our chosen vacuum is not only one of many but it is totally
unnatural. Had we chosen a generic point in moduli space of
$G_2$-manifolds, it would not be very difficult to account for
it, as one point has to be chosen after all. However, we do not
want a generic point, but we choose a certain limit lying on the
boundary of the moduli space. In a sense, after a lot of
engineering and pulling and squashing the spaces in a way that
they satisfy the conditions we want, we just manage to stay at
the brink of moduli space, a result which might not be very
satisfying.\\
It remains an open question whether this criticism is justified.
One certainly hopes to find a mechanism in M-theory that singles
out the specific vacuum we constructed. Yet, it should be
emphasized that it is really amazing that it is possible at all
to construct a realistic theory in the given setup. Certainly,
there remain a lot of open questions, which we leave for future
exploration.

\vspace{2cm}
{\bf \Large Acknowledgements}\\
\\
This thesis was written under the joint supervision of Adel Bilal
at the École Normale Supérieure in Paris and Julius Wess at the
Ludwig-Maximilians-Universität in Munich. I would like to thank
both of them for their guidance and support and Adel Bilal for his
never ceasing patience, endless discussions and a fruitful
collaboration.\\
Furthermore, I would like to thank Jean-Pierre Derendinger, Yacine
Dolivet, Jean Iliopoulos, Dan Israel, Branislav Jurco, Thomas
Klose, Ruben Minasian, Chris Pope and Ivo Sachs for discussions
and useful comments.

\begin{appendix}

\chapter{General Notation}

\bigskip
The number of space-time dimensions is denoted by $d$.\\

\bigskip
{\bf Indices:}\footnote{These are the conventions chosen for
compactifications of M-theory to four dimensions. When we talk
about M2- and M5-branes index conventions will be different.}
\begin{center}
\begin{tabular}{l|l|l|c}
                & World                                & Tangent space  &
                Coordinates\\\hline
$d=11$ space-time & $M,N,\ldots\in\lbrace 0,1,\ldots10\rbrace$ & $A,B,\ldots\in\lbrace 0,1,\ldots10\rbrace$ & $z^M$ \\
$d=4$ space-time  & $\mu,\nu,\ldots\in\lbrace0,1,2,3\rbrace$& $a,b,\ldots\in\lbrace 0,1,2,3\rbrace$    & $x^{\mu}$\\
$d=7$ internal space& $m,n,\ldots\in\lbrace 1,2,\ldots,7\rbrace$&$\hat{m}, \hat{n},\ldots\in\lbrace1,2,\ldots,7\rbrace$&  $y^m$\\
\end{tabular}
\end{center}
\bigskip
The metric on flat space is given by
\begin{equation}
\eta :={\rm diag}(-1,1,\ldots1),
\end{equation}
curved metrics are denoted by $g_{MN}$, with\footnote{The symbol $g$ is used to denote
both the metric tensor and the modulus of its determinant. It should always be clear from
the context which one is meant.} $g:=|{\rm det}\ (g_{MN})|$.\\
The anti-symmetric tensor is defined as
\begin{eqnarray}
\widetilde{\epsilon}_{012\ldots d-1}&:=&{\widetilde{\epsilon}}^{\ 012\ldots d-1}:=+1,\\
\epsilon_{M_1\ldots M_{d}}&:=&\sqrt{g}\
\widetilde{\epsilon}_{M_1\ldots M_{d}}.
\end{eqnarray}
That is, we define $\widetilde\epsilon$ to be the tensor density
and $\epsilon$ to be the tensor. We obtain
\begin{eqnarray}
\epsilon_{012\ldots d-1}&=&\sqrt{g}=e:=|{\rm det} \ e^A_{\ \ M}|, \\
\epsilon^{M_1\ldots M_{d}}&=&{\rm sig}(g){1\over {\sqrt{g}}}\
\widetilde{\epsilon}^{\ M_1\ldots M_{d}},\mbox{\
\ and}\\
\widetilde{\epsilon}^{\ M_1\ldots M_rP_1\ldots
P_{d-r}}\widetilde{\epsilon}_{N_1\ldots N_rP_1\ldots
P_{d-r}}&=&r!(d-r)!\delta^{[M_1\ldots M_r]}_{N_1\ldots N_r}.
\end{eqnarray}
\begin{eqnarray}
A_{(M_1\ldots M_l)}&:=&{1 \over l!}\sum_{\pi}A_{M_{\pi(1)}\ldots M_{\pi(l)}}   \\
A_{[M_1\ldots M_l]}&:=&{1 \over l!}\sum_{\pi} {\rm
sig}(\pi)A_{M_{\pi(1)}\ldots M_{\pi(l)}}
\end{eqnarray}
p-forms are defined with a factor of $p!$, e.g.
\begin{equation}
\omega:={1\over p!}\omega_{M_1\ldots M_p}dz^{M_1}\wedge\ldots
\wedge dz^{M_p}.
\end{equation}
The Hodge dual is defined as
\begin{equation}
*\omega = {1\over p! (d-p)!}\, \omega_{M_1\ldots M_p}\
\epsilon^{M_1\ldots M_p}_{\phantom{M_1\ldots M_p} M_{p+1}\ldots
M_d} \ d z^{M_{p+1}}\wedge\ldots\wedge d z^{M_d}\ .
\end{equation}
\\
$[n]$ denotes the integer part of $n$.

\chapter{Spinors}

\section{Clifford Algebras and their Representation}

{\bf Definition B.1}\\
A {\em Clifford algebra} in $d$ dimensions is defined as a set
containing $d$ elements $\Gamma^{A}$ which satisfy the
relation\footnote{Note that Clifford algebras are defined using
flat indices. If one wants to write down $\Gamma$-matrices in
curved spaces one has to use the vielbeins defined in appendix D
to convert from one basis to the other. Sometimes ambiguities
occur for expressions like $\Gamma^0$, $\Gamma^1$, etc. In places
where the difference between the two frames is important we will
put hats on indices in tangent space and write $\Gamma^{\hat0}$,
$\Gamma^{\hat1}$, etc.}
\begin{equation}
\lbrace\Gamma^A,\Gamma^B\rbrace=2\eta^{AB}\opone.\label{Clifford
algebra}
\end{equation}
Under multiplication this set generates a finite group, denoted
$C_d$, with elements
\begin{equation}
C_d=\lbrace\pm1,\pm\Gamma^A,
\pm\Gamma^{A_1A_2},\ldots,\pm\Gamma^{A_1\ldots A_d}\rbrace,
\end{equation}
where $\Gamma^{A_1...A_l}:=\Gamma^{[A_1}\ldots \Gamma^{A_l]}$. The
order of this group is
\begin{equation}
{\rm ord}(C_d)=2\sum_{p=0}^d {d \choose p}=2\cdot2^d=2^{d+1}.
\end{equation}
{\bf Definition B.2}\\
Let $G$ be a group. Then the {\em conjugacy class} $[a]$ of $a\in
G$ is defined as
\begin{equation}
[a]:=\lbrace gag^{-1}|g\in G\rbrace.
\end{equation}
{\bf Proposition B.3}\\
Let $G$ be a finite dimensional group. Then the number of its
irreducible representations equals the number of its conjugacy
classes.\\
\\
{\bf Definition B.4}\\
Let $G$ be a finite group. Then the {\em commutator group}
$Com(G)$ of $G$ is defined as
\begin{equation}
{\rm Com}(G):=\lbrace ghg^{-1}h^{-1}|g,h\in G\rbrace.
\end{equation}
{\bf Proposition B.5}\\
Let G be a finite group. Then the number of inequivalent
one-dimensional representations is equal to the order of $G$
divided by the order of the commutator group of $G$.\\
{\bf Proposition B.6}\\
Let $G$ be a finite group with inequivalent irreducible
representations of dimension $n_p$, where $p$ labels the
representation. Then we have
\begin{equation}
{\rm ord}(G)=\sum_p(n_p)^2.\label{order}
\end{equation}
{\bf Proposition B.7}\\
Every class of equivalent representations of a finite group $G$
contains a unitary representation.\\
\\
For the unitary choice we get
$\Gamma^A{\Gamma^A}^{\dagger}=\opone$. From (\ref{Clifford
algebra}) we infer (in Minkowski space)
${\Gamma^0}^{\dagger}=-\Gamma^0$ and
$(\Gamma^A)^{\dagger}=\Gamma^A$ for $A\neq0$. This can be
rewritten as
\begin{equation}
{\Gamma^A}^{\dagger}=\Gamma^0\Gamma^A\Gamma^0.\label{Gammadagger}
\end{equation}

\subsection{Clifford Algebras in even Dimensions}
{\bf Theorem B.8}\\
For $d=2k+2$ even the group $C_d$ has $2^d+1$ inequivalent
representations. Of these $2^d$ are one-dimensional and the
remaining representation has (complex) dimension $2^{d\over2}=2^{k+1}$.\\
\\
This can be proved by noting that for even $d$ the conjugacy
classes of $C_d$ are given by
\begin{equation*}
\left\lbrace
[+1],[-1],[\Gamma^A],[\Gamma^{A_1A_2}],\ldots,[\Gamma^{A_1\ldots
A_d}]\right\rbrace,
\end{equation*}
hence the number of inequivalent irreducible representations of
$C_d$ is $2^d+1$. The commutator of $C_d$ is ${\rm
Com}(C_d)=\lbrace \pm1\rbrace$ and we conclude that the number of
inequivalent one-dimensional representations of $C_d$ is $2^d$.
From (\ref{order}) we read off that the dimension of the remaining
representation has
to be $2^{d\over2}$.\\
Having found irreducible representations of $C_d$ we turn to the
question whether we also found representations of the Clifford
algebra. In fact, for elements of the Clifford algebra we do not
only need the group multiplication but the addition of two
elements must be well-defined as well, in order to make sense of
(\ref{Clifford algebra}). It turns out that the one-dimensional
representation of $C_d$ do not extend to representations of the
Clifford algebra, as they do not obey the rules for addition and
subtraction. Hence, we found that for $d$ even there is a unique
irreducible representation of the Clifford algebra of dimension
$2^{d\over2}=2^{k+1}$.\\
\\
Given an irreducible representation $\lbrace \Gamma^A\rbrace$ of a
Clifford algebra, it is clear that $\pm\lbrace
{\Gamma^{A}}^*\rbrace$ and $\pm\lbrace {\Gamma^{A}}^{\tau}\rbrace$
form irreducible representations as well. As there is a unique
representation in even dimensions, these have to be related by
similarity transformations,

\bigskip
\parbox{14cm}{
\begin{eqnarray}
{\Gamma^{A}}^*&=&\pm(B_{\pm})^{-1}\Gamma^A B_{\pm},\nonumber\\
{\Gamma^A}^{\tau}&=&\pm (C_{\pm})^{-1}\Gamma^A C_{\pm}.\nonumber
\end{eqnarray}}\hfill\parbox{8mm}{\begin{eqnarray}\label{defBC}\end{eqnarray}}
The matrices $C_{\pm}$ are known as {\em charge conjugation
matrices}. Iterating this definition gives conditions for
$B_{\pm}, C_{\pm}$,
\begin{eqnarray}
(B_{\pm})^{-1}&=&b_{\pm}{B_{\pm}}^*,\\
C_{\pm}&=&c_{\pm}{C_{\pm}}^{\tau},
\end{eqnarray}
with $b_{\pm}$ real, $c_{\pm}\in\lbrace\pm1\rbrace$ and $C_{\pm}$
symmetric or anti-symmetric.

\subsection{Clifford Algebras in odd Dimensions}
{\bf Theorem B.9}\\
For $d=2k+3$ odd the group $C_d$ has $2^d+2$ inequivalent
representations. Of these $2^d$ are one-dimensional\footnote{As
above these will not be considered any longer as they are
representations of $C_d$ but not of the Clifford algebra.} and the
remaining two representation have (complex) dimension $2^{d-1\over2}=2^{k+1}$.\\
\\
As above we note that for odd $d$ the conjugacy classes of $C_d$
are given by
\begin{equation*}
\left\lbrace[+1],[-1],[\Gamma^A],[\Gamma^{A_1A_2}],\ldots,[\Gamma^{A_1\ldots
A_d}],[-\Gamma^{A_1\ldots A_d}]\right\rbrace
\end{equation*}
and the number of inequivalent irreducible representations of
$C_d$ is $2^d+2$. Again we find the commutator ${\rm
Com}(C_d)=\lbrace \pm1\rbrace$, hence, the number of inequivalent
one-dimensional representations of $C_d$ is $2^d$. Now define the
matrix
\begin{equation}
\Gamma^d:=\Gamma^0\Gamma^1\ldots\Gamma^{d-1},
\end{equation}
which commutes with all elements of $C_d$. By Schur's lemma this
must be a multiple of the identity, $\Gamma^d=a^{-1}\opone$, with
some constant $a$. Multiplying by $\Gamma^{d-1}$ we find
\begin{equation}
\Gamma^{d-1}=a\Gamma^0\Gamma^1\ldots\Gamma^{d-2}.\label{Gammaodd}
\end{equation}
Furthermore,
$(\Gamma^{0}\Gamma^1\ldots\Gamma^{d-2})^2=-(-1)^{k+1}$. As we know
from (\ref{Clifford algebra}) that $(\Gamma^{d-1})^2=+1$ we
conclude that $a=\pm1$ for $d=3$ (mod 4) and $a=\pm i$ for $a=5$
(mod 4). The matrices
$\lbrace\Gamma^0,\Gamma^1,\ldots,\Gamma^{d-2}\rbrace$ generate an
even-dimensional Clifford algebra the dimension of which has been
determined to be $2^{k+1}$. Therefore, the two inequivalent
irreducible representations of $C_d$ for odd $d$ must coincide
with this irreducible representation when restricted to $C_{d-1}$.
We conclude that the two irreducible representations for $C_d$ and
odd $d$ are generated by the unique irreducible representation for
$\lbrace\Gamma^0,\Gamma^1,\Gamma^{d-2}\rbrace$, together with the
matrix $\Gamma^{d-1}=a\Gamma^0\Gamma^1\ldots\Gamma^{d-2}$. The two
possible choices of $a$ correspond to the two inequivalent
representations.
The dimension of these representation is $2^{k+1}$.\\

\section{Dirac, Weyl and Majorana Spinors}
\subsection{Dirac Spinors}
Let $(M,g)$ be an oriented pseudo-Riemannian manifold of dimension
$d$, which is identified with $d$-dimensional space-time, and let
$\lbrace\Gamma^A\rbrace$ be a $d$-dimensional Clifford algebra.
The metric and orientation induce a unique
$SO(d-1,1)$-structure\footnote{The definition of a $G$-structure
is given in section 2.2.} $P$ on $M$. A {\em spin structure}
$(\widetilde{P},\pi)$ on $M$ is a principal bundle $\widetilde{P}$
over $M$ with fibre ${\rm Spin}(d-1,1)$, together with a map of
bundles $\pi:\widetilde{P}\rightarrow P$.
${\rm Spin}(d-1,1)$ is the universal covering group of $SO(d-1,1)$.\\
Spin structures do not exist on every manifold. An oriented
pseudo-Riemannian manifold $M$ admits a spin structure if and only
if $w_2(M)=0$, where $w_2(M)\in H^2(M,\mathbb{Z})$ is the {\em
second Stiefel-Whitney class} of $M$. In that case we call $M$ a
{\em spin
manifold}.\\
Define the anti-Hermitian generators
\begin{equation}
\Sigma^{AB}:={1\over2}\Gamma^{AB}={1\over4}[\Gamma^A,\Gamma^B].
\end{equation}
Then the $\Sigma^{AB}$ form a representation of
\verb"so"$(d-1,1)$, the Lie algebra of $SO(d-1,1)$,
\begin{equation}
[\Sigma^{AB},\Sigma^{CD}]=-\Sigma^{AC}g^{BD}+\Sigma^{AD}g^{BC}+\Sigma^{BC}g^{AD}-\Sigma^{BD}g^{AC}.
\end{equation}
In fact, $\Sigma^{AB}$ are generators of ${\rm Spin}(d-1,1)$. Take
$\Delta^d$ to be the natural representation of ${\rm
Spin}(d-1,1)$. We define the {\em(complex) spin bundle}
$S\rightarrow M$ to be\footnote{$\widetilde{P}\times_{{\rm
Spin}(d-1,1)}\Delta^d$ is the fibre bundle which is associated to
the principal bundle $\widetilde{P}$ in a natural way. Details of
this construction can be found in any textbook on differential
geometry, see for example \cite{Na90}.}
$S:=\widetilde{P}\times_{{\rm Spin}(d-1,1)}\Delta^d$. Then $S$ is
a complex vector bundle over $M$, with fibre $\Delta^d$ of
dimension $2^{[d/2]}$. A {\em Dirac spinor} $\psi$ is defined as a
section
of the spin bundle $S$. \\
Under a local Lorentz transformation with infinitesimal parameter
$\a_{AB}=-\a_{BA}$ a Dirac spinor transforms as
\begin{equation}
\psi'=\psi+\delta\psi=\psi-{1\over2}\a_{AB}\Sigma^{AB}\psi.\label{spinortrafo}
\end{equation}
The {\em Dirac
conjugate} $\bar\psi$ of the spinor $\psi$ is defined as
\begin{equation}
\bar\psi:=i\psi^{\dagger}\Gamma^0.
\end{equation}
With this definition we have $\delta(\bar\psi\eta)=0$ and
$\bar\psi\psi$ is Hermitian,
$(\bar\psi\psi)^{\dagger}=\bar\psi\psi$.

\subsection{Weyl Spinors}
In $d=2k+2$-dimensional space-time we can construct the
matrix\footnote{This definition of the $\Gamma$-matrix in
Minkowski space agrees with the one of \cite{Pol}. Sometimes it is
useful to define a Minkowskian $\Gamma$-matrix as
$\Gamma=i^{k}\Gamma^0\ldots\Gamma^{d-1}$ as in \cite{BM03c}.
Obviously the two conventions agree in 2, 6 and 10 dimensions and
differ by a sign in dimensions 4 and 8.}
\begin{equation}
\Gamma_{d+1}:=(-i)^k\Gamma^0\Gamma^1\ldots\Gamma^{d-1},\label{Gamma}
\end{equation}
which satisfies
\begin{eqnarray}
(\Gamma_{d+1})^2&=&\opone,\\
\lbrace\Gamma_{d+1},\Gamma^A\rbrace&=&0,\\
\ [\Gamma_{d+1},\Sigma^{AB}]&=&0.
\end{eqnarray}
Then, we can define the chirality projectors
\begin{eqnarray}
P_L\equiv P_-:={1\over2}(\opone-\Gamma_{d+1}),\\
P_R\equiv P_+:={1\over2}(\opone+\Gamma_{d+1}),
\end{eqnarray}
satisfying
\begin{eqnarray}
P_L+P_R&=&\opone,\\
P_{L,R}^2&=&P_{L,R},\\
P_LP_R&=&P_RP_L=0,\\
\ [P_{L,R},\Sigma^{AB}]&=&0.
\end{eqnarray}
A {\em Weyl spinor} in even-dimensional spaces is defined as a
spinor satisfying the Weyl condition,
\begin{equation}
P_{L,R}\psi=\psi.
\end{equation}
Note that this condition is Lorentz invariant, as the projection
operators commute with $\Sigma^{AB}$. Spinors satisfying
$P_L\psi_L=\psi_L$ are called {\em left-handed} Weyl spinors and
those satisfying $P_R\psi_R=\psi_R$ are called {\em
right-handed}. The Weyl condition reduces the number of complex
components of a
spinor to $2^{k}$. \\
Obviously, under the projections $P_{L,R}$ the space $\Delta^d$
splits into a direct sum $\Delta^d=\Delta^d_+\oplus\Delta^d_-$ and
the spin bundle is given by the Whitney sum $S=S_+\oplus S_-$.
Left- and right-handed Weyl spinors are sections of $S_-$ and
$S_+$, respectively.

\subsection{Majorana Spinors}
In (\ref{defBC}) we defined the matrices $B_{\pm}$ and $C_{\pm}$.
We now want to explore these matrices in more detail. For
$d=2k+2$ we define {\em Majorana spinors} as those spinors that
satisfy
\begin{equation}
\psi=B_+\psi^*,\label{Majorana}
\end{equation}
and pseudo-Majorana spinors as those satisfying
\begin{equation}
\psi=B_-\psi^*.\label{pseudoMajorana}
\end{equation}
As in the case of the Weyl conditions, these conditions reduce the
number of components of a spinor by one half. The definitions
imply\\
\parbox{14cm}{
\begin{eqnarray}
B_+^*B_+&=&\opone,\nonumber\\
B_-^*B_-&=&\opone\nonumber,
\end{eqnarray}}\hfill\parbox{8mm}{\begin{eqnarray}\label{Majimpl}\end{eqnarray}}
which in turn would give $b_+=1$ and $b_-=1$. These are
non-trivial conditions since $B_\pm$ is fixed by its definition
(\ref{defBC}). The existence of (pseudo-) Majorana spinors relies
on the possibility to construct matrices $B_+$ or $B_-$ which
satisfy (\ref{Majimpl}).\\
It turns out\footnote{See for example \cite{JPD02} for a detailed
analysis.} that Majorana conditions can be imposed in 2 and 4 (mod
8) dimensions. Pseudo-Majorana conditions are possible in 2 and 8
(mod 8) dimensions.

\bigskip
Finally, we state that for $d=2k+3$ Majorana spinors can be
defined in dimensions 1 and 3 (mod 8).

\subsection{Majorana-Weyl Spinors}
For $d=2k+2$ dimensions one might try to impose both the Majorana
and the Weyl condition. This certainly leads to spinors with
$2^{k-1}$ components. From (\ref{Gamma}) we get
$(\Gamma_{d+1})^*=(-1)^k B_{\pm}^{-1}\Gamma_{d+1} B_{\pm}$ and
therefore for $d=2$ (mod 4)
\begin{equation}
P_{L,R}^*=B_{\pm}^{-1}P_{L,R}B_{\pm}
\end{equation}
and for $d=4$ (mod 4)
\begin{equation}
P_{L,R}^*=B_{\pm}^{-1}P_{R,L}B_{\pm}.
\end{equation}
But this implies that imposing both the Majorana and the Weyl
condition is consistent only in dimensions $d=2$ (mod 4), as we
get
\begin{equation*}
B_{\pm}(P_{L,R}\psi)^*=P_{L,R}B_{\pm}\psi^*,
\end{equation*}
for $d=2$ (mod 4), but
\begin{equation}
B_{\pm}(P_{L,R}\psi)^*=P_{R,L}B_{\pm}\psi^*
\end{equation}
for $d=4$ (mod 4). We see that in the latter case the operator
$B_{\pm}$ is a map between states of different chirality, which is
inconsistent with the Weyl condition. As the Majorana condition
can be imposed only in dimensions 2, 4 and 8 (mod 8) we conclude
that Majorana-Weyl spinors can only exist in dimensions 2 (mod 8).

\bigskip
We summarize the results on spinors in various dimensions
in the following table.

\bigskip
\begin{center}
\begin{tabular}{|c|c|c|c|c|c|}\hline
d&Dirac& Weyl& Majorana&Pseudo-Majorana&Majorana-Weyl\\\hline
2&4&2&2&2&1\\
3&4& &2& & \\
4&8&4&4&&\\
5&8&&&&\\
6&16&8&&&\\
7&16&&&&\\
8&32&16&&16&\\
9&32&&&16&\\
10&64&32&32&32&16\\
11&64&&32&&\\
12&128&64&64&&\\\hline
\end{tabular}

\bigskip
The numbers indicate the real dimension of a spinor, whenever it
exists.\\
\end{center}

\chapter{Gauge Theory}

Gauge theories are formulated on principal bundles $P\rightarrow
M$ on a base space $M$ with fibre $G$ known as the {\em gauge
group}. Any group element $g$ of the connected component of $G$
that contains the unit element can be written as $g:=e^{\Lambda}$,
with $\Lambda:=\Lambda_a T_a$ and $T_a$ basis vectors of the Lie
algebra $\verb"g":=Lie(G)$. We always take $T_a$ to be
anti-Hermitian, s.t. $T_a=:-i t_a$ with $t_a$ Hermitian. The
elements of a Lie algebra satisfy commutation relations
\begin{eqnarray}
[T_a,T_b]&=&C^c_{\ ab}T_c,\\
\ [t_a,t_b]&=& iC^c_{\ ab} t_c,
\end{eqnarray}
with the real valued {\em structure coefficients} $C^c_{\
ab}$.\\
Of course, the group $G$ can come in various representations. The
{\em adjoint representation}
\begin{equation}
(T^{Ad}_{\ a})^b_{\ c}:=-C^b_{\ ca}
\end{equation}
is particularly important. Suppose a connection is given on the
principal bundle. This induces a local Lie algebra valued
connection form $A=A_aT_a$ and the corresponding local form of the
curvature, $F=F_aT_a$. These forms are related by\footnote{The
commutator of Lie algebra valued forms $A$ and $B$ is understood
to be
$[A,B]:=[A_M,B_N]\ dz^M\wedge dz^N$.}\\
\parbox{14cm}{
\begin{eqnarray}
F&:=&dA+{1\over2}[A, A],\nonumber\\
F_{\mu\nu}&=&\partial_{\mu}
A_{\nu}-\partial_{\nu} A_{\mu}+[A_{\mu},A_{\nu}],\nonumber\\
F_{\mu\nu}^a&=&\partial_{\mu} A^a_{\nu}-\partial_{\nu}
A^a_{\mu}+C^a_{\ bc}A^b_{\mu}A^c_{\nu}.\nonumber
\end{eqnarray}}\hfill\parbox{8mm}{\begin{eqnarray}\end{eqnarray}}\\
In going from one chart to another they transform as\\
\parbox{14cm}{
\begin{eqnarray}
A^g&:=&g^{-1}(A+d)g,\nonumber\\
F^g&:=&dA^g+{1\over2}[A^g, A^g]=g^{-1}Fg.\nonumber
\end{eqnarray}}\hfill\parbox{8mm}{\begin{eqnarray}\end{eqnarray}}\\
For $g=e^{\epsilon}=e^{\epsilon_aT_a}$ with $\epsilon$
infinitesimal we get $A^g=A+D\epsilon$.\\
For any object on the manifold which transforms under some
representation $\widetilde{T}$ (with $\widetilde{T}$
anti-Hermitian) of the gauge group $G$ we define a
gauge covariant derivative\\
\parbox{14cm}{
\begin{eqnarray}
D&:=&d+A,\nonumber\\
A&:=&A_a\widetilde{T}_a.\nonumber
\end{eqnarray}}\hfill\parbox{8mm}{\begin{eqnarray}\label{covder}\end{eqnarray}}\\
When acting on Lie algebra valued fields the covariant derivative
is understood to be $D:=d+[A,\ ]$.  An interesting example are
spinor fields which transform as
\begin{equation}
\psi^{g}(x):=g^{-1}\psi(x)\nonumber.
\end{equation}
Here $g=e^{\Lambda_aT^S_a}$ with $T^S_a$ a representation of $G$
acting on the spinor $\psi$. Its covariant derivative is $D\psi$,
and we verify
\begin{eqnarray}
(D\psi)'&=&(d+A)'\psi'=(d+g^{-1}Ag+g^{-1}(dg))g^{-1}\psi\nonumber\\
&=&d(g^{-1})\psi+g^{-1}d\psi+g^{-1}Agg^{-1}\psi+g^{-1}(dg)g^{-1}\psi\nonumber\\
&=&-g^{-1}(dg)g^{-1}\psi+g^{-1}(d+A)\psi+g^{-1}(dg)
g^{-1}\psi\nonumber\\
&=&g^{-1}D\psi,
\end{eqnarray}
as expected.\\
For fields $\phi$ which are not Lie algebra valued we get the
rather general operator formula
\begin{equation}
DD\phi=F\phi,
\end{equation}
which reads in components
\begin{equation}
[D_M,D_N]\phi=F_{MN}\phi.\label{curvPB}
\end{equation}
Finally, we note that the curvature satisfies the Bianchi
identity,
\begin{eqnarray}
DF=0.
\end{eqnarray}

\chapter{Curvature}
Usually general relativity on a manifold $M$ of dimension $d$ is
formulated in a way which makes the invariance under
diffeomorphisms, $Diff(M)$, manifest. Its basic objects are
tensors which transform covariantly under $GL(d,\mathbb{R})$.
However, as $GL(d,\mathbb{R})$ does not admit a spinor
representation, the theory has to be reformulated if we want to
couple spinors to a gravitational field. This is done by choosing
an orthonormal basis in the tangent space $TM$ which is different
from the one induced by the coordinate system. From that procedure
we get an additional local Lorentz invariance of the theory. As
$SO(d-1,1)$ does have a spinor representation we can couple
spinors to this reformulated theory. The basic features of the
Lorentz invariant theory are
listed in this appendix.\\
At a point $x$ on a pseudo-Riemannian manifold $(M,g)$ we define
the vielbeins $e_A(x)$ as
\begin{equation}
e_A(x):=e_A^{\ \ M}(x)\partial_M,
\end{equation}
with coefficients $e_A^{\ \ M}(x)$ such that $\lbrace e_A\rbrace$
are orthogonal,
\begin{equation}
g(e_A,e_B)=e_A^{\ \ M}e_B^{\ \ N}g_{MN}=\eta_{AB}.
\end{equation}
Define the inverse coefficients via $e_A^{\ \ M}e_{\ \
M}^B=\delta_A^B$ and $e^A_{\ \ M}e_A^{\ \ N}=\delta_M^N$, which
gives
\begin{equation}
g_{MN}(x)=e^A_{\ \ M}(x)e^B_{\ \ N}(x)\eta_{AB}.
\end{equation}
The dual basis $\lbrace e^A \rbrace$ is defined as,
\begin{equation}
e^A:=e^A_{\ \ M}dz^M.
\end{equation}
The commutator of two vielbeins defines the {\em anholonomy
coefficients} $\Omega_{AB}^{\ \ \ \ C}$,
\begin{equation}
[e_A,e_B]:=[e^{\ \ M}_A\partial_M,e^{\ \
N}_B\partial_N]=\Omega_{AB}^{\ \ \ \ C}e_C.
\end{equation}
The definition of $e_A$ gives
\begin{equation}
\Omega_{AB}^{\ \ \ \ C}(x)=e^C_{\ \ N}[e_A^{\ \ K}(\partial_K
e_B^{\ \ N})-e_B^{\ \ K}(\partial_K e_A^{\ \ N})](x).
\end{equation}

When acting on tensors expressed in the orthogonal basis, the
covariant derivative has to be rewritten using the {\em spin
connection coefficients} $\omega_{M\ \ B}^{\ \ A}$,
\begin{equation}
\nabla_M^S V^{AB\ldots}_{CD\ldots}:=\partial_M
V^{AB\ldots}_{CD\ldots}+\omega_{M\ \ E}^{\ \ A}
V^{EB\ldots}_{CD\ldots}+\ldots-\omega_{M\ \ C}^{\ \ E}
V^{AB\ldots}_{ED\ldots}.\label{tensortrafo}
\end{equation}
The object $\nabla^S$ is called the {\em spin
connection}\footnote{Physicists usually use the term "spin
connection" for the connection coefficients.}. Its action can be
extended to objects transforming under an arbitrary
representation of the Lorentz group. Take a field $\phi$ which
transforms as
\begin{equation}
\delta\phi^i=-{1\over2}\epsilon_{AB}(T^{AB})^i_{\ j}\phi^j
\end{equation}
under the infinitesimal Lorentz transformation $\Lambda^A_{\ \
B}(x)=\delta^A_B+\epsilon^A_{\ \
B}=\delta^A_B+{1\over2}\epsilon_{CD}(T^{CD}_{vec})^A_{\ \ B}$,
with the vector representation $(T^{CD}_{vec})^A_{\ \
B}=(\eta^{CA}\delta^D_B-\eta^{DA}\delta^C_B)$. Then its covariant
derivative is defined as
\begin{equation}
\nabla^S_M\phi^i:=\partial_M\phi^i+{1\over2}\omega_{MAB}(T^{AB})^i_{\
j}\phi^j.\label{covder2}
\end{equation}
We see that the spin connection coefficients can be interpreted as
the gauge field corresponding to local Lorentz invariance.
Commuting two covariant derivatives gives the general formula
\begin{equation}
[\nabla^S_M,\nabla^S_N]\phi={1\over2}R_{MNAB}T^{AB}\phi.\label{curvgeneral}
\end{equation}
In particular we can construct a connection on the spin bundle $S$
of $M$. As we know that for $\psi\in C^{\infty}(S)$ the
transformation law reads (with $\S^{AB}$ as defined in appendix B)
\begin{equation}
\delta\psi=-{1\over2}\epsilon_{AB}\Sigma^{AB}\psi,
\end{equation}
we find
\begin{equation}
\nabla^S_M\psi=\partial_M\psi+{1\over2}\omega_{MAB}\Sigma^{AB}\psi=\partial_M\psi+{1\over4}\omega_{MAB}\Gamma^{AB}\psi.
\end{equation}
If we commute two spin connections acting on spin bundles we get
\begin{equation}
[\nabla^S_M,\nabla^S_N]\psi={1\over4}R_{MNAB}\Gamma^{AB}\psi,\label{curv}
\end{equation}
where $R$ is the curvature corresponding to $\omega$, i.e.
$R^A_{\ \ B}=D(\omega_{M\ \ B}^{\ \ A}dz^M)$.

\bigskip
In the vielbein formalism the property $\nabla
g_{MN}=0$ translates to
\begin{equation}
\nabla^S_N e^A_{\ M}=0.
\end{equation}
In the absence of torsion this gives the dependence of
$\omega_{MAB}$ on the vielbeins. It can be expressed most
conveniently using the anholonomy coefficients
\begin{equation}
\omega_{MAB}(e)={1\over2}(-\Omega_{MAB}+\Omega_{ABM}-\Omega_{BMA}).
\end{equation}
If torsion does not vanish one finds
\begin{equation}
\omega_{MAB}=\omega_{MAB}(e)+\kappa_{MAB},
\end{equation}
where $\kappa_{MAB}$ is the contorsion tensor. It is related to
the torsion tensor by
\begin{equation}
\kappa_{MAB}=\mathcal{T}_{MN}^L(e_{AL}e_B^{\ \ N}-e_{BL}e_A^{\ \ N
})+g_{ML}\mathcal{T}^L_{NR}e^{\ \ N}_Ae^{\ \ R}_B.
\end{equation}
Defining $\omega^A_{\ \ B}:=\omega^{\ \ A}_{M\ \ B}dz^M$ one can
derive the {\em Maurer-Cartan structure equations},
\begin{eqnarray}
de^A+\omega^A_{\ \ B}\wedge e^B&=&\mathcal{T}^A, \label{MC1}\\
d\omega^A_{\ \ B}+\omega^A_{\ \ C}\wedge \omega^C_{\ \
B}&=&R^A_{\ \ B},\label{MC2}
\end{eqnarray}
where
\begin{eqnarray}
\mathcal{T}^A&=&{1\over 2}\mathcal{T}^A_{\ \ MN}dz^M\wedge dz^N,\\
R^A_{\ \ B}&=&{1\over2}R^A_{\ \ BMN}dz^M\wedge dz^N,
\end{eqnarray}
and
\begin{eqnarray}
\mathcal{T}^A_{\ \ MN}&=&e^A_{\ \ P}T^P_{\ \ MN},\\
R^A_{\ \ BMN}&=&e^A_{\ \ Q}e_B^{\ \ P}R^Q_{\ \ PMN}.
\end{eqnarray}
These equations tell us that the curvature corresponding to
$\nabla$ and the one corresponding to $\nabla^S$ are basically
the same.\\
The Maurer-Cartan structure equations can be rewritten
as
\begin{eqnarray}
\mathcal{T}&=&De,\\
R&=&D\omega,
\end{eqnarray}
where $D=d+\omega$. $\mathcal{T}$ and $R$ satisfy the Bianchi
identities
\begin{eqnarray}
D\mathcal{T}&=&Re,\\
DR&=&0.
\end{eqnarray}
The Ricci tensor $\mathcal{R}_{MN}$ and the Ricci scalar
$\mathcal{R}$ are given by
\begin{eqnarray}
\mathcal{R}_{MN}&:=&R_{MPNQ}g^{PQ},\\
\mathcal{R}&:=&\mathcal{R}_{MN}g^{MN}.
\end{eqnarray}
Finally, we note that general relativity is a gauge theory in the
sense of appendix C. If we take the induced basis as a basis for
the tangent bundle the relevant group is $GL(d,\mathbb{R})$. If on
the other hand we use the vielbein formalism the gauge group is
$SO(d-1,1)$. The gauge fields are $\Gamma$ and $\omega$,
respectively. The curvature of these one-forms is the Riemann
curvature tensor and the curvature two-form, respectively.
However, general relativity is a very special gauge theory, as its
connection coefficients can be constructed from another basic
object on the manifold, namely the metric tensor $g_{MN}$ or the
vielbein $e_A^{\ \ M}$.

\chapter{Index Theorems}
It turns out that anomalies are closely related to the index of
differential operators. A famous theorem found by Atiyah and
Singer tells us how to determine the index of these operators from
topological quantities. In this chapter we collect important index
theory results which are needed to calculate the anomalies.\\
\cite{Na90} gives a rather elementary introduction to index
theorems. Their relation to anomalies is explained in \cite{AG85}
and \cite{AGG84}.

\bigskip
{\bf Theorem (Atiyah-Singer index theorem)}\footnote{In its
general form the Atiyah-Singer index theorem can be formulated for
elliptic complexes over compact manifolds. We only consider the
special
cases of the twisted spin complex.}\\
Let $M$ be a manifold of even dimension, $d=2n$, $G$ a Lie group,
$P(M,G)$ the principal bundle of $G$ over $M$ and let $E$ the
associated vector bundle with $k:={\rm dim}(E)$. Let $A$ be the
gauge potential corresponding to a connection on $E$ and let
$S_{\pm}$ be the positive and negative chirality part of the spin
bundle. Define the Dirac operators $D_{\pm}:S_{\pm}\otimes
E\rightarrow S_{\mp}\otimes E$ by
\begin{equation}
D_{\pm}:=i\Gamma^M\left(\partial_M+{1\over4}\omega_{MAB}\Gamma^{AB}+A_M\right)P_{\pm}\label{Diracoperator}.
\end{equation}
Then ${\rm ind}(D_+)$ with
\begin{equation}
{\rm ind}(D_+):={\rm dim}({\rm ker}D_+)-{\rm dim}({\rm ker}D_-)
\end{equation}
is given by
\begin{eqnarray}
{\rm ind}(D_+)&=&\int_M[{\rm ch}(F)\hat A(M)]_{{\rm vol}}\label{ASI}\\
\hat A(M)&:=&\prod_{j=1}^n{x_j/2\over \sinh(x_j/2)}\label{Ahat}\\
{\rm ch}(F)&=&{\rm tr}\ \exp\left({iF\over2\pi}\right).
\end{eqnarray}
The $x_j$ are defined as
\begin{equation}
p(E):={\rm det}\left(1+{R\over
2\pi}\right)=\prod_{j=1}^{[n/2]}(1+x_j^2)=1+p_1+p_2+\ldots
\end{equation}
where $p(E)$ is the {\em total Pontrjagin class} of the bundle E.
The $x_j$ are nothing but the skew eigenvalues of $R/2\pi$,
\begin{equation}
{R\over2\pi}=\left(\begin{matrix}
                    0&x_1&0&0&\ldots&\\
                    -x_1&0&0&0&\ldots&\\
                    0&0&0&x_2&\ldots\\
                    0&0&-x_2&0&\ldots\\
                    \vdots&\vdots&\vdots&\vdots&
                   \end{matrix}\right).
\end{equation}
$\hat A(M)$ is known as the {\em Dirac genus} and ${\rm ch}(F)$ is
the {\em total Chern character}. The subscript ${\rm vol}$ means
that one has to extract the form whose
degree equals the dimension of $M$.\\
\\
To read off the volume form both $\hat A(M)$ and ${\rm ch}(F)$
need to be expanded. We get \cite{AG85}, \cite{AGG84}
\begin{eqnarray}
\hat A(M)&=&1+{1\over
(4\pi)^2}{1\over12}{\rm tr}R^2+{1\over(4\pi)^4}\left[{1\over288}({\rm tr}R^2)^2+{1\over360}{\rm tr}R^4\right]\nonumber\\
&&+{1\over(4\pi)^6}\left[{1\over
10368}({\rm tr}R^2)^3+{1\over4320}{\rm tr}R^2{\rm tr}R^4+{1\over5670}{\rm tr}R^6\right]\nonumber\\
&&+{1\over(4\pi)^8}\left[{1\over
497664}({\rm tr}R^2)^4+{1\over103680}({\rm tr}R^2)^2{\rm tr}R^4+\right.\nonumber\\
&&+\left.{1\over68040}{\rm tr}R^2{\rm tr}R^6+{1\over259200}({\rm tr}R^4)^2+{1\over75600}{\rm tr}R^8\right]+\ldots\\
{\rm ch}(F)&:=&{\rm
tr}\exp\left({iF\over2\pi}\right)=k+{i\over2\pi}{\rm
tr}F+{i^2\over2(2\pi)^2}{\rm tr}F^2+\ldots+{i^s\over
s!(2\pi)^s}{\rm tr}F^s+\ldots\nonumber\\ \label{ch(F)}
\end{eqnarray}
From these formulae we can determine the index of the Dirac
operator on arbitrary manifolds, e.g. for $d=4$ we get
\begin{equation}
{\rm ind}(D_+)={1\over(2\pi)^2}\int_M\left({i^2\over2}{\rm
tr}F^2+{k\over48}{\rm tr}R^2\right).
\end{equation}

\bigskip
The Dirac operator (\ref{Diracoperator}) is not the only operator
we need to calculate anomalies. We also need the analogue of
(\ref{ASI}) for spin-$3/2$ fields which is given by \cite{AG85},
\cite{AGG84}
\begin{eqnarray}
{\rm ind}(D_{3/2})&=&\int_M[ \hat A(M)({\rm tr}
\exp(iR/2\pi)-1){\rm
ch}(F)]_{{\rm vol}}\nonumber\\
&=&\int_M[ \hat A(M)({\rm tr}(\exp(iR/2\pi)-\opone)+d-1){\rm
ch}(F)]_{{\rm vol}}.
\end{eqnarray}
Explicitly,
\begin{eqnarray}
\hat
A(M){\rm tr}\,({\rm exp}(R/2\pi)-\opone)&=&-{1\over(4\pi)^2}\ 2\ {\rm tr}R^2\nonumber\\
&&+{1\over(4\pi)^4}\left[-{1\over6}({\rm tr}R^2)^2+{2\over3}{\rm tr}R^4\right]\nonumber\\
&&+{1\over(4\pi)^6}\left[-{1\over144}({\rm tr}R^2)^3+{1\over20}{\rm tr}R^2{\rm tr}R^4-{4\over45}{\rm tr}R^6\right]\nonumber\\
&&+{1\over (4\pi)^8}\left[-{1\over5184}({\rm tr}R^2)^4+{1\over540}({\rm tr}R^2)^2{\rm tr}R^4-\right.\nonumber\\
&&-\left.{22\over2835}{\rm tr}R^2{\rm tr}R^6+{1\over540}({\rm tr}R^4)^2+{2\over315}{\rm tr}R^8\right]\nonumber\\
&&+\ldots
\end{eqnarray}

\bigskip
Finally, in 4k+2 dimensions there are anomalies related to forms
with (anti-)self-dual field strength. The relevant index is given
by \cite{AG85}, \cite{AGG84}
\begin{equation}
{\rm ind}(D_A)={1\over4}\int_M[L(M)]_{2n},
\end{equation}
where the subscript $A$ stands for anti-symmetric tensor. $L(M)$
is known as the {\em Hirzebruch L-polynomial} and is defined as
\begin{equation}
L(M):=\prod_{j=1}^n {x_j/2\over \tanh(x_j/2)}.
\end{equation}
For reference we present the expansion
\begin{eqnarray}
L(M)&=&1-{1\over(2\pi)^2}{1\over6}{\rm tr}R^2+{1\over(2\pi)^4}\left[{1\over72}({\rm tr}R^2)^2-{7\over180}{\rm tr}R^4\right]\nonumber\\
&&+ {1\over(2\pi)^6}\left[-{1\over1296}({\rm tr}R^2)^3+{7\over1080}{\rm tr}R^2{\rm tr}R^4-{31\over2835}{\rm tr}R^6\right]\nonumber\\
&&+ {1\over(2\pi)^8}\left[{1\over31104}({\rm tr}R^2)^4-{7\over12960}({\rm tr}R^2)^2{\rm tr}R^4+\right.\nonumber\\
&&+\left.{31\over17010}{\rm tr}R^2{\rm tr}R^6+{49\over64800}({\rm
tr}R^4)^2-{127\over37800}{\rm tr}R^8\right]+\ldots \,.
\end{eqnarray}

\chapter{A-D-E Singularities}
In this appendix we list the finite subgroups of $SU(2)$,
which have an $ADE$ classification, and give the definition of an $ADE$ singularity.\\
Let $SU(2)$ act on $\mathbb{C}^2$ in the standard way and let
$\Gamma$ be a finite subgroup of $SU(2)$. Then
$\mathbb{C}^2/\Gamma$ is an orbifold and its singularities are the
points in $\mathbb{C}^2$ which are fixed under $\Gamma$. Thus the
singularities can be characterized by $\Gamma$. The finite
subgroups of $SU(2)$ have a classification in terms of the simply
laced semi-simple Lie algebras, $A_n$, $D_k$, $E_6$, $E_7$ and
$E_8$. This is why the singularities in the orbifold
$\mathbb{C}^2/\Gamma_{ADE}$ is called an {\em $ADE$ singularity.}
We will describe the corresponding subgroups $\Gamma_{ADE}$
explicitly.
\begin{equation}
\Gamma_{A_{n-1}}:=\left\langle\left(\begin{matrix}
               e^{2\pi i\over n} &       0\\
               0                 & e^{-{2\pi i\over n}}
        \end{matrix}\right)\right\rangle
\end{equation}
Obviously, $\Gamma_{A_{n-1}}$ is isomorphic to $\mathbb{Z}_n$.
The other groups have the following structure
\begin{equation}
\Gamma_{D_k}:=\left\langle\left(\begin{matrix}
               e^{\pi i\over {k-2}} &       0\\
               0                 & e^{-{\pi i\over {k-2}}}
        \end{matrix}\right),
        \left(\begin{matrix}
               0 &       1\\
               -1     &   0
        \end{matrix}\right)
        \right\rangle
\end{equation}
\begin{equation}
\Gamma_{E_6}:=\left\langle\left(\begin{matrix}
               e^{\pi i\over 2} &       0\\
               0                 & e^{-{\pi i\over 2}}
        \end{matrix}\right),
        {1\over \sqrt{2}}\left(\begin{matrix}
               e^{2\pi i 7\over 8} &  e^{2\pi i 7\over 8}     \\
               e^{2\pi i 5\over 8}    &   e^{2\pi i \over 8}
        \end{matrix}\right),
        \left(\begin{matrix}
               0 &       1\\
               -1     &   0
        \end{matrix}\right)
        \right\rangle
\end{equation}
\begin{equation}
\Gamma_{E_7}:=\left\langle\left(\begin{matrix}
               e^{\pi i\over 2} &       0\\
               0                 & e^{-{\pi i\over 2}}
        \end{matrix}\right),
        {1\over \sqrt{2}}\left(\begin{matrix}
               e^{2\pi i 7\over 8} &  e^{2\pi i 7\over 8}     \\
               e^{2\pi i 5\over 8}    &   e^{2\pi i \over 8}
        \end{matrix}\right),
        \left(\begin{matrix}
               e^{2\pi i\over 8} &       0\\
               0                 & e^{{2\pi i 7\over 8}}
        \end{matrix}\right),
        \left(\begin{matrix}
               0 &       1\\
               -1     &   0
        \end{matrix}\right)
        \right\rangle
\end{equation}
\begin{equation}
\Gamma_{E_8}:=\left\langle-\left(\begin{matrix}
               e^{2\pi i 3\over 5} &       0\\
               0                 & e^{{2\pi i2 \over 5}}
        \end{matrix}\right),
{1\over {e^{2\pi i2\over 5}}-e^{2\pi i
3\over5}}\left(\begin{matrix}
               e^{2\pi i \over 5}+ e^{{2\pi i 4\over 5}}&  1     \\
               1& -e^{2\pi i \over 5}- e^{{2\pi i 4\over 5}}
        \end{matrix}\right)
        \right\rangle.
\end{equation}

\end{appendix}

\end{document}